\documentclass[aps, prd, final, superscriptaddress,
               amsfonts, amssymb, amsmath, preprintnumbers,
               showpacs, showkeys, floatfix, nofootinbib]{revtex4}

\pdfoutput=1
\usepackage[latin1]{inputenc}                    
\usepackage{graphicx}                            
\usepackage{amsmath}
\usepackage{rotating}

\graphicspath{{.}{Figures/}}                     
\newcommand{\mps}{m_\pi}                         
\newcommand{\fps}{f_\pi}                         
\newcommand{\imag}{{\rm i}}                      
\newcommand{\msbar}{$\overline{\mbox{\rm MS}}$}  
\newcommand{\fslash}[1]{\slash\!\!\!{#1}}        
\newcommand{\Dlr}{\stackrel{\leftrightarrow}{D}} 
\newcommand{\dlangle}{\langle\!\langle}          
\newcommand{\drangle}{\rangle\!\rangle}          

\newcommand{\la}{\label}
\newcommand{\be}{\begin{equation}}
\newcommand{\ee}{\end{equation}}
\newcommand{\ba}{\begin{eqnarray}}
\newcommand{\ea}{\end{eqnarray}}
\newcommand{\eq}{Eq.~}
\newcommand{\fig}{Fig.~}
\newcommand{\sect}{Sec.~}
\newcommand{\nn}{\nonumber \\}
\newcommand{\tab}{Tab.~}

\renewcommand{\>}{\rangle}
\newcommand{\MeV}{\text{MeV}}

\newcommand{\MSbar}{\overline{\text{MS}}}

\begin{document}

\title[Mixed-action nucleon structure]{Nucleon structure from mixed
  action calculations using 2+1 flavors of asqtad sea and domain wall
  valence fermions}

\author{J.D.~Bratt} \affiliation{Center for Theoretical Physics,
  Massachusetts Institute of Technology, Cambridge, MA 02139}
\author{R.G.~Edwards} \affiliation{Thomas Jefferson National
  Accelerator Facility, Newport News, VA 23606}
\author{M.~Engelhardt} \affiliation{Department of Physics, New Mexico
  State University, Las Cruces, NM 88003-0001}
\author{Ph.~H{\"a}gler} \affiliation{Physik-Department der TU
  M\"unchen, James-Franck-Stra\ss{}e, D-85748 Garching, Germany}
\author{H.W.~Lin} \affiliation{Thomas Jefferson National Accelerator
  Facility, Newport News, VA 23606} \affiliation{Department of
  Physics, University of Washington, Seattle, WA 98195-1560}
\author{M.F.~Lin} \affiliation{Center for Theoretical Physics,
  Massachusetts Institute of Technology, Cambridge, MA 02139}
\author{H.B.~Meyer} \affiliation{Center for Theoretical Physics,
  Massachusetts Institute of Technology, Cambridge, MA 02139}
\affiliation{CERN Physics Department, 1211 Geneva 23, Switzerland}
\author{B.~Musch} \affiliation{Physik-Department der TU M\"unchen,
  James-Franck-Stra\ss{}e, D-85748 Garching, Germany}
\affiliation{Thomas Jefferson National Accelerator Facility, Newport
  News, VA 23606}
\author{J.W.~Negele} \affiliation{Center for Theoretical Physics,
  Massachusetts Institute of Technology, Cambridge, MA 02139}
\author{K.~Orginos} \affiliation{Department of Physics, College of
  William and Mary,P.O. Box 8795, Williamsburg VA 23187-8795}
\author{A.V.~Pochinsky} \affiliation{Center for Theoretical Physics,
  Massachusetts Institute of Technology, Cambridge, MA 02139}
\author{M.~Procura} \affiliation{Center for Theoretical Physics,
  Massachusetts Institute of Technology, Cambridge, MA 02139}
\affiliation{Physik-Department der TU M\"unchen,
  James-Franck-Stra\ss{}e, D-85748 Garching, Germany}
\author{D.G.~Richards} \affiliation{Thomas Jefferson National
  Accelerator Facility, Newport News, VA 23606}
\author{W.~Schroers} \altaffiliation{Current address: NuAS,
  Stubenrauchstr.~3, 12357 Berlin} \affiliation{Institute of Physics,
  Academia Sinica, Taipei 115, Taiwan, R.O.C.}
\author{S.N.~Syritsyn} \affiliation{Center for Theoretical Physics,
  Massachusetts Institute of Technology, Cambridge, MA 02139}
\collaboration{LHPC} \noaffiliation \date{\today}

\begin{abstract}
  We present high statistics results for the structure of the nucleon
  from a mixed-action calculation using 2+1 flavors of asqtad sea and
  domain wall valence fermions. We perform extrapolations of our data
  based on different chiral effective field theory schemes and compare
  our results with available information from phenomenology. We
  discuss vector and axial form factors of the nucleon, moments of
  generalized parton distributions, including moments of forward
  parton distributions, and implications for the decomposition of the
  nucleon spin.
\end{abstract}

\preprint{TUM/T39-10-01}
\preprint{TUM-EFT 6/10}
\preprint{CERN-PH-TH/2010-005}



\maketitle

%
%

\section{\label{sec:introduction}Introduction}
Determining the structure of the nucleon in terms of quarks and gluons
is central to our goal of understanding baryonic matter at the level
of its smallest constituents.  While the theory describing the strong
interactions of quarks and gluons, Quantum Chromodynamics, was
identified thirty-five years ago, its predictions at low energies have
been notoriously hard to derive ab initio. The modern approach to
calculate the properties of hadrons is based on the Euclidean path
integral representation of QCD discretized on a space-time lattice,
i.e.~lattice QCD\@. Importance sampling methods, implemented on
massively parallel computers, make it possible to extract, in
particular, many properties of the nucleon.

In recent years, advances both in algorithms and in computer
technology made a series of remarkable calculations possible that had
a large impact on our understanding of nucleon structure. Among the
quantities calculated we would like to mention the quark contribution
to the nucleon spin~\cite{Gockeler:2003jfa, Hagler:2003jd}, the
nucleon transverse structure~\cite{LHPC:2003is}, and the nucleon axial
charge~\cite{Edwards:2005ym, Khan:2006de}. Recently, the nucleon
electromagnetic and axial form factors have received special attention
in Refs.~\cite{Alexandrou:2006ru} and~\cite{Alexandrou:2007xj} using
dynamical Wilson and asqtad fermions. Another important milestone is
the advent of full domain-wall calculations, see
Refs.~\cite{Lin:2008uz, Syritsyn:2009mx}, and of dynamical
twisted-mass fermions~\cite{Alexandrou:2009ng}. Disconnected diagrams
play a key role in an ongoing study of the strange quark content of
the nucleon~\cite{Takeda:2009ga}. For reviews and progress reports on
the current state of the field, see Refs.~\cite{Zanotti:2008zm,
  Renner:2009lat, Gockeler:2009pe, Hagler:2009ni}.

Over the past years several of us have reported on hadron structure
measurements using mixed action calculations with 2+1 flavors of
dynamical asqtad sea quarks~\cite{Bernard:2001av, Bazavov:2009bb} ---
corresponding to degenerate $u,d$ quarks + the strange quark --- and
domain wall valence quarks~\cite{Negele:2005za, Schroers:2007qf,
  Hagler:2007xi, WalkerLoud:2008bp, Syritsyn:2009np}. A significant
milestone was reached in Ref.~\cite{Hagler:2007xi}, which summarized
our findings for higher moments of generalized form factors. The
current paper represents a major update of that work: it includes the
observables presented previously with higher statistics, as well as an
additional, lower pion mass calculation. Beyond that, it covers form
factors and chiral extrapolations of the forward moments that were not
shown previously. The propagators and technology underlying these
calculations have not only successfully been applied to nucleon
structure, but have also turned out to be enormously valuable to other
studies, see e.g.~\cite{Alexandrou:2007dt, Detmold:2008fn,
  Beane:2008dv} and references therein. Mixed action calculations have
also been studied in the framework of effective field theory,
see~\cite{Golterman:2005xa, Orginos:2007tw, Chen:2007ug}. It is the
purpose of the current paper to report our final results including all
improvements we have made over the years on the method and technology
of our computations.

The layout of this paper is as follows. We give an overview of our
notation and conventions and present the observables that we study in
Sec.~\ref{sec:overv-phys-observ}. The discussion of our technology and
of the improvements we have made over our previous calculations
together with consistency checks takes place in
Sec.~\ref{sec:new-latt-calc}. Our results are presented in
Sec.~\ref{sec:results} which is divided into the following
subsections: Section~\ref{sec:gA} is dedicated to the axial charge
$g_A$. The discussion of the electromagnetic form factors and the
axial form factors takes place in Sec.~\ref{sec:form-factors}
and~\ref{sec:axial-form-factors}, respectively. The generalized form
factors and their extrapolation using different schemes of chiral
effective theory are presented in Sec.~\ref{sec:gener-form-fact}.

The generalized form factors of the energy-momentum tensor provide
vital information as to how spin is apportioned within the nucleon ---
a long-standing puzzle of hadron physics.  In particular, they enable
a first-principles calculation in lattice QCD of $J_q$, the total
angular momentum carried by the quarks~\cite{Mathur:1999uf,
  Gockeler:2003jfa, Hagler:2003jd, Hagler:2007xi, Gadiyak:2001fe}, and
hence have a crucial role in resolving this puzzle.
Section~\ref{sec:decomp-nucl-spin} is dedicated to this topic. Our
summary and outlook for future work are given in
Sec.~\ref{sec:summary-conclusions}. Since the summary includes
cross-references to the most significant tables and figures, the
reader mainly interested in new results might find it useful to use
that section as a guide to the highlights of our calculations.

%
%

\section{\label{sec:overv-phys-observ}Overview of physical
  observables}
The observables we are reporting on are defined via matrix elements of
bi-local light cone quark operators in nucleon states. They can be
systematically discussed in the framework of so-called generalized
parton distributions~\cite{Mueller:1998fv, Ji:1996nm,
  Radyushkin:1997ki, Diehl:2003ny}. The relevant bi-local operator is
given by
\begin{equation}
  \label{eq:gpd-op-def}
  {\cal O}_{q,\Gamma}(x) = \int\, \frac{d\lambda}{4\pi}
  e^{{\rm i}\lambda x} \bar{q}(\frac{-\lambda
    n}{2})\,\Gamma_\mu n^\mu \, {\cal P} e^{-{\rm i}
    g\int_{-\lambda / 2}^{\lambda / 2} d\alpha\, n\cdot A(\alpha n)}
  \, q(\frac{\lambda n}{2})\,,
\end{equation}
with $x$ being the momentum fraction, $n$ a light-cone vector, and
$\Gamma$ representing any gamma matrix from the basis $\gamma_\mu$,
$\gamma_\mu\gamma_5$, and $\sigma_{\mu j}$, $j=1,2$.

The quark field $q$ in Eq.~\eqref{eq:gpd-op-def} can carry any of the
up, down or strange flavors, however here we restrict ourselves to
$q=u,d$. Since in our lattice calculations the up- and down-quarks
have degenerate masses, isospin symmetry is built in by
construction. Note that our matrix elements always refer to the case
of a proton, i.e.~the nucleon sources used in the construction of the
matrix elements in Eqs.~\eqref{eq:two-pt} and~\eqref{eq:three-pt}
below always contain two $u$- and one $d$-quark. When comparing
lattice matrix elements to experiment, we need to choose appropriate
flavors or flavor combinations. The isovector combination ${\cal
  O}_q\equiv{\cal O}_u-{\cal O}_d$, where we subtract the down-quark
contribution from that of the up-quarks, can be constructed from the
difference between proton and neutron observables obtained in
experiment. The isoscalar combination ${\cal O}_{u+d}$, on the other
hand, corresponds to the sum of proton and neutron observables. To
compute proton observables as measured by probing a proton with a
photon, one needs to take into account the charge weighting factors of
the quark-photon vertex and thus consider the combination
$q=2/3u-1/3d$. In all but the isovector case, the matrix elements will
in principle receive contributions from both the connected and the
disconnected diagrams. In this work we have neglected the disconnected
contributions since they are very costly to obtain. For recent studies
of disconnected contributions in the framework of calculations of form
factors and moments of parton distribution functions (PDFs) we refer
to, e.g.~Refs.~\cite{Deka:2008xr, Doi:2009sq}.

The matrix elements of the operator Eq.~\eqref{eq:gpd-op-def} between
nucleon states with momentum $\vec{p}$ and polarization $\lambda$ can
be parameterized generically by exploiting their Lorentz tensor
structure. This has been discussed in detail in the
literature~\cite{Diehl:2003ny} and we merely present the results here.

For the case $\Gamma_\mu=\gamma_\mu$, the nucleon matrix element
adopts the form
\begin{equation}
  \label{eq:nucleon-vector-gpd}
  \langle p',\lambda'\vert {\cal O}_{q,\gamma}(x) \vert
  p,\lambda\rangle = \dlangle \fslash{n} \drangle H(x,\xi,t) +
  \frac{n_\mu \Delta_\alpha} {2m}
  \dlangle\imag\sigma^{\mu\alpha}\drangle E(x,\xi,t)\,,
\end{equation}
where we have introduced the notation $\dlangle {\cal X}\drangle=
\bar{u}(p',\lambda') {\cal X} u(p,\lambda)$ and the parameters
$\Delta=p'-p$, $t\equiv -Q^2= -(p'-p)^2$ and $\xi=-n\cdot\Delta/2$. In
the framework of form factors in Secs.~\ref{sec:form-factors}
and~\ref{sec:axial-form-factors}, we will denote the squared momentum
transfer by $Q^2$ since this is a common and wide-spread convention.
The unpolarized generalized parton distributions (GPDs) $H(x,\xi,t)$
and $E(x,\xi,t)$ are Lorentz scalars and thus frame-independent
functions parameterizing the matrix element. We point out that the
matrix element also depends implicitly on a renormalization scale,
$\mu^2$, and scheme.

In the case $\Gamma_\mu=\gamma_\mu\gamma_5$ we obtain a
Lorentz-covariant parameterization in terms of the polarized GPDs
$\tilde{H}(x,\xi,t)$ and $\tilde{E}(x,\xi,t)$:
\begin{equation}
  \label{eq:nucleon-axvector-gpd}
  \langle p',\lambda'\vert {\cal O}_{q,\gamma\gamma_5}(x) \vert
  p,\lambda\rangle = \dlangle \fslash{n}\gamma_5\drangle
  \tilde{H}(x,\xi,t) + \frac{n\cdot\Delta}{2m} \dlangle \gamma_5
  \drangle \tilde{E}(x,\xi,t)\,.
\end{equation}
The case $\Gamma_{\mu,j}=\sigma_{\mu j}$ will be discussed in a
separate publication~\cite{Bratt:2009xx}. The kinematic parameter $x$
is the average longitudinal momentum fraction of the struck quark and
$\xi$ and $t$ are the longitudinal and the total squared momentum
transfer to the nucleon, respectively. The GPDs are defined over the
full intervals $x=-1\dots +1$, and $\xi=-1\dots +1$. Depending on
whether $|x|>|\xi|$ or vice versa they have the interpretation of
amplitudes for the emission and absorption of a quark or for the
emission of a quark-antiquark pair, respectively.

An attractive feature of the generalized parton distributions is that
they occur in a range of different processes, e.g.~deeply virtual
Compton scattering, wide-angle Compton scattering and exclusive meson
production, in addition to the classic processes that probe the
forward parton distributions and form factors.  The challenge of GPDs
lies in their more complex structure--- each generalized parton
distribution is a function of three parameters rather than just one,
and the different experimental processes provide different constraints
on their form. Typically only convolutions of these functions in the
$x$ variable are experimentally accessible.

Since lattice calculations deal with operators and matrix elements in
Euclidean space, a direct computation of non-local light-cone elements
is not possible. To facilitate the lattice calculations, one takes
$x^{n-1}$-moments of Eqs.~\eqref{eq:nucleon-vector-gpd}
and~\eqref{eq:nucleon-axvector-gpd}, yielding a tower of local
operators whose matrix elements can be related to the corresponding
moments of $H$, $E$, $\tilde{H}$ and $\tilde{E}$. In this study, we
will compute matrix elements of the following local generalized
currents,
\begin{equation}
  \label{eq:gen-currents}
  {\cal O}_{q,\Gamma}^{\{\mu_1\ldots\mu_n\}} = \bar{q}(0) \Gamma^{\{\mu_1}
  \imag \Dlr\!\phantom{}^{\mu_{2}}\!\cdots
  \imag\Dlr\!\phantom{}^{\mu_{n}\}} q(0)\,,
\end{equation}
where, again, $\Gamma^\mu$ can refer to either $\gamma^\mu$ or
$\gamma^\mu\gamma_5$. Curly braces around indices represent a
symmetrization and the subtraction of traces of the indices and the
derivative is defined via $\Dlr=1/2
(\stackrel{\rightarrow}{D}-\stackrel{\leftarrow}{D})$.

Taking the moments w.r.t.~$x$ of the GPDs we define
\begin{eqnarray}
  \label{eq:mellin-def}
   H^n(\xi, t) & \equiv & \int_{-1}^{1} dx\, x^{n-1} H(x, \xi, t)\,,
   \nonumber \\
   E^n(\xi, t) & \equiv & \int_{-1}^{1} dx\, x^{n-1} E(x, \xi, t)\,,
   \nonumber \\
   \tilde{H}^n(\xi, t) & \equiv & \int_{-1}^{1} dx\, x^{n-1}
   \tilde{H}(x, \xi, t)\,, \nonumber \\
   \tilde{E}^n(\xi, t) & \equiv & \int_{-1}^{1} dx\, x^{n-1}
   \tilde{E}(x, \xi, t)\,.
\end{eqnarray}
The non-forward nucleon matrix elements of the local operators,
Eq.~\eqref{eq:gen-currents}, can in turn be parametrized according to
their Lorentz structure in terms of generalized form factors (GFFs)
$A_{nm}(t)$, $\tilde{A}_{nm}(t)$, $B_{nm}(t)$, $\tilde{B}_{nm}(t)$,
and $C_{nm}(t)$,
\begin{eqnarray}
  \label{eq:lattice-vec-gff}
  \langle p',\lambda'\vert {\cal O}^{\mu_1} \vert p,\lambda \rangle
  & = & \dlangle\gamma^{\mu_1}\drangle A_{10}(t) + \frac{\imag}{2
    m}\dlangle\sigma^{\mu_1\alpha}\drangle \Delta_{\alpha}
  B_{10}(t)\,, \nonumber \\
  \langle p',\lambda\vert {\cal O}^{\lbrace\mu_1\mu_2\rbrace}\vert
  p,\lambda\rangle & = & \bar{p}^{\lbrace\mu_1}\dlangle
  \gamma^{\mu_2\rbrace}\drangle A_{20}(t) + \frac{\imag}{2 m}
  \bar{p}^{\lbrace\mu_1} \dlangle
  \sigma^{\mu_2\rbrace\alpha}\drangle \Delta_{\alpha} B_{20}(t)
  + \frac{1}{m}\Delta^{\lbrace\mu_1} \Delta^{\mu_2\rbrace}
  C_{20}(t)\,, \nonumber \\
  \langle p',\lambda'\vert {\cal
    O}^{\lbrace\mu_1\mu_2\mu_3\rbrace}\vert p,\lambda \rangle 
  & = & \bar{p}^{\lbrace\mu_1}\bar{p}^{\mu_2} \dlangle
  \gamma^{\mu_3\rbrace} \drangle A_{30}(t)
  + \frac{\imag}{2 m} \bar{p}^{\lbrace\mu_1}\bar{p}^{\mu_2}
  \dlangle \sigma^{\mu_3\rbrace\alpha} \drangle
  \Delta_{\alpha} B_{30}(t) \nonumber \\
  && \quad + \Delta^{\lbrace \mu_1}\Delta^{\mu_2} \dlangle
  \gamma^{\mu_3\rbrace}\drangle A_{32}(t)
  + \frac{\imag}{2 m} \Delta^{\lbrace\mu_1}\Delta^{\mu_2}
  \dlangle \sigma^{\mu_3\rbrace\alpha}\drangle
  \Delta_{\alpha} B_{32}(t)\,,
\end{eqnarray}
for the vector operators and
\begin{eqnarray}
  \label{eq:lattice-ax-gff}
  \langle p',\lambda'\vert {\cal O}_{\gamma_5}^{\mu_1} \vert p,\lambda
  \rangle & = & \dlangle \gamma^{\mu_1 }\gamma_5\drangle \tilde A_{10}(t)
  + \frac{1}{2 m}\Delta^{\mu_1}\dlangle \gamma_5 \drangle
  \tilde B_{10}(t)\,, \nonumber \\
  \langle p',\lambda'\vert {\cal
    O}_{\gamma_5}^{\lbrace\mu_1\mu_2\rbrace} \vert p,\lambda \rangle
  & = & \bar{p}^{\lbrace\mu_1}\dlangle
  \gamma^{\mu_2\rbrace}\gamma_5\drangle \tilde A_{20}(t)
  + \frac{1}{2 m} \Delta^{\lbrace\mu_1} \bar{p}^{\mu_2\rbrace} \dlangle
  \gamma_5 \drangle \tilde B_{20}(t) \,, \nonumber \\
  \langle p',\lambda'\vert {\cal
    O}_{\gamma_5}^{\lbrace\mu_1\mu_2\mu_3\rbrace} \vert p,\lambda \rangle
  & = & \bar{p}^{\lbrace\mu_1}\bar{p}^{\mu_2} \dlangle
  \gamma^{\mu_3\rbrace}\gamma_5 \drangle \tilde A_{30}(t)
  + \frac{1}{2 m}\Delta^{\lbrace\mu_1} \bar{p}^{\mu_2}
  \bar{p}^{\mu_3\rbrace} \dlangle \gamma_5\drangle \tilde B_{30}(t)
  \nonumber \\ && \quad
  + \Delta^{\lbrace \mu_1}\Delta^{\mu_2} \dlangle
  \gamma^{\mu_3\rbrace}\gamma_5\drangle \tilde A_{32}(t)
  + \frac{1}{2 m} \Delta^{\lbrace\mu_1}\Delta^{\mu_2}\Delta^{\mu_3\rbrace}
  \dlangle\gamma_5\drangle \tilde B_{32}(t)\,,
\end{eqnarray}
for the axial vector operators. Here we have defined the average
nucleon momentum $\bar{p}=(p'+p)/2$. By comparing this with the
$x^{n-1}$-moments of Eqs.~\eqref{eq:nucleon-vector-gpd},
\eqref{eq:nucleon-axvector-gpd}, and using Eq.~\eqref{eq:mellin-def},
one finds that the $\xi$ dependence of the moments of the GPDs is
merely polynomial,
\begin{eqnarray}
  \label{eq:lattice-vec-gff-def}
  H^{n=1}(\xi, t) & = & A_{10}(t)\,, \nonumber \\
  H^{n=2}(\xi, t) & = & A_{20}(t)+(2\xi)^2 C_{20}(t)\,, \nonumber \\
  H^{n=3}(\xi, t) & = & A_{30}(t)+(2\xi)^2 A_{32}(t)\,, \nonumber \\
  E^{n=1}(\xi, t) & = & B_{10}(t)\,, \nonumber \\
  E^{n=2}(\xi, t) & = & B_{20}(t)-(2\xi)^2 C_{20}(t)\,, \nonumber \\
  E^{n=3}(\xi, t) & = & B_{30}(t)+(2\xi)^2 B_{32}(t)\,, \nonumber \\
  \dots
\end{eqnarray}
and
\begin{eqnarray}
  \label{eq:lattice-ax-gff-def}
  \tilde H^{n=1}(\xi, t) & = & \tilde{A}_{10}(t)\,, \nonumber \\
  \tilde H^{n=2}(\xi, t) & = & \tilde{A}_{20}(t)\,, \nonumber \\
  \tilde H^{n=3}(\xi, t) & = & \tilde{A}_{30}(t)+(2\xi)^2
  \tilde{A}_{32}(t)\,, \nonumber \\
  \tilde E^{n=1}(\xi, t) & = & \tilde B_{10}(t)\,, \nonumber \\
  \tilde E^{n=2}(\xi, t) & = & \tilde B_{20}(t)\,, \nonumber \\
  \tilde E^{n=3}(\xi, t) & = & \tilde B_{30}(t)+(2\xi)^2\tilde
  B_{32}(t)\,, \nonumber \\
  \dots \,.
\end{eqnarray}
In the forward limit of Eqs.~\eqref{eq:nucleon-vector-gpd}
and~\eqref{eq:nucleon-axvector-gpd} with $\vec{p}=\vec{p}^{\,\prime}$,
we obtain the well-known parton distribution functions,
\begin{eqnarray}
  \label{eq:forward-pd}
  q(x) & = & H(x, \xi=0, t=0)\,, \nonumber\\
  \Delta q(x) & = & \tilde{H}(x, \xi=0, t=0)\,.
\end{eqnarray}
Note that in the case $\xi=0$ the GPDs --- and also the GFFs,
including the form factors for $n=1$ --- admit a probability
interpretation~\cite{Burkardt:2000za} and that this property holds
even in the case $t\ne 0$. Taking together Eqs.~\eqref{eq:mellin-def},
\eqref{eq:lattice-vec-gff-def}, \eqref{eq:lattice-ax-gff-def}
and~\eqref{eq:forward-pd} and setting $t=0$ will similarly yield the
moments
\begin{eqnarray}
  \label{eq:forward-mellin}
  \langle x^{n-1}\rangle_q & = & H^n(0,0) = A_{n0}(0)\,, \nonumber \\
  \langle x^{n-1}\rangle_{\Delta q} & = & \tilde{H}^n(0,0) =
  \tilde{A}_{n0}(0)\,.
\end{eqnarray}
The matrix elements are obtained on the lattice from the two point
functions
\begin{equation}
  \label{eq:two-pt}
  C^{\mbox{\tiny 2pt}}(T,\vec{p}) = \sum_{\vec{x}} e^{-{\imag}
    \vec{p}\cdot\vec{x}} \;\;\mbox{Tr} \left( \Gamma_{\mbox{\tiny pol}}
    \langle n(\vec{x},T) \bar{n}(\vec{0},0)\rangle\right)\,,
\end{equation}
and the three-point functions
\begin{equation}
  \label{eq:three-pt}
  C^{\mbox{\tiny 3pt}}_{{\cal O}}(T,T_0,\vec{p},\vec{p}^{\,\prime}) =
  \sum_{\vec{x},\vec{y}} e^{-{\imag} \vec{p}^{\,\prime}\cdot\vec{x}
    +\imag (\vec{p}^{\,\prime}-\vec{p})\cdot\vec{y}} \;\;
  \mbox{Tr}  \left( \Gamma_{\mbox{\tiny pol}} \langle n(\vec{x},T_0)
    {\cal O}(\vec{y},T) \bar{n}(\vec{0},0)\rangle\right)\,.
\end{equation}
We have introduced the lattice proton operators, $n(\vec{x},T)$ and
$\bar{n}(\vec{x},T)$. In order to maximize overlap with the ground
state, we use the smeared sources defined in
Ref.~\cite{Dolgov:2002zm}. This overlap can be parameterized by a
function $Z(\vec{p})$ according to $\langle\Omega\vert n(x) \vert
\vec{p},\lambda\rangle = \sqrt{Z(\vec{p})} u(p,\lambda) e^{-{\imag}
  p\cdot x}$. We also use the projection operator,
$\Gamma_{\mbox{\tiny pol}}=\frac{1}{2}(1+\gamma_4) \frac{1}{2}(1-\imag
\gamma_3\gamma_5)$. ${\cal O}=\bar{q}(0){\cal J}q(0)$ denotes the
operator with all appropriate indices in which we are interested.

Applying the transfer matrix formalism yields the following expression
for the behavior of the two- and three-point functions
\begin{eqnarray}
  \label{eq:two-three-ground}
  C^{\mbox{\tiny 2pt}}(T,\vec{p}) &=& \frac{Z(\vec{p}) e^{-E T}} {2E}
  \;\;\mbox{Tr}\left( \Gamma_{\mbox{\tiny pol}} (\imag \fslash{p} + m_N)
  \right) + \mbox{excited states}\,. \nonumber \\
  C^{\mbox{\tiny 3pt}}_{{\cal O}}(T,T_0,\vec{p},\vec{p}^{\,\prime}) &=&
  \frac{\sqrt{Z(\vec{p}) Z(\vec{p}^{\,\prime})} e^{-E'(T_0-T) -E T}}
  {2E' 2E} \;\;\mbox{Tr} \left( \Gamma_{\mbox{\tiny pol}} (\imag
    \fslash{p}^{\,\prime}+m_N) {\cal J} (\imag \fslash{p} +
    m_N)\right) + \mbox{excited states}\,.
\end{eqnarray}
In order to cancel the exponential factors and wave-function
normalizations, we construct the ratio
\begin{eqnarray}
  \label{eq:two-three-ratio}
  R_{{\cal O}}(T,T_0) &=& \frac{ C^{\mbox{\tiny 3pt}}_{\cal O}(T,
    T_0,\vec{p}, \vec{p}^{\,\prime})} {\sqrt{C^{\mbox{\tiny 2pt}}(T,
      \vec{p}) C^{\mbox{\tiny 2pt}}(T,\vec{p}^{\,\prime})}}
  \sqrt{\frac{C^{\mbox{\tiny 2pt}}(T_0-T,\vec{p})
      C^{\mbox{\tiny 2pt}}(T,\vec{p}^{\,\prime})} 
    {C^{\mbox{\tiny 2pt}}(T_0-T,\vec{p}^{\,\prime})
      C^{\mbox{\tiny 2pt}}(T,\vec{p})}} \nonumber \\
  & \buildrel{T_0\gg T\gg 1}\over\to & \sum_{\lambda,\lambda'} \frac{
    \bar{u}(\vec{p},\lambda) \Gamma_{\mbox{\tiny pol}}
    u(\vec{p}^{\,\prime},\lambda') } {\sqrt{2E(E+m_N) 2E'(E'+m_N)}}
  \langle \vec{p}^{\,\prime},\lambda'\vert {\cal O}\vert
  \vec{p},\lambda\rangle\,,
\end{eqnarray}
which becomes proportional to the desired matrix elements for
sufficiently large source-sink separations, $T_0$, and with the
operator insertion sufficiently far from both source and sink, $T\gg
1$ and $T_0-T\gg 1$. Typically, we find a plateau region
$[T_{\mbox{\tiny min}}, T_{\mbox{\tiny max}}]$ over which we average
the resulting value of the operator.

Since we operate with a finite lattice of extent $aL$, the momentum
values which we can choose are discrete and are given by
$\vec{p}=2\pi/(aL) \vec{n}_p$ with $\vec{n}_p$ a vector whose
components are integers ranging from $-L/2$ to $L/2$.  For the nucleon
sink we choose the two values $\vec{p}^{\,\prime}=2\pi/(aL) \vec{0}$
and $\vec{p}^{\,\prime}=2\pi/(aL) (-1,0,0)$ and for the source we
choose $\vec{p}$ such that the absolute value of the integer momentum
vector, $|\vec{n}_p|$ is smaller than five. This defines the set of
$t$ values accessible in our calculation. Note that the (generalized)
form factors will receive contributions from several different
momentum and index combinations at any fixed value of $t$. By
constructing an overdetermined system of equations from all those
combinations we make optimal use of the available data. This procedure
has been discussed in detail in Ref.~\cite{Hagler:2003jd}.

The energy of a state at momentum $\vec p$ is related to its mass
through the dispersion relation. In our analysis we use the continuum
dispersion relation.  We have verified that the resulting energy
agrees, for the spatial momenta employed in our calculation, with the
energy of a nucleon at non-zero momentum actually calculated on the
lattice.

In the case of the electromagnetic current,
$\bar{\psi}\gamma_\mu\psi$, the generalized form factors correspond to
the electromagnetic form factors of the nucleon. For the axial
current, $\bar{\psi}\gamma_\mu\gamma_5\psi$, the form factors
correspond to the axial and the pseudoscalar form factors. These will
be covered in detail in Secs.~\ref{sec:form-factors}
and~\ref{sec:axial-form-factors}.

%
%

\section{\label{sec:new-latt-calc}New lattice calculations}
We now present the methods and technologies we have used for our
calculations. As discussed previously in Ref.~\cite{Hagler:2007xi}, we
continue to employ the asqtad action for the sea quarks and the domain
wall (DWF) action for the valence quarks. In addition, we also add one
lighter mass to our data set. The data sets used in this paper are
summarized in Tab.~\ref{tab:data-sets}. The columns show the bare
asqtad quark mass for $N_f=2+1$ dynamical fermions, the corresponding
bare DWF mass, the volume in lattice units, the number of gauge field
configurations used and the number of measurements included in the
analysis. The bare valence DWF masses have been tuned such that the
physical pion masses agree with those obtained from the purely asqtad
calculation. The size of the 5th dimension has been set to
$L_5=16$. The choice and tuning of these parameters has been discussed
in detail in Ref.~\cite{Hagler:2007xi}.
\begin{table}[htb]
  \centering
  \begin{tabular}[c]{*{4}{c|}c}
    \hline\hline
    $m_{\mbox{\tiny sea}}^{\mbox{\tiny asqtad}}\,, N_f=2+1$ &
    $m_{\mbox{\tiny val}}^{\mbox{\tiny DWF}}$ & \strut Volume & \#
    confs & \# meas \\ \hline
    0.007/0.050 & 0.0081 & $20^3\times 64$ & 463 & 3704 \\
    0.010/0.050 & 0.0138 & $28^3\times 64$ & 274 & 2192 \\
    0.010/0.050 & 0.0138 & $20^3\times 64$ & 631 & 5048 \\
    0.020/0.050 & 0.0313 & $20^3\times 64$ & 486 & 3888 \\
    0.030/0.050 & 0.0478 & $20^3\times 64$ & 563 & 4504 \\
    0.040/0.050 & 0.0644 & $20^3\times 32$ & 350 & 350 \\
    0.050       & 0.0810 & $20^3\times 32$ & 425 & 425 \\
    \hline\hline
  \end{tabular}
  \caption{Summary of our data sets.}
  \label{tab:data-sets}
\end{table}

We work in a mass independent scheme. The lattice spacing is therefore
independent of the bare quark mass, and its value has been determined
to be $a=0.1241(25)$fm, corresponding to \mbox{$a^{-1}=1.591(32)$GeV}
in Ref.~\cite{WalkerLoud:2008bp}, taken from heavy quark
spectroscopy~\cite{Aubin:2004wf}. This yields a physical volume of
$V=(aL)^3=(2.5$ fm$)^3$ on the $L^3=20^3$ lattices and of
$V=(aL)^3=(3.5$ fm$)^3$ on the $L^3=28^3$ lattice. The physical values
of the nucleon masses, pion masses, and pion decay constants are
needed for our computation of hadron structure. These have been
previously determined in Ref.~\cite{WalkerLoud:2008bp}. We have listed
them in Tab.~\ref{tab:phys-pars}. The columns show the bare asqtad sea
quark mass, the lattice size, and the resulting lattice pion mass,
pion decay constant and nucleon mass. Finally, they are converted to
physical values in MeV\footnote{Note that unlike in
  Ref.~\cite{WalkerLoud:2008bp}, in this paper we use the
  normalization of $\fps$ such that $\fps=92.4$MeV}.
\begin{table}[htb] \centering
  \begin{tabular}[c]{*{7}{c|}c}
    \hline\hline
    Light \strut $m_{\mbox{\tiny sea}}^{\mbox{\tiny asqtad}}$ & Volume
    $\Omega$ & $(am)_\pi$ & $(af)_\pi$ & $(am)_N$ &  $\mps$ [MeV] &
    $\fps$ [MeV] & $m_N$ [MeV] \\ \hline 
    0.007 & $20^3\times 64$ & 0.1842(7)  & 0.0657(3) & 0.696(7) & 292.99(111) & 104.49(45) & 1107.1(111) \\
    0.010 & $28^3\times 64$ & 0.2238(5)  & 0.0681(2) & 0.726(5) & 355.98(80)  & 108.31(34) & 1154.8(80)  \\
    0.010 & $20^3\times 64$ & 0.2238(5)  & 0.0681(2) & 0.726(5) & 355.98(80)  & 108.31(34) & 1154.8(80)  \\
    0.020 & $20^3\times 64$ & 0.3113(4)  & 0.0725(1) & 0.810(5) & 495.15(64)  & 115.40(23) & 1288.4(80)  \\
    0.030 & $20^3\times 64$ & 0.3752(5)  & 0.0761(2) & 0.878(5) & 596.79(80)  & 121.02(34) & 1396.5(80)  \\
    \hline
    0.040 & $20^3\times 32$ & 0.4325(12) & 0.0800(5) & 0.941(6) & 687.94(191) & 127.21(78) & 1496.8(95)  \\
    0.050 & $20^3\times 32$ & 0.4767(10) & 0.0822(4) & 0.991(5) & 758.24(159) & 130.70(67) & 1576.3(80)  \\
    \hline\hline
  \end{tabular}
  \caption{Hadron masses and decay constants in physical units.}
  \label{tab:phys-pars}
\end{table}

In contrast with our previous publication~\cite{Hagler:2007xi}, in
this work we did not use Dirichlet boundary conditions on the first
and the middle time-slice, forming so-called chopped
lattices. Instead, the technology that has been employed consists in
computing multiple source/sink pairs on a single gauge field
configuration; it is discussed in more detail in the next
section~\ref{sec:improved-statistics}.  We find this technique both
more convenient to use and more powerful in making optimal use of the
existing resources. The quality of the results is superior since it
allows us to process eight source/sink pairs instead of just two as
before. We have recalculated our results on the lattices with asqtad
sea quark masses of $m_{\mbox{\tiny sea}}^{\mbox{\tiny
    asqtad}}=0.007$--$0.030$. The higher masses are included in some
plots to guide the eye, but have never been included in the chiral
fits. Section~\ref{sec:source-sink-separ} discusses possible
systematic errors of our nucleon matrix elements.

We also take all possible sources of correlations into account
carefully by performing all fits using the error correlation matrix
among all data points at fixed pion mass and the ``super jackknife''
technique for combining data from different pion masses in a single
fit. These techniques are discussed in
\sect\ref{sec:super-jackkn-analys}. Section~\ref{sec:renorm-latt-oper}
specifies our renormalization procedure for the lattice operators we
use and lists all relevant renormalization constants. Finally,
Sec.~\ref{sec:finite-volume-effect} discusses potential influences of
finite-volume effects on our lattice data.

\subsection{\label{sec:improved-statistics}Improved statistics}
In order to improve the statistical quality of our data set, we
employ a method we call ``coherent sink technique''. This method
proves to be a substantial improvement over previous methods employed
for the extraction of three-point functions from lattice
data. Effectively, we obtain eight measurements of matrix elements per
configuration. Hence, this allows us to make optimal use of the rather
expensive configurations generated with dynamical quarks.

On every other configuration we place sources at space-time positions
$(\vec{0},0)$, $(\vec{L}/2,16)$, $(\vec{0},32)$, and $(\vec{L}/2,48)$,
where $\vec{L}$ denotes a spatial vector with components $(L,L,L)$.
We then perform twelve inversions of the Dirac operator,
corresponding to the four spin and three color indices of the quarks,
and feed them into the construction of the forward propagators.
Using these forward propagators, we create a
momentum projected nucleon sink a temporal distance $T_0$ away 
from the source, i.e.~we end up
with four sinks located at Euclidean times $T_0$, $T_0+16$,
$T_0+32$, and $T_0+48$.

If we did a conventional calculation, we would consider each
source-sink pair completely separately, constructing a set of backward
propagators from the sink and evaluating the three-point
function. Thus, each measurement would require a separate set of
backward propagator calculations. With our new approach, however, we
calculate a single set of coherent backward propagators in the
simultaneous presence of all four sinks! Combining these coherent
backward propagators with the forward propagators yields the physical
matrix elements plus terms that vanish due to gauge invariance when
computing expectation values.

Additionally, we also create a momentum projected antinucleon sink a
temporal distance $-T_0$ away from each source and perform an
analogous calculation for coherent antinucleon propagators. It is then
straightforward to relate the matrix elements of our twist-two quark
operators in an antinucleon to the desired results in a nucleon. To
summarize, given a set of forward propagators, we obtain eight
measurements of $C^{\mbox{\tiny 3pt}}_{\cal O}$ at the cost of two
rather than eight sets of inversions for the backward propagators.
\begin{figure}[htb]
  \centerline{\includegraphics[width=6cm,angle=0]{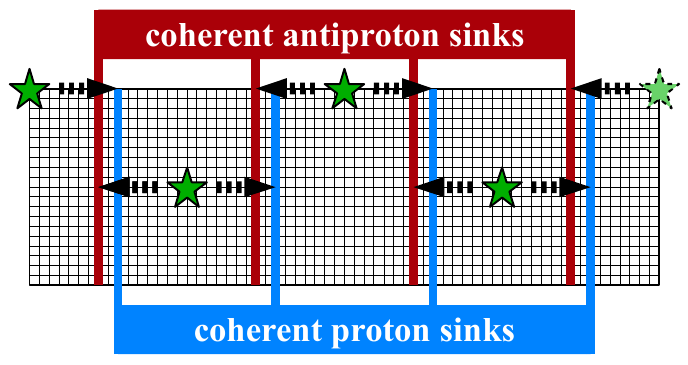}
    \hspace{1.4cm}
    \includegraphics[width=6cm,angle=0]{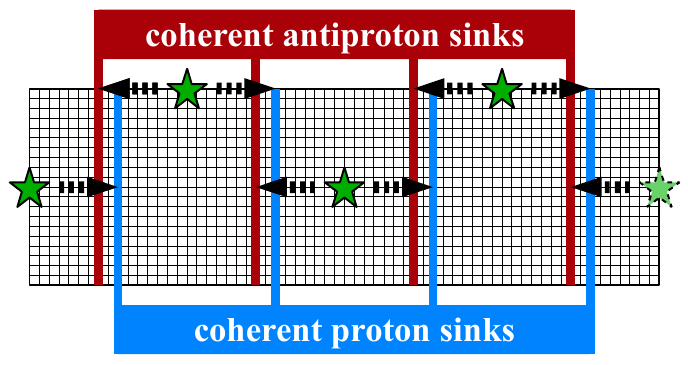}}
  \caption{Layout of smeared nucleon sources (stars) and coherent,
    fixed momentum sinks (vertical lines) on our lattices: (left) even
    configurations; (right) odd configurations.}
  \la{fig:latlayout}
\end{figure}
To minimize correlations, we alter the source locations on every other
lattice to be $(\vec{L}/2,0)$, $(\vec{0},16)$, $(\vec{L}/2,32)$, and
$(\vec{0},48)$. Thus, each source is shifted by a displacement of
$\vec{L}/2$. This layout of the sources is illustrated in
\fig\ref{fig:latlayout}.

We find that the individual measurements only exhibit minimal
correlations and thus this strategy provides a valuable increase of
statistics, see Ref.~\cite{Bratt:2008uf} for a discussion of
autocorrelations. We consistently apply binning with a bin-size above
ten which eliminates residual autocorrelations from our data. Below we
study whether this technique introduces any systematic errors into our
calculation.

\subsection{\label{sec:source-sink-separ}Tests for systematic errors}
We have performed tests concerning three possible sources of
systematic errors: The coherent sink scheme described in the previous
section, the choice of boundary conditions in the temporal direction,
and the choice of the source-sink separation, $T_0$, which is an
important input parameter entering any three-point function
calculation\footnote{The issue of finite size effects is addressed
  separately in \sect\ref{sec:finite-volume-effect}}. Following
Eq.~\eqref{eq:two-three-ground} it is advisable to pick $T_0$ as large
as possible. However, if $T_0$ becomes too large the two-point
function in the ratio Eq.~\eqref{eq:two-three-ratio} introduces an
increasing noise that will eventually wipe out the signal. Thus, $T_0$
should be chosen such that we are still able to project out the
ground-state in a suitably chosen plateau-region, but not so large
that the signal to noise ratio becomes too bad.

\begin{figure}[htb]
  \centering
  \includegraphics[scale=0.27,clip=true]{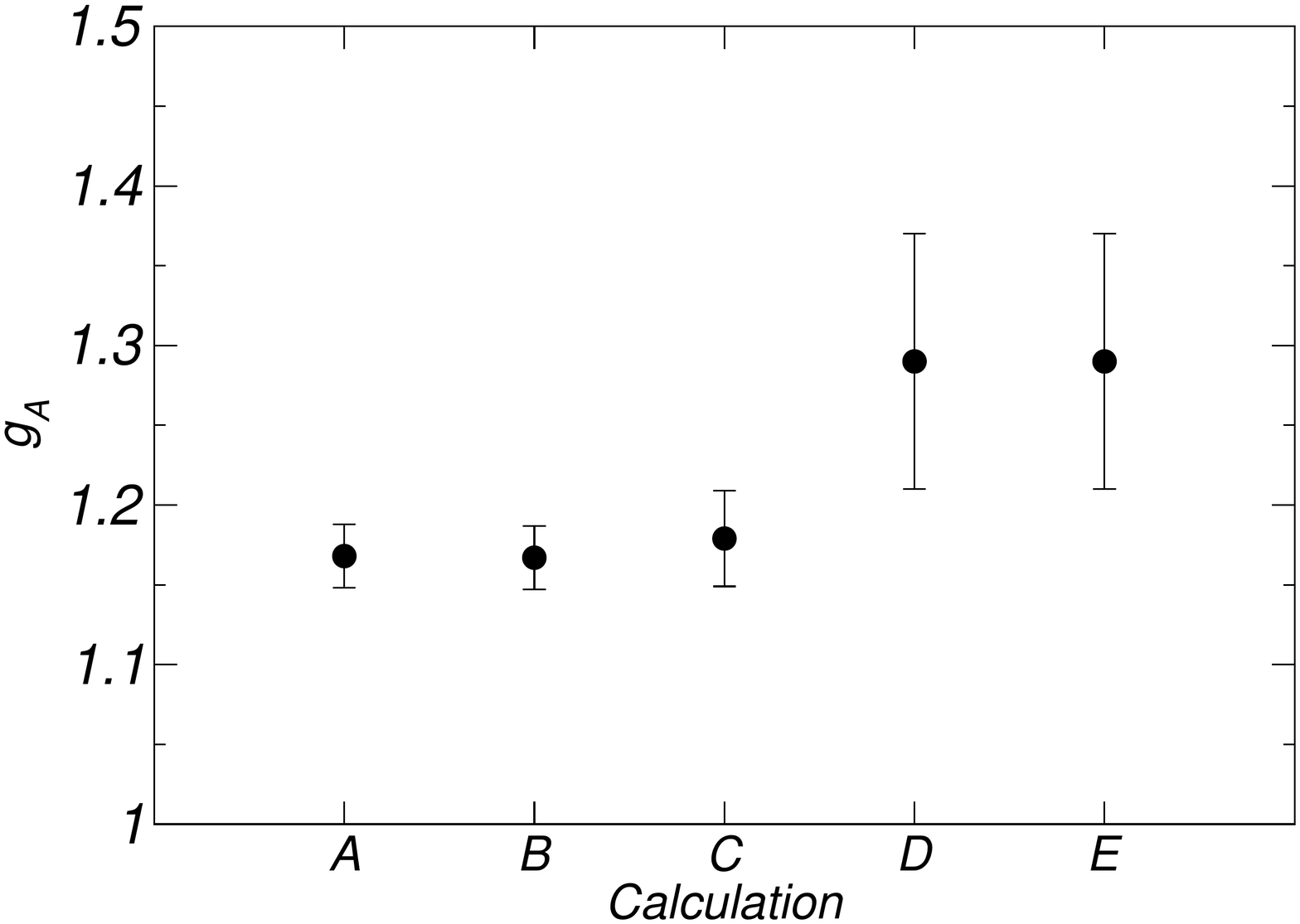} \qquad
  \includegraphics[scale=0.27,clip=true]{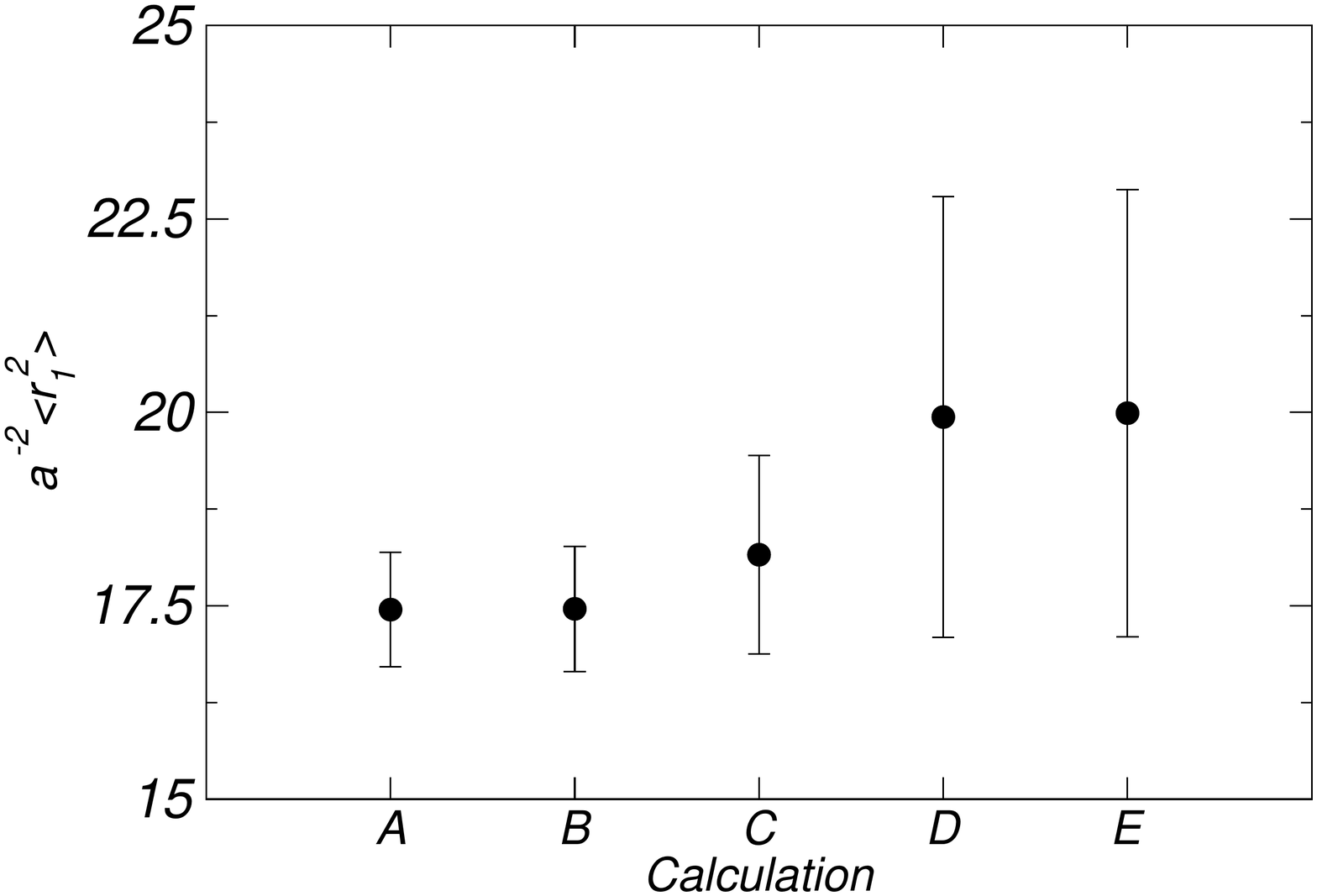}
  \caption{Comparison of coherent sink technique and two source-sink
    separations with our previous calculation. The observables are the
    axial charge, $g_A$ in the left panel, see \sect\ref{sec:gA}, and
    the isovector charge radius, $a^{-2}\langle r_1^2\rangle$ in the
    right panel, see \sect\ref{sec:form-factors}.}
  \label{fig:tech-compare}
\end{figure}

From previous calculations, cf.~e.g.~\cite{Hagler:2007xi}, we know
that a separation of $T_0\simeq 1.2$ fm is a reasonable choice. Our
tests of this assumption, together with the other characteristics of
our calculation mentioned above, are summarized in
\fig\ref{fig:tech-compare}. The labels for the calculations denoted on
the abscissa are explained in Tab.~\ref{tab:tech-compare}. The gauge
fields are a sample of 448 configurations at the working point
$m_{\mbox{\tiny sea}}^{\mbox{\tiny asqtad}}=0.010/0.050$ on the $20^3$
lattice, cf.~Tab.~\ref{tab:data-sets}.
\begin{table}[htb]
  \centering
  \begin{tabular}[c]{c|c|r|l}
    \hline\hline
    Label & \# meas & $T_0$ & Technology \\ \hline
    A & $448\cdot 8$ & 9 & Independent backward propagators, nucleon/antinucleon \\
    B & $448\cdot 8$ & 9 & Coherent backward propagators, nucleon/antinucleon \\
    C & $448\cdot 8$ & 10 & Coherent backward propagators, nucleon/antinucleon \\
    D & $448$ & 10 & Single source/conf., nucleon only, unchopped $L_t=64$ lattice \\
    E & $448$ & 10 & Single source/conf., nucleon only, chopped $L_t=32$ lattice \\
    \hline\hline
  \end{tabular}
  \caption{Calculation techniques used for the comparison of the
    coherent sink techniques and two source-sink separations with
    our previous calculation.}
  \label{tab:tech-compare}
\end{table}
The technique employed in our previous publication,
Ref.~\cite{Hagler:2007xi}, corresponds to the label ``E'',
cf.~Tab.~\ref{tab:tech-compare}. To address whether the chopping
prescription--- to cut the lattice into two halves, impose Dirichlet
boundary conditions, and compute observables on both halves
separately--- employed in that publication introduces a systematic
error, we have repeated the calculation on the unchopped lattice. This
case is denoted by label ``D''. It is evident that the results are
essentially identical both for the axial charge and the mean squared
radius. We thus conclude that the chopping prescription did not
introduce any systematic error.

Next, using the same sample of configurations we performed eight
independent calculations of propagators on the lattice with locations
identical to those chosen for the coherent propagators,
cf.~\sect\ref{sec:improved-statistics}. This case is denoted by
``A''. It is evident that the statistical improvement is remarkable,
indicating that the results are sufficiently decorrelated to warrant
the extra effort.

The true power of the coherent sink technique is demonstrated by case
``B''. The computational requirements for the backward propagators are
just 1/4th of those of case ``A'', yet the result is almost identical
and completely consistent. This gives us further confidence that our
technology is indeed correct.

Finally, we also compare the two source-sink separations, $T_0=9$ and
$T_0=10$, the latter denoted by case ``C''. There is an increase of
about 50\% in the error bar between case ``B'' and case ``C'', but the
results are fully compatible within error bars. This indicates that a
separation of $T_0=9$ is already sufficient to extract the ground
state and higher-state contaminations are negligibly small. We thus
proceed to use the method denoted by ``B'' for the remainder of this
publication.

\begin{figure}[htb]
  \centering
  \includegraphics[scale=0.3,clip=true]{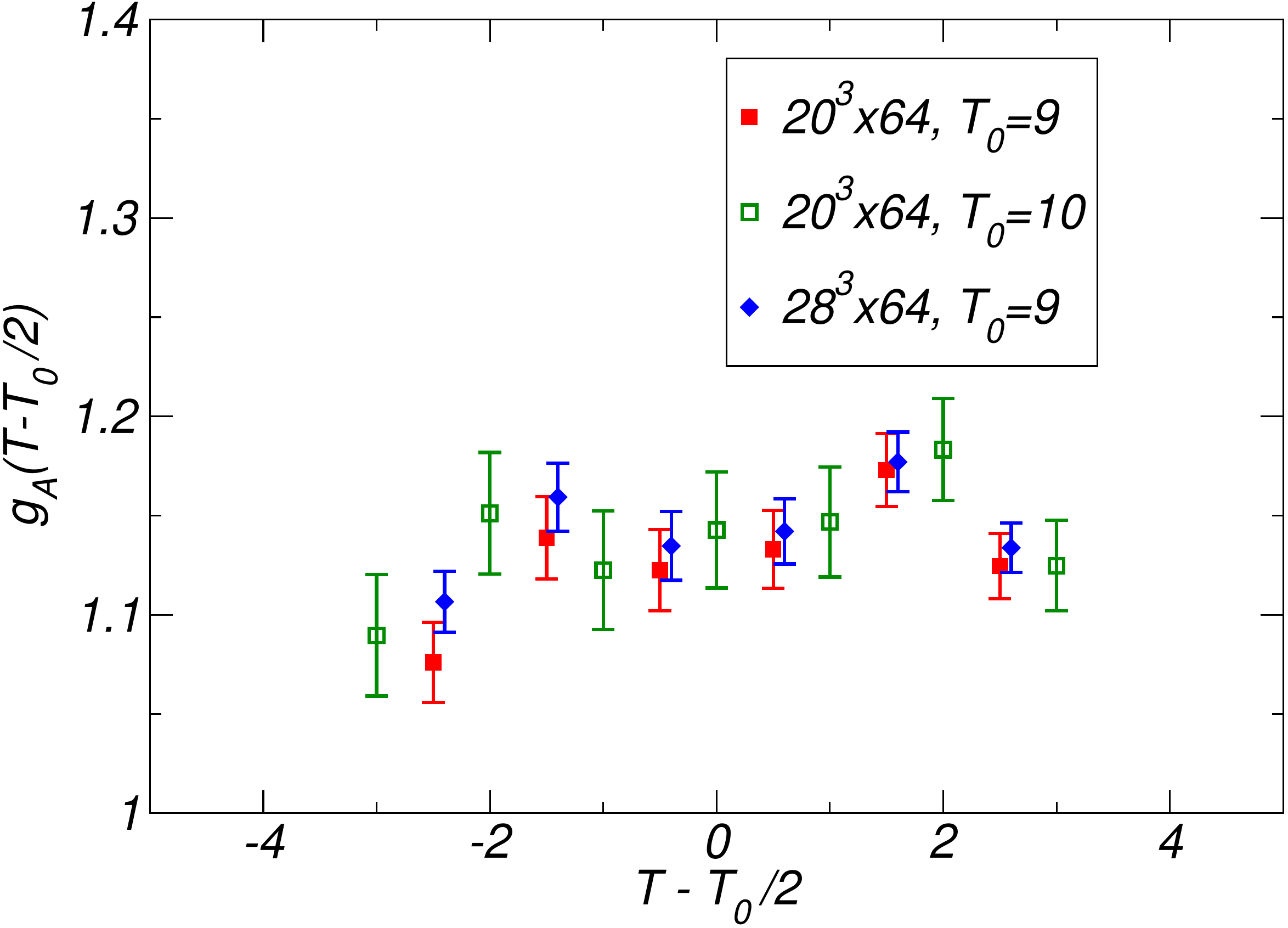}
  \caption{Comparison of two separations on two volumes for the
    isovector axial charge, $g_A$. This plot shows the resulting
    plateau plot, i.e.~$g_A(T)$ as a function of the location of the
    operator insertion at fixed source-sink separation.}
  \label{fig:ga-plateau}
\end{figure}

To illustrate the result from different source-sink separations and to
make sure that excited state contributions are small, we also show a
plateau plot for one of our observables. Figure~\ref{fig:ga-plateau}
shows the isovector axial charge, $g_A$, as a function of the location
of the operator insertion prior to averaging over the plateau
region. The plateau is flat up to ${\cal O}(a^2)$ cutoff effects for
any $T_0$ since the operator corresponds to a conserved charge. If we
had to suspect contamination from excited state contributions, the
overall level of the plateau would be systematically different from
the corresponding ones using the $T_0=9$ separation.  We do not
observe such a systematic effect and thus conclude that going to the
smaller separation does not introduce systematic uncertainties from
excited states.

In summary, the new technology does indeed constitute a major
improvement over the one previously employed in terms of statistical
accuracy, but does not lead to a detectable increase in systematic
error.

\subsection{\label{sec:super-jackkn-analys}Super jackknife analysis
  and error correlation}
When calculating physical observables and their uncertainties from
different lattice data sets we have to take into account the
correlations of different data points with each other. It turns out
that there are three possible cases we need to consider: (a)
Observables calculated from data points computed on a single ensemble
of lattice data. (b) Observables calculated from data points computed
on several statistically independent ensembles of lattice data, where
each ensemble contributes only a single data point. (c) Observables
calculated from data points from different ensembles, where each
ensemble contributes more than one data point. An example for case (a)
are dipole fits to form factors, where each fit is done to a set of
different $Q^2$ values at a fixed pion mass. Case (b) occurs for
chiral fits to hadron masses~\cite{WalkerLoud:2008bp} or moments of
forward parton distributions~\cite{Dolgov:2002zm}. Case (c) is the
most complicated and shows up in simultaneous fits to (generalized)
form factors as functions of both $Q^2$ (or of $t$ in case of the
GFFs) and $m_\pi$.

In case (a) we are dealing with data that may have correlations among
data points. Case (b) has no correlation among different data points
since the underlying gauge field configurations are entirely
independent. Case (c) has data points that are correlated (those
points obtained on the same sample of gauge field configurations) and
others that are not correlated (those obtained on different samples of
configurations). Although in case (c) the error correlation matrix
will be strictly block-diagonal, a straightforward numerical
estimation may not take this property into account. In particular, the
off-diagonal entries in the error correlation matrix may have large
uncertainties themselves and thus introduce numerical instabilities.

In this paper, we have decided to consistently adopt the jackknife
method to compute uncertainties and to use the error correlation
matrix in the function $\chi^2$ where it does not introduce numerical
instabilities.  The standard jackknife error prescription is
well-known and discussed widely in the literature, see
e.g.~\cite{Efron_The_jackknife_1982,Schroers:2001fw}. We briefly
summarize the method for an observable with exact statistical mean $A$
as follows: given a series of measurements $\lbrace a_i\rbrace$,
$i=1\dots N$, with $N$ being the total number of measurements, $a_i$
being the $i^{\rm th}$ measurement we define the $i^{\rm th}$
jackknife average via
\begin{equation}
  \label{eq:jackknife-block-def}
  \bar{a}_i = \frac{1}{N-1} \sum_{j=1, i\ne j}^N a_j\,.
\end{equation}
Thus, we obtain a new set of $N$ jackknife averages,
$\lbrace\bar{a}_i\rbrace$.  The natural estimator for $A$ reads $\bar
a = \frac{1}{N}\sum_{i=1}^N \bar a_i$ in terms of the
$\lbrace\bar{a}_i\rbrace$.  More generally, for any function $f(A)$ of
the observable, the set of $N$ jackknife blocks provides an estimate
of the mean $\bar{F}$,
\begin{equation}
  \label{eq:jackknife-av}
 \bar{F} = \frac{1}{N} \sum_{i=1}^N f(\bar{a}_i)\,,
\end{equation}
as well as an estimator for its uncertainty,
\begin{equation}
  \label{eq:jackknife-err}
  \sigma^2_{F} = \frac{N-1}{N}\sum_{i=1}^N \left(
    f(\bar{a}_i) - f(\bar a)\right)^2\,.
\end{equation}
Note that this method also generalizes to the case of a function of
several observables, labeled by a Greek index, with expectation values
$A_\alpha$, $\alpha=1\dots n$. Again, this topic has been widely
discussed in the literature, including in our previous studies, see
e.g.~\cite{Syritsyn:2009np}. As an important example, we describe the
case of a fit to several observables. The function that we minimize in
a nonlinear fit to $n$ different lattice measurements is given by
\begin{equation}
  \label{eq:chisq-def}
  \chi^2_i = \sum_{\alpha=1}^n\sum_{\beta=1}^n (y_\alpha(\lbrace
  x\rbrace)-\bar{a}_{\alpha,i}) (y_\beta(\lbrace
  x\rbrace)-\bar{a}_{\beta,i})
  C^{-1}_{\alpha\beta}\,,
\end{equation}
where $y_\alpha(\lbrace x\rbrace)$ is a model function with parameters
$\lbrace x\rbrace$ and a possible dependence on the index,
$\alpha$. The covariance matrix $C_{\alpha\beta}$ is defined by
\begin{equation}
  \label{eq:errcorrmat-def}
  C_{\alpha\beta} = \frac{N-1}{N}\sum_{i=1}^N
  (\bar{a}_\alpha-\bar{a}_{\alpha,i})
  (\bar{a}_\beta-\bar{a}_{\beta,i})\,.
\end{equation}
For each configuration with number $i$ we substitute the jackknife
averages $\bar{a}_{\alpha,i}$ and minimize $\chi^2_i$ w.r.t.~the
parameters $\lbrace x\rbrace$. The parameter values which minimize
Eq.~\eqref{eq:chisq-def} are thus implicit functions of the original
data set. Finally, we use Eqs.~\eqref{eq:jackknife-av}
and~\eqref{eq:jackknife-err} on the resulting set of parameters to
obtain estimates of their central values and of their statistical
uncertainties.

As a special case, we consider the case of a dipole fit at fixed pion
mass, where each $a_{\alpha,i}$ denotes the form factor measurement on
the $i^{\rm th}$ jackknife block at momentum transfer $Q^2_\alpha$,
and the function $y_\alpha(M_d,A_0)=A_0/(1+Q^2_\alpha/M_d^2)^2$ in
Eq.~\eqref{eq:chisq-def} is the dipole model function,
cf.~Eq.~\eqref{eq:dipole} below, the parameters being the dipole mass,
$M_d$, and the overall normalization, $A_0$. The best fit parameters
are obtained from a minimization of the associated function $\chi^2$
for all jackknife blocks in the sample and we get the results by using
Eqs.~\eqref{eq:jackknife-av} and~\eqref{eq:jackknife-err}, where the
function $f(\lbrace A_{\alpha}\rbrace)$ denotes either $M_d$ or $A_0$.

As has been mentioned previously, it may happen that the off-diagonal
elements of $C_{\alpha\beta}$ are only poorly determined and the
resulting inverse has zero or negative eigenvalues. In that case, the
function $\chi^2$ is not positive-definite and the problem is
ill-defined. If this happens, we have to resort to simply using the
diagonal matrix elements of $C_{\alpha\beta}$ and the method reduces
to the regular, uncorrelated fit. We point out that even in this case
the statistical error associated with this procedure is correctly
estimated by the jackknife method, Eq.~\eqref{eq:jackknife-err}.

We also want to point out that each measurement can be taken from a
single gauge field configuration or obtained as an average of a small
set of $B$ subsequent configurations. In the latter case, the
jackknife procedure corresponds to what is called ``jackknife with
block-size $B$'' in Ref.~\cite{Schroers:2001fw}. To summarize, the
jackknife method allows us to compute functions of averages obtained
on a given ensemble. The estimation of the uncertainty in
Eq.~\eqref{eq:jackknife-err} correctly takes into account correlations
among the input data.
\begin{figure}[htb]
  \centering
  \includegraphics[scale=0.25,clip=true]{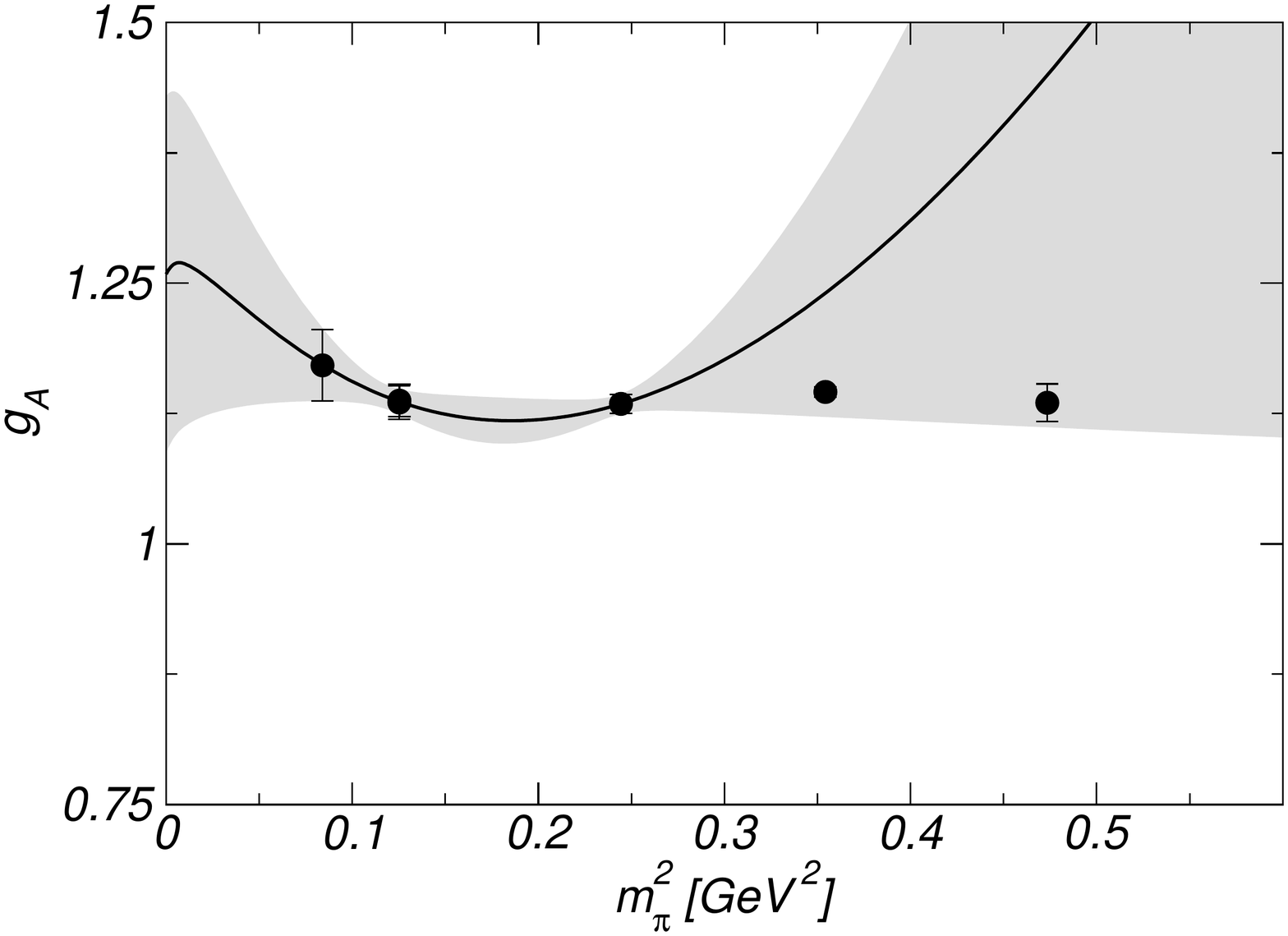} \qquad
  \includegraphics[scale=0.25,clip=true]{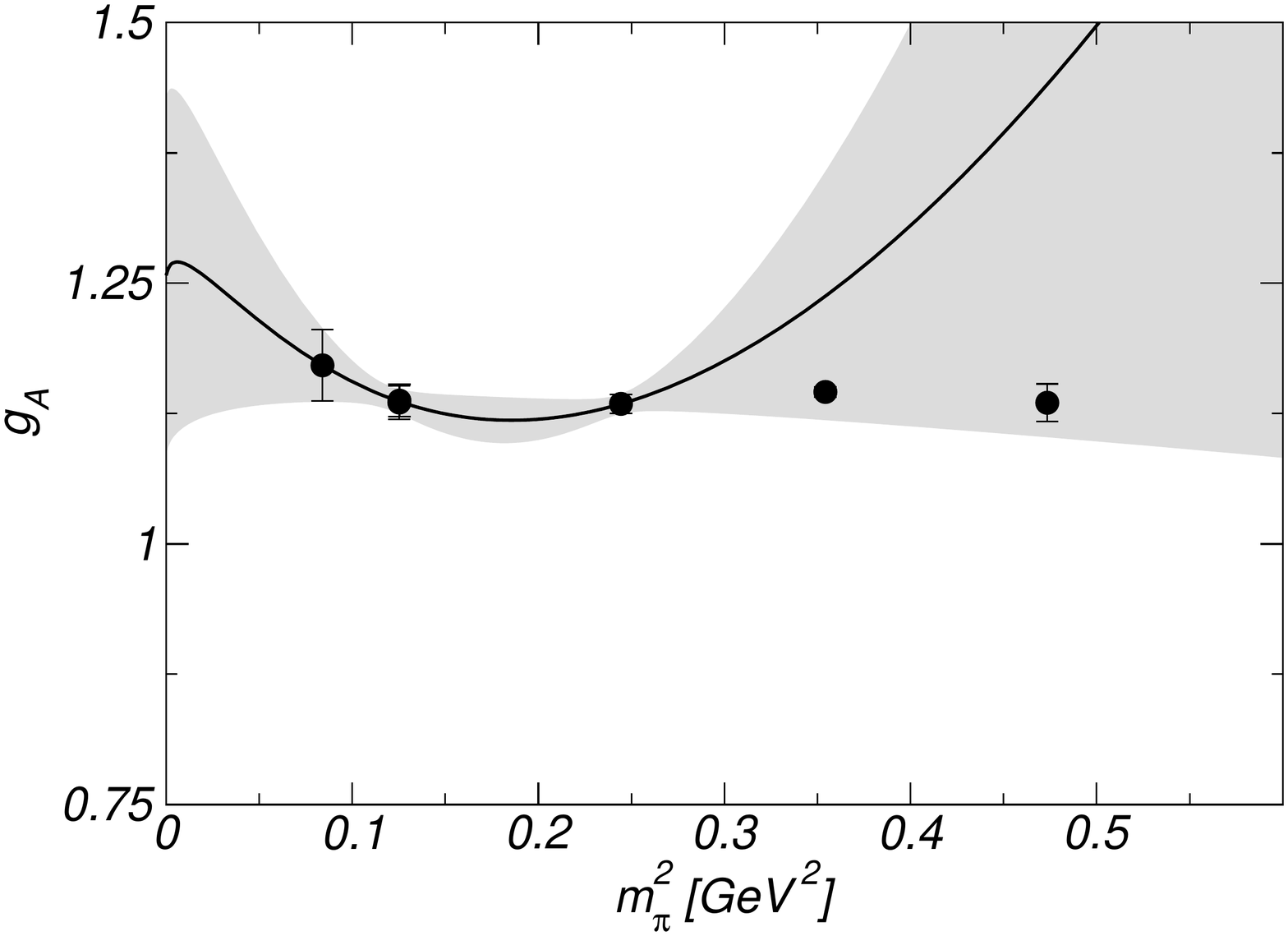} \\
  \includegraphics[scale=0.25,clip=true]{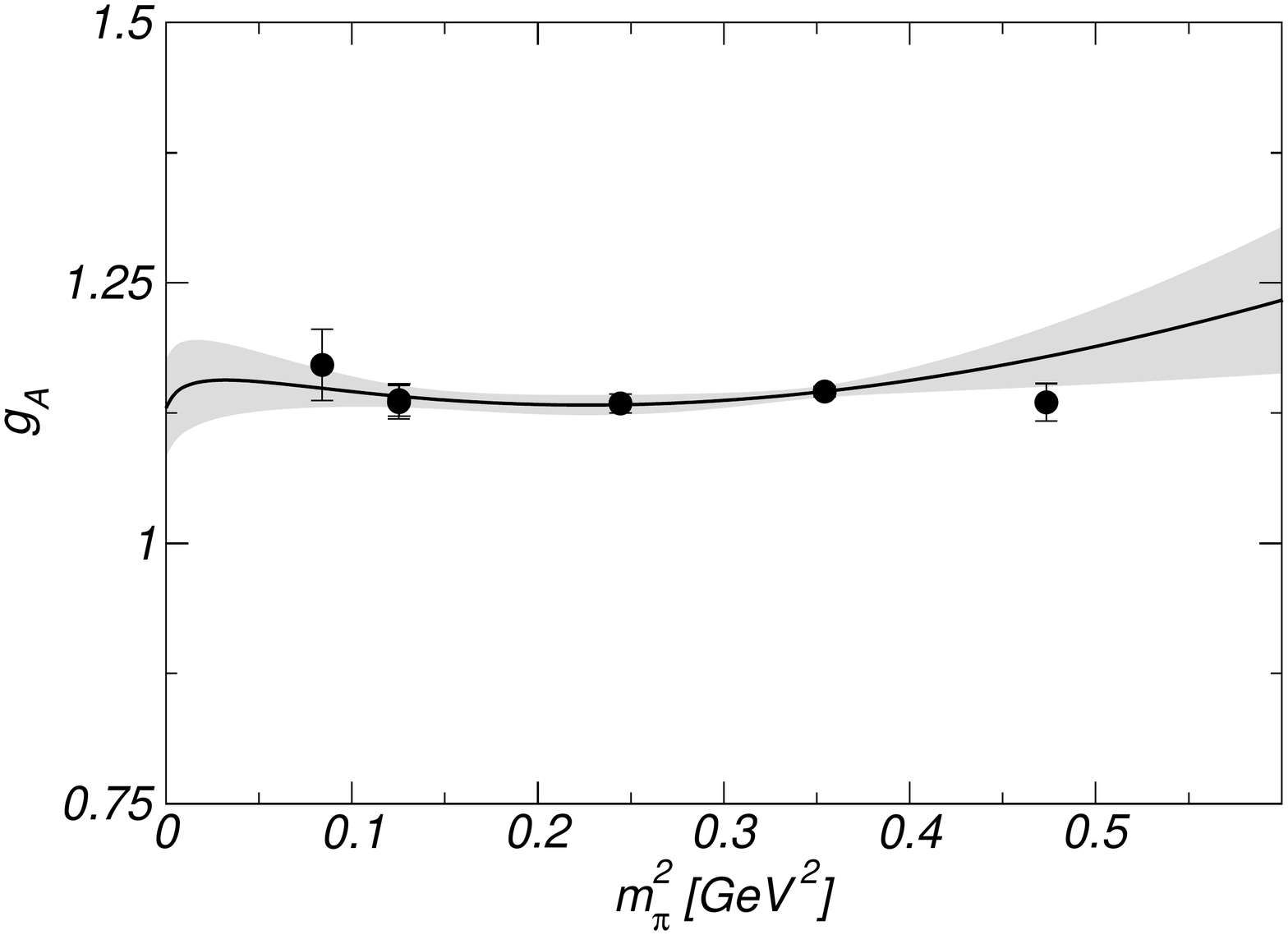} \qquad
  \includegraphics[scale=0.25,clip=true]{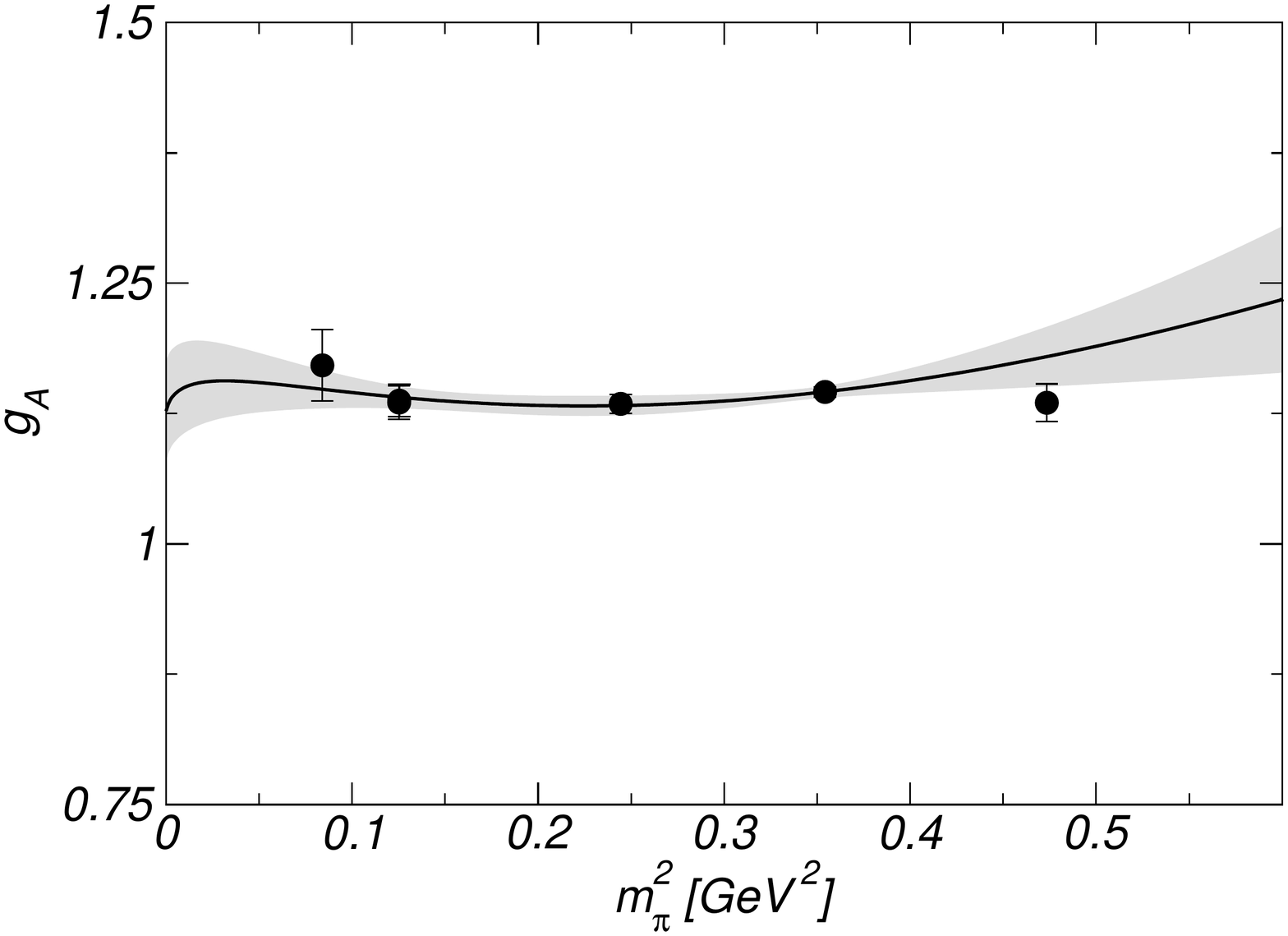}
  \caption{Comparison of bootstrap (left panels) and super jackknife
    (right panels) resampling plans for chiral fits to the nucleon
    axial charge, $g_A$. The upper two plots show fits to the lowest
    three masses, the lower two show fits to the lowest four masses.}
  \label{fig:bs-sjk-comp}
\end{figure}

When we are dealing with a function of several observables computed on
different ensembles, the method needs to be generalized further. In
order to take the particular features of such a ``super-sample''
containing both correlated and uncorrelated data into account, we use
an extension of the jackknife method, called ``super
jackknife''~\cite{Blum:2009pr, AliKhan:2001tx}. The idea is to define
generalized (super) jackknife blocks, with which averages and errors
can still be estimated using Eqs.~\eqref{eq:jackknife-av}
and~\eqref{eq:jackknife-err}. To illustrate this procedure we start
with $M$ distinct, uncorrelated ensembles with $N_k$ samples available
in the $k^{\rm th}$ ensemble. On these ensembles we again have a set
of measurements, $\lbrace a_{\alpha,i}^{(k)}\rbrace$, which denotes
the measurement on the $i^{\rm th}$ sample of the $k^{\rm th}$
ensemble. The averages $\bar{a}_{\alpha}^{(k)}=1/N_k \sum_{i=1}^{N_k}
a_{\alpha,i}^{(k)}$ and the jackknife blocks
$\bar{a}_{\alpha,i}^{(k)}$ are introduced as in
Eq.~\eqref{eq:jackknife-block-def} above. The total number of
super-jackknife blocks is defined by $N=\sum_{k=1}^M N_k$. With
appropriate super-jackknife blocks
$\lbrace\tilde{a}^{(k)}_{\alpha,i}\rbrace$ we can generalize
Eq.~\eqref{eq:chisq-def} to
\begin{equation}
  \label{eq:chisq-sjn-def}
  \chi^2_i = \sum_{k=1}^M\sum_{\alpha=1}^{n^{(k)}}
  \sum_{\beta=1}^{n^{(k)}} \left(y^{(k)}_\alpha(\lbrace
    x\rbrace)-\tilde{a}_{\alpha,i}^{(k)}\right)
  \left(y^{(k)}_\beta(\lbrace x\rbrace)-\tilde{a}_{\beta
      i}^{(k)}\right) (C^{(k)})^{-1}_{\alpha\beta}\,,
\end{equation}
where we introduce the notation $n^{(k)}$ for the number of
observables in the $k^{\rm th}$ ensemble. The index $i$ now denotes
the number of the super-jackknife block, $i=1\dots N$. The $N$
super-jackknife blocks $\tilde{a}_{\alpha,i}^{(k)}$, i.e.~the $N$ sets
of arguments at which the function is to be evaluated are constructed
as follows. The first $N_1$ blocks consist of
\begin{equation}
  \label{eq:sjblock1}
  \tilde{a}^{(k)}_{\alpha,i} = \left\lbrace
    {\renewcommand\arraystretch{1.5}
      \begin{array}{r l}
        \bar{a}_{\alpha,i}^{(k)} & : k = 1 \\
        \bar{a}_{\alpha}^{(k)} & : k \ne 1
      \end{array}}\right.
  \qquad\mbox{with } i=1\dots N_1, k=1\dots M,
  \alpha=1\dots n^{(k)}\,.
\end{equation}
The following $N_2$ blocks consist of
\begin{equation}
  \label{eq:sjblock2}
  \tilde{a}^{(k)}_{\alpha,i+N_1} = \left\lbrace
    {\renewcommand\arraystretch{1.5}
      \begin{array}{r l}
        \bar{a}_{\alpha,i}^{(k)} & : k = 2 \\
        \bar{a}_{\alpha}^{(k)} & : k \ne 2
      \end{array}} \right.
  \qquad\mbox{with } i=1\dots N_2, k=1\dots M,
  \alpha=1\dots n^{(k)}\,,
\end{equation}
and so on. This generalization takes the correlations within each
ensemble correctly into account, while at the same time implicitly
sets correlations among different ensembles to zero. It is evident
that it reduces to the regular jackknife method,
Eq.~\eqref{eq:jackknife-block-def}, in the case of a single ensemble.

A typical case that the super-jackknife method can be applied to is
the fit of a form factor which has been expanded simultaneously in
$\mps$ and $Q^2$ to all available lattice data, see e.g.~the SSE
formula in Eq.~\eqref{eq:f1v-sse}. The function
$y_\alpha^{(k)}(c_A,B_{10}^r(\lambda))=y(c_A, B_{10}^r(\lambda);
\mps^{(k)}, (Q^{(k)}_\alpha)^2)$ in Eq.~\eqref{eq:chisq-sjn-def} is
the form factor model function and Eqs.~\eqref{eq:sjblock1}
and~\eqref{eq:sjblock2} collect all form factor data on the ensembles
with different pion masses, $\mps^{(k)}$, and at different
$(Q^{(k)}_{\alpha})^2$ values. Note that the values of $Q^2$ on each
ensemble are different in general since both the lattice volumes and
the nucleon masses are different. After $N$ minimizations to the super
samples, we obtain all parameters --- in our example the low-energy
constants $c_A$ and $B_{10}^r(\lambda)$ --- by virtue of
Eqs.~\eqref{eq:jackknife-av} and~\eqref{eq:jackknife-err} from the
resulting set of best fit parameters.

In order to ascertain that results obtained with the super jackknife
method are compatible with those obtained with competitor schemes,
like the bootstrap resampling plan, which has, e.g., been employed in
Ref.~\cite{WalkerLoud:2008bp}, we have made a detailed comparison
between the two methods based on fits to our data for the nucleon
axial charge, $g_A$. Figure~\ref{fig:bs-sjk-comp} shows a comparison
of two fits to $g_A$, one based on the three smallest pion masses and
one based on the four smallest pion masses, corresponding respectively
to the upper panels and the lower panels of the figure. The two left
plots show error bands determined from the bootstrap resampling
method, and the two right plots show results from the super jackknife
prescription. The technical details and results will be discussed
later in \sect\ref{sec:gA}, we use this fit merely as a test to verify
that the bootstrap and the super jackknife methods give
indistinguishable error bands in the two situations. We point out that
the case with three masses is a ``bad'' fit with large uncertainties
and the case with four masses has substantially smaller uncertainties
and provides a much better fit. We conclude that the two resampling
schemes give essentially identical error estimates and are thus
equally applicable, both in fits with large and in fits with small
uncertainties.

The error correlation matrix also allows us to study a phenomenon
about lattice data that looks puzzling when one assumes that data
points at distinct momentum transfers $Q^2$ are independent. In this
paper we will plot a couple of those cases, see
e.g.~Fig.~\ref{fig:at10vsqQ}. A couple of data points are
systematically higher, they are off by more than one standard
deviation from the central fit. If these points were independent
measurements, it would be a highly significant deviation. Thus, the
question is whether these six points are highly correlated so that
only one or two degrees of freedom have fluctuated randomly. Using the
technology of error correlations presented in this section we have
addressed this phenomenon in detail in Ref.~\cite{Bratt:2008uf}. We
have found that the correlation of such data points is very high,
typically $(80-90)\%$. On the other hand, the resulting $\chi^2/$dof
is still around $1$ which proves that such ``outliers'' are not
statistically significant, although a naive visual analysis would lead
to the erroneous conclusion that the data is incompatible with the fit
function. In this way the treatment of error correlations as
implemented in this work is absolutely necessary to derive conclusive
statements about the statistical quality of fits to lattice data
points.

\subsection{\label{sec:renorm-latt-oper}Renormalization of lattice
  operators}
The matrix elements we compute numerically are in the lattice
regularization at a scale of the lattice cutoff, $\mu^2=a^{-2}$. In
order to compare them to experiment, we need to convert them to a
commonly used scheme. When the renormalization is multiplicative, the
conversion is done via
\begin{equation}
  \label{eq:renormapply}
  \langle {\cal O}^{\rm cont}\rangle = Z_{\cal O} \langle {\cal
    O}^{\rm latt}\rangle\,,
\end{equation}
$\langle {\cal O}^{\rm cont}\rangle$ being the matrix element in the
\msbar-scheme and $\langle {\cal O}^{\rm latt}\rangle$ the bare
lattice operator. The factor $Z_{\cal O}$ is the renormalization
constant which depends on the details of the lattice action and the
operator, but not the external states of the matrix element.

Since on the lattice all operators are representations of the finite
hypercubic group $H(4)$, there will necessarily be fewer operators
than in the continuum group, $O(4)$. This means that in general,
continuum operators correspond to linear combinations of a finite set
of lattice operators and the r.h.s.~of Eq.~\eqref{eq:renormapply} is
replaced by a sum containing operators of different dimensions. In
such a situation, computing renormalization coefficients will be
impractical since subtracting power-law divergences non-perturbatively
is required. It is possible, however, to compute the coefficients if
operators of different dimensions do not mix --- in the case of the
operators appearing in Eq.~\eqref{eq:gen-currents}, this requires
choosing distinct indices, which is only possible for operators with
at most four indices. Hence, our lattice technology limits us to the
computation of only the lowest moments of (generalized) parton
distributions.

We apply the perturbative renormalization constants computed in
Ref.~\cite{Bistrovic:2005ph}. We have employed them previously
in~\cite{Hagler:2007xi}.  All the perturbative coefficients relevant
to this work are listed in Tab.~\ref{tab:renormcoeff}, together with
the representation of $H(4)$. In addition, we also obtain
renormalization constants for the electromagnetic and axial currents
non-perturbatively, see Tab.~\ref{tab:renorm-const} in
\sect\ref{sec:form-factors} and Tab.~\ref{tab:renorm-const-ax} in
\sect\ref{sec:axial-form-factors}. In particular, the wave function
renormalization encoded in the axial current renormalization factor
$Z_A$ implicitly enters most of the operators we study, and is not
small; therefore, it is desirable to determine this one common factor
non-perturbatively and employ the same recipe as
in~\cite{Hagler:2007xi}, namely,
\begin{equation}
  \la{eq:Z_O}
  Z_{\cal O} = \frac{Z_{{\cal O},{\rm pert}}}{Z_{A,{\rm pert}}} \cdot
  Z_{A,{\rm nonpert}}\,,
\end{equation}
with $Z_{A,{\rm pert}}=0.964$ for all but the vector and axial
currents. The numerical results in this paper are all transformed to
the \msbar-scheme at the scale of $\mu^2=4$GeV$^2$.
\begin{table}[htb]
  \centering
  \begin{tabular}[c]{c|c|c}
    \hline\hline
    Operator & $H(4)$ & $Z_{\cal O}^{\rm pert}$ \\
    \hline
    $\bar{q}[\gamma_5]\gamma_{\lbrace\mu}D_{\nu\rbrace}q$ & $\tau_1^{(3)}$ & 0.962 \\
    $\bar{q}[\gamma_5]\gamma_{\lbrace\mu}D_{\nu\rbrace}q$ & $\tau_1^{(6)}$ & 0.968 \\
    $\bar{q}[\gamma_5]\gamma_{\lbrace\mu}D_\nu D_{\rho\rbrace}q$ & $\tau_1^{(4)}$ & 0.980 \\
    $\bar{q}[\gamma_5]\gamma_{\lbrace\mu}D_\nu D_{\rho\rbrace}q$ & $\tau_1^{(8)}$ & 0.982 \\
    \hline\hline
  \end{tabular}
  \caption{Perturbative renormalization constants to convert bare
    matrix elements to the scale $\mu = a^{-1}$ in the \msbar-scheme.}
  \label{tab:renormcoeff}
\end{table}

We note that an ongoing analysis of \emph{non-perturbative}
renormalization constants employing the
Rome-Southampton-scheme~\cite{Martinelli:1994ty} turns out to be
challenging due to discretization effects and the restricted range of
applicability of the perturbative scale evolution. From our
preliminary results~\cite{Syritsyn:2010xy}, we can however obtain a
numerical estimate of the potential systematic uncertainty due to the
renormalization of the one-derivative operators, which is $\approx
7\%$. We include this systematic uncertainty in our final results in
Table~\ref{TabSpinDecomp}.

\subsection{\label{sec:finite-volume-effect}Finite-volume effects}
It is important to assess the potential influence of finite-volume
effects on our observables. Our calculations are done in a fixed
physical volume, hence we expect the finite-volume effects to increase
as the pion mass decreases. Since at the working point $m_{\mbox{\tiny
    sea}}^{\mbox{\tiny asqtad}}=0.010/0.050$ we have two volumes at
our disposal, $20^3$ and $28^3$, corresponding respectively to
$(2.5\text{ fm})^3$ and $(3.5\text{ fm})^3$, we have a way to estimate
the size of these effects at a fixed pion mass of $356$MeV.

\begin{figure}[htb]
  \centerline{\includegraphics[scale=0.3,clip=true,angle=0]{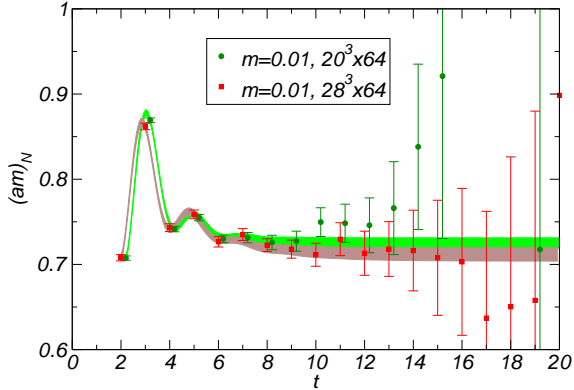}}
  \caption{Mass plateau at $am=0.010$ on the $20^3$ and the $28^3$
    lattices. The band corresponds to Eq.~(\ref{eq:c2pt-fit-func}).}
  \label{fig:MNplateau2028}
\end{figure}

We start with the simplest quantity, namely the mass of the nucleon.
Figure~\ref{fig:MNplateau2028} displays the mass plateau on the two
volumes, together with the band from fitting the two-point correlators
with the function~\cite{Syritsyn:2009mx}
\begin{equation}
\label{eq:c2pt-fit-func}
C_\text{2pt}(t) = Z_0^{-m_Nt} + Z_1 e^{-m_{N,\text{exc}}t} 
+ (-1)^t Z_\text{osc} e^{-E_\text{osc}t}
\end{equation}
in range $2\le (t/a) \le 15$.
The same source is used on both volumes, and the effective
mass is extremely similar.  Fits to the mass plateau yield the result
\be m_\pi=356{\rm MeV}: \qquad | 1- m_N(L=2.5{\rm fm}) / m_N(L=3.5{\rm
  fm})| < 1.8\% \qquad (95\% {\rm~conf.~lev.}), \la{eq:mbound} \ee As
a consequence, in extrapolations carried out throughout this paper, we
will use a common value of the nucleon mass for both volumes.
\begin{figure}[htb]
  \centerline{\includegraphics[scale=0.3,clip=true]{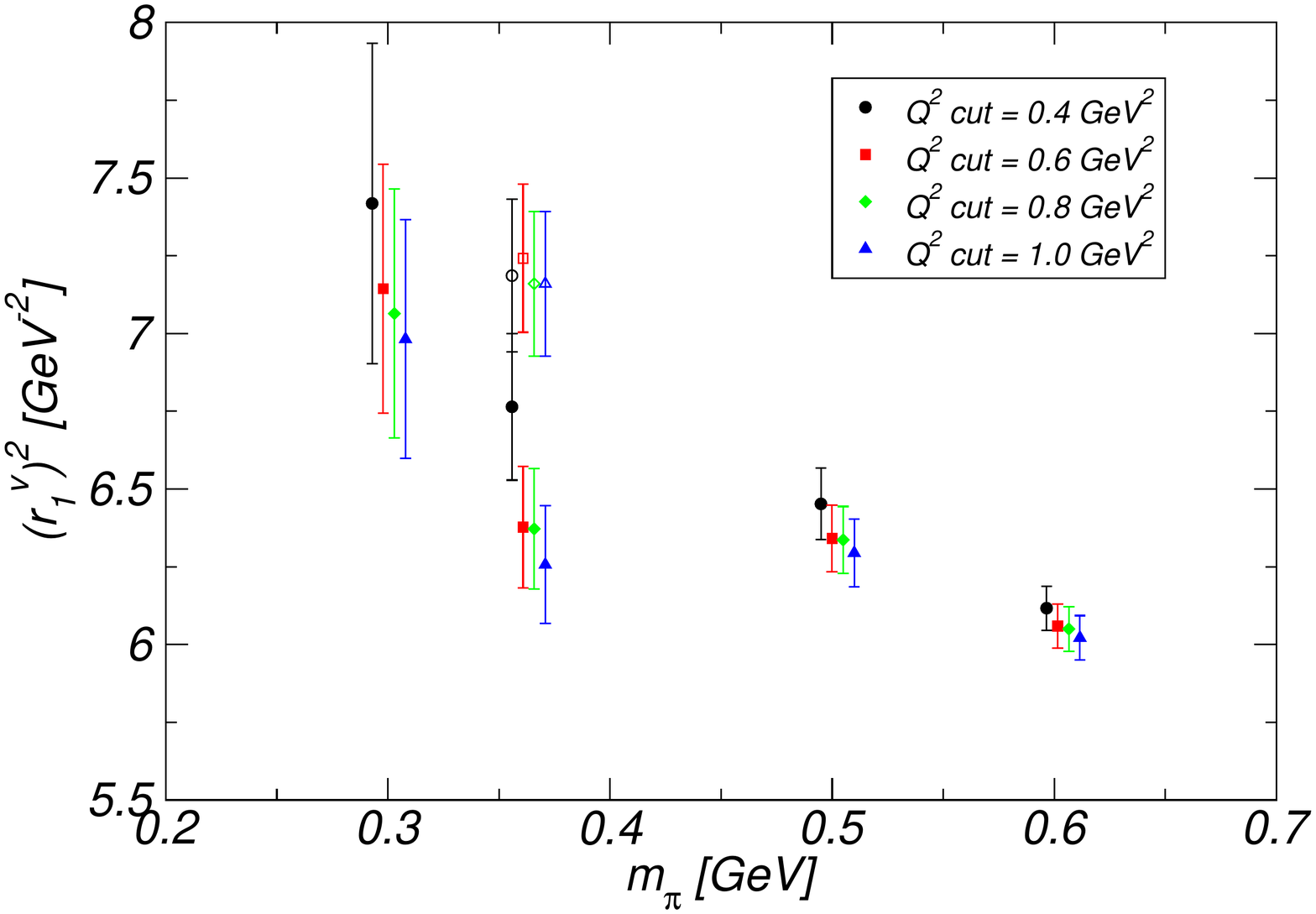}}
  \caption{Isovector Dirac radius $\langle r_1^2\rangle$ as a function
    of the pion mass for all ensembles.}
  \label{fig:r1v-varq}
\end{figure}
Secondly, Fig.~\ref{fig:ga-plateau}, used earlier to illustrate the
source-sink separation dependence, also compares the plateau plot for
$g_A$ on the $20^3$ and the $28^3$ lattices. The whole function
$g_A(T)$, where $T$ is the source-operator separation, is strikingly
similar between the two volumes.  We will return to this fact in
Sec.~\ref{sec:gA} dedicated to $g_A$.

Figure~\ref{fig:r1v-varq} shows the isovector Dirac radius, $\langle
r_1^2\rangle$ extracted from a dipole fit to the form factor
$F_1^v(Q^2)$, as a function of the pion mass, $m_\pi$. These
observables will be discussed in detail in
\sect\ref{sec:form-factors}; here we only wish to exhibit the finite
size effects. The upper cut-off in $Q^2$ for the dipole fit has been
varied and the results for different cut-offs have been drawn with a
slight displacement for clarity. At the pion mass of $m_\pi=356$MeV,
there is a discrepancy outside the error bars between the two
volumes. However, when reducing the data set to $Q^2$ values below
$0.4$GeV$^2$, we find that the two Dirac radii are actually
compatible. The discrepancy only becomes apparent at data points
beyond $Q^2>0.4$GeV$^2$.

\begin{figure}[htb]
  \centering
  \includegraphics[scale=0.27,clip=true]{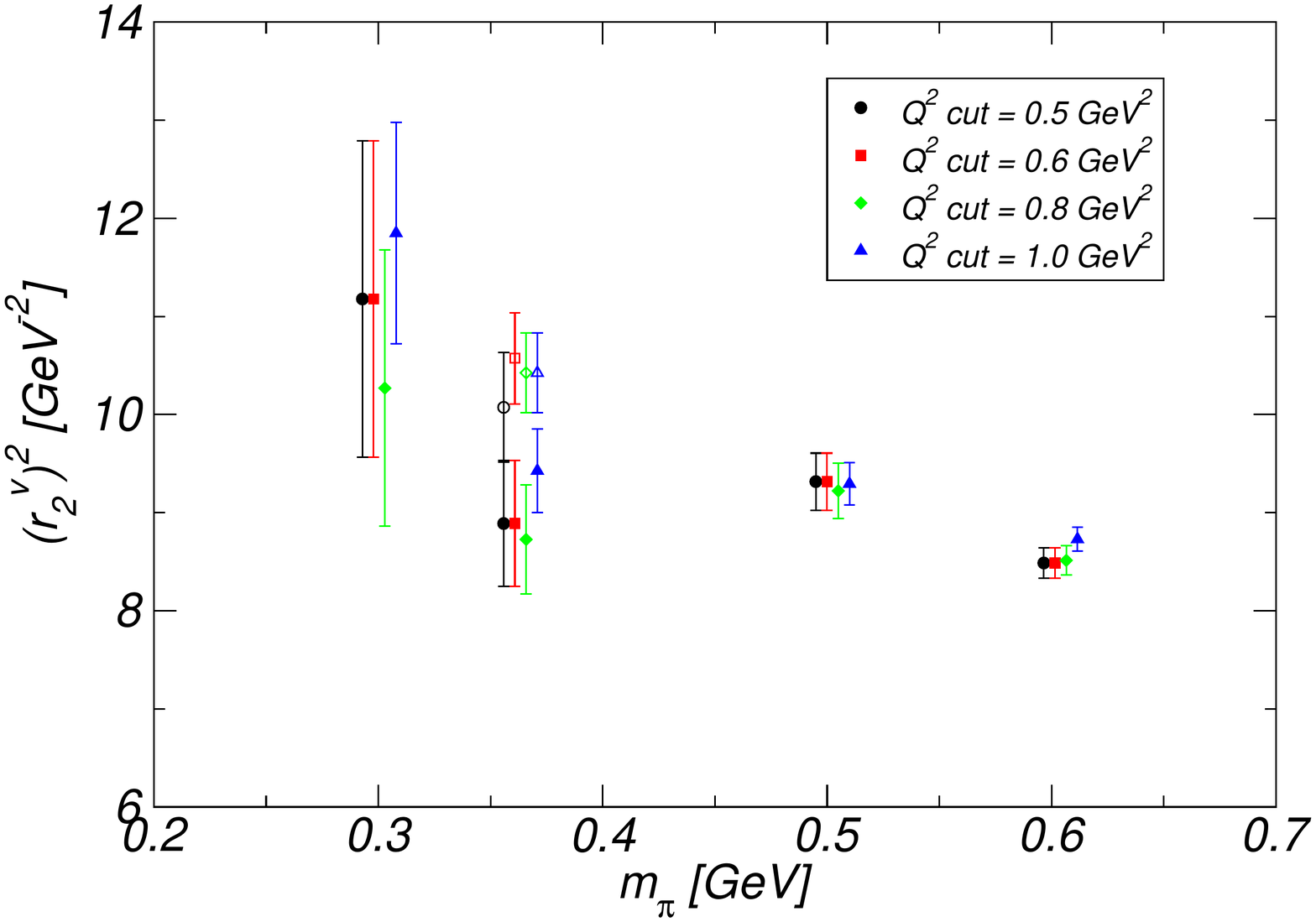} \qquad
  \includegraphics[scale=0.27,clip=true]{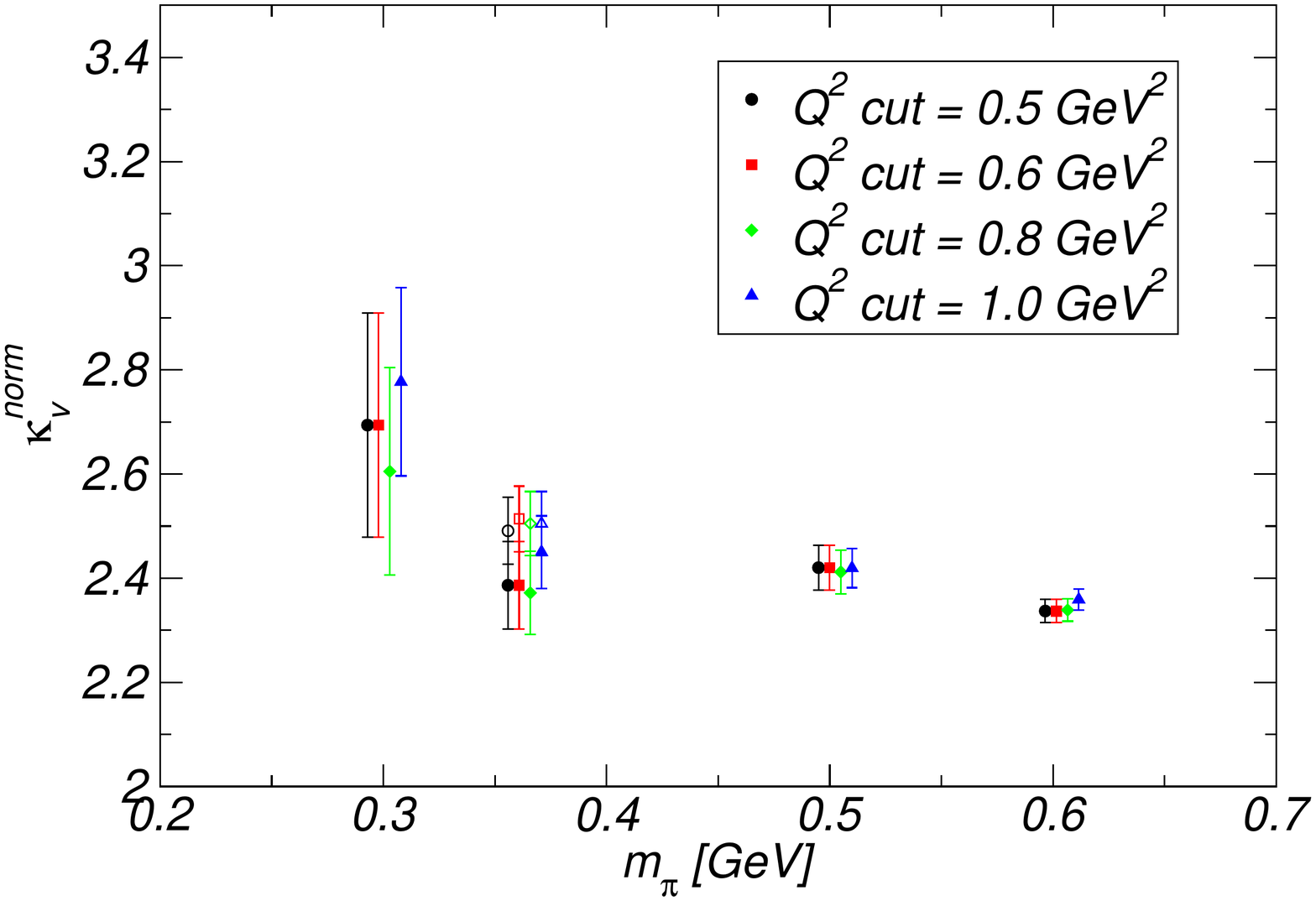}
  \caption{Isovector Pauli radius, $\langle r_2^2\rangle$, and
    anomalous magnetic moment, $\kappa^{\rm norm}_v\equiv
    \frac{m_N^{\rm phys}}{m_N(m_\pi)} F_2^v(0)$, as a function of the
    pion mass for all ensembles.}
  \label{fig:r2vkv-varq}
\end{figure}

When studying the isovector form factor $F_2^v(Q^2)$, we obtain the
Pauli radii, $\langle r_2^2\rangle$, and the anomalous magnetic
moments, $\kappa_v$, shown in Fig.~\ref{fig:r2vkv-varq}. Also these
observables will be discussed in detail in
\sect\ref{sec:form-factors}; here we again only wish to exhibit the
finite size effects. As in the previous plot, we have applied dipole
fits with varying upper cut-off in $Q^2$.
Again, we find that there is a notable discrepancy for $\langle
r_2^2\rangle$ when data points at $Q^2>0.4$GeV$^2$ are included in the
plot. For $\kappa_v$ we find that the results on the two volumes
deviate systematically, but are still compatible taking into account
their statistical uncertainty.

\begin{figure}[htb]
  \centering
  \includegraphics[scale=0.27,clip=true]{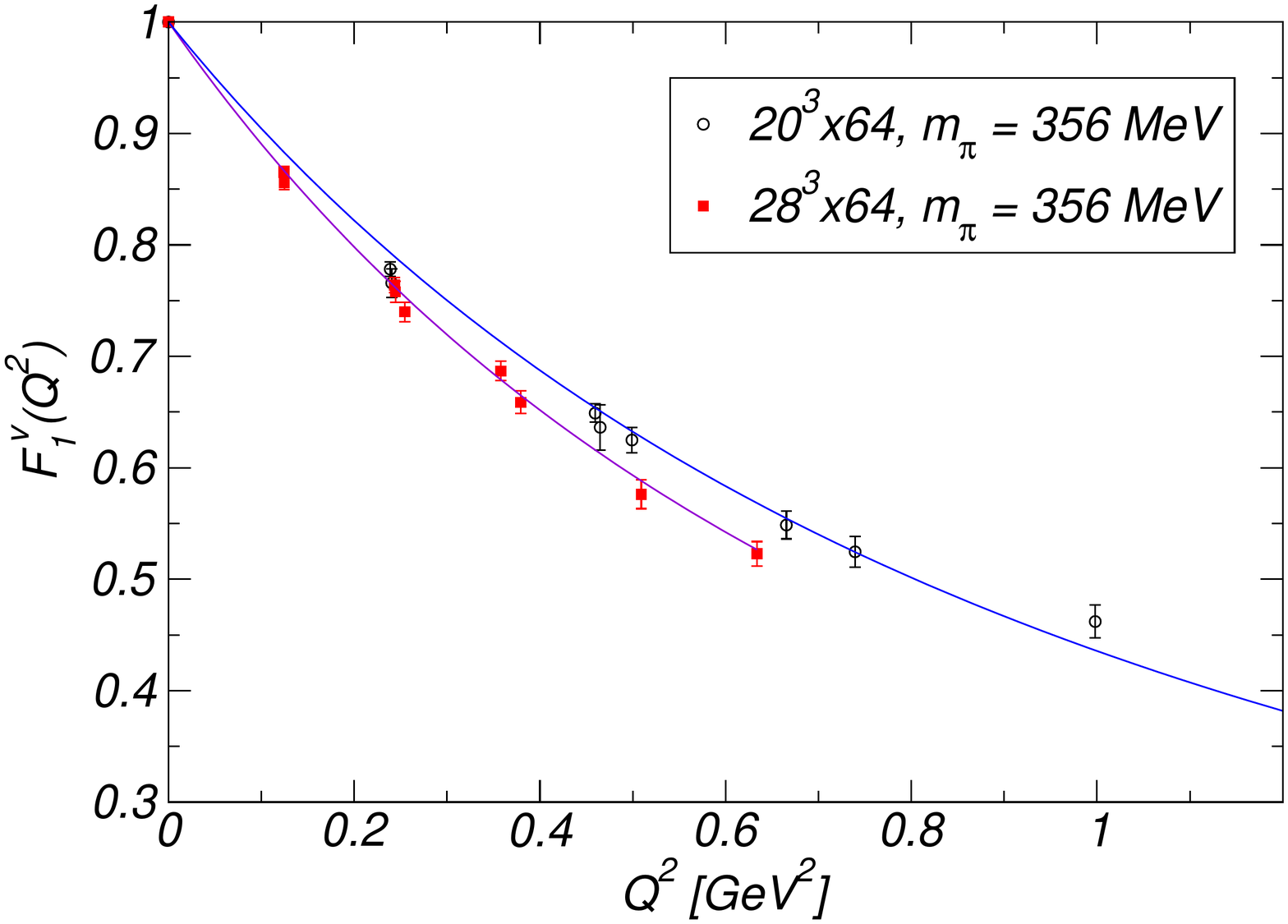} \qquad
  \includegraphics[scale=0.27,clip=true]{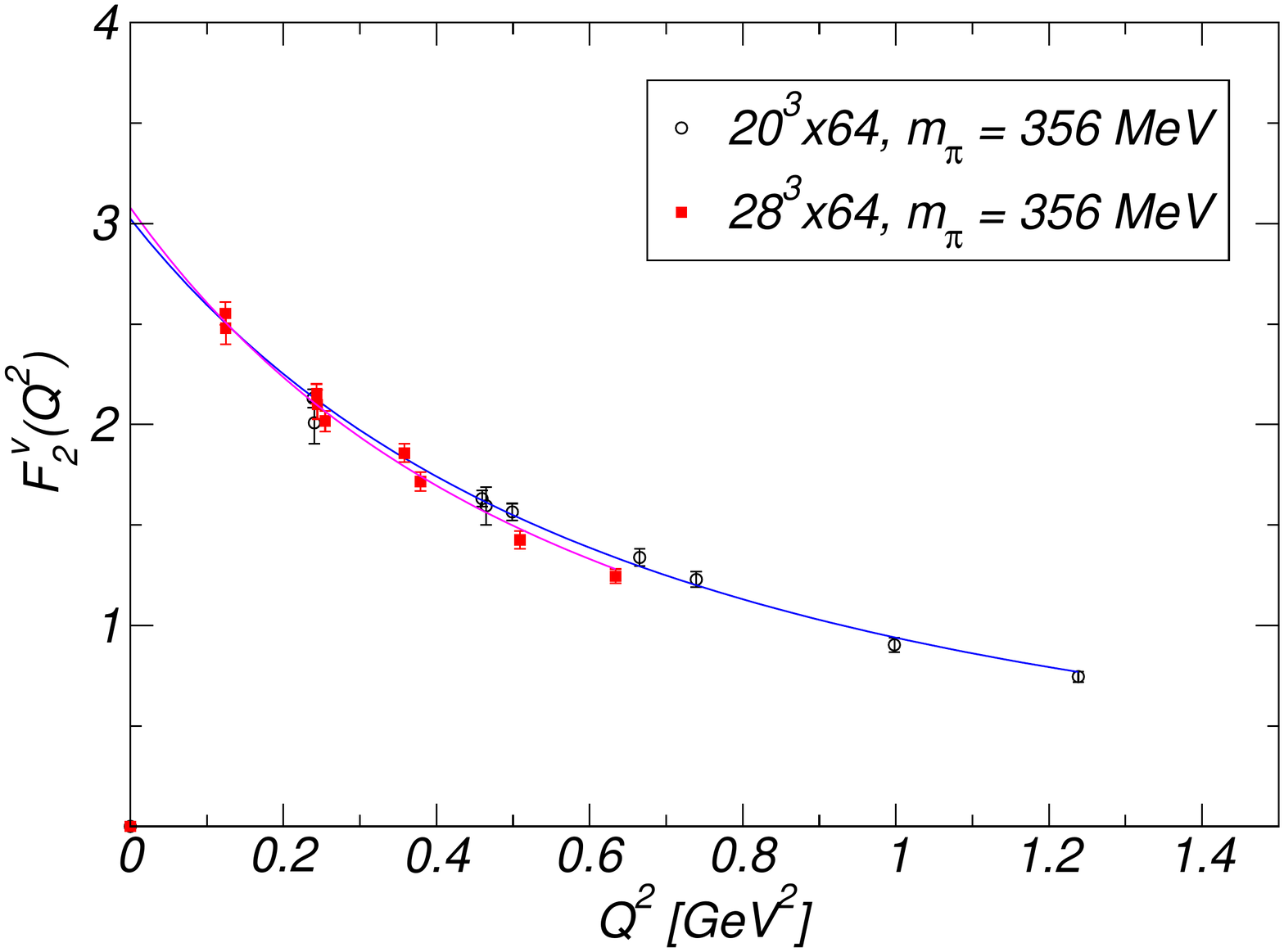}
  \caption{Direct comparison of finite-volume effects for the
    isovector form factors, $F_1^v(Q^2)$ and $F_2^v(Q^2)$, at
    $m_\pi=356$MeV on the two volumes $\Omega=28^3\times 64$ and
    $\Omega=20^3\times 64$.}
  \label{fig:f12v-finvol}
\end{figure}

We can also scrutinize this behavior by looking at the form factors as
a function of $Q^2$ directly and compare the location of the points on
the two volumes. Figure~\ref{fig:f12v-finvol} shows this comparison.
The solid curves are dipole fits to all available data points. It is
evident that their dipole masses are different, but the data points at
small $Q^2$ are identical. The difference in curvature is only caused
by points beyond that.

We conclude that we observe finite-size effects in our
lattice form-factor data. However, the finite-size effects are only
significant for intermediate values of $Q^2$. Based on the
smallest values of $Q^2$, the Dirac and Pauli radii may be identical
on both volumes. We did not find a satisfactory explanation based on
chiral expansions or models at this point and leave this matter for
future investigations.

As a final remark, in Sec.~\ref{sec:gener-form-fact} we will compare
the radii defined by the generalized form factors. There we will also
find suggestive evidence that the mass radius is larger on the larger
volume. Here our intention was to illustrate our observation of
finite-volume effects with the data that is most accurate.

%
%

\section{\label{sec:results}Results}

\subsection{\label{sec:gA}The axial charge}
In this section we present new data on the nucleon axial charge
$g_A$. Although axial form factors are discussed in more depth below
in \sect\ref{sec:axial-form-factors}, the fundamental phenomenological
importance of the axial charge warrants highlighting our new results
for this observable already at this point. Aside from the changes in
technology described in Sec.~\ref{sec:new-latt-calc}, our calculation
follows closely the methods of Ref.~\cite{Edwards:2005ym}. In
particular we use the local axial current for the calculation, and the
five-dimensional axial current is used to determine the normalization
of the local current, as described in detail at the beginning of
\sect\ref{sec:axial-form-factors}.

The new data are displayed in Fig.~\ref{fig:gA-SSE-600}. The value of
$g_A$ is remarkably independent of the pion mass, and lies at a value
$(8-10)\%$ lower than the experimental value of $1.2695(29)$, while
statistical errors are less than $2\%$.  A naive extrapolation linear
in $m_\pi^2$ of the $m_\pi<500$MeV data leads to
$g_A(m_\pi)=1.153(28)$.  We will discuss below what difference more
sophisticated chiral effective theory fits make.

It is worth describing to what extent the situation has changed since
the calculation~\cite{Edwards:2005ym}. In the latter, less accurate
calculation, the lattice data also showed a very mild pion-mass
dependence. Using a 3-parameter fit based on the leading one-loop pion
mass dependence in the Small Scale Expansion (SSE) at finite volume
leads to the value $g_A(m_\pi=140{\rm MeV})=1.226(84)$.  The
finite-volume effects predicted by the formula at the simulation
points were found to be negligible compared to the statistical errors.
The largest pion mass included in the fit was $760$MeV, and the
lightest $356$MeV.

Thanks to the new, higher statistics data, we control the
finite-volume effects to a higher level of accuracy.  Indeed, fitting
the $20^3$ and $28^3$ $g_A$ plateau at $am=0.010$, see
Fig.~\ref{fig:ga-plateau}, leads to the bound \be m_\pi=356{\rm MeV}:
\qquad | g_A(L=3.5{\rm fm}) - g_A(L=2.5{\rm fm}) | < 0.045 \qquad
(95\% {\rm~conf.~lev.}).  \la{eq:deltagA} \ee To further tighten this
statement, we want to constrain the possibility that the plateau for
$g_A$ could be affected by different excited state contributions on
the two volumes.  Indeed, even if the nucleon mass has a weak volume
dependence for $L\geq 2.5$fm, see Eq.~\eqref{eq:mbound}, the energy of
the first excited state in that symmetry channel could \emph{a priori}
have a significant volume dependence: in large volume we expect it to
be a nucleon and a pion with a non-vanishing relative momentum.
However, comparing the local effective mass on the $20^3$ and $28^3$
lattices, Fig.~\ref{fig:MNplateau2028}, we see good agreement between
them (the same nucleon interpolating operator is used on both
volumes).  We conclude that the contamination of the first excited
state does not increase significantly with the volume.  In particular,
the bound \eqref{eq:deltagA} on the $g_A$ finite-size effects is
robust.  While a non-monotonic volume-dependence of $g_A$ that would
make the difference in Eq.~\eqref{eq:deltagA} accidentally small
cannot be excluded, this bound strongly constrains how much of the
discrepancy between the lattice data and the experimental value of
$g_A$ can be attributed to finite-volume effects.
\begin{figure}[t]
  \centerline{\includegraphics[width=8.5cm,angle=0]{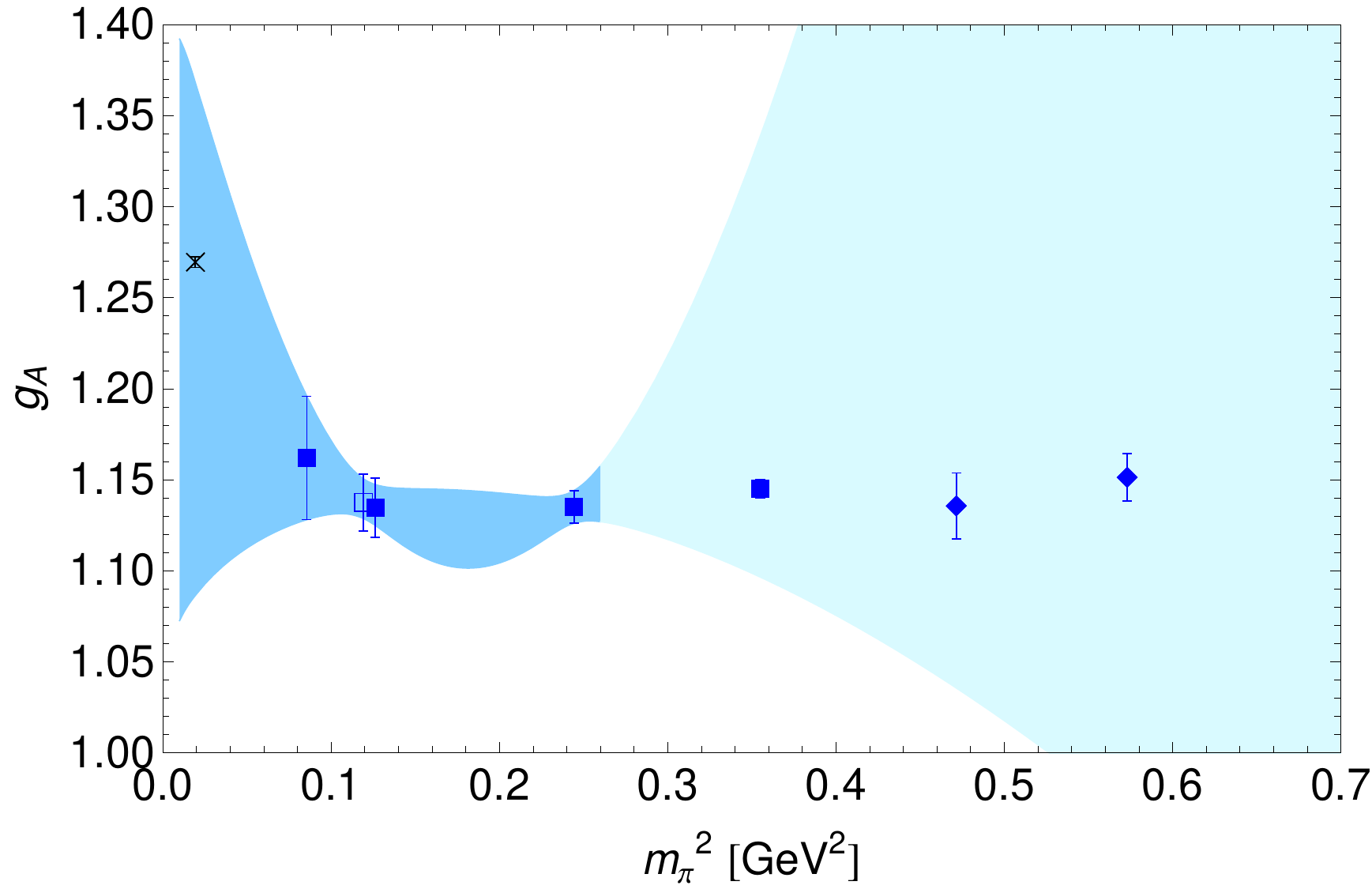}
    \qquad\includegraphics[width=8.5cm,angle=0]{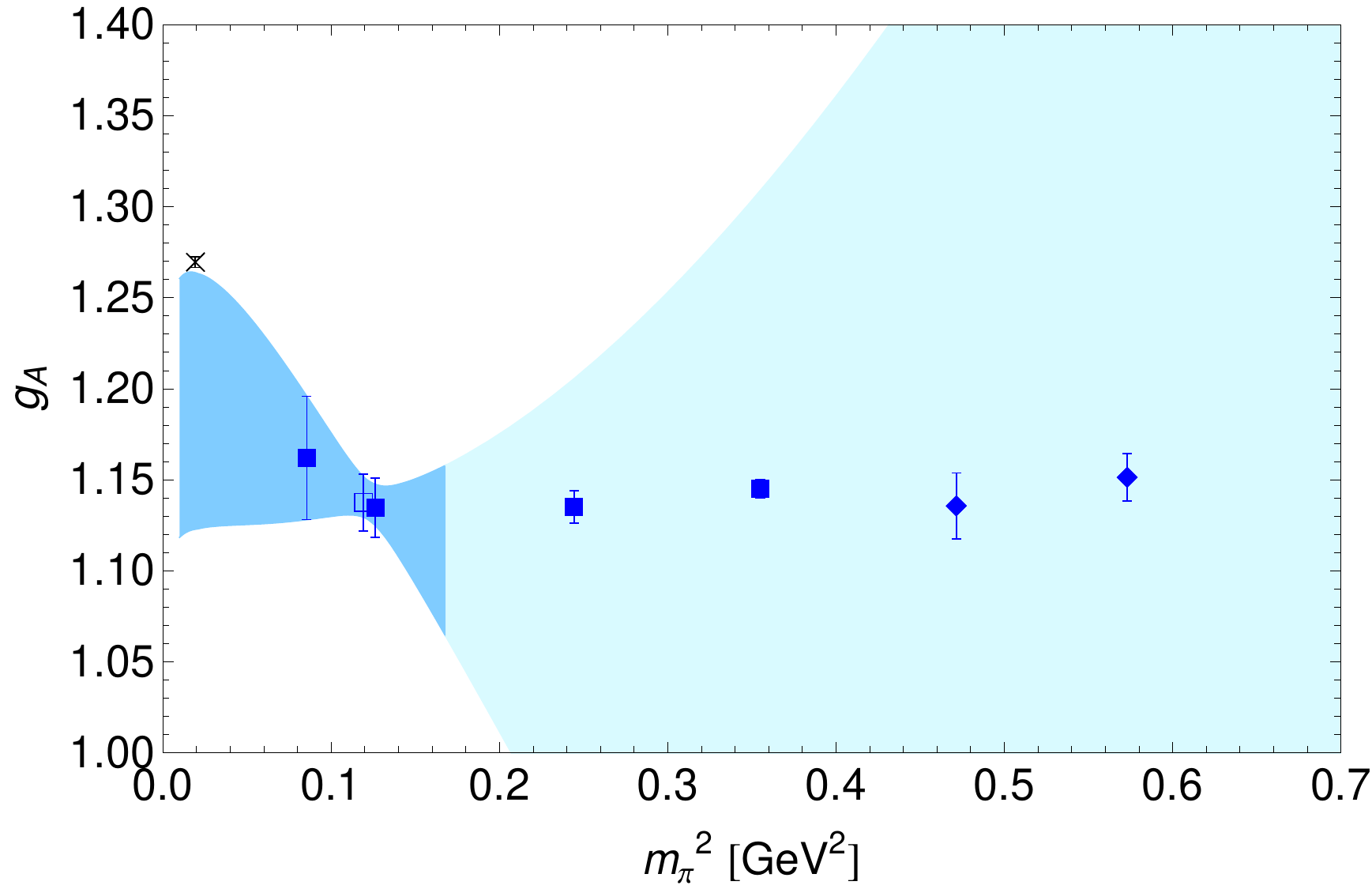}}
  \caption{SSE fit to the axial charge. Left: three-parameter fit with
    $m_\pi<500$MeV. Right: two-parameter fit with $m_\pi<360$MeV.}
  \la{fig:gA-SSE-600}
\end{figure}

As mentioned above, a naive extrapolation linear in $m_\pi^2$ of our
lattice data leads to values of $g_A(m_\pi)$ about $10\%$ lower than
the phenomenological value.  We now proceed with the fit ansatz
provided by the small-scale expansion (SSE)
framework~\cite{Hemmert:2003cb},
\begin{eqnarray}
  g_A(m_\pi) &=& g_A-\frac{g_A^3m_\pi^2}{16\pi^2 f_\pi^2}
  +4m_\pi^2\Big\{C(\lambda)
  +\frac{c_A^2}{4\pi^2f_\pi^2}[{\textstyle\frac{155}{972}}g_1-
  {\textstyle\frac{17}{36}}g_A]
  +\gamma\log\frac{m_\pi}{\lambda}\Big\} \nonumber
  \\ && \quad +\frac{4c_A^2g_A}{27\pi f_\pi^2\Delta}m_\pi^3 
  + \frac{8c_A^2g_Am_\pi^2}{27\pi^2f_\pi^2}
  [1-{\textstyle\frac{m_\pi^2}{\Delta^2}}]^{\frac{1}{2}}\log R(m_\pi)
  \nonumber \\ && \quad
  +\frac{c_A^2\Delta^2}{81\pi^2f_\pi^2}(25g_1 - 57g_A)
  \Big\{\log{\textstyle\frac{2\Delta}{m_\pi}}-
  [1-{\textstyle\frac{m_\pi^2}{\Delta^2}}]^{\frac{1}{2}}\log
  R(m_\pi)\Big\}\,,
  \label{eq:axial-expand}
\end{eqnarray}
with $g_1$ the axial-delta-delta coupling, $c_A$ the
axial-nucleon-delta coupling and $\Delta$ denoting the delta-nucleon
mass splitting, in the chiral limit. Following~\cite{Hemmert:2003cb},
we define the function
\begin{equation}
  R(m)=\frac{\Delta}{m}+\sqrt{\frac{\Delta^2}{m^2}-1} \,.
  \label{eq:capr-def}
\end{equation}
When the $\Delta$ baryon is below threshold, as is the case in our
lattice calculations, $\sqrt{\Delta^2-m_\pi^2}\log R(m_\pi)$ is
substituted by $-\sqrt{m_\pi^2-\Delta^2}{\rm arccos}(\Delta/m_\pi)$.
A three-parameter SSE fit to our data at pion masses below $500$MeV
with a fixed value of $c_A=1.5$ --- a motivation for this choice is
given in Ref.\cite{Syritsyn:2009mx} --- gives, see left panel of
Fig.~\ref{fig:gA-SSE-600},
\begin{equation}
  g_A^0 = 1.22(17),\qquad g_1=3.9(3.0)\,.
\end{equation}
The result thus does increase as the pion mass is lowered, but only
becomes consistent with the phenomenological value by virtue of its
uncertainty also rising significantly. As an alternative, we perform a
two-parameter fit to $m_\pi < 360$MeV, where we fix the value of $g_1$
to 2.5, close to the $SU(4)$ spin-flavor quark symmetry prediction
$9/5 g_A$. This fit is illustrated on the right panel of
Fig.~\ref{fig:gA-SSE-600}. Here the result for $g_A$ is slightly
lower than in the three-parameter fit, and the error bar barely
extends to the phenomenological value.

As has already been mentioned above, our previous calculation of $g_A$
in Ref.~\cite{Edwards:2005ym} included data points at larger pion
masses in the fit --- as a result of which it has a smaller
statistical uncertainty for the extrapolated value. Another
calculation in Ref.~\cite{Khan:2006de} contains only pion masses
larger than $500$MeV. A discussion of the range of validity of one
chiral expansion scheme in Ref.~\cite{Bernard:2006te} concludes that
lattice data below pion masses of $300$MeV are necessary for a
reliable prediction. Reference~\cite{Yamazaki:2008py} observes a
bending down of the extrapolation due to the data point at the
smallest available pion mass in that calculation, $m_\pi=331$MeV. This
is above our smallest mass. However, our smallest pion mass data point
has a larger error bar and is just consistent with the one from
Ref.~\cite{Yamazaki:2008py}; it may be the case that at this parameter
the data is already affected by finite-volume effects -- a possibility
also mentioned in that paper. This interpretation is supported by the
observation that the other data points at larger pion masses tend to
be systematically higher than our data points. However, at our data
point at $m_\pi=356$MeV we do not find any evidence of finite-volume
effects which indicates that at lighter pion mass these effects would
have to set in rather quickly.

While in the present work we find no significant evidence for a pion
mass dependence of $g_A$, our data is simultaneously compatible with
the possibility that the functional form predicted by the small-scale
expansion applies below $\mps=350$MeV and with the phenomenological
value of $g_A$.

\subsection{\label{sec:form-factors}Electromagnetic form factors}
The matrix element of the electromagnetic current between nucleon
states can be parameterized in terms of two form factors. Common
choices are the Dirac and Pauli form factors, $F_1(Q^2)$ and
$F_2(Q^2)$, and the electric and magnetic Sachs form factors,
$G_E(Q^2)$ and $G_M(Q^2)$. The former directly correspond to the form
factors $A_{10}(Q^2)$ and $B_{10}(Q^2)$ from
Eq.~\eqref{eq:lattice-vec-gff}.  The latter are related by a simple
linear transformation to the isovector Dirac and Pauli form factors,
$F_1^v(Q^2)$ and $F_2^v(Q^2)$:
\begin{eqnarray}
  \label{eq:sachs-ff-def}
  G_E(Q^2) &=& F_1^v(Q^2) - \frac{Q^2}{(2 m_N)^2} F_2^v(Q^2)\, \\
  G_M(Q^2) &=& F_1^v(Q^2) + F_2^v(Q^2)\,.
\end{eqnarray}
We will also use the standard notation for the anomalous magnetic
moment of the nucleon in units of $e/2m_N(m_\pi)$,
\begin{equation}
  \kappa_v = F_2^v(0)\,.
\end{equation}
When performing chiral fits we will work with 
\begin{equation}
  \label{eq:kappanorm-def}
  \kappa_v^{\rm norm} = \frac{m_N^{\rm phys}}{m_N(m_\pi)} \kappa_v\,,
\end{equation}
which represents the isovector anomalous magnetic moment in units of
the physical Bohr magneton, $e/2m_N^{\rm phys}$.

We use the ultra-local discretizations of the dimension three quark
bilinear operators, i.e.~their support is a single lattice site. Due
to quantum effects the matrix elements of these lattice operators are
not trivially renormalized, and we have to apply renormalization
constants to them. Since the forward matrix element $\langle
p,\lambda\vert \bar{\psi}\gamma_\mu\psi\vert p,\lambda\rangle$ counts
the total number of quarks of type $\psi$ and this number is known by
construction, we obtain $Z_V$ by dividing the unrenormalized isovector
current in the forward case. We point out that in the forward case the
disconnected contribution is exactly zero since the disconnected
operator cannot change the total number of quarks of any type. Thus,
the value for $Z_V$ obtained this way will be exact also if we
consider disconnected contributions in future work. The resulting
renormalization constants $Z_V$ for the vector current are listed in
Tab.~\ref{tab:renorm-const}. The renormalization constants of the
axial current are discussed later in
\sect\ref{sec:axial-form-factors}.
\begin{table}[htb]
  \centering
  \begin{tabular}[c]{*{2}{c|}c}
    \hline\hline
    $m_{\mbox{\tiny sea}}^{\mbox{\tiny asqtad}}$ & Volume & $Z_V$ \\
    \hline
    0.007/0.050 & $20^3\times 64$ & 1.1159 \\
    0.010/0.050 & $28^3\times 64$ & 1.1169 \\
    0.010/0.050 & $20^3\times 64$ & 1.1206 \\
    0.020/0.050 & $20^3\times 64$ & 1.1351 \\
    0.030/0.050 & $20^3\times 64$ & 1.1464 \\
    \hline\hline
  \end{tabular}
  \caption{Renormalization constant of the vector currents.}
  \label{tab:renorm-const}
\end{table}

To study the charge distribution of the nucleon at large distances, it
makes sense to consider the leading contribution of the form factors
at small values of $Q^2$~\cite{LHPC:2003is}. The linear coefficient of
the small-$Q^2$ expansion can serve as a measure of the nucleon size
and is known as the mean squared radius, $\langle r^2_i\rangle$, where
$i$ labels the different Lorentz and flavor structures one may
consider:
\begin{equation}
  \label{eq:rms-radius-def}
  F_i(Q^2) = F_i(0) \left( 1 - \frac{1}{6}Q^2\cdot \langle
    r^2_i\rangle + {\cal O}(Q^4)\right)\,.
\end{equation}
The radii, $\langle r^2_i\rangle$, can also be extracted from
experiment. For a recent review see Ref.~\cite{Perdrisat:2006hj}.
Although this is straightforward for the proton isovector $F_1^v(Q^2)$
form factor, a determination from fits to the
experiment~\cite{Friedrich:2003iz, Arrington:2007ux} turns out to be
inconsistent with an analysis based on dispersion
theory~\cite{Hohler:1976ax, Mergell:1995bf, Belushkin:2006qa}. The
latter radii are systematically larger than the former. To resolve
this discrepancy, a dedicated experiment is currently being
performed~\cite{Bernauer:2008zz}. For the proton isovector
$F_2^v(Q^2)$ a different discrepancy has been found in recent
spin-transfer measurements~\cite{Milbrath:1997de, Pospischil:2001pp,
  Gayou:2001qd, Gayou:2001qt, Punjabi:2005wq}. The source of this
mismatch is generally believed to be two-photon exchange
processes~\cite{Arrington:2007ux}, which is challenging to verify. On
the lattice, we can study these observables without any two-photon
contamination and thus make a significant contribution towards
resolving the discrepancy.

This chapter discusses the results for the form factors of the
electromagnetic current. First, we study the isovector Dirac form
factor $F_1^v(Q^2)$ in \sect\ref{sec:isovector-dirac-form} and the
isovector Pauli form factor $F_2^v(Q^2)$ in
\sect\ref{sec:isovector-pauli-form}. The scaling behavior of form
factors at larger values of $Q^2$ is shown in
\sect\ref{sec:asymptotic-scaling}. Section~\ref{sec:isovector-sachs-form}
discusses the Sachs parameterization of form factors.
Section~\ref{sec:flavor-dependence} discusses the slope of the ratio
$F_1^d/F_1^u(Q^2)$ to learn about the flavor dependence of the form
factors. The isoscalar form factors are shown in
\sect\ref{sec:isosc-form-fact}. Section~\ref{sec:summ-electr-form}
summarizes our findings. Where applicable, we compare the chirally
extrapolated results to experiment.

\subsubsection{\label{sec:isovector-dirac-form}Isovector Dirac form
  factor $F_1^v(Q^2)$}
This section covers the isovector Dirac form factor,
$F_1^v(Q^2)$. Phenomenologically, this form factor is commonly fit
using a dipole form at fixed pion mass. We will thus first attempt to
fit $F_1^v(Q^2)$ using the dipole form and study the stability of this
fit as a function of the $Q^2$ range. Next, we will perform chiral
fits using the small scale expansion (SSE), Ref~\cite{Hemmert:1997ye},
which includes explicit $\Delta\,(1232)$ degrees of
freedom~\cite{Gockeler:2003ay, Bernard:1998gv}. We will first compare
the expansion applied to the Dirac radii, $\langle r_1^2\rangle$,
obtained from the previous dipole fits. We will then study the
covariant baryon chiral perturbation theory expansion (BChPT) for the
same quantity, see Ref.~\cite{Gail:2009ph, Syritsyn:2009mx}.

Finally, we will present SSE fits to the simultaneous $Q^2$ and $\mps$
dependence of our lattice data. The latter method has the strong
advantage that no reliance on the applicability of the dipole form is
assumed. For these fits we apply the super jackknife and error
correlation matrix methods discussed in
\sect\ref{sec:super-jackkn-analys}. Thus, we believe that this fit
strategy is superior to the ones previously employed.

\paragraph{Dipole fits to isovector $F_1^v(Q^2)$}
In this section we discuss the $Q^2$-dependence of the form factors at
fixed values of the pion mass, $\mps$. The function we will use
throughout this section is the dipole formula,
\begin{equation}
  \label{eq:dipole}
  F_1^v(Q^2) = A_0/(1+Q^2/M_d^2)^2\,,
\end{equation}
with $A_0$ fixing the overall normalization and $M_d$ being the dipole
mass. From Eqs.~\eqref{eq:dipole} and \eqref{eq:rms-radius-def} it is
immediately obvious that the dipole mass is related to the Dirac
radius of $F_1^v(Q^2)$ via
\begin{equation}
  \label{eq:dipole-mass-rms}
  \langle r_1^2\rangle = \frac{12}{M_d^2}\,.
\end{equation}
In order to verify if the functional form indeed allows for a
meaningful application of the dipole formula, we have performed a
series of fits in which we varied the fit interval $[Q^2_{\mbox{\tiny
    min}}, Q^2_{\mbox{\tiny max}}]$ and listed the variation of the
fit parameters. We have restricted ourselves to the $28^3$ lattice
with pion mass $\mps=356$MeV. Results are summarized in
Tab.~\ref{tab:f1v-dip-vary-t}. The table shows the fit interval used,
the resulting value of $\chi^2$/dof (degrees of freedom), the
normalization $A_0$ --- which must be equal to one within precision
due to the conservation of the vector current --- and the dipole mass,
$M_d$, together with the resulting Dirac radius, $\langle
r_1^2\rangle$. All error estimates have been obtained by applying the
Jackknife method to the minimization of the $\chi^2$ including the
error correlation matrix, as discussed in
\sect\ref{sec:super-jackkn-analys}.
\begin{table}[htb]
  \centering
  \begin{tabular}[c]{*{4}{c|}c}
    \hline\hline
    $[Q^2_{\mbox{\tiny min}}, Q^2_{\mbox{\tiny max}}]$ [GeV$^2$] &
    $\chi^2$/dof & $A_0$ & $M_d$ [GeV] & $\langle r_1^2\rangle$
    [fm$^2$] \\ \hline
    $[0,1.5]$   & 1.22 & 1.0004(20) & 1.299(21) & 0.2770(92) \\
    $[0,0.5]$   & 1.09 & 9.9931(21) & 1.298(23) & 0.2774(97) \\
    $[0,0.4]$   & 1.27 & 9.9956(22) & 1.293(23) & 0.2797(99) \\
    $[0,0.3]$   & 1.52 & 9.9932(22) & 1.287(22) & 0.2820(98) \\
    $[0,0.2]$   & 2.43 & 1.0006(23) & 1.276(24) & 0.2870(11) \\
    \hline
    $[0,1.5]$   & 1.22 & 1.0004(20) & 1.299(21) & 0.2770(92) \\
    $[0.1,1.5]$ & 1.23 & 0.9965(42) & 1.305(22) & 0.2744(93) \\
    $[0.2,1.5]$ & 0.72 & 0.9780(80) & 1.338(25) & 0.2611(97) \\
    $[0.3,1.5]$ & 0.90 & 0.9765(148) & 1.339(31) & 0.2605(120) \\
    $[0,4,1.5]$ & 1.21 & 0.9445(354) & 1.384(59) & 0.2440(207) \\
    $[0.5,1.5]$ & 1.41 & 0.9469(692) & 1.378(105) & 0.2462(376) \\
    \hline
    $[0.3,0.5]$ & 0.65 & 1.0098(499) & 1.288(85) & 0.2819(370) \\
    $[0.2,0.4]$ & 0.77 & 0.9749(117) & 1.351(38) & 0.2558(146) \\
    $[0.1,0.3]$ & 1.59 & 0.9948(47) & 1.297(24) & 0.2778(104) \\
    \hline\hline
  \end{tabular}
  \caption{Dipole fits to isovector $F_1^v(Q^2)$ with varying fit
    intervals on the $28^3$ lattice with $\mps=356$MeV.}
  \label{tab:f1v-dip-vary-t}
\end{table}
The table is divided into three blocks --- first, the large-$Q^2$
cut-off is varied, next the small-$Q^2$ cut-off is varied and finally
the fit-interval window is moved along the available data set. Note
that when one leaves out the small $Q^2$ values, the data point
$F_1^v(0)$ is no longer included in the fit interval and $A_0$ can
vary more.

The overall conclusion is that $A_0$ is always compatible with one
within error bars and all results for $M_d$ are consistent over the
entire table. The former is an important internal consistency check
and the latter allows us to conclude that the dipole function is
indeed an excellent description of the $F_1^v(Q^2)$ form factor over
the entire range of available $Q^2$ values.

After performing similar fits at all available pion masses, we
obtained the numbers compiled in Tab.~\ref{tab:f1v-dip-allmpi}. We
have taken all available $Q^2$ values for each fit at fixed pion
mass. Again, we have performed a combined error analysis with error
correlation matrix and jackknife.
\begin{table}[htb]
  \centering
  \begin{tabular}[c]{*{4}{c|}c}
    \hline\hline
    $\mps$ [MeV] & $\chi^2$/dof & $A_0$ & $M_d$ [GeV] & $\langle
    r_1^2\rangle$ [fm$^2$] \\ \hline
    293 & 0.89 & 0.9983(73) & 1.307(35) & 0.2734(147) \\
    356 on $28^3$ & 1.22 & 1.0004(20) & 1.299(21) & 0.2770(92) \\
    356 on $20^3$ & 1.95 & 0.9999(15) & 1.382(20) & 0.2447(70) \\
    495 & 1.57 & 0.9994(9) & 1.3829(12) & 0.2444(42) \\
    597 & 3.76 & 0.9998(5) & 1.4144(8)  & 0.2336(27) \\
    \hline\hline
  \end{tabular}
  \caption{Dipole fits to isovector $F_1^v(Q^2)$ for all data sets.}
  \label{tab:f1v-dip-allmpi}
\end{table}
We find that the resulting values for the Dirac radii, $\langle
r_1^2\rangle$ are systematically smaller than the experimental value,
$\langle r_1^2\rangle^{\mbox{\tiny exp}}=0.637(12)$fm$^2$ from
Ref.~\cite{Amsler:2008zzb}. We discuss possible resolutions of this
discrepancy in the following by discussing chiral fits to our lattice
data.

\begin{figure}[htb]
  \centering
  \includegraphics[scale=0.3, clip=true]{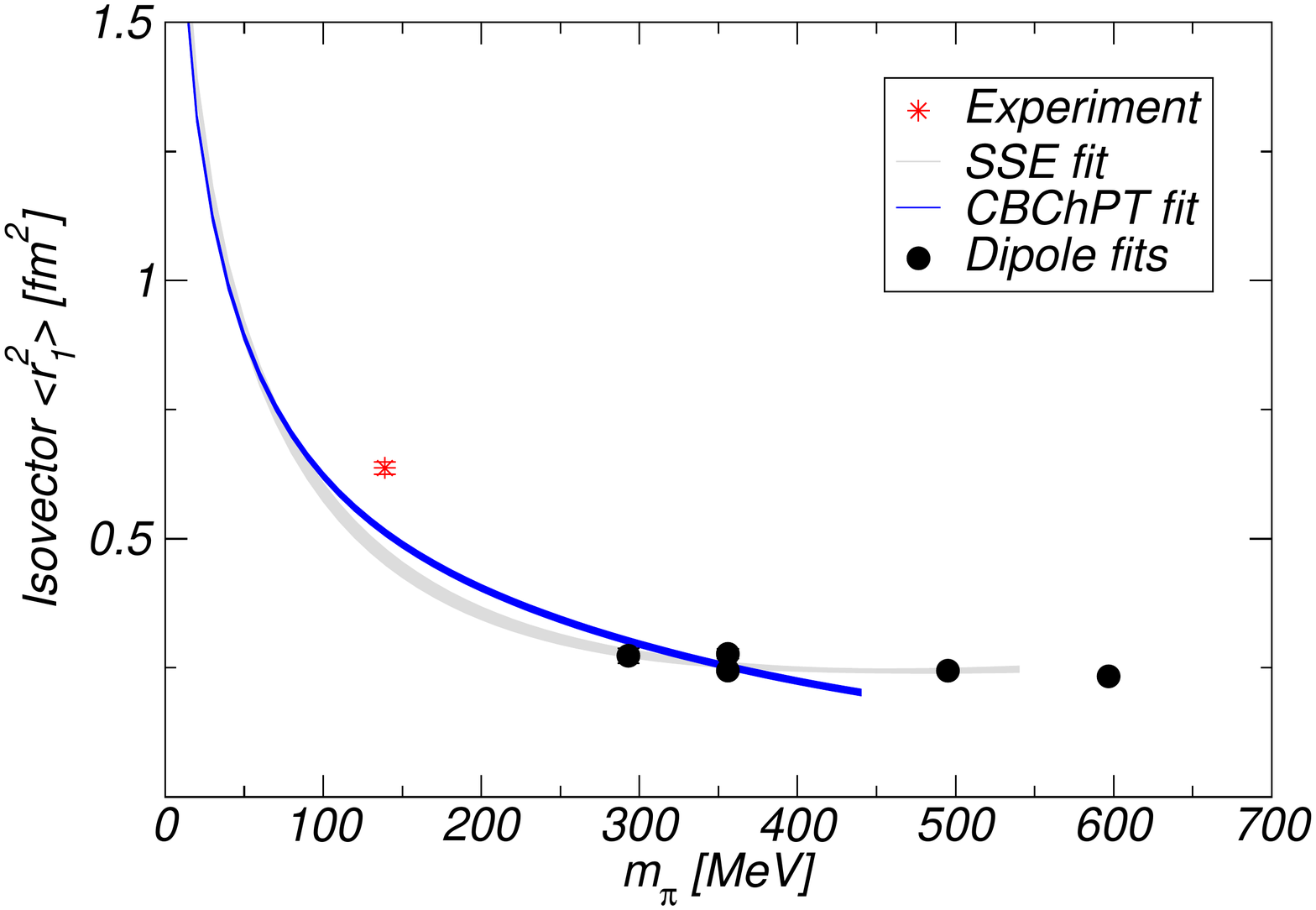}
  \caption{Chiral fits to isovector Dirac radii, $\langle
    r_1^2\rangle$, for $\mps<500$MeV (SSE) and $\mps<400$MeV (BChPT).}
  \label{fig:r1v-varmpi}
\end{figure}

\paragraph{SSE fits to isovector $\langle r_1^2\rangle$}
As we have seen, the Dirac radius --- and, consequently, the size of
the nucleon --- is lower than experiment at the pion masses we use.
Hence, for a meaningful comparison the Dirac radii $\langle
r_1^2\rangle$ obtained in Tab.~\ref{tab:f1v-dip-allmpi} need to be
extrapolated as a function of the pion mass, $\mps$. In this chapter
we perform the chiral extrapolation using the small-scale expansion
(SSE). The pion mass dependence to next-to-leading order (NLO) is
given by:
\begin{eqnarray}
  \label{eq:r1v-sse}
  \langle r_1^2\rangle &=& -\frac{1}{(4\pi f_\pi)^2} \left( 1+ 7g_A^2
    + (10g_A^2+2) \log\left( \frac{\mps}{\lambda}\right) \right)
  \nonumber \\
  && \quad - \frac{12 B_{10}^r(\lambda)}{(4\pi f_\pi)^2} +
  \frac{c_A^2}{54\pi^2 f_\pi^2} \left(
    26+30\log\left(\frac{\mps}{\lambda}\right) + 30 
    \frac{\Delta} {\sqrt{\Delta^2-\mps^2}}
    \log\left(\frac{\Delta}{\mps} +
      \sqrt{ \frac{\Delta^2}{\mps^2}-1}\right) \right)\,.
\end{eqnarray}
This expansion has a logarithmic divergence at $\mps\to 0$. Since the
proton and the neutron are linear combinations of the isovector and
isoscalar operators, the radius $\langle r_1^2\rangle$ of either the
proton or the neutron will similarly diverge in the chiral limit. In
our fits we fix some of the parameters involved in the chiral
expressions, see Tab.~\ref{tab:sse-pars}. The axial coupling $g_A$ in
the chiral limit has been set equal to $1.2$ according to
Refs.~\cite{Edwards:2005ym, Hemmert:2003cb, Procura:2006gq}. The
chiral limit values of $\fps$ and the nucleon mass have been
determined in Ref.~\cite{Colangelo:2003hf} and
Refs.~\cite{WalkerLoud:2008bp, Procura:2003ig, Procura:2006bj},
respectively. The delta-nucleon mass splitting $\Delta$ is taken equal
to its physical value from the position of the delta resonance
pole. Without any loss of generality, we set the regularization scale
$\lambda$ equal to $1$GeV.
\begin{table}[htb]
  \centering
  \begin{tabular}[c]{cc}
    \hline\hline
    Parameter & Value \\ \hline
    $g_A$ & 1.2 \\
    $c_A$ & 1.5 or free parameter \\
    $f_\pi$ [MeV] & 86.2 \\
    $m_N$ [GeV] & 0.8900 \\
    $\Delta$ [GeV] & 0.2711 \\
    $g_1$           & 2.5   \\
    \hline\hline
  \end{tabular}
  \caption{Input parameters used for the chiral expansion.}
  \label{tab:sse-pars}
\end{table}
As input for our fits of $\langle r_1^2\rangle$ we take the Dirac
radii obtained from the dipole fits in
Tab.~\ref{tab:f1v-dip-allmpi}. We perform a two parameter fit with the
counterterm $B_{10}^r(\lambda=1{\rm GeV})$ and the coupling $c_A$. We
vary the upper cut-off in $\mps$ and collect the resulting fits in
Tab.~\ref{tab:r1v-varmpi-sse}. The quality of the fits is not so good;
we will see in \fig\ref{fig:r1v-varmpi} that the curvature of the SSE
curve is stronger than that of the lattice data. Furthermore, the
value of $c_A$ tends to be larger than $1.5$. One interpretation is
that the range of validity of the SSE does not extend to our lattice
data. On the other hand, if the smallest data point suffers from
finite-size effects, it is still possible that the SSE is consistent
with lattice data on a very large volume.
\begin{table}[htb]
  \centering
  \begin{tabular}[c]{*{3}{c|}c}
    \hline\hline
    $\mps$ max [MeV] & $\chi^2$/dof & $c_A$ & $B_{10}^r(1{\rm GeV})$ \\
    \hline
    500 & 8.0 & 1.951(36) & 1.713(78) \\
    600 & 7.2 & 1.873(20) & 1.557(48) \\
    \hline\hline
  \end{tabular}
  \caption{NLO SSE fits to the isovector Dirac radii,
    $\langle r_1^2\rangle$, obtained from dipole fits.}
  \label{tab:r1v-varmpi-sse}
\end{table}

\paragraph{BChPT fits to isovector $\langle r_1^2\rangle$}
In the covariant baryon chiral perturbation theory (BChPT) scheme we
use, the $\Delta\,(1232)$ degrees of freedom are not explicitly
included while recoil corrections to the non-relativistic Heavy Baryon
results which correspond to kinetic insertions in the nucleon
propagator are systematically resummed~\cite{Becher:1999he}. The
chiral expansion of the $\langle r_1^2\rangle$ Dirac radius, is given
by~\cite{Gail:2009ph}:
\begin{equation}
  \label{eq:r1v-hbchpt}
  \langle r_1^2\rangle = B_{c1} + (\langle r_1^2\rangle)^{(3)} +
  (\langle r_1^2\rangle)^{(4)} + {\cal O}(p^6)\,,
\end{equation}
The notation $(\langle\cdot\rangle)^{(M)}$ denotes the contribution of
a quantity to the $M$th order in the expansion in $p^M$. The specific
expressions are
\begin{eqnarray}
  \label{eq:r1v-hbchpt-allorder}
  B_{c1} & = & -12d_6^r(\lambda)\,, \nonumber \\
  (\langle r_1^2\rangle)^{(3)} & = & -\frac{1}{16\pi^2 f_\pi^2 M_N^4}
  \Bigl( 7g_A^2M_N^4 + 2(5g_A^2+1)M_N^4\log\frac{\mps}{\lambda} +
  M_N^4 \nonumber \\ && \qquad\qquad -
  15g_A^2 \mps^2 M_N^2 + g_A^2\mps^2(15\mps^2 -
  44M_N^2)\log\frac{\mps}{M_N} \Bigr) \nonumber \\
  && \quad + \frac{g_A^2 \mps}{16\pi^2 f_\pi^2 M_N^4 \sqrt{4M_N^2
      -\mps^2}} \left(15 \mps^4 -74\mps^2M_N^2+70M_N^4\right)
  \arccos\frac{\mps}{2M_N}\,, \nonumber \\
  (\langle r_1^2\rangle)^{(4)} & = & -\frac{3c_6g_A^2\mps^2}
  {16\pi^2f_\pi^2 M_0^4\sqrt{4M_0^2-\mps^2}} \Bigl[ \mps (\mps^2 -
  3M_0^2) \arccos\frac{\mps}{2M_0} \nonumber \\
  && \qquad\qquad\qquad\qquad\qquad +
  \sqrt{4M_0^2-\mps^2}\left(M_0^2  +
    (M_0^2-\mps^2)\log\frac{\mps}{M_0}\right) \Bigr]\,.
\end{eqnarray}
The expression up to order ${\cal O}(p^4)$ introduces two fit
parameters, $d_6^r(\lambda)$ and $c_6$. We set the scale of
dimensional regularization $\lambda$ equal to the value of the nucleon
mass in the chiral limit, $M_0$. We point out that when only the third
order in expansion~\eqref{eq:r1v-hbchpt} is considered, one should
replace $M_N$ by $M_0$ in Eqs.~\eqref{eq:r1v-hbchpt-allorder}. In our
fits, however, we only consider the full expansion at order ${\cal
  O}(p^4)$. Consequently, we are always using a pion-mass dependent
form, $M_N(\mps)$, for $M_N$. Similar to Ref.~\cite{Syritsyn:2009mx}
we use the expansion from Ref.~\cite{Dorati:2007bk} to model the
functional form $M_N(\mps)$:
\begin{eqnarray}
  \label{eq:mn-expand}
  M_N(\mps) &=& M_0 - 4 c_1 \mps^2 \nonumber \\
  && \quad + \frac{3g_A^2 \mps^3}{32\pi^2
    f_\pi^2\sqrt{4-\mps^2/M_0^2}}\left( -4+\frac{\mps^2}{M_0^2} +4c_1
    \frac{\mps^4}{M_0^3}\right) \arccos\frac{\mps}{2M_0} \nonumber \\
  && \quad -\frac{3\mps^4}{128\pi^2f_\pi^2}\left(\left(
      \frac{6g_A^2}{M_0} -c_2\right) + 4\left(\frac{g_A^2}{M_0} -8c_1
      + c_2 + 4c_3\right) \log\frac{\mps}{\lambda}\right) \nonumber \\
  && \quad +4e_1^r(\lambda)\mps^4 -
  \frac{3c_1g_A^2\mps^6}{8\pi^2f_\pi^2 M_0^2}\log\frac{\mps}{M_0}\,,
\end{eqnarray}
with $c_1$, $c_2$, $c_3$, and $e_1^r(\lambda)$ being parameters that
need to be fixed. In order to fix these constants we follow
Ref.~\cite{Syritsyn:2009mx} again and pick the parameters $c_2$,
$c_3$, and $c_4$ from the literature, cf.~Refs.~\cite{Bernard:1996gq,
  Fettes:1998ud, Entem:2002sf, Procura:2006bj, Khan:2003cu}. For $M_0$
we could adopt the chiral value $m_N$ listed in
Tab.~\ref{tab:sse-pars} which we have used for the SSE\@. However,
since the functional dependence on $M_0$ is highly complicated and all
expressions can be sensitive to small changes in $M_0$, we decided to
consider it a free parameter and determine both $M_0$ and the
remaining parameters $c_1$ and $e_1^r(\lambda)$ from a fit to the data
points listed in Tab.~\ref{tab:phys-pars} with pion masses
$\mps<500$MeV. The physical nucleon mass is included in the fit. The
resulting $\chi^2$/dof=2.93 is acceptable and the parameters are in
agreement with those reported in~\cite{Syritsyn:2009mx}. If we include
all nucleon masses up to $\mps<600$MeV, the $\chi^2$/dof=4.15
increases slightly. Since we do not know the exact range of validity,
we adopt the more conservative choice $\mps<500$MeV. When the
experimental point is not included, the uncertainty of $M_0$ increases
dramatically and becomes inconsistent with
Ref.~\cite{Syritsyn:2009mx}. The results are summarized in
Tab.~\ref{tab:mn-expand-pars}.  As an alternative to
Eq.~\eqref{eq:mn-expand} we may use an expression from
Ref.~\cite{WalkerLoud:2008bp}. It turns out, however, that the
influence of different expansions of $M_N(\mps)$ on the radii is quite
small.
\begin{table}[htb]
  \centering
  \begin{tabular}{*{5}{c|}c}
    \hline\hline
    $M_0$ [MeV] & $e_1^r(\lambda=1{\rm GeV})$ [GeV$^{-3}$] & $c_1$
    [GeV$^{-1}$] & $c_2$ [GeV$^{-1}$] &  $c_3$ [GeV$^{-1}$] &  $c_4$
    [GeV$^{-1}$] \\
    \hline
    876.6(17) & 1.27(12) & -0.983(22) & 3.2 & -3.4 & 3.5 \\
    \hline\hline
  \end{tabular}
  \caption{Low energy constants involved in the chiral expansion of
    the nucleon mass up to NNLO\@. $M_0$, $e_1^r(1{\rm GeV})$,
    and $c_1$ are determined from a fit to lattice data and
    experiment.}
  \label{tab:mn-expand-pars}
\end{table}

There is one more subtlety that prevents a straightforward BChPT fit
in the way it was possible for the SSE: The parameter $c_6$ in
Eq.~\eqref{eq:r1v-hbchpt-allorder} appears only at ${\cal
  O}(p^4)$. The physical meaning of this low-energy constant is that
of the chiral limit of the isovector anomalous magnetic moment. In
order to gain statistics and better constrain our free parameters, we
follow the prescriptions devised in Ref.~\cite{Syritsyn:2009mx} and
perform a simultaneous fit to $\langle r_1^2\rangle$, $\langle
r_2^2\rangle$, and $\kappa_v$ in the BChPT scheme. This is done later
in the BChPT part of \sect\ref{sec:isovector-pauli-form} and the
results are summarized in Tab.~\ref{tab:hbchpt-r1r2kv-varmpi}. Note
that we restrict the joint fit to pion masses $\mps<400$MeV. The
reason is that both the $\langle r_1^2\rangle$ and the $\langle
r_2^2\rangle$ have a stronger curvature at larger pion masses than our
lattice data and the $\chi^2$/dof would be unacceptably large when
pion masses beyond $400$MeV are included. We cannot exclude that the
range of validity of the BChPT is smaller than that of the
SSE\@. Figure~\ref{fig:r1v-varmpi} displays the results from the fits
in Tabs.~\ref{tab:r1v-varmpi-sse} and~\ref{tab:hbchpt-r1r2kv-varmpi}
with $\mps<500$MeV (for the SSE) and $\mps<400$MeV (for the BChPT)
graphically. The experimental value in the graph is taken from
Ref.~\cite{Amsler:2008zzb}.
Both curves are quite similar near the physical point. However, the
lattice data is still quite flat for the available pion masses. Our
lattice results do not give any indication of the divergence we should
see approaching the chiral limit. We also remark that we observe a
finite-volume dependence at $\mps=356$MeV of almost $12\%$ if we
perform dipole fits to all available $Q^2$, as discussed in
\sect\ref{sec:finite-volume-effect}. This effect could potentially
increase rapidly at lower pion masses, and therefore the smallest
$\mps=293$MeV data point could be too low. Thus, it is not excluded
that the fit works well below $400$MeV if one had data at larger
volume. At this point we observe that the effect goes in the right
direction and is qualitatively consistent. For a conclusive statement,
we still require more data at smaller pion masses and at larger
volumes.

\paragraph{Simultaneous expansion fit to isovector $F_1^v(Q^2,\mps)$}
The dipole fits to $F_1^v(Q^2)$ discussed previously indeed provide an
excellent description of the data over the entire range of available
data. However, applying dipole fits suffers from the disadvantage that
there is little fundamental justification for their usage. They are
simply employed because they appear to work well. As an alternative
strategy, we consider now the simultaneous chiral expansions in both
$\mps$ and $Q^2$. These expansions are only expected to hold for small
values of $Q^2$ and $ m_\pi$, but we can apply them to our entire data
set, i.e.~by combining data from several different ensembles at
different $\mps$ values in the same chiral fit without model-dependent
assumptions on the functional form.

In the following, we present results from the application of the
expression from Ref.~\cite{Bernard:1998gv} to the isovector form
factor $F_1^v(Q^2,\mps)$:
\begin{eqnarray}
  \label{eq:f1v-sse}
  F_1^v(Q^2,\mps) &=& 1 + \frac{1}{(4\pi f_\pi)^2}\Bigl( Q^2\left[
    \frac{68}{81} c_A^2 - \frac{2}{3}g_A^2 -2B_{10}^r(\lambda) + \left(
      \frac{40}{27} c_A^2 -\frac{5}{3}g_A^2 -\frac{1}{3}\right)
    \log\left( \frac{\mps}{\lambda}\right) \right] \nonumber \\
  && \quad + \int_0^1 dx \left[ \frac{16}{3} \Delta^2 c_A^2 + \mps^2
    \left( 3g_A^2 + 1 - \frac{8}{3}c_A^2\right) - Q^2 x(1-x) \left(
      5g_A^2+1 -\frac{40}{9}c_A^2\right) \right] \log\left(
    \frac{\tilde{m}^2} {\mps^2}\right) \nonumber \\
  && \quad + \int_0^1 dx \left[ \frac{32}{9} c_A^2 Q^2 x(1-x)
    \frac{\Delta \log R(\tilde{m})}
    {\sqrt{\Delta^2-\tilde{m}^2}}\right]
  \nonumber \\
  && \quad - \int_0^1 dx\, \frac{32}{3} c_A^2\Delta
  \left[\sqrt{\Delta^2 -\mps^2} \log R(\mps) -\sqrt{\Delta^2
      -\tilde{m}^2} \log R(\tilde{m}) \right] \Bigr)\,,
\end{eqnarray}
where we use the function $R(z)$ defined in Eq.~\eqref{eq:capr-def}
and introduce
\begin{equation}
  \label{eq:mtilde-def}
  \tilde{m}^2 = \mps^2 - Q^2 x (1-x)\,.
\end{equation}
Note that the expansion in Eq.~\eqref{eq:f1v-sse} is finite at
$Q^2=0$, i.e.~$F_1^v(Q^2=0,\mps)=1$ for any value of $\mps$ including
zero, but its derivative w.r.t.~$Q^2$ at the origin will diverge
logarithmically as $\mps\to 0$, cf.~Eq.~\eqref{eq:r1v-sse}. Again, the
fit parameters are $B_{10}^r(1{\rm GeV})$ and $c_A$. The other
parameters are fixed at their chiral values, see
Tab.~\ref{tab:sse-pars}.

As we will also point out in \sect\ref{sec:asymptotic-scaling}, our
comparatively large lattice volume puts us at a disadvantage when
studying the ratio $F_2(Q^2)/F_1(Q^2)$ for large values of
$Q^2>1$GeV$^2$. However, when using chiral expansions, this choice
turns out to work to our advantage since we have sufficiently many
data points at lower values of $Q^2$ to meaningfully apply the chiral
expansion directly to the $Q^2$-dependence. In the chiral fits we will
certainly need to apply cuts in $Q^2$. Hence, every fit will have
fewer data points contributing to it than in the dipole case. Thus, in
the end the uncertainty may very well turn out to be larger. However,
this is offset by the advantage that we do not make any
phenomenological assumptions on the $Q^2$-dependence and the results
can truly be considered ``first principle'' results.

As the first step, we fix the interval in $Q^2$ to be $Q^2\in [0,
0.5]$ GeV$^2$. This choice is a reasonable guess based on the
discussion in Ref.~\cite{Gockeler:2003ay}. With this interval, we vary
the cut representing the upper value of the pion mass, $\mps$. The
resulting values for $\chi^2$/dof, the fit parameters and the
extrapolated Dirac radii, $\langle r_1^2\rangle$ at the physical pion
mass, are listed in Tab.~\ref{tab:F1v-varmpi}.
\begin{table}[htb]
  \centering
  \begin{tabular}[c]{*{4}{c|}c}
    \hline\hline
    $\mps$ max [MeV] & $\chi^2$/dof & $B_{10}^r(1{\rm GeV})$ & $c_A$ &
    $\langle r_1^2\rangle(\mps^{\rm phys})$ [fm$^2$] \\ \hline
    300 & 0.59 & 0.60(39) & 1.15(33) & 0.623(68) \\
    400 & 1.86 & 0.35(11) & 1.009(97) & 0.686(21) \\
    500 & 12.74 & 1.195(46) & 1.623(23) & 0.5355(98) \\
    600 & 30.74 & 1.181(29) & 1.631(13) & 0.5446(65) \\
    \hline\hline
  \end{tabular}
  \caption{Fits to isovector $F_1^v(Q^2,\mps)$ at fixed interval of
    $Q^2=[0,0.5]$ GeV$^2$ with different pion mass cuts. The
    experimental form factors are not included in the fit.}
  \label{tab:F1v-varmpi}
\end{table}
From this table we conclude that it is necessary to apply a rather
conservative cut and restrict ourselves to $\mps < 400$MeV. The value
of $\chi^2$/dof becomes extremely large beyond this point which
implies that the function fails to properly describe the $\mps$
dependence of the data. This fact is not surprising and has been
reported also in~\cite{Gockeler:2003ay, Gockeler:2007ir,
  Gockeler:2007hj}. We have also observed it above when applying the
SSE and BChPT formulae to the radii obtained from dipole fits. In all
those cases it became evident that the pion mass dependence of the
lattice data is weaker than NLO SSE demands. Although the values of
the fit parameters in Tab.~\ref{tab:F1v-varmpi} stabilize when
ensembles at $\mps > 400$MeV are included, the resulting parameters
are incompatible with those obtained in the region $\mps <
400$MeV. Next, we need to ascertain that the cut in $Q^2$ can be
justified. We vary the upper end of the fit interval in $Q^2$ by
keeping the cut for $\mps < 400$MeV in place. The results are shown in
Tab.~\ref{tab:F1v-vart}.
\begin{table}[htb]
  \centering
  \begin{tabular}[c]{*{4}{c|}c}
    \hline\hline
    $Q^2$ max [GeV$^2$] & $\chi^2$/dof & $B_{10}^r(1{\rm GeV})$ & $c_A$ &
    $\langle r_1^2\rangle(\mps^{\rm phys})$ [fm$^2$] \\
    \hline
    0.7 & 1.84 & 0.483(71) & 1.127(54) & 0.661(14) \\
    0.5 & 1.86 & 0.35(11) & 1.009(97) & 0.686(21) \\
    0.3 & 1.82 & 0.60(21) & 1.241(158) & 0.645(37) \\
    \hline\hline
  \end{tabular}
  \caption{Fits to isovector $F_1^v(Q^2,\mps)$ at varying intervals of
    $Q^2$ with fixed pion mass cut, $\mps<400$MeV.}
  \label{tab:F1v-vart}
\end{table}
It turns out that the fit quality only mildly depends on the cut we
apply in $Q^2$. This situation is quite different from the pion mass
dependence we encountered previously. The SSE expression
Eq.~\eqref{eq:f1v-sse} provides an excellent fit to the $Q^2$
dependence for essentially the entire range of data points, but fails
to describe the pion mass dependence for all except the smallest
masses.

\begin{figure}[htb]
  \centering
  \includegraphics[scale=0.3,clip=true]{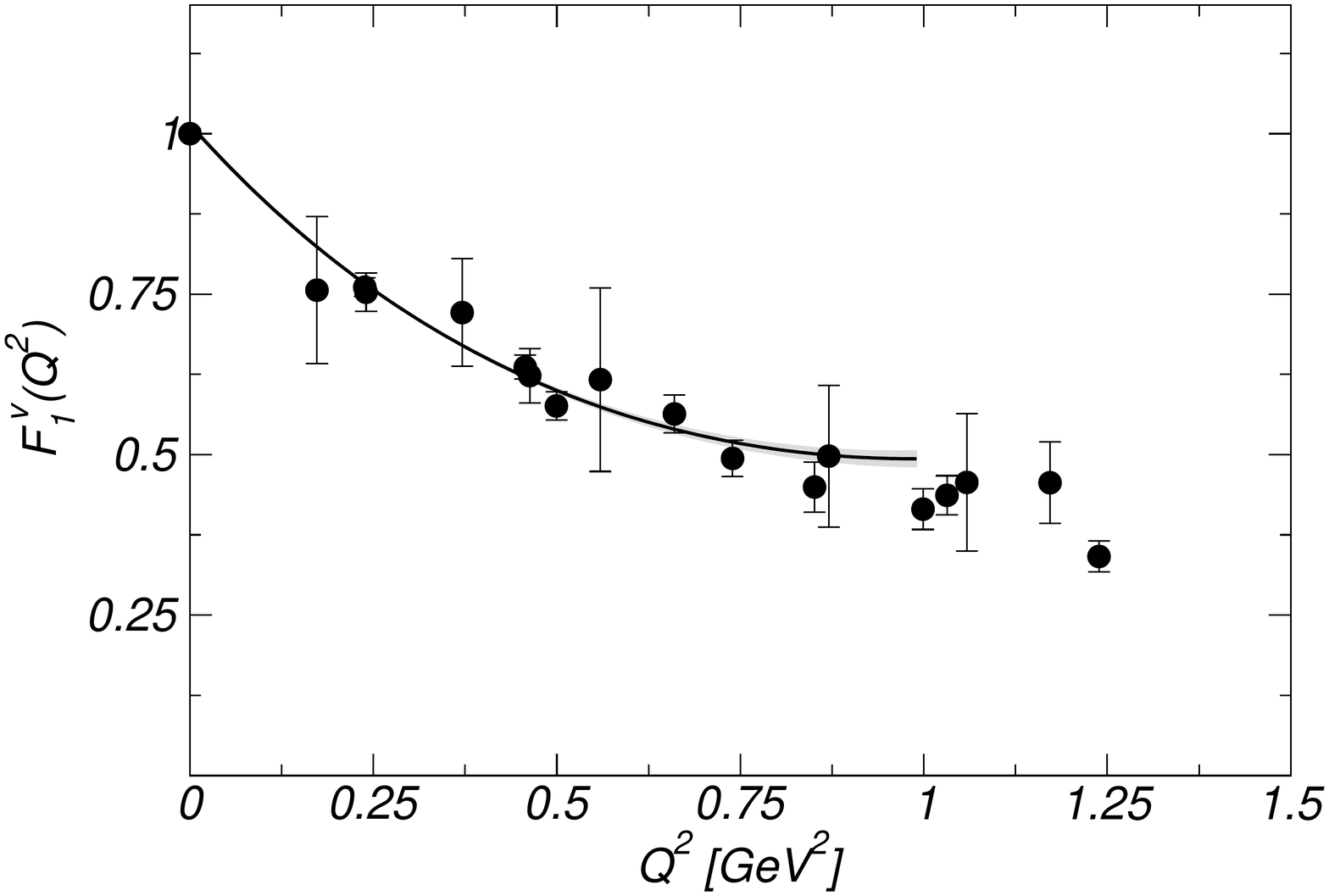}
  \caption{SSE best fit to isovector form factor
    $F_1^v(Q^2,\mps=293{\rm MeV})$ with lattice data for that
    ensemble.}
  \label{fig:F1v-comb-final}
\end{figure}

We now discuss the resulting curves for the chiral expansion with the
cuts $Q^2<0.5$GeV$^2$ and
$\mps<400$MeV. Figure~\ref{fig:F1v-comb-final} shows the data for the
ensemble at $\mps=293$MeV and the curve based on the best fit
parameters listed in Tab.~\ref{tab:F1v-vart} applied to the same pion
mass. Regarding the $Q^2$-dependence in \fig\ref{fig:F1v-comb-final}
we observe the surprising feature, that the resulting curve appears to
fit the data very well over a very large range of $Q^2$
values. However, we do not believe this to be of physical significance
since there is no reason to believe in the validity of the chiral
expansion at $Q^2$ values as large as $1$GeV$^2$!  Hence, we consider
this feature to be merely accidental and the upper cut in $Q^2$
necessary on theoretical grounds.

\begin{figure}[htb]
  \centering
  \includegraphics[scale=0.3,clip=true]{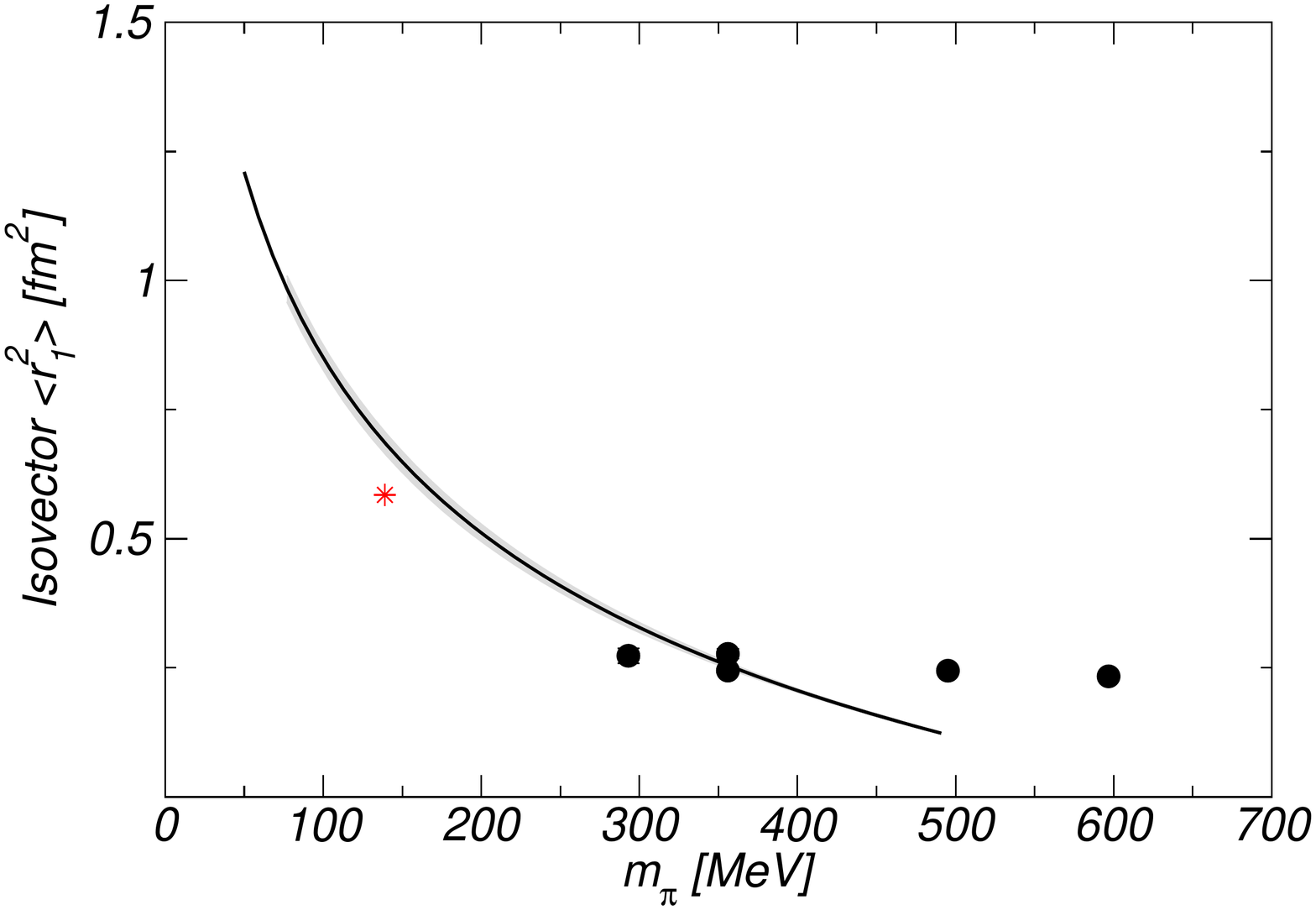}
  \caption{Isovector Dirac radii $\langle r_1^2\rangle$ with best fit
    using kinematic cuts $Q^2<0.5$GeV$^2$ and $\mps<400$MeV. The data
    points shown are dipole radii and experiment, which serve to
    compare this method with the previous one. They have not been
    included in the fit.}
  \label{fig:r1v-comb-final}
\end{figure}

Figure~\ref{fig:r1v-comb-final} shows the resulting chiral
extrapolation of the Dirac radii as a function of the pion mass,
$\mps$. For illustration purposes, we have included the radii obtained
from the dipole fits, cf.~Tab.~\ref{tab:f1v-dip-allmpi}, in the
plot. These data points have no influence on the curve presented and
just serve as a comparison of the two fitting methods. The red star in
the plot shows the experimental value taken from
Ref.~\cite{Belushkin:2006qa}.
When studying the resulting Dirac radii in
\fig\ref{fig:r1v-comb-final} we find that the fit slightly overshoots
the experiment and that the lattice data may very well account for the
physical value of $\langle r_1^2\rangle$. Finally, we point out that
the fitted parameters $c_A$ and $B_{10}^r(1{\rm GeV})$ from the
simultaneous strategy are different from those obtained with the
previous strategy, cf.~Tab.~\ref{tab:r1v-varmpi-sse}.

From the resulting $\chi^2/$dof we do not see that the SSE at NLO
fails to describe the functional form of $F_1^v(Q^2,\mps)$ if we
consider the parameter region $Q^2<0.5$GeV$^2$ and $\mps<400$MeV. The
simultaneous fit yields larger uncertainties than an SSE or BChPT fit
to Dirac radii from dipole fits and prefers a smaller value of
$c_A$. In general a combined fit has the advantage that no
phenomenological assumption on the functional behavior w.r.t.~$Q^2$ is
needed. Therefore this approach should become the method of choice as
sufficiently small $\mps$ and $Q^2$ are reached.

\subsubsection{\label{sec:isovector-pauli-form}Isovector Pauli form
  factor $F_2^v(Q^2)$}
Since according to the definition Eq.~\eqref{eq:lattice-vec-gff-def},
$F_2^v(Q^2)$ involves a spin-flip, this form factor may be better
described with an additional suppression of $Q^2$. On the other hand,
as we will discuss in the section on asymptotic scaling,
\sect\ref{sec:asymptotic-scaling}, the ratio of form factors does not
follow the quark-counting rules one would expect from the leading
perturbative expansion. Hence, one could use either the dipole
expression already employed for the form factor $F_1^v(Q^2)$,
Eq.~\eqref{eq:dipole}, or a tripole expression via
\begin{equation}
  \label{eq:tripole}
    F_2^v(Q^2) = A_0/(1+Q^2/M_t^2)^3\,,
\end{equation}
with the tripole mass, $M_t$. The tripole mass is related to the Pauli
radius via
\begin{equation}
  \label{eq:tripole-mass-rms}
  \langle r_2^2\rangle = \frac{18}{M_t^2}\,.
\end{equation}
In the following we will first fit $F_2^v(Q^2)$ using the dipole form
at fixed pion mass and study the stability of this fit as a function
of cut-offs. Next, we also apply the tripole form to $F_2^v(Q^2)$ and
study whether the data favors one of the two forms.

\begin{figure}[htb]
  \centering
  \includegraphics[scale=0.3,clip=true]{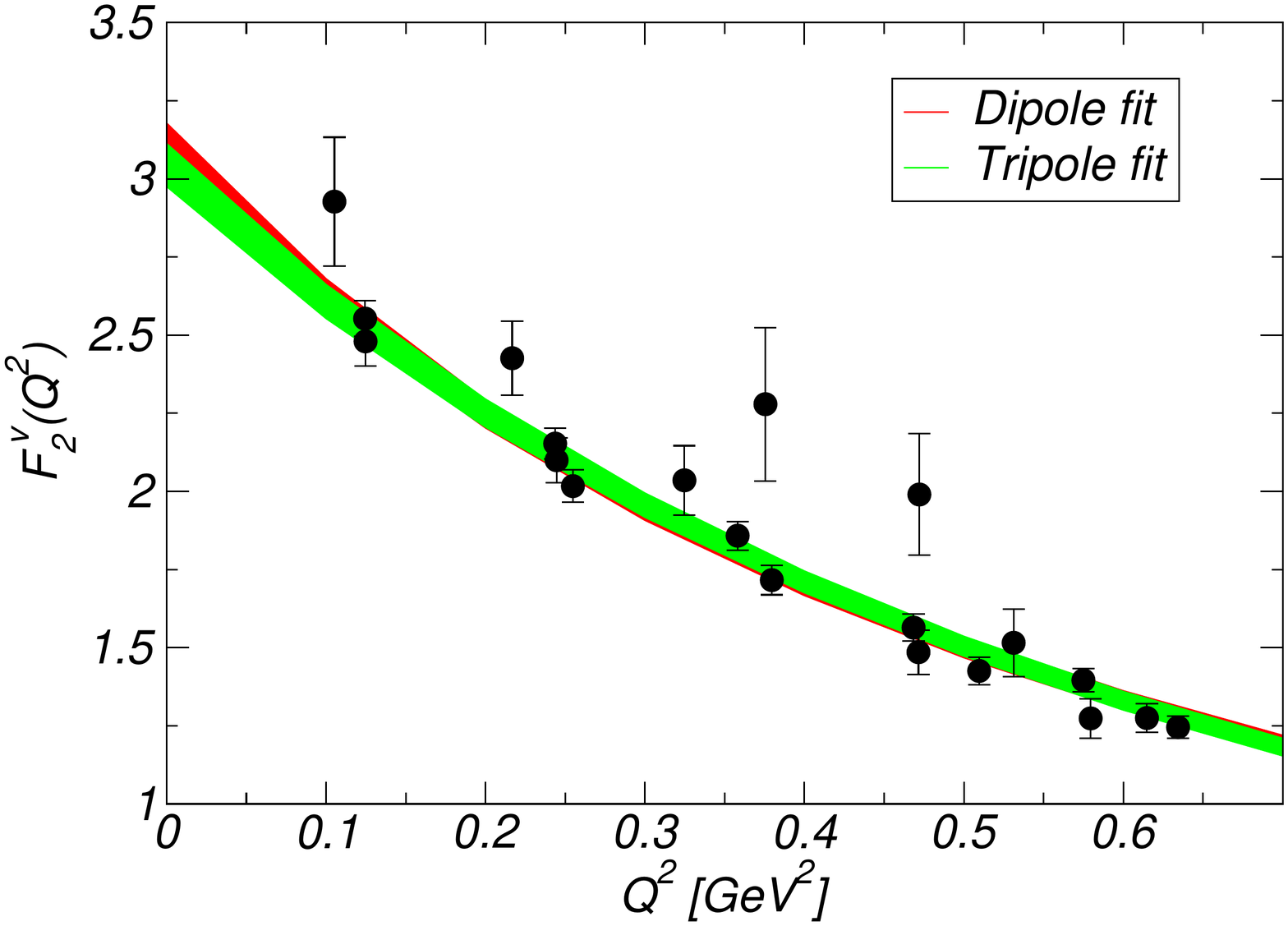}
  \caption{Comparison of dipole and tripole fit for isovector
    $F_2^v(Q^2)$ on the $28^3$ lattice at $\mps=356$MeV.}
  \label{fig:dip-tri-comp}
\end{figure}

Again, we first consider the lattice results at the large volume,
$28^3$, at pion mass $\mps=356$MeV. At this point as well as in the
remainder of this section we determine the uncertainty by applying the
jackknife method to the minimization of $\chi^2$ including the error
correlation matrix, cf.~\sect\ref{sec:super-jackkn-analys}. When
comparing dipole and tripole fits for the entire range of available
data, we find the results listed in Tab.~\ref{tab:dip-tri-comp}. The
fits are applied to the entire range of available $Q^2$ values. They
are shown graphically in \fig\ref{fig:dip-tri-comp}.
\begin{table}[htb]
  \centering
  \begin{tabular}[c]{*{3}{c}}
    \hline\hline
    & Dipole & Tripole \\ \hline
    $\chi^2 $/dof   & 1.04      & 1.15    \\
    $A_0$             & 3.107(71) & 3.044(68) \\
    $M_{d/t}$ [GeV]    & 1.067(17) & 1.374(20) \\
    $\langle r_2^2\rangle$ [fm$^2$] & 0.411(13) & 0.371(11) \\
    \hline\hline
  \end{tabular}
  \caption{Comparison of dipole and tripole fit for isovector
    $F_2^v(Q^2)$ on the $28^3$ lattice at $\mps=356$MeV.}
  \label{tab:dip-tri-comp}
\end{table}

Both results agree within error bars for $A_0$. The resulting
$\chi^2$/dof is almost identical, but the resulting Pauli radii
disagree by two sigma. Based on the quality of the fits it is not
possible to favor either choice since it appears that the distinction
only becomes important outside of the range of the available data.

In order to study if both fit strategies remain stable over the entire
range of data we perform the same variation of the fitting interval as
previously for $F_1^v(Q^2)$. The results for the dipole fit are shown
in Tab.~\ref{tab:f2-dip-vary-t} and the corresponding tripole fit
results in Tab.~\ref{tab:f2-tri-vary-t}.
\begin{table}[htb]
  \centering
  \begin{tabular}[c]{*{4}{c|}c}
    \hline\hline
    $[Q^2_{\mbox{\tiny min}}, Q^2_{\mbox{\tiny max}}]$ [GeV]$^2$ &
    $\chi^2$/dof & $A_0$ & $M_d$ [GeV] & $\langle r_2^2\rangle$
    [fm$^2$] \\ \hline
    $[0,1.5]$   & 1.04 & 3.107(71) & 1.067(17) & 0.411(13) \\
    $[0,0.5]$   & 1.15 & 3.111(74) & 1.067(20) & 0.411(16) \\
    $[0,0.4]$   & 1.19 & 3.066(78) & 1.095(30) & 0.389(21) \\
    $[0,0.3]$   & 1.54 & 3.065(82) & 1.086(34) & 0.396(25) \\
    \hline
    $[0,1.5]$   & 1.04 & 3.107(71) & 1.067(17) & 0.411(13) \\
    $[0.2,1.5]$ & 1.16 & 3.113(75) & 1.064(18) & 0.413(14) \\
    $[0.3,1.5]$ & 1.54 & 3.145(126) & 1.052(33) & 0.422(27) \\
    $[0.4,1.5]$ & 1.57 & 2.824(239) & 1.139(76) & 0.360(48) \\
    $[0.5,1.5]$ & 2.05 & 2.843(312) & 1.130(96) & 0.366(63) \\
    \hline
    $[0.3,0.5]$ & 1.83 & 3.866(444) & 0.893(75) & 0.586(98) \\
    $[0.2,0.4]$ & 1.59 & 3.054(92)  & 1.098(37) & 0.387(26) \\
    $[0.1,0.3]$ & 1.54 & 3.065(82)  & 1.086(34) & 0.396(25) \\
    \hline\hline
  \end{tabular}
  \caption{Dipole fits to $F_2^v(Q^2)$ with varying fit intervals.}
  \label{tab:f2-dip-vary-t}
\end{table}

\begin{table}[htb]
  \centering
  \begin{tabular}[c]{*{4}{c|}c}
    \hline\hline
    $[Q^2_{\mbox{\tiny min}}, Q^2_{\mbox{\tiny max}}]$ [GeV]$^2$ &
    $\chi^2$/dof & $A_0$ & $M_t$ [GeV] & $\langle r_2^2\rangle$
    [fm$^2$] \\ \hline
    $[0,1.5]$   & 1.15 & 3.044(68) & 1.374(20) & 0.371(11) \\
    $[0,0.5]$   & 1.15 & 3.072(72) & 1.358(24) & 0.380(14) \\
    $[0,0.4]$   & 1.25 & 3.034(76) & 1.386(36) & 0.365(19) \\
    $[0,0.3]$   & 1.53 & 3.046(80) & 1.364(40) & 0.377(22) \\
    \hline
    $[0,1.5]$   & 1.15 & 3.044(68) & 1.374(20) & 0.371(11) \\
    $[0.2,1.5]$ & 1.24 & 3.029(71) & 1.377(21) & 0.370(11) \\
    $[0.3,1.5]$ & 1.60 & 2.998(108) & 1.384(38) & 0.366(20) \\
    $[0.4,1.5]$ & 1.57 & 2.709(204) & 1.490(88) & 0.316(37) \\
    $[0.5,1.5]$ & 2.04 & 2.714(261) & 1.483(11) & 0.319(47) \\
    \hline
    $[0.3,0.5]$ & 1.84 & 3.643(361) & 1.183(85) & 0.501(72) \\
    $[0.2,0.4]$ & 1.59 & 3.004(86) &  1.400(43) & 0.357(22) \\
    $[0.1,0.3]$ & 1.53 & 3.046(80) &  1.364(40) & 0.377(22) \\
    \hline\hline
  \end{tabular}
  \caption{Tripole fits to $F_2^v(Q^2)$ with varying fit intervals.}
  \label{tab:f2-tri-vary-t}
\end{table}

When varying the upper cut-off it is apparent that the variation is
minimal and the resulting parameters only weakly depend on the
cut-off. The uncertainty increases as expected, but the effect is
small. On the other hand, when varying the lower cut-off the error bar
increases notably. However, the central values remain stable and the
data is still well described by the dipole fit over the entire range.
When shifting the fit interval, the inclusion of data points at
smaller values of $Q^2$ improves the quality of the fit and reduces
the error bars notably. In the case of the tripole fit, the value of
$A_0$ is systematically lower, but still within error bars. We
conclude that $F_2^v(Q^2)$ is well described by either functional form
over the entire kinematic range, but the strongest influence on fixing
the parameters of the fit comes from the region of smaller $Q^2$
values, in particular from $Q^2 < 0.3$GeV$^2$. Compared to the fits to
$F_1^v(Q^2)$ we thus find a qualitatively similar picture, although
the sensitivity of the parameters to the inclusion of the data at
small values of $Q^2$ is larger.

Finally, we perform a series of tripole fits to our entire data set at
all pion masses. Results are summarized in
Tab.~\ref{tab:f2-tri-allmpi}.
\begin{table}[htb]
  \centering
  \begin{tabular}[c]{*{4}{c|}c}
    \hline\hline
    $\mps$ [MeV] & $\chi^2$/dof & $A_0$ & $M_t$ [GeV] & $\langle
    r_2\rangle$ [fm] \\ \hline
    293 & 1.31 & 2.896(162) & 1.389(53) & 0.363(28) \\
    356 on $28^3$ & 1.15 & 3.044(68) & 1.374(20) & 0.371(11) \\
    356 on $20^3$ & 2.09 & 2.958(76) & 1.436(25) & 0.340(12) \\
    495 & 1.39 & 3.210(44) & 1.482(14) & 0.319(6) \\
    597 & 1.93 & 3.402(27) & 1.529(94) & 0.300(4) \\
    \hline\hline
  \end{tabular}
  \caption{Tripole fits to isovector $F_2^v(Q^2)$ for all data sets.}
  \label{tab:f2-tri-allmpi}
\end{table}
We again observe finite-size effects for the Pauli radius $\langle
r_2\rangle$. On top of this, the Pauli radius increases as the pion
mass decreases on the lattices with fixed physical volume. Hence, we
have reason to expect that finite-size effects may be non-negligible
for the smallest value of $\mps$.

\paragraph{SSE chiral fits to isovector $\langle r_2^2\rangle$ and
  $\kappa^v$}
Similar to what we did for the isovector Dirac form factor, we now fit
the mean squared Pauli radius $\langle r_2^2\rangle$ and the isovector
anomalous magnetic moment, $\kappa_v$, using chiral formulae from the
SSE expansion at NLO\@. In this case, the expressions as functions of
$\mps$ are~\cite{Bernard:1998gv, Hemmert:2002uh}:
\begin{eqnarray}
  \label{eq:r2kv-kappav}
  \langle r_2^2\rangle &=& \frac{g_A^2 m_N} {8 f_\pi^2\kappa_v(\mps)
    \pi \mps} + \frac{c_A^2 m_N} {9f_\pi^2 \kappa_v(\mps) \pi^2
    \sqrt{\Delta^2-\mps^2}} \log\left( \frac{\Delta}{\mps} +
    \sqrt{\frac{\Delta^2}{\mps^2} - 1}\right) + \frac{24m_N}
  {\kappa_v(\mps)} B_{c2}\,, \nonumber \\
  \kappa_v(\mps) &=& \kappa_v^0 - \frac{g_A^2 \mps m_N}{4\pi
    f_\pi^2} + \frac{2c_A^2\Delta m_N}{9\pi^2f_\pi^2} \left(
    \sqrt{1-\frac{\mps^2}{\Delta}} \log R(\mps) +
    \log\left(\frac{\mps} {2\Delta}\right) \right) \nonumber \\
  && \quad - 8 E_1^r(\lambda) m_N \mps^2 + \frac{4 c_A c_V g_A m_N 
    \mps^2}{9\pi^2 f_\pi^2} \log\left( \frac{2\Delta}{\lambda} \right)
  + \frac{4c_A c_V g_A m_N \mps^3} {27\pi f_\pi^2 \Delta} \nonumber \\
  && \quad - \frac{8 c_A c_V g_A \Delta^2 m_N}{27\pi^2 f_\pi^2}
  \left(\left( 1-\frac{\mps^2}{\Delta^2}
    \right)^{3/2} \log R(\mps) + \left( 1-\frac{3 \mps^2}
      {2\Delta^2} \right) \log\left( \frac{\mps}{2\Delta} \right)
  \right)\,,
\end{eqnarray}
where $R(m)$ was defined in Eq.~\eqref{eq:capr-def}. The pion mass
dependence of the product $\kappa_v(m_\pi) \langle r_2^2\rangle$
contains a single unknown parameter, $B_{c2}$, whose effect is similar
to the counterterm $B_{10}^r(\lambda)$ in $\langle r_1^2\rangle$. In
addition to the expression above, we need to correct for the fact that
Eq.~\eqref{eq:r2kv-kappav} assumes the nucleon mass to be constant
when computing the $\mps$ dependence of $\kappa_v$. This assumption is
certainly not justified, so we need to correct the expression by
working with the normalized anomalous magnetic moment, $\kappa_v^{\rm
  norm}$, as defined in Eq.~\eqref{eq:kappanorm-def}. The SSE pion
mass dependence for $\kappa_v$ involves three additional low-energy
constants, $\kappa_v^0$, $c_V$, and $E_1^r(\lambda)$. Since
$\kappa_v(\mps)$ is also part of the expression for $\langle
r_2^2\rangle$ and is always obtained from the same tripole fit to
$F_2^v(Q^2)$ it makes sense to fit $\kappa_v$ separately at first and
then perform a single-parameter fit to $\kappa_v \langle
r_2^2\rangle$. In this way, the best fit stability is guaranteed.

We employ the same parameters for the expression as listed previously
in Tab.~\ref{tab:sse-pars}. In this case we fix $c_A$ to the value of
$1.5$. Furthermore, since $\kappa_v$ has three fit parameters, we can
only work with a pion mass cut of $\mps<600$MeV. The result is shown
in Tab.~\ref{tab:kv-varmpi}. The subsequent fit to $\kappa_v \langle
r_2^2\rangle$ with $\kappa_v$ being canceled out in the next step is
shown in Tab.~\ref{tab:r2v-varmpi}.
\begin{table}[htb]
  \centering
  \begin{tabular}[c]{*{4}{c|}c}
    \hline\hline
    $\mps$ max [MeV] & $\chi^2$/dof & $\kappa_v^0$ & $c_V$
    [GeV$^{-1}$] & $E_1^r(1{\rm GeV})$ [GeV$^{-3}$] \\ \hline
    600 & 1.18 & 4.68(24) & -2.84(37) & -6.08(41) \\
    \hline\hline
  \end{tabular}
  \caption{SSE fit to the isovector anomalous magnetic moment,
    $\kappa_v^{\rm norm}$, obtained from tripole fits.}
  \label{tab:kv-varmpi}
\end{table}
\begin{table}[htb]
  \centering
  \begin{tabular}[c]{*{2}{c|}c}
    \hline\hline
    $\mps$ max [MeV] & $\chi^2$/dof & $B_{c2}$ [GeV$^{-3}$] \\
    \hline
    600 & 22.2 & -6.1(12) $10^{-2}$ \\
    \hline\hline
  \end{tabular}
  \caption{SSE fit to the isovector Pauli radius, $\langle
    r_2^2\rangle$, obtained from tripole fits.}
  \label{tab:r2v-varmpi}
\end{table}
Unfortunately, the resulting $\chi^2$/dof is not very good, evidently
due to the stronger curvature of the NLO SSE form. The curves will be
plotted below in Fig.~\ref{fig:r2vkv-final}.

\paragraph{BChPT fits to isovector $\langle r_2^2\rangle$ and
  $\kappa^v$}
The expansions of the isovector anomalous magnetic moment, $\kappa_v$,
and the isovector mean squared Pauli radius, $\langle r_2^2\rangle$,
up to the order ${\cal O}(p^4)$ are given by:
\begin{eqnarray}
  \label{eq:r2vkv-hbchpt}
  \langle r_2^2\rangle & = & \frac{M_N}{M_0} \left(
    \frac{1}{\kappa_v} B_{c2} + (\langle r_2^2\rangle)^{(3)} +
    (\langle r_2^3\rangle)^{(4})\right) + {\cal O}(p^6) \,,
  \nonumber \\
  \kappa_v &=& \frac{M_N}{M_0}\left(c_6 - 16 M_0\mps^2
    e_{106}^r(\lambda) + \delta\kappa_v^{(3)} +
    \delta\kappa_v^{(4)}\right) + {\cal O}(p^6)\,,
\end{eqnarray}
with the individual contributions at each order:
\begin{eqnarray*}
  B_{c2} & = & 24M_0e_{74}^r(\lambda)\,, \nonumber \\
  (\langle r_2^2\rangle)^{(3)} & = & \frac{1}{\kappa_v} \frac{g_A^2
    M_0}{16\pi^2f_\pi^2M_N^5(\mps^2-4M_N^2)} \Bigl[ -124M_N^6 +
  105\mps M_N^4 -18\mps^4 M_N^2 + \nonumber \\
  && \quad\qquad 6(3\mps^6-22M_N^2 \mps^4 +
  44M_N^4 \mps^2 -16M_N^6)\log\frac{\mps}{M_N}\Bigr]
  \nonumber \\
  && \quad + \frac{1}{\kappa_v} \frac{g_A^2 M_0}{8\pi^2f_\pi^2M_N^5
    \mps(4M_N^2 - \mps^2)^{3/2}} \bigl[ 9\mps^8-84M_N^2 \mps^6
  \nonumber \\ && \qquad\qquad\qquad\qquad
  +246 M_N^4 \mps^4 -216M_N^6\mps^2 + 16M_N^8\bigr]
  \arccos\frac{\mps}{2M_N}\,, \nonumber
\end{eqnarray*}

\begin{eqnarray}
  \label{eq:hbchpt-highorder}
  (\langle r_2^2\rangle)^{(4)} & = & -\frac{1}{\kappa_v} \frac{c_6
    g_A^2\mps^3}{16\pi^2f_\pi^2M_0^4(4M_0^2-\mps^2)^{3/2}}\left(
    4\mps^4 -27 \mps^2M_0^2 +42M_0^4\right)\arccos\frac{\mps}{2M_0}
  \nonumber \\
  && \quad + \frac{1}{\kappa_v}
  \frac{1}{16\pi^2f_\pi^2M_0^4(\mps^2-4M_0^2)} \Biggl[ 16c_4M_0^7 + 52
  g_A^2M_0^6 \nonumber \\
  && \qquad\quad- 4c_4 \mps^2 M_0^5 - 14c_6g_A^2 \mps^2 M_0^4 -
  13g_A^2\mps^2M_0^4 \nonumber \\
  && \qquad\quad +
  8(3g_A^2-c_4M_0)(\mps^2-4M_0^2)M_0^4\log\frac{\mps}{M_0}
  \log\frac{\mps}{M_0} +4c_6 g_A^2\mps^4M_0^2 \nonumber \\
  && \qquad\quad - g_A^2(\mps^2-4M_0^2) (4c_6 \mps^4 -
  3c_6\mps^2M_0^2+24M_0^4) \log\frac{\mps}{M_0}\Biggr]\,, \nonumber \\
  \delta\kappa_v^{(3)} & = & \frac{g_A^2M_0}{16\pi^2\fps^2M_N^5(\mps^2-4
    M_N^2)}\Bigl[ -124 M_N^6 + 105 \mps^2 M_N^4 - 18\mps^4M_N^2
  \nonumber \\ && \qquad\quad
  +6(3\mps^6-22M_N^2\mps^4+44M_N^4\mps^2-16M_N^6)
  \log\frac{\mps}{M_N}\Bigr] \nonumber \\ &&  \quad +
  \frac{g_A^2M_0}{8\pi^2\fps^2M_N^5\mps (4M_N^2-\mps^2)^{3/2}} \bigl[
  9\mps^8 -84 M_N^2\mps^6 +246M_N^4\mps^4 \nonumber \\ && \qquad\quad
  - 216M_N^6\mps^2 + 16M_N^8\bigr] \arccos\frac{\mps}{2M_N}\,,
  \nonumber \\
  \delta\kappa_v^{(4)} & = & - \frac{g_A^2 c_6 \mps^3} {16\pi^2 \fps^2
    M_0^4 (4M_0^2-\mps^2)^{3/2}} \left[4\mps^4 - 27\mps^2M_0^2 +42
    M_0^4\right] \arccos\frac{\mps}{2M_0} \nonumber \\ && \quad +
  \frac{1}{16\pi^2\fps^2M_0^4(\mps^2-4M_0^2)}\Bigl[ 16c_4 M_0^7 +
  52g_A^2M_0^6 - 4c_4\mps^2M_0^5 - 14c_6g_A^2\mps^2M_0^4 \nonumber \\
  && \qquad\quad -13g_A^2\mps^2M_0^4 + 8(3g_A^2 - c_4M_0)(\mps^2-4
  M_0^2)M_0^4\log\frac{\mps}{M_0} + 4c_6g_A^2\mps^4M_0^2 \nonumber \\
  && \qquad\quad -g_A^2(\mps^2-4M_0^2)(4c_6\mps^4 - 3c_6\mps^2M_0^2 +
  24M_0^4)\log\frac{\mps}{M_0}\Bigr] \,.
\end{eqnarray}
In the Pauli radius, $c_6$ only shows up in the ${\cal O}(p^4)$ part
which is of higher order. Thus, we cannot use the mean squared radius
$\langle r_2^2\rangle$ for a determination of $c_6$ in isolation. We
therefore settle for a simultaneous fit to all three quantities
$\langle r_1^2\rangle$, $\langle r_2^2\rangle$ and $\kappa_v$. To do
this, we insert the expansion from Eq.~\eqref{eq:mn-expand} for the
nucleon mass $M_N$ in Eqs.~\eqref{eq:r1v-hbchpt},
\eqref{eq:r1v-hbchpt-allorder}, \eqref{eq:r2vkv-hbchpt}
and~\eqref{eq:hbchpt-highorder}. The parameter $c_4$ is again taken
from Tab.~\ref{tab:mn-expand-pars}. We take correlations into account
by using both the error-correlation matrix and the super-jackknife
prescriptions as outlined in \sect\ref{sec:super-jackkn-analys}. For
these fits we also set the scale to $\lambda=M_0$. The resulting fits
are stable, however the $\chi^2$/dof is relatively large. The reason
is that the BChPT curves for both $\langle r_1^2\rangle$ and $\langle
r_2^2\rangle$ show a stronger curvature than our lattice data. If we
included data at higher pion masses, $\chi^2$/dof would increase even
more. The results are summarized in
Tab.~\ref{tab:hbchpt-r1r2kv-varmpi}.
\begin{table}[htb]
  \centering
  \begin{tabular}[c]{*{4}{c|}c}
    \hline\hline
    $\chi^2$/dof & $c_6$ & $d_6^r(M_0)$ [GeV$^{-2}$] &
    ${e_{74}^r(M_0)}$ [GeV$^{-3}$] & $e_{106}^r(M_0)$ [GeV$^{-3}$] \\
      \hline
    10.23 & 4.31(11) & 0.924(12) & 1.201(34) & -0.08(11) \\
    \hline\hline
  \end{tabular}
  \caption{Simultaneous fit of the BChPT expression to the anomalous
    magnetic moment, $\kappa_v$, and the radii $\langle r_1^2\rangle$
    and $\langle r_2^2\rangle$, as obtained from tripole fits.}
  \label{tab:hbchpt-r1r2kv-varmpi}
\end{table}
Both the (poor) quality of the fit as well as the parameters agree
with Ref.~\cite{Syritsyn:2009mx}. We finally display the fits with
$\mps<600$MeV (SSE) and $\mps<400$MeV (BChPT) for the normalized
isovector anomalous magnetic moment, $\kappa_v^{\rm norm}$, on the
left panel of \fig\ref{fig:r2vkv-final} and the results from the fits
with $\mps<600$MeV (SSE) and $\mps<400$ MeV(BChPT) for the Pauli
radius, $\langle r_2^2\rangle$, in the right panel of said figure.
\begin{figure}[htb]
  \centering
  \includegraphics[scale=0.27,clip=true]{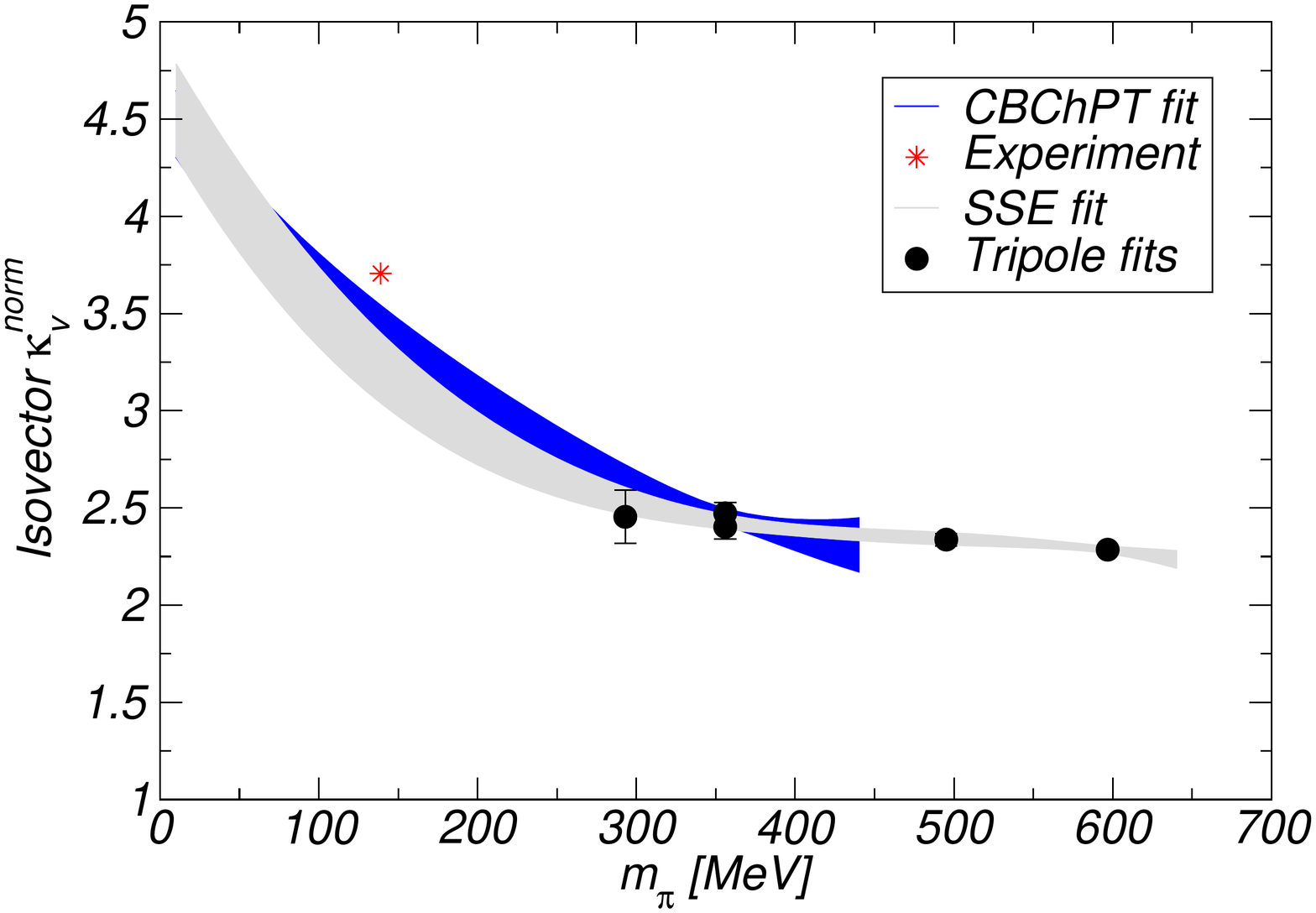}\qquad
  \includegraphics[scale=0.27,clip=true]{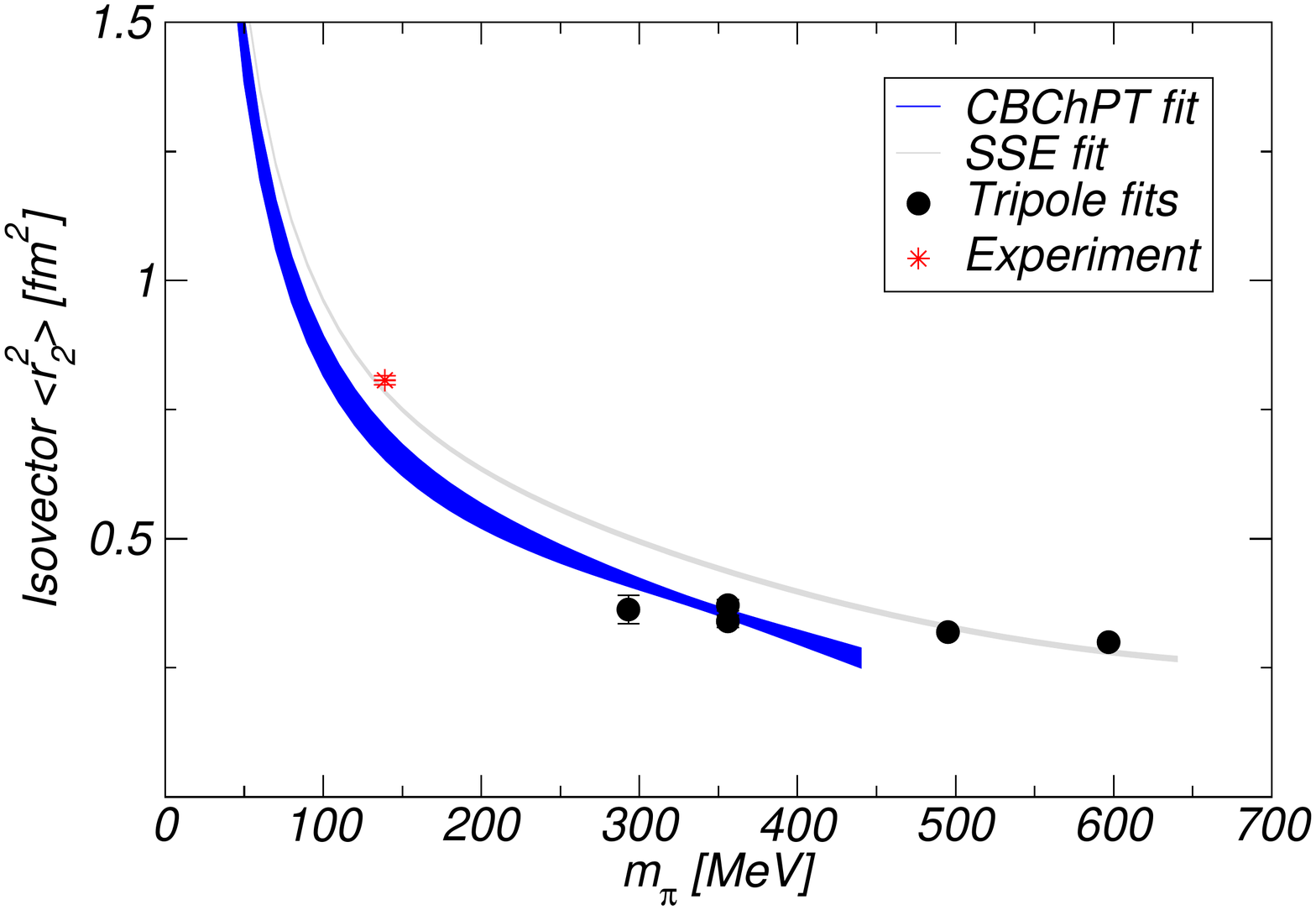}
  \caption{Results of SSE and BChPT fits to the normalized anomalous
    magnetic moment $\kappa_v^{\rm norm}$ (left panel) and to the
    isovector mean squared radius $\langle r_2^2\rangle$ (right
    panel).}
  \label{fig:r2vkv-final}
\end{figure}

We find that the fit quality for $\kappa_v^{\rm norm}$ is good and the
fits to $\langle r_2^2\rangle$ are similar to those for $\langle
r_1^2\rangle$ --- the lattice data does not yet display the feature of
strong divergence in the chiral limit and tends to be flatter as a
function of the pion mass than the associated curves from chiral
perturbation theory. Again, it is possible that the fits work already
well below $\mps=400$MeV for calculations carried out in sufficiently
large spatial volumes.

\paragraph{Simultaneous fits to $F_2^v(Q^2,\mps)$}
For the chiral expansion in $F_2^v(Q^2,\mps)$ we use the simultaneous
expansion based on SSE in $Q^2$ and $\mps$ from
Ref.~\cite{Bernard:1998gv,Hemmert:2002uh}:
\begin{eqnarray}
  \label{eq:f2v-sse}
  F_2^v(Q^2,\mps) &=& \kappa_v(\mps) - g_A^2 \frac{4\pi m_N} {(4\pi
    f_\pi)^2} \int_0^1 dx \left( \tilde{m} - \mps\right)
  \nonumber \\
  && \quad + \frac{32 c_A^2 m_N\Delta} {9(4\pi f_\pi)^2} \int_0^1 dx
  \Biggl[ \frac{1}{2} \log\left( \frac{\tilde{m}^2} {4\Delta^2} \right)
  - \log\left( \frac{\mps}{2\Delta}\right) \nonumber \\
  && \qquad\qquad\qquad\qquad\qquad + \frac{\sqrt{\Delta^2 -
      \tilde{m}^2}} {\Delta} \log R(\tilde{m}) - \frac{\sqrt{\Delta^2
      - \mps^2}} {\Delta} \log R(\mps)\Biggr]\,,
\end{eqnarray}
with $R(x)$ and $\tilde{m}$ as defined previously in
Eq.~\eqref{eq:capr-def} and~\eqref{eq:mtilde-def}. Note that this
expansion also includes the function for $\kappa_v(\mps)$ defined
previously in Eq.~\eqref{eq:r2kv-kappav}. The input parameters are the
same as used previously, Tab.~\ref{tab:sse-pars}. The parameters are
$c_V$, $E_1^r(\lambda=1{\rm GeV})$, and $\kappa_v^0$ in addition to
$c_A$. The number of parameters is larger than for $F_1^v(Q^2)$, but
we decided nevertheless not to add the experimental data point in this
fit.

\begin{figure}[htb]
  \centering
  \includegraphics[scale=0.3,clip=true]{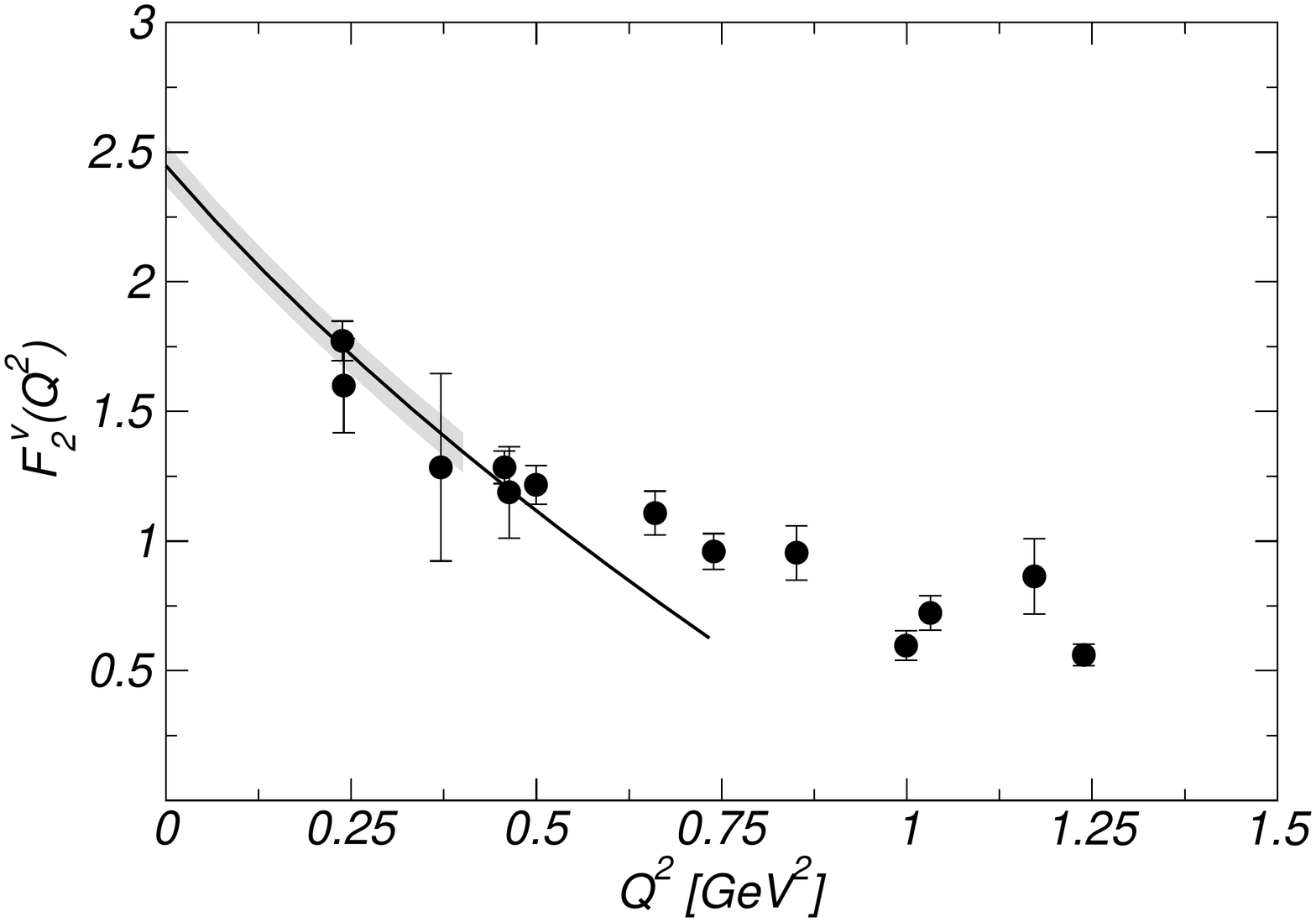}
  \caption{Isovector form factor $F_2^v(Q^2,\mps)$ lattice data with
    best fit SSE at $\mps=293$MeV.}
  \label{fig:F2v-comb-final}
\end{figure}

Similar to our study for the form factor $F_1^v(Q^2,\mps)$ in
\sect\ref{sec:isovector-dirac-form}, we also vary the cuts in $Q^2$
and $\mps$ to find an acceptable fit range for $F_2^v(Q^2,\mps)$. We
start out by keeping the cut $Q^2<0.4$GeV$^2$ in place and varying the
cut in the pion mass. Table~\ref{tab:F2v-varmpi} summarizes our
findings. Note, that we cannot use $\mps<400$MeV since we would have
insufficient data points to constrain all three parameters in
$\kappa_v$, cf.~Eq.~\eqref{eq:r2kv-kappav}.
\begin{table}[htb]
  \centering
  \begin{tabular}[c]{*{5}{c|}c}
    \hline\hline
    $\mps$ max [MeV] & $\chi^2$/dof & $c_V$ [GeV$^{-1}$] &
    $E_1^r(1{\rm GeV})$ [GeV$^{-3}$] & $\kappa_v^0$ & $c_A$ \\ \hline
    500 & 1.82 & $-4.1(20)$  & $-4.7(11) $ & 4.47(50) & 0.852(98) \\
    600 & 1.61 & $-3.54(46)$ & $-4.40(31)$ & 4.62(14) & 0.851(97) \\
    \hline\hline
  \end{tabular}
  \caption{Fits to isovector $F_2^v(Q^2,\mps)$ at fixed interval of
    $Q^2=[0,0.4]$ GeV$^2$ with different pion mass cuts.}
  \label{tab:F2v-varmpi}
\end{table}
As seen from the resulting values of $\chi^2/$dof the quality of the
fits is good for the entire range of pion masses. This feature is
distinct from the corresponding case of $F_1^v(Q^2,\mps)$. Next, we
vary the upper cut in $Q^2$ and keep $\mps<500$MeV. The results are
collected in Tab.~\ref{tab:F2v-vart}.
\begin{table}[htb]
  \centering
  \begin{tabular}[c]{*{5}{c|}c}
    \hline\hline
    $Q^2$ max [GeV$^2$] & $\chi^2$/dof & $c_V$ [GeV$^{-1}$] &
    $E_1^r(1{\rm GeV})$ [GeV$^{-3}$] & $\kappa_v^0$ & $c_A$ \\ \hline
    0.3 & 1.54 & $-6.6(68)$ & $-3.8(11) $ & 4.67(51) & 0.40(32)  \\
    0.4 & 1.82 & $-4.1(20)$ & $-4.7(11) $ & 4.47(50) & 0.852(98) \\
    0.5 & 2.03 & $-4.0(14)$ & $-4.57(78)$ & 4.63(36) & 0.851(42) \\
    0.6 & 2.77 & $-4.4(13)$ & $-5.11(77)$ & 4.44(36) & 0.941(35) \\
    0.7 & 4.87 & $-4.0(11)$ & $-5.40(74)$ & 4.40(35) & 1.100(25) \\
    \hline\hline
  \end{tabular}
  \caption{Fits to isovector $F_2^v(Q^2,\mps)$ with varying cut in
    $Q^2$ at fixed pion mass cut, $\mps<500$MeV.}
  \label{tab:F2v-vart}
\end{table}
When increasing the number of data points, the $\chi^2/$dof increases
and we conclude that the selection $Q^2<0.4$GeV$^2$ is
preferred. Apparently, the $Q^2$ dependence is not described well
beyond that point. We also remind the reader that we find
finite-volume effects to be reduced when only considering
$Q^2<0.4$GeV$^2$ in \sect\ref{sec:finite-volume-effect}. Hence, we
finally settle for the cuts $Q^2=[0,0.4]$ GeV$^2$ and $\mps<500$MeV
for our final plots. This choice is compatible with our choice for
$F_1^v(Q^2,\mps)$ and it gives both acceptable fits and minimizes the
unknown higher-order contributions in the SSE\@. We notice, however,
that the parameter $c_A$ is systematically lower than in the previous
SSE fits and not compatible with its estimate from the $\Delta\to N
\pi$ decay width~\cite{Syritsyn:2009mx}. Graphically, the resulting
curve is shown for a single ensemble at $\mps=293$MeV in
\fig\ref{fig:F2v-comb-final}.
The graph makes it apparent that $F_2^v(Q^2,\mps)$ should be cut below
$Q^2<0.4$GeV$^2$.

\begin{figure}[htb]
  \centering
  \includegraphics[scale=0.27,clip=true]{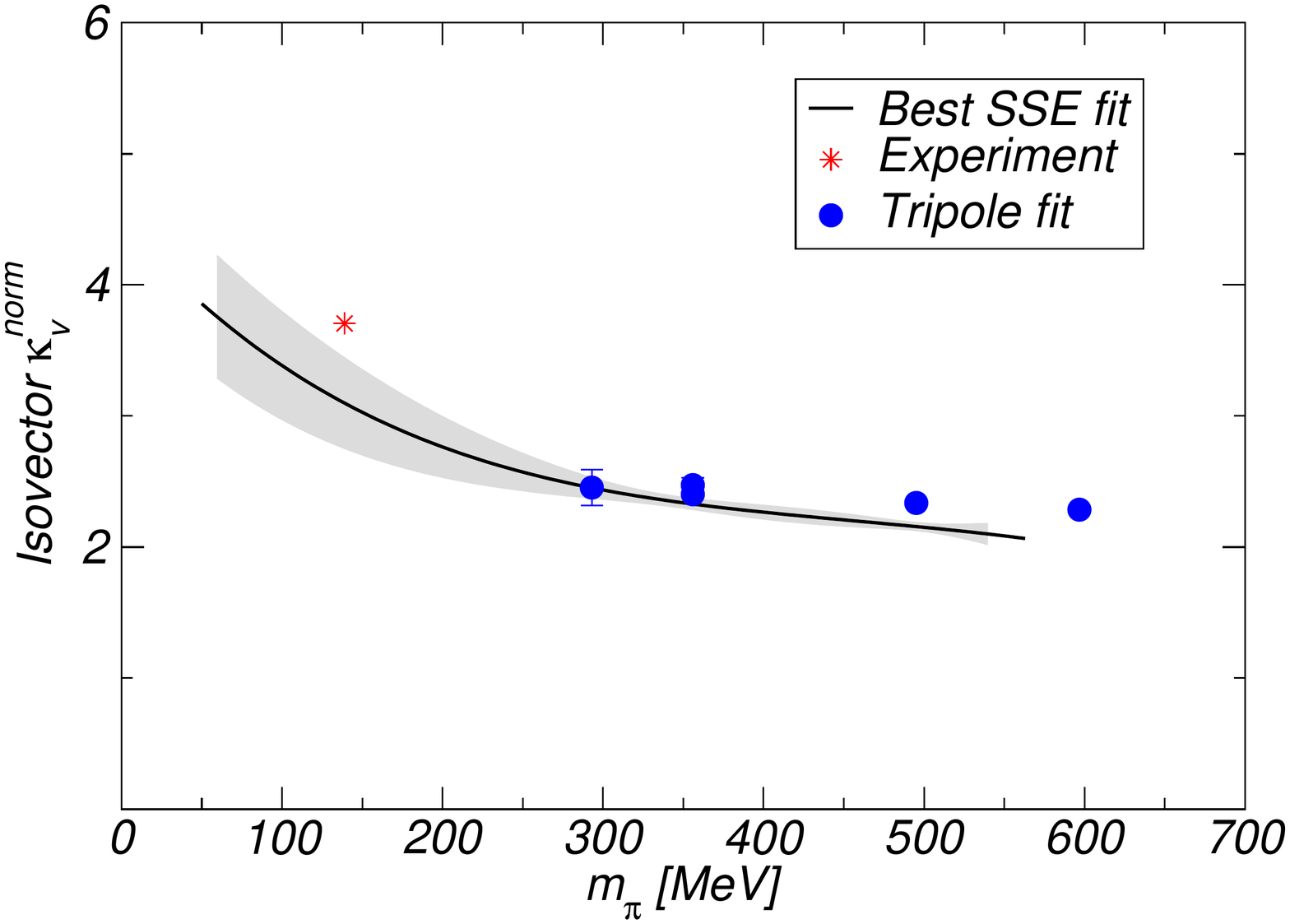}\qquad
  \includegraphics[scale=0.27,clip=true]{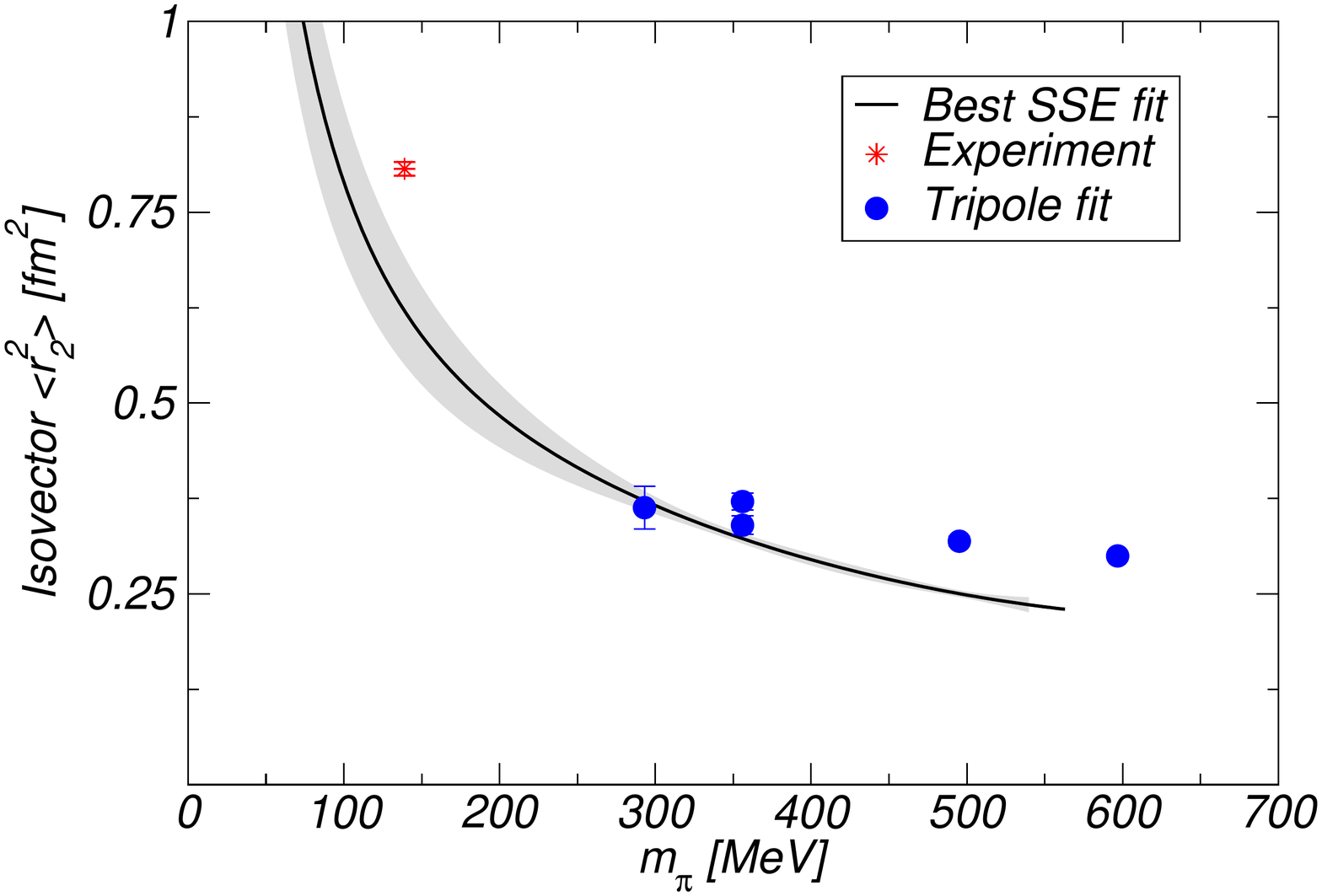}
  \caption{Isovector normalized anomalous magnetic moment
    $\kappa_v^{\rm norm}$ (left panel) and mean squared radius
    $\langle r_2^2\rangle$ (right panel) as a function of the pion
    mass as obtained from an SSE fit to the chiral expansion of
    $F_2^v(Q^2, \mps)$. The data points shown are tripole radii and
    experiment, which serve to compare this method to the previous one
    and to phenomenology. They have not been included in the fit.}
  \label{fig:F2v-final}
\end{figure}

We finally plot curves of $\langle r_2^2\rangle$ and $\kappa_v^{\rm
  norm}$ derived from the above fit as functions of $\mps$ in
Fig.~\ref{fig:F2v-final}. The tripole fit data points and the
experimental result are included in the graph to provide a comparison
to the former method. They have not been used in the fit.
The qualitative picture is similar to that from the SSE and BChPT fits
to the tripole radii and magnetic moments. The curves slightly
underestimate the results at the physical pion mass. The advantage of
this scheme is the same as in the simultaneous fits to
$F_1^v(Q^2,\mps)$, namely the absence of phenomenological assumptions
on the functional behavior w.r.t.~$Q^2$.

\subsubsection{\label{sec:asymptotic-scaling}Asymptotic scaling}
Although form factors have been studied experimentally for several
decades and using perturbative QCD many qualitative and quantitative
features have been understood very well, for the proton $F_2^p(Q^2)$ a
notable discrepancy has been found in recent spin-transfer
measurements~\cite{Milbrath:1997de, Pospischil:2001pp, Gayou:2001qd,
  Gayou:2001qt, Punjabi:2005wq}. From quark-counting rules one expects
the ratio $F_2^p(Q^2)/F_1^p(Q^2)$ to scale proportionally to $Q^{-2}$
which is consistent with experimental measurements using the
Rosenbluth method. The recent spin-transfer experiments, on the other
hand, found a scaling of $F_2^p(Q^2)/F_1^p(Q^2)$ proportional to
$Q^{-1}$, instead. The source of the discrepancy is now generally
believed to be two-photon exchange processes, see
Ref.~\cite{Arrington:2007ux}. On the lattice we are in a unique
position to study form factors using exactly single-photon exchanges
without contamination from other processes. This analysis can proceed
in a fully model-independent way.

The downside of the lattice technology is the limitation to rather
small virtualities, $Q^2$, since the external momenta,
$\vec{p}^{\,\prime}$ and $\vec{p}$ in
Eq.~\eqref{eq:nucleon-vector-gpd}, cannot be chosen too large. For
larger values of the external momenta the exponential in
Eqs.~\eqref{eq:two-pt} and \eqref{eq:three-pt} introduces large
fluctuations which quickly deteriorate the signal-to-noise ratio. A
quantitative analysis of this phenomenon has been given in
Ref.~\cite{Edwards:2005kw}. Also, controlling the cutoff effects
requires $|\vec q| \ll \pi/a$.

\begin{figure}[htb]
  \centerline{\includegraphics[scale=0.5,clip=true]{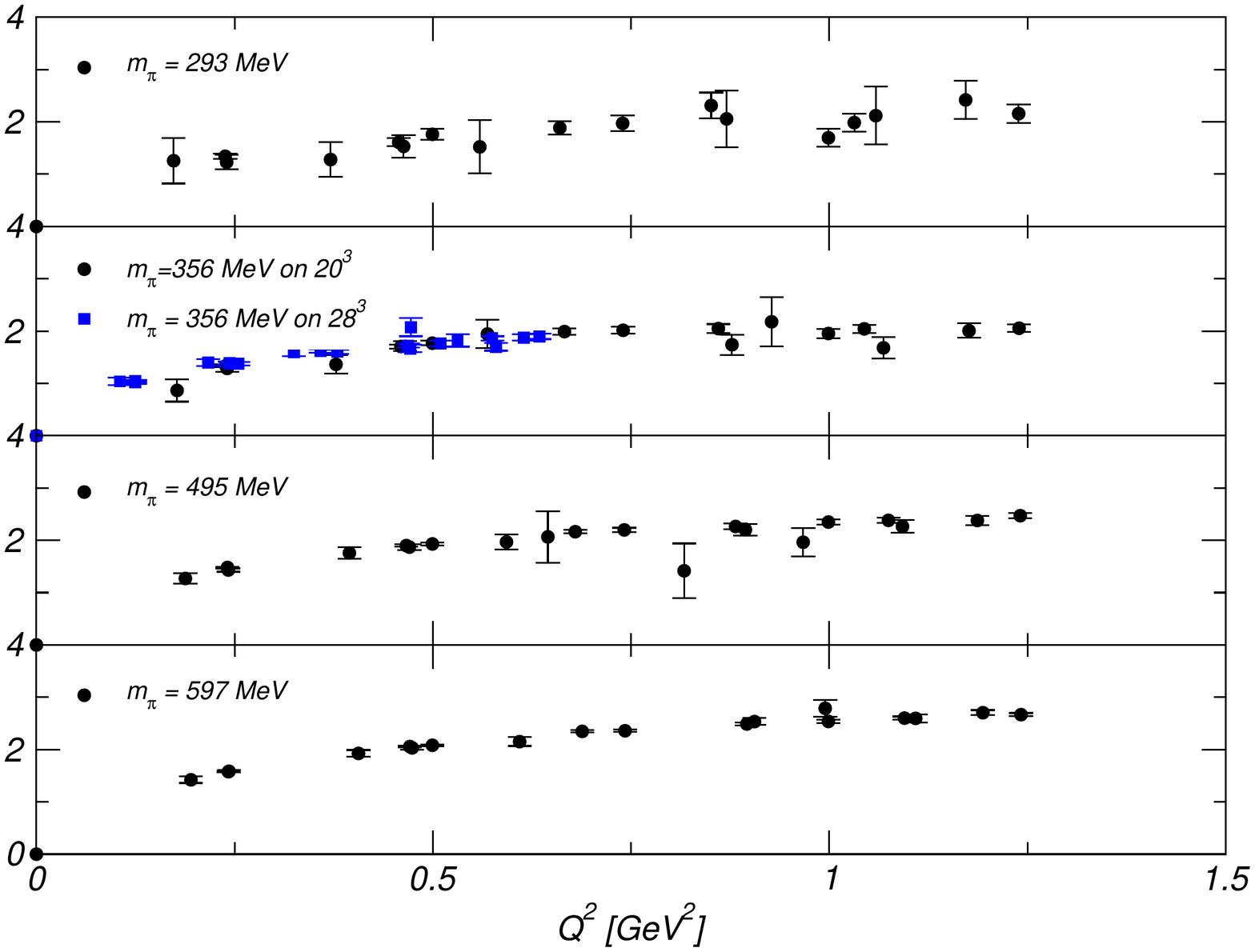}}
  \caption{Data for the ratio of $Q F_2^v(Q^2)/F_1^v(Q^2)$.}
  \label{fig:asymp-ratio}
\end{figure}

Figure~\ref{fig:asymp-ratio} shows our results for the ratio $Q
F_2^v(Q^2)/F_1^v(Q^2)$ for all available ensembles. The quantities
displayed show signs of saturation beyond $Q^2=1$GeV$^2$ for the
largest pion masses. For the smaller pion masses saturation is not
achieved conclusively, but we observe evidence of a pion-mass
dependence of the ratio. Unfortunately, we do not have data at
sufficiently high momentum transfer on the $28^3$ lattice at
$\mps=356$MeV to reach the scaling region of interest.

To conclude, we find that the picture obtained from spin-transfer
experiments is qualitatively reproduced in our lattice data. The
residual pion mass dependence prohibits a more precise quantitative
analysis. New techniques are probably needed to establish the relative
scaling of the form factors with confidence.

\subsubsection{\label{sec:isovector-sachs-form}Isovector Sachs form
  factors}
In the previous Secs.~\ref{sec:isovector-dirac-form}
and~\ref{sec:isovector-pauli-form} we have discussed the Dirac and
Pauli form factors, $F_1^v(Q^2)$ and $F_2^v(Q^2)$ and found them to
obey dipole forms very well over their entire parameter range. The
Sachs form factors, $G_E(Q^2)$ and $G_M(Q^2)$, are linear combinations
thereof and phenomenologically are usually fit using also dipole
forms. It is therefore important to see whether the lattice data
prefers the one or the other dipole fit scheme since the sum of
dipoles can only be approximately another dipole form.

To facilitate this study, we have subjected both $G_E(Q^2)$ and
$G_M(Q^2)$ to the same tests as previously applied to $F_1^v(Q^2)$
and $F_2^v(Q^2)$. The results are summarized in
Tabs.~\ref{tab:ge-dip-vary-t} and~\ref{tab:gm-dip-vary-t}. They show
fit results obtained by applying various cuts in $Q^2$ to dipole fits
to $G_E(Q^2)$ and $G_M(Q^2)$, respectively. All tables correspond to
the case $\mps=356$MeV on the $28^3$ lattice.
\begin{table}[htb]
  \centering
  \begin{tabular}[c]{*{4}{c|}c}
    \hline\hline
    $[Q^2_{\mbox{\tiny min}}, Q^2_{\mbox{\tiny max}}]$ [GeV$^2$] &
    $\chi^2$/dof & $A_0$ & $M_d$ [GeV] & $\langle r_E^2\rangle$
    [fm$^2$] \\ \hline
    $[0,1.5]$   & 1.00 & 0.9992(20) & 1.029(11) & 0.4406(96) \\
    $[0,0.5]$   & 0.69 & 0.9994(22) & 1.035(12) & 0.4362(103) \\
    $[0,0.4]$   & 0.80 & 0.9997(22) & 1.035(12) & 0.4361(104) \\
    $[0,0.3]$   & 1.23 & 0.9999(22) & 1.035(12) & 0.4364(104) \\
    $[0,0.2]$   & 1.75 & 1.0005(23) & 1.041(13) & 0.4311(111) \\
    \hline
    $[0,1.5]$   & 1.00 & 0.9992(20) & 1.029(11) & 0.4406(96) \\
    $[0.1,1.5]$ & 0.93 & 1.0048(42) & 1.022(12) & 0.4474(108) \\
    $[0.2,1.5]$ & 0.82 & 1.0175(108) & 1.008(16) & 0.4600(150) \\
    $[0.3,1.5]$ & 0.81 & 1.0500(243) & 0.983(23) & 0.4835(222) \\
    $[0.4,1.5]$ & 1.23 & 1.0636(625) & 0.975(41) & 0.4915(412) \\
    \hline
    $[0.3,0.5]$ & 0.39 & 1.0295(649) & 1.003(58) & 0.4649(534) \\
    $[0.2,0.4]$ & 0.40 & 1.0061(151) & 1.023(23) & 0.4461(205) \\
    $[0.1,0.3]$ & 1.06 & 1.0064(51) & 1.024(14) & 0.4455(125) \\
    \hline\hline
  \end{tabular}
  \caption{Dipole fits to $G_E(Q^2)$ with varying fit intervals.}
  \label{tab:ge-dip-vary-t}
\end{table}

\begin{table}[htb]
  \centering
  \begin{tabular}[c]{*{4}{c|}c}
    \hline\hline
    $[Q^2_{\mbox{\tiny min}}, Q^2_{\mbox{\tiny max}}]$ [GeV$^2$] &
    $\chi^2$/dof & $A_0$ & $M_d$ [GeV] & $\langle r_M^2\rangle$
    [fm$^2$] \\ \hline
    $[0,1.5]$   & 1.07 & 4.068(69) & 1.127(15) & 0.368(10) \\
    $[0,0.5]$   & 1.21 & 4.082(73) & 1.126(18) & 0.369(12) \\
    $[0,0.4]$   & 1.33 & 4.041(77) & 1.148(26) & 0.355(16) \\
    $[0,0.3]$   & 1.79 & 4.051(81) & 1.134(29) & 0.363(19) \\
    \hline
    $[0.2,1.5]$ & 1.15 & 4.064(73) & 1.127(16) & 0.368(11) \\
    $[0.3,1.5]$ & 1.55 & 4.085(121) & 1.120(29) & 0.372(20) \\
    $[0.4,1.5]$ & 1.51 & 3.716(234) & 1.208(66) & 0.320(35) \\
    \hline
    $[0.3,0.5]$ & 1.79 & 4.837(415) & 0.966(68) & 0.501(70) \\
    $[0.2,0.4]$ & 1.70 & 4.010(915) & 1.158(33) & 0.348(20) \\
    $[0.1,0.3]$ & 1.79 & 4.051(807) & 1.134(29) & 0.363(19) \\
    \hline\hline
  \end{tabular}
  \caption{Dipole fits to $G_M(Q^2)$ with varying fit intervals.}
  \label{tab:gm-dip-vary-t}
\end{table}

\begin{figure}[htb]
  \centering
  \includegraphics[scale=0.3,clip=true]{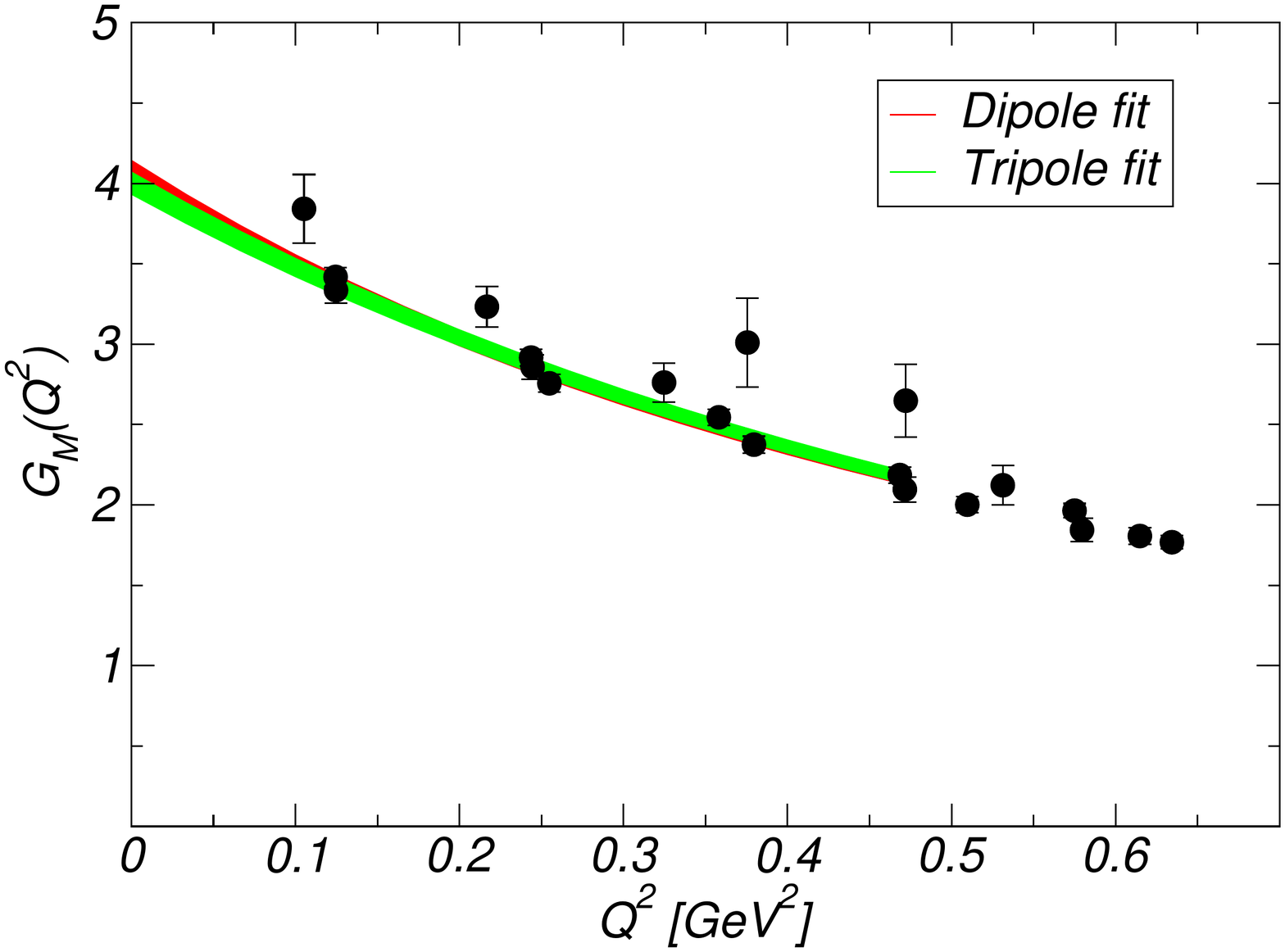}
  \caption{Comparison between dipole and tripole fit for isovector
    $G_M(Q^2)$.}
  \label{fig:sachs-dip-tri-comp}
\end{figure}

We also compared the dipole vs.~the tripole form for $G_M(Q^2)$. The
results from the fits are shown in
Tab.~\ref{tab:sachs-dip-tri-comp}. Figure~\ref{fig:sachs-dip-tri-comp}
compares the results graphically. Similar to the case of the Pauli
form factor we do not find a favorite fitting function.
\begin{table}[htb]
  \centering
  \begin{tabular}[c]{*{3}{c}}
    \hline\hline
                      & Dipole      & Tripole \\ \hline
    $\chi^2$/dof    & 1.07        & 1.28 \\
    $A_0$             & 4.0682(694) & 3.9976(671) \\
    $M_{d/t}$ [GeV]    & 1.127(15)   & 1.444(18) \\
    $\langle r_M^2\rangle$ [fm$^2$] & 0.368(10) & 0.336(86) \\
    \hline\hline
  \end{tabular}
  \caption{Comparison between dipole and tripole fit for isovector
    $G_M(Q^2)$.}
  \label{tab:sachs-dip-tri-comp}
\end{table}

Finally, we apply the dipole fits to all available data sets and
extract the radii, $\langle r_E\rangle$ and $\langle r_M\rangle$. The
results are shown in Tabs.~\ref{tab:ge-dip-allmpi}
and~\ref{tab:gm-dip-allmpi}.
\begin{table}[htb]
  \centering
  \begin{tabular}[c]{*{4}{c|}c}
    \hline\hline
    $\mps$ [MeV] & $\chi^2$/dof & $A_0$ & $M_d$ [GeV] & $\langle
    r_E^2\rangle$ [fm$^2$] \\ \hline
    293 & 0.77 & 1.0008(74) & 1.014(20) & 0.4544(180) \\
    356 on $28^3$ & 1.00 & 0.9992(20) & 1.029(11) & 0.4406(96) \\
    356 on $20^3$ & 1.66 & 0.9998(15) & 1.060(11) & 0.4158(86) \\
    495 & 1.43 & 0.9999(9) & 1.106(7) & 0.3818(46) \\
    597 & 4.02 & 0.9999(5) & 1.145(4) & 0.3562(28) \\
    \hline\hline
  \end{tabular}
  \caption{Dipole fits to $G_E(Q^2)$ for all data sets.}
  \label{tab:ge-dip-allmpi}
\end{table}
\begin{table}[htb]
  \centering
  \begin{tabular}[c]{*{4}{c|}c}
    \hline\hline
    $\mps$ [MeV] & $\chi^2$/dof & $A_0$ & $M_d$ [GeV] & $\langle
    r_M^2\rangle$ [fm$^2$] \\ \hline
    293 & 0.93 & 3.93831(1842) & 1.133(44) & 0.364(28) \\
    356 on $28^3$ & 1.07 & 4.0682(694) & 1.127(15) & 0.368(10) \\
    356 on $20^3$ & 2.11 & 3.9854(826) & 1.180(20) & 0.336(11) \\
    495 & 1.13 & 4.2913(471) & 1.196(11) & 0.3266(62) \\
    597 & 1.14 & 4.4684(291) & 1.236(8) & 0.3060(38) \\
    \hline\hline
  \end{tabular}
  \caption{Dipole fits to $G_M(Q^2)$ for all data sets.}
  \label{tab:gm-dip-allmpi}
\end{table}
It turns out that a dipole fit to $G_E(Q^2)$ has a slightly smaller
$\chi^2$/dof compared to a fit to $F_1^v(Q^2)$. On the other hand,
fitting $G_M(Q^2)$ with a dipole or a tripole is no better than
fitting $F_2^v(Q^2)$. In both cases the fits do not exhibit a clear
preference for either a dipole or a tripole function. Similar to the
fits to $F_1^v(Q^2)$ and the case of $F_2^v(Q^2)$ we observe
finite-size effects when comparing the two volumes at $20^3$ and
$28^3$. However, in the case of $G_E(Q^2)$ they appear to be less
pronounced, even though the $\chi^2$/dof is slightly smaller.

\subsubsection{\label{sec:flavor-dependence}Flavor dependence}
\begin{figure}[htb]
  \centering
  \includegraphics[scale=0.5,clip=true]{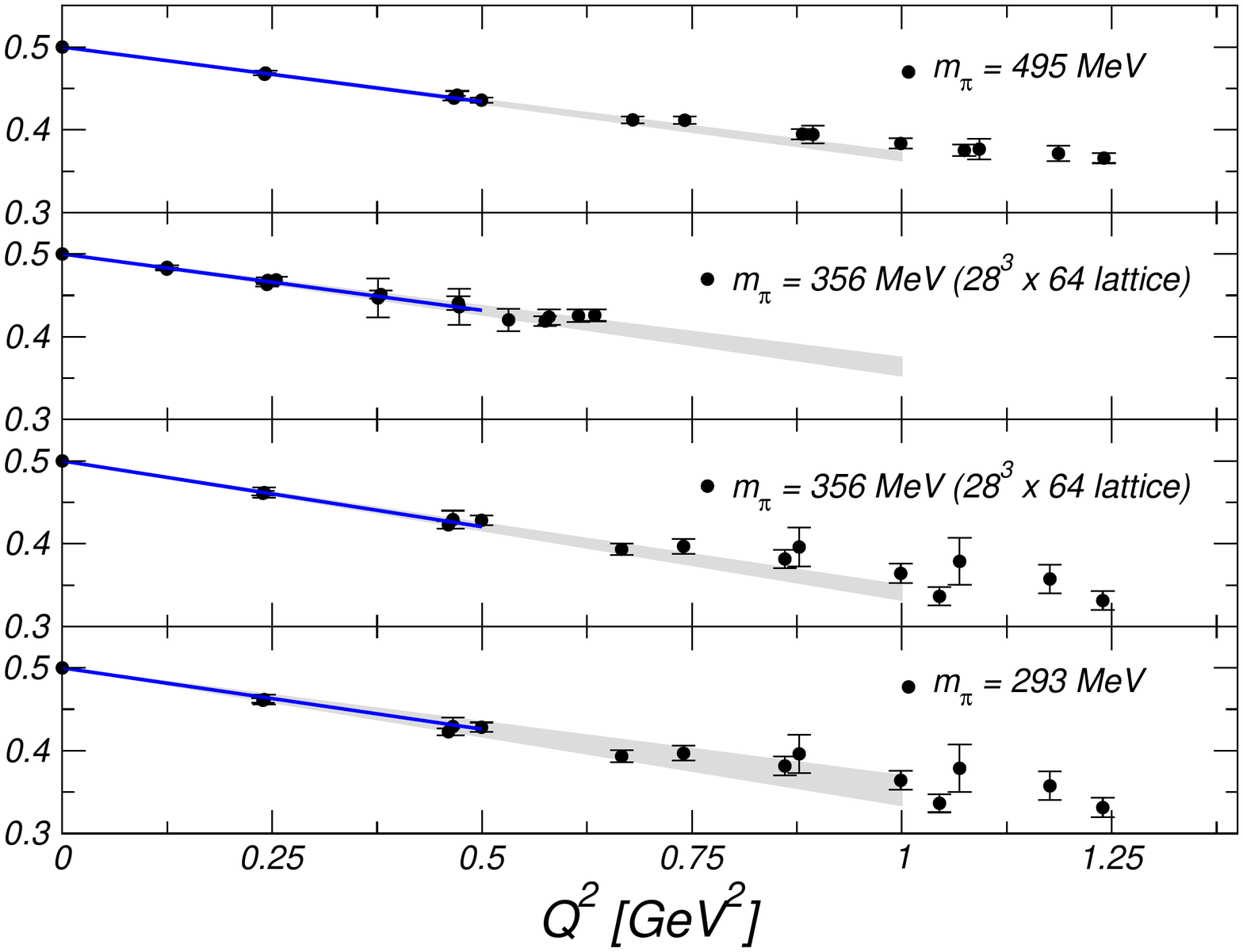}
  \caption{Linear fits to $F_1^d/F_1^u(Q^2)$ ratio.}
  \label{fig:a10-ratio-linear}
\end{figure}

We investigated the flavor dependence of the connected part of the
$F_1(Q^2)$ form factor by studying the ratio of
$F_1^d(Q^2)/F_1^u(Q^2)$. Experimentally, this ratio is being
scrutinized currently, see Ref.~\cite{JLABE02013:2009lnk} for the
website of the experiment. Since the forward value of the ratio is
trivially determined by the number of quarks, the interesting quantity
is the slope, $s(\mps)$, w.r.t~$Q^2$ of the ratio. Before comparing to
experiment, this quantity needs to be chirally extrapolated and we
have adopted the form~\cite{Bratt:2009ph}
\begin{equation}
  \label{eq:slope-extrap}
  s(\mps) = k_1 - k_2 \log\left(\frac{\mps^2}{(4\pi\fps)^2}\right)\,,
\end{equation}
with two generic parameters, $k_1$ and
$k_2$. Figure~\ref{fig:a10-ratio-linear} shows the lattice
calculations of the ratios at the three lightest quark masses. The
resulting slopes obtained from linear fits to these data are
summarized in Tab.~\ref{tab:a10-ratio-linear}.
\begin{table}[htb]
  \centering
  \begin{tabular}[c]{c|c}
    \hline\hline
    $\mps$ [MeV] & Slope [GeV$^{-2}$] \\ \hline
    495 & -0.132(5) \\
    356 on $28^3$ & -0.136(11) \\
    356 on $20^3$ & -0.159(9) \\
    293 & -0.148(18) \\ \hline
    $\mps^{\rm phys}=139$ & -0.190(26) \\
    \hline\hline
  \end{tabular}
  \caption{Slopes at $Q^2=0$GeV$^2$ of $F_1^d/F_1^u(Q^2)$ from linear
    fits. The last line shows the chirally extrapolated value.}
  \label{tab:a10-ratio-linear}
\end{table}

\begin{figure}[htb]
  \centering
  \includegraphics[scale=0.3,clip=true]{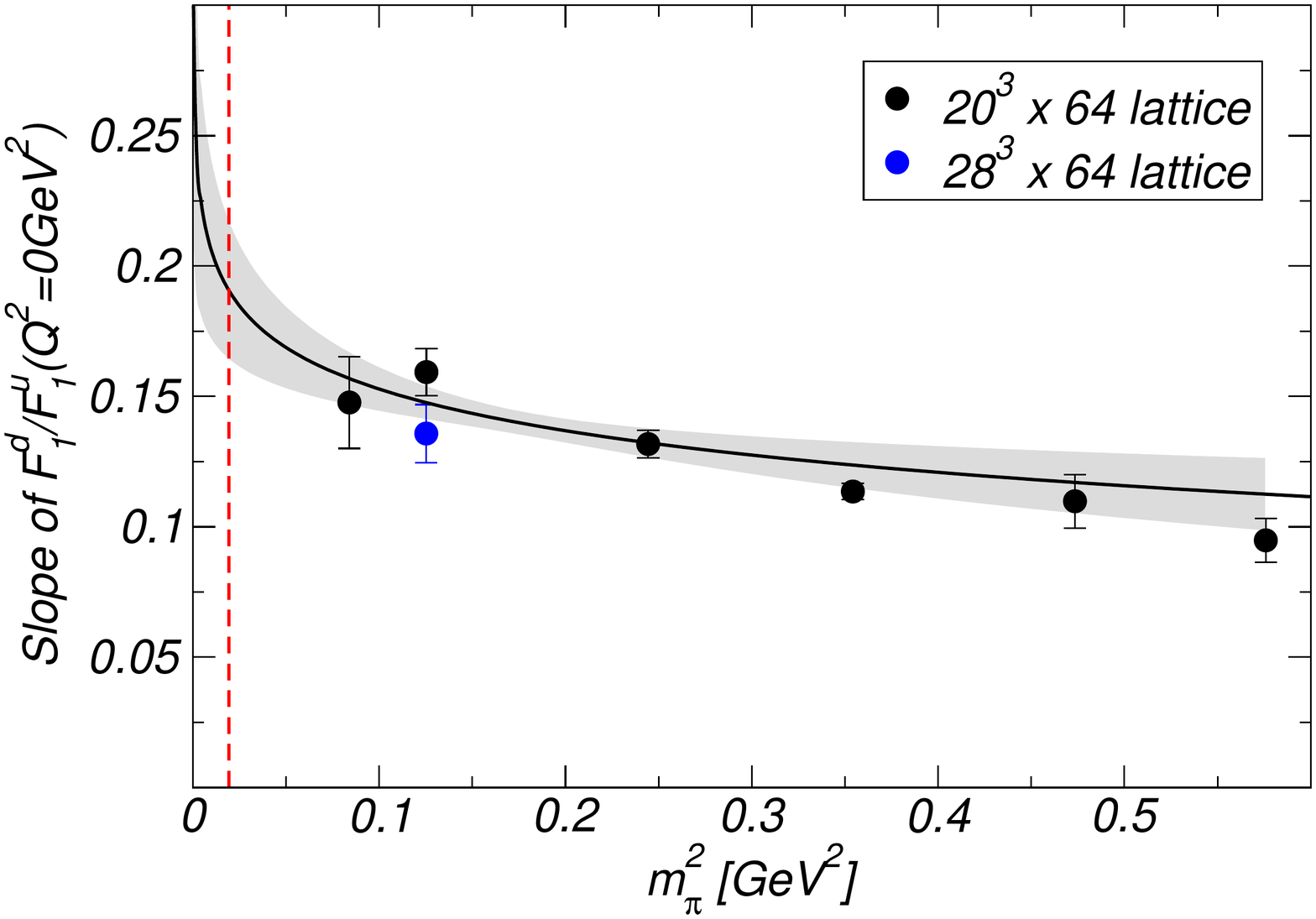}
  \caption{Chiral fit to the slope of the ratio $F_1^d/F_1^u(Q^2=0{\rm
      GeV}^2)$. The physical pion mass is marked by the dashed red
    line.}
  \label{fig:fit-ratio-slope}
\end{figure}

We observe that the lattice data are quite linear for $Q^2<0.5$GeV$^2$
and approximately linear up to $1$GeV$^2$. Since the leading chiral
expansion Eq.~\eqref{eq:slope-extrap} only tells us how to extrapolate
the slope to the physical pion mass, we only fit the slope at low
$Q^2$ values in these fits. Figure~\ref{fig:fit-ratio-slope} shows the
chiral extrapolation of the slope displayed in
\fig\ref{fig:a10-ratio-linear} to the physical pion mass, denoted by
the vertical dashed line. The leading chiral singularity is $\log
\mps$, cf.~\sect\ref{sec:isovector-dirac-form}, which explains the
divergence of the curve as the pion mass goes to zero. The error band
propagates the statistical errors to the physical mass, and yields the
final result for the slope. This final number is shown in the last row
of Tab.~\ref{tab:a10-ratio-linear}. The fit parameters are summarized
in Tab.~\ref{tab:fit-ratio-slope}.
\begin{table}[htb]
  \centering
  \begin{tabular}[c]{cc}
    \hline\hline
    $\chi^2$/dof & 1.5 \\
    $k_1$ & 0.093(23) GeV$^{-2}$ \\
    $k_2$ & 0.023(11) GeV$^{-2}$ \\ \hline
    $s(\mps^{\rm phys})$ & -0.190(26) GeV$^{-2}$ \\
    \hline\hline
  \end{tabular}
  \caption{Fit parameters to the slope of the ratio $F_1^d/F_1^u(Q^2)$
    as a function of the pion mass.}
  \label{tab:fit-ratio-slope}
\end{table}

\begin{figure}[htb]
  \centering
  \includegraphics[scale=0.3,clip=true]{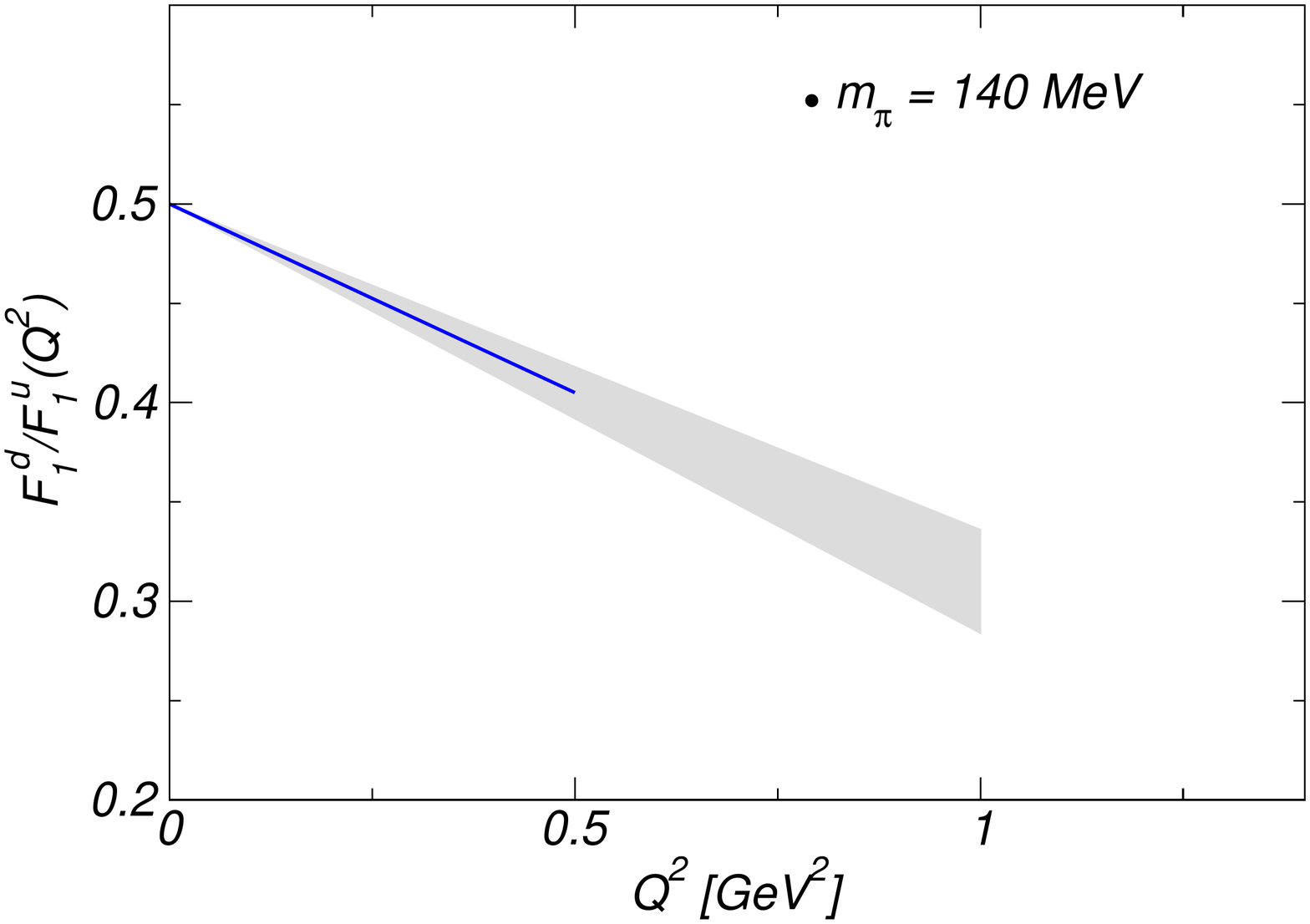}
  \caption{Linear slope of $F_1^d/F_1^u(Q^2)$ at the physical pion mass.}
  \label{fig:ratio-extrap}
\end{figure}

The prediction that can be meaningfully compared to experiment is
shown in \fig\ref{fig:ratio-extrap}. The linear behavior of
$F_1^d/F_1^u$ as a function of $Q^2$ departing from $Q^2=0$,
extrapolated to the physical pion mass is plotted, including the
one-sigma error band.

\subsubsection{\label{sec:isosc-form-fact}Isoscalar form factors}
We now turn our attention to the isoscalar form factors. We remind the
reader that we have only computed the connected Wick contractions, and
that the disconnected diagrams remain to be calculated in the
future. In the following we discuss the isoscalar form factors,
$F_1^s(Q^2)$ and $F_2^s(Q^2)$ separately.

\paragraph{Isoscalar form factor $F_1^s(Q^2)$}
We first apply the well-known dipole fits again to the isoscalar form
factor $F_1^s(Q^2)$. First, we study the stability of the dipole fits
to variations in the fit interval like we did before. We focus again
on the ensemble with $\mps=356$MeV on the $28^3$
lattice. Table~\ref{tab:f1s-dip-vart} summarizes our findings. It is
evident that the dipole provides an excellent, stable fit over the
entire available data range.
\begin{table}[htb]
  \centering
  \begin{tabular}[c]{*{4}{c|}c}
    \hline\hline
    $[Q^2_{\mbox{\tiny min}}, Q^2_{\mbox{\tiny max}}]$ [GeV$^2$] &
    $\chi^2$/dof & $A_0$ & $M_d$ [GeV] & $\langle r_1^2\rangle$
    [fm$^2$] \\ \hline
    $[0,1.5]$   & 1.20 & 3.0511(27) & 1096.9(62) & 0.3883(27) \\
    $[0,0.5]$   & 0.77 & 3.0518(27) & 1098.6(65) & 0.3871(46) \\
    $[0,0.4]$   & 0.90 & 3.0521(28) & 1099.0(66) & 0.3868(46) \\
    $[0,0.3]$   & 0.91 & 3.0516(29) & 1070.0(67) & 0.3883(47) \\
    $[0,0.2]$   & 0.76 & 3.0520(29) & 1098.1(68) & 0.3875(48) \\
    \hline
    $[0,1.5]$   & 1.20 & 3.0511(27) & 1096.9(62) & 0.3883(27) \\
    $[0.1,1.5]$ & 1.18 & 3.0587(66) & 1090.3(81) & 0.3931(59) \\
    $[0.2,1.5]$ & 1.09 & 3.0356(14) & 1103.4(109) & 0.3838(76) \\
    $[0.3,1.5]$ & 0.57 & 3.0141(306) & 1107.9(151) & 0.3807(104) \\
    $[0.4,1.5]$ & 0.64 & 3.0522(799) & 1097.8(248) & 0.3877(175) \\
    \hline
    $[0.3,0.5]$ & 0.30 & 3.0309(963) & 1102.9(375) & 0.3842(260) \\
    $[0.2,0.4]$ & 1.04 & 3.0430(204) & 1102.2(135) & 0.3853(98) \\
    $[0.1,0.3]$ & 0.85 & 3.0589(73) & 1090.7(88) & 0.3928(63) \\
    \hline\hline
  \end{tabular}
  \caption{Dipole fits to isoscalar $F_1^s(Q^2)$ with varying fit
    intervals.}
  \label{tab:f1s-dip-vart}
\end{table}

Applying dipole fits to all available pion masses yields the results
shown in Tab.~\ref{tab:f1s-dip-allmpi}.
\begin{table}[htb]
  \centering
  \begin{tabular}[c]{*{4}{c|}c}
    \hline\hline
    $\mps$ [MeV] & $\chi^2$/dof & $A_0$ & $M_d$ [GeV] & $\langle
    r_1^2\rangle$ [fm$^2$] \\ \hline
    293 & 1.08 & 3.0421(90) & 1086.6(129) & 0.3958(94) \\
    356 on $28^3$ & 1.20 & 3.0511(27) & 1096.9(62) & 0.3883(27) \\
    356 on $20^3$ & 2.46 & 3.0538(22) & 1107.3(68) & 0.3811(47) \\
    495 & 1.78 & 3.0541(15) & 1152.7(41) & 0.3516(25) \\
    597 & 1.27 & 3.0550(9) & 1201.3(29) & 0.3238(16) \\
    \hline\hline
  \end{tabular}
  \caption{Dipole fits to isoscalar $F_1^s(Q^2)$ for all data sets.}
  \label{tab:f1s-dip-allmpi}
\end{table}
We find that even at the lightest available pion mass the obtained
radii underestimate the empirical value $\langle r_1^2\rangle$ =
(0.782fm)$^2$ quoted from Ref.~\cite{Bernard:1998gv} substantially by
a factor of two. A similar discrepancy also appeared in the isovector
case, \sect\ref{sec:isovector-dirac-form}, but it seems to be even
more pronounced here. Another noteworthy fact is the inconsistency of
the forward value, $F_1^s(Q^2=0)$, with the expected value of
three. This is most likely a cutoff effect. Although the disagreement
is only mild (about $1\%$), it is outside the relative error bars. We
will comment on this below when discussing the chiral expansion.

Reference~\cite{Bernard:1998gv} discusses the chiral expansion in
$\mps$ and $Q^2$. The next-to-leading order SSE expression
\begin{equation}
  \label{eq:f1s-chiral}
  F_1^s(Q^2,\mps) = 3 - 12 \tilde{B}_1 \frac{Q^2}{(4\pi f_\pi)^2}\,,
\end{equation}
has only a linear $Q^2$ dependence and no $\mps$ dependence. As we
have pointed out before, in comparison with the isovector form factor
$F_1^v(Q^2,\mps)$, the derivative w.r.t.~$\mps$ of the isoscalar one
does not diverge in the limit $\mps\to 0$. It merely approaches a
constant value. Since the $Q^2$ dependence only has a linear part, we
need to again restrict ourselves to the regime of small values of
$Q^2$. Table~\ref{tab:f1s-dip-allmpi} suggests a non-trivial
$\mps$-dependence. However, we observe that the results for $\langle
r_1^2\rangle$ are still constant within statistics when we restrict
ourselves to the region $\mps<400$MeV. This cut is consistent with
what we have found before in the isovector case.

With the restriction of $\mps<400$MeV we perform a chiral fit based on
Eq.~\eqref{eq:f1s-chiral}. However, when covering the entire fit
interval, $Q^2=[0,0.5]$ GeV$^2$, we find an unacceptably large
$\chi^2$/dof=128. We traced the problem back to the value at the
origin, $F_1^s(Q^2=0)$. Due to the systematic shift with a small
relative error we end up with such a huge discrepancy. Hence, we
restrict ourselves to the fitting interval, $Q^2>0.01$GeV$^2$, which
excludes this data point. Varying the upper cut-off in $Q^2$ yields
the results tabulated in Tab.~\ref{tab:f1s-vart}.
\begin{table}[htb]
  \centering
  \begin{tabular}[c]{*{3}{c|}c}
    \hline\hline
    $Q^2$ max [GeV$^2$] & $\chi^2$/dof & $\tilde{B}_{1}$ & $\langle
    r_1^2\rangle$ [fm$^2$] \\
    \hline
    0.4 & 42.7 & 0.3304(23) & 0.2631(18) \\
    0.3 & 17.9 & 0.3493(25) & 0.2782(20) \\
    0.2 & 2.1  & 0.3878(47) & 0.3089(38) \\
    \hline\hline
  \end{tabular}
  \caption{Fits to isoscalar $F_1^s(Q^2,\mps)$ at varying intervals of
    $Q^2$ with fixed pion mass cut, $\mps<400$MeV.}
  \label{tab:f1s-vart}
\end{table}
Surprisingly, the radii are even smaller than those obtained from the
dipole expression. Furthermore, the overall quality of the fits is
poor until $Q^2 < 0.2$GeV$^2$ is imposed, at which point the resulting
radius is still about $20\%$ smaller than the one obtained from the
dipole form. We conclude that applying the NLO
expression~\eqref{eq:f1s-chiral} does not give new insight into the
problem.

\begin{figure}[htb]
  \centering
  \includegraphics[scale=0.3,clip=true]{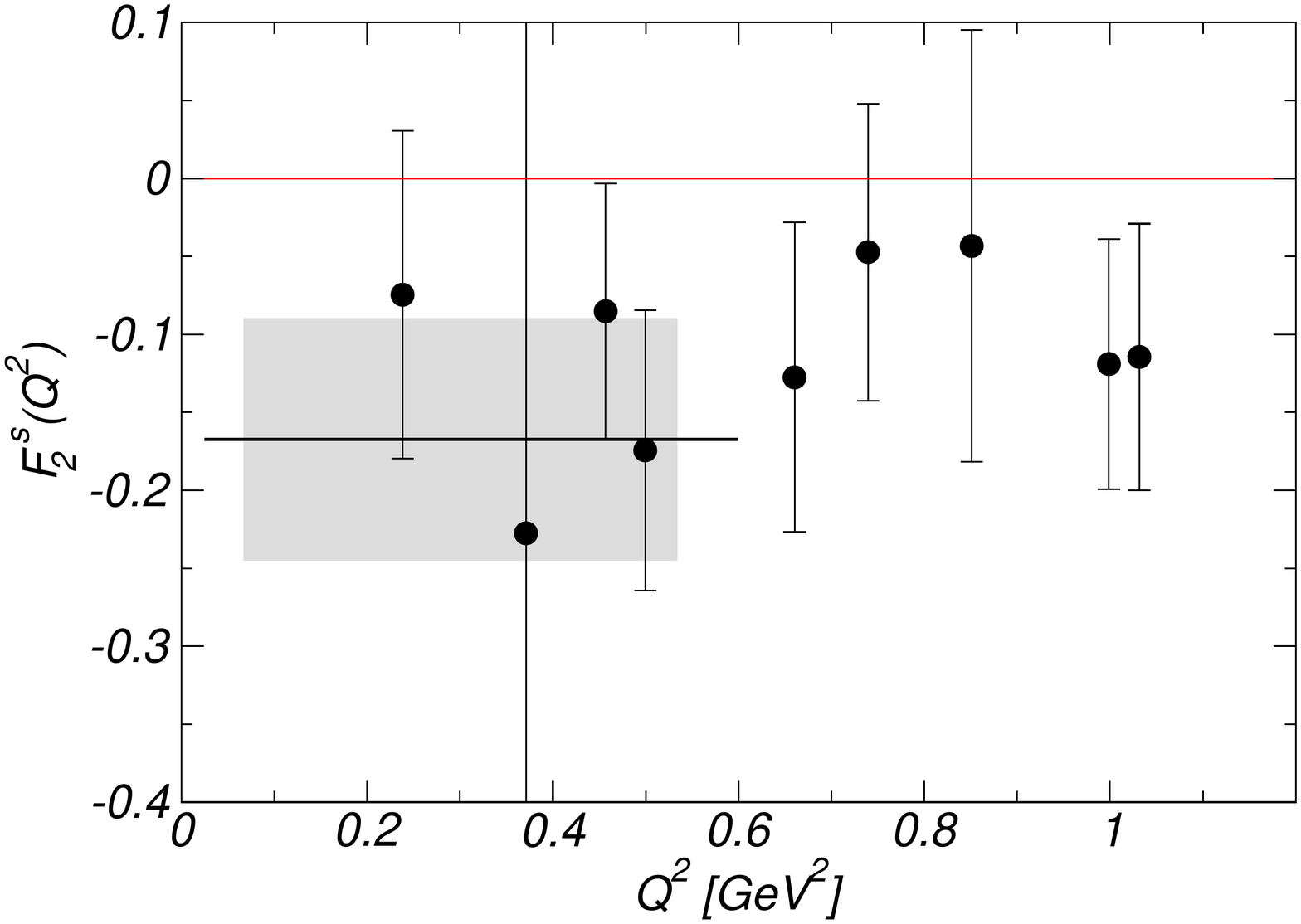}
  \caption{Isoscalar form factor $F_2^s(Q^2)$ as a function of $Q^2$
    with SSE fit at $\mps=293$MeV.}
  \label{fig:f2s-fixed}
\end{figure}

\paragraph{Isoscalar form factor $F_2^s(Q^2)$}
The chiral expansion from Ref.~\cite{Hemmert:2002uh} yields for the
isoscalar spin-flip form factor, $F_2^s(Q^2,\mps)$, an expansion of
the form
\begin{equation}
  \label{eq:f2s-sse}
  F_2^s(Q^2,\mps) = \kappa_s^0 - 8 m_N \mps^2 \tilde{E}_2\,,
\end{equation}
i.e., the leading $Q^2$ dependence vanishes and the $\mps$ dependence
is quadratic. Table~\ref{tab:f2s-sse} shows the result of a fit of
Eq.~\eqref{eq:f2s-sse} to our lattice data with the cuts $Q^2=[0,0.5]$
GeV$^2$ and $\mps<400$MeV. The experimental value for
$F_2^s(Q^2=0,\mps=\mps^{\rm phys})$ taken from
Ref.~\cite{Amsler:2008zzb} is listed on the last line.
\begin{table}[htb]
  \centering
  \begin{tabular}[c]{*{2}{c}}
    \hline\hline
    & Fit result \\ \hline
    $\chi^2$/dof & 1.34 \\
    $\kappa_s^0$ & -0.36(25) \\
    $\tilde{E}_2$ [GeV$^{-3}$] & $-0.32(29)$ \\
    Extrap.~$F_2^s(Q^2=0,\mps^{\rm phys})$ & -0.32(21) \\ \hline
    Exp.~$F_2^s(Q^2=0,\mps^{\rm phys})$ & -0.360586110(14) \\
    \hline\hline
  \end{tabular}
  \caption{Fit results to isoscalar form factor $F_2^s(Q^2)$ for the
    interval $Q^2=[0,0.5]$ GeV$^2$ and $\mps<400$MeV. The next-to-last
    line is our extrapolated result at the physical pion mass and the
    last line lists the experimental value for $F_2^s(Q^2=0,\mps^{\rm
      phys})$.}
  \label{tab:f2s-sse}
\end{table}
It is evident that the data suggests $\kappa_s^0<0$, but we can only
fix the order of magnitude since the relative error is about $50\%$.
When plotting the data for the ensemble at $\mps=293$MeV, we find that
the data is indeed flat, albeit with a large relative
error. Figure~\ref{fig:f2s-fixed} shows this graph with the fit
function from Tab.~\ref{tab:f2s-sse} for $F_2^s(Q^2,\mps=293{\rm
  MeV})$, which is constant w.r.t.~$Q^2$.

Note that the parameters are poorly determined, although the fit
quality as indicated by $\chi^2$/dof is acceptable. We can extract the
order of magnitude of both $\kappa_s^0$ and $\tilde{E}_2$, but with
large statistical uncertainty. On the other hand, we have found the
extrapolated value of $F_2^s(Q^2=0,\mps^{\rm phys})$ to be fully
compatible with experiment, having both the correct order of magnitude
and the correct sign. We point out again that the forward value will
not receive disconnected contributions, thus the value quoted
corresponds directly to the experiment.

\subsubsection{\label{sec:summ-electr-form}Summary of electromagnetic
  form factors}
For the isovector form factors of the vector current we find that the
lattice mean squared radii at present pion masses are significantly
below the empirical ones. This finding is qualitatively compatible
with chiral perturbation theory, though, since the latter predicts a
sharp increase of the mean squared radii as the pion mass approaches
the chiral limit. Although our data is consistent with one-loop
expressions from both SSE and BChPT, we still require data at smaller
pion masses and larger volumes to successfully compare with
phenomenology and make predictions in the chiral limit with negligible
systematic errors. For the isoscalar form factors of the vector
currents we find our results to be in agreement with expectations from
chiral expansions --- the connected part of $F_1^s(Q^2)$ has a
non-vanishing $Q^2$-dependence, while the $Q^2$-dependence of
$F_2^s(Q^2)$ was not measurable. The overall magnitude of
$F_2^s(Q^2=0)$ is well in agreement with empirical information, albeit
with large uncertainty.

\subsection{\label{sec:axial-form-factors}Axial form factors}
The nucleon axial current matrix element is also expressed in terms of
two form factors: The axial form factor, $G_A(Q^2)$, and the induced
pseudoscalar form factor, $G_P(Q^2)$. They correspond to the
generalized form factors $\tilde{A}_{10}(Q^2)$ and
$\tilde{B}_{10}(Q^2)$ in Eq.~\eqref{eq:lattice-ax-gff-def}. For a
review of the current experimental and theoretical understanding of
the axial structure of the nucleon, we refer to
Ref.~\cite{Bernard:2001rs}. For more details on the chiral effective
field theory expressions that we will use in our fits, see
Ref.~\cite{Bernard:1998gv}. For lattice results from other groups, see
Ref.~\cite{Gockeler:2007hj}.

Like the vector current, the axial current needs to be
renormalized. The renormalization coefficient can be computed from the
conserved axial current, see Ref.~\cite{Blum:2000kn}. Defining the
two-point functions $C(T)$ and $L(T)$ of the conserved and local
currents via
\begin{eqnarray}
  \label{eq:axial-curr}
  C(T+1/2) &=& \sum_{\vec{x}} \langle {\cal A}_0(\vec{x},T+1/2)
  \pi(\vec{0},0)\rangle\,, \nonumber \\
  L(T) &=& \sum_{\vec{x}} \langle {A}_0(\vec{x},T)
  \pi(\vec{0},0)\rangle\,,
\end{eqnarray}
the axial renormalization constant, $Z_A$, can then be computed from
\begin{equation}
  \label{eq:axial-za}
  Z_A(T) = \frac{1}{2} \left( \frac{C(T+1/2)+C(T-1/2)} {2L(T)} +
    \frac{2C(T+1/2)} {L(T)+L(T+1)}\right)\,.
\end{equation}
For large $T\gg 1$, the ratio in Eq.~\eqref{eq:axial-za} becomes the
axial renormalization constant, $Z_A\equiv
\lim_{T\to\infty}Z_A(T)$. Based on the ensembles listed in
Tab.~\ref{tab:data-sets} we have used the ranges $T=[12,29]$ and
$T=[35,52]$ to obtain our final values of the renormalization
constants. Table~\ref{tab:renorm-const-ax} summarizes our findings.
\begin{table}[htb]
  \centering
  \begin{tabular}[c]{*{2}{c|}c}
    \hline\hline
    $m_{\mbox{\tiny sea}}^{\mbox{\tiny asqtad}}$ & Volume & $Z_A$ \\
    \hline
    0.007/0.050 & $20^3\times 64$ & 1.0816 \\
    0.010/0.050 & $28^3\times 64$ & 1.0850 \\
    0.010/0.050 & $20^3\times 64$ & 1.0849 \\
    0.020/0.050 & $20^3\times 64$ & 1.0986 \\
    0.030/0.050 & $20^3\times 64$ & 1.1090 \\
    \hline\hline
  \end{tabular}
  \caption{Renormalization constant of the axial currents.}
  \label{tab:renorm-const-ax}
\end{table}
We thus obtain a quark-mass dependent $Z_A$ factor. Since all the
ensembles considered here are at the same value of the lattice spacing
$a$, the differences between the $Z_A$ values in
Tab.~\ref{tab:renorm-const-ax} are entirely due to quark mass
effects. To leading order these effects are expected to be linear in
$am_{\rm q}$~\cite{Luscher:1996sc}, and it is important to keep at
least this leading quark-mass dependence to avoid introducing ${\cal
  O}(a)$ discretization errors in the matrix elements of $A_0$.

We will now review the isovector axial form factor, $G_A(Q^2)$, in
\sect\ref{sec:isovector-axial-form}, the induced pseudoscalar form
factor, $G_P(Q^2)$, in \sect\ref{sec:isov-pseud-form}, and the two
isoscalar axial form factors $\tilde{A}_{10}(Q^2)$ and
$\tilde{B}_{10}(Q^2)$ in
\sect\ref{sec:isoscalar-axial-form}. Section~\ref{sec:summary-axial-form}
summarizes our findings.

\subsubsection{\label{sec:isovector-axial-form}Isovector axial form
  factor $G_A(Q^2)$}
The isovector axial form factor is usually fit by a dipole form. The
forward value, $g_A = G_A(Q^2=0)$ has been discussed in detail in
\sect\ref{sec:gA}. The dipole mass and its relevance to experiment has
been studied in detail in
Ref.~\cite{Bernard:2001rs}. Table~\ref{tab:ga-dip-vart} shows results
of a series of fits to $G_A(Q^2)$ on the $28^3$ lattice at
$\mps=356$MeV. It is evident that the dipole form provides a good fit
to the lattice data at all available values of the virtuality, $Q^2$.
\begin{table}[htb]
  \centering
  \begin{tabular}[c]{*{4}{c|}c}
    \hline\hline
    $[Q^2_{\mbox{\tiny min}}, Q^2_{\mbox{\tiny max}}]$ [GeV$^2$] &
    $\chi^2$/dof & $g_A$ & $M_d$ [GeV] & $\langle r_A^2\rangle$
    [fm$^2$] \\ \hline
    $[0,1.5]$   & 1.70 & 1.1245(147) & 1.587(29) & 0.1856(67) \\
    $[0,0.5]$   & 1.77 & 1.1334(158) & 1.545(31) & 0.1956(79) \\
    $[0,0.4]$   & 2.22 & 1.1343(160) & 1.536(34) & 0.1980(89) \\
    $[0,0.3]$   & 1.71 & 1.1331(161) & 1.550(39) & 0.1945(97) \\
    $[0,0.2]$   & 2.25 & 1.1308(163) & 1.508(47) & 0.2056(13) \\
    \hline
    $[0,1.5]$   & 1.70 & 1.1245(147) & 1.587(29) & 0.1856(67) \\
    $[0.1,1.5]$ & 1.79 & 1.1247(150) & 1.587(29) & 0.1856(68) \\
    $[0.2,1.5]$ & 1.43 & 1.1303(168) & 1.584(33) & 0.1862(78) \\
    $[0.3,1.5]$ & 0.53 & 1.1123(228) & 1.598(50) & 0.1830(115) \\
    \hline
    $[0.3,0.5]$ & 0.54 & 1.1005(642) & 1.617(17) & 0.1787(369) \\
    $[0.2,0.4]$ & 1.71 & 1.1533(209) & 1.496(49) & 0.2088(136) \\
    $[0.1,0.3]$ & 2.01 & 1.1305(171) & 1.563(49) & 0.1913(120) \\
    \hline\hline
  \end{tabular}
  \caption{Dipole fits to isovector $G_A(Q^2)$ with varying fit
    intervals.}
  \label{tab:ga-dip-vart}
\end{table}
It turns out that the axial mass at $\mps=356$MeV is about $50\%$
larger than the phenomenological value from neutrino scattering, $M_d
= (1.026\pm 0.021)$ GeV, and the one from electro-production, $M_d =
(1.069\pm 0.016)$ GeV, see Ref.~\cite{Bernard:2001rs}. Hence, the
situation is qualitatively similar to the case of the form factors of
the vector current.

For the other ensembles, we obtain the results from the dipole fits at
fixed pion mass summarized in Tab.~\ref{tab:ga-dip-allmpi}. The data
has been fit to all available $Q^2$. The large offset between lattice
and phenomenology exhibited by the dipole masses, and thus also by the
radii, remains to be explained.
\begin{table}[htb]
  \centering
  \begin{tabular}[c]{*{4}{c|}c}
    \hline\hline
    $\mps$ [MeV] & $\chi^2$/dof & $A_0$ & $M_d$ [GeV] & $\langle
    r_A^2\rangle$ [fm$^2$] \\ \hline
    293 & 0.80 & 1.154(26) & 1.577(56) & 0.1879(134) \\
    356 on $28^3$ & 1.70 & 1.125(15) & 1.587(29) & 0.1856(67) \\
    356 on $20^3$ & 1.30 & 1.144(15) & 1.661(33) & 0.1694(68) \\
    495 & 1.25 & 1.142(77) & 1.654(16) & 0.1708(34) \\
    597 & 1.05 & 1.146(46) & 1.686(11) & 0.1644(21) \\
    \hline\hline
  \end{tabular}
  \caption{Dipole fits to isovector $G_A(Q^2)$ for all data sets.}
  \label{tab:ga-dip-allmpi}
\end{table}

We also study $G_A(Q^2,\mps)$ using the chiral expansion from
Ref.~\cite{Bernard:1998gv}. At this order, the $Q^2$-dependence is
linear and contains a counterterm which is directly related to the
mean squared axial radius, $\langle r_A^2\rangle$. The pion mass
dependence is encoded in $g_A(\mps)$:
\begin{equation}
  \label{eq:ga-sse}
  G_A(Q^2,\mps) = g_A(\mps) - \frac{Q^2}{(4\pi f_\pi)^2} \tilde{B}_3\,.
\end{equation}
For the expression of $g_A(\mps)$ see Eq.~\eqref{eq:axial-expand}. In
the following we have performed fits of the $Q^2$-dependence of $G_A$
at fixed pion masses and extracted the mean squared axial radii by
taking
\begin{equation}
  \label{eq:ra-extract}
  \langle r_A^2\rangle = \frac{-6}{g_A(\mps)} \left. \frac{{\rm d}
      G_A(Q^2, \mps)}{{\rm d} Q^2} \right\vert_{Q^2=0} \,.
\end{equation}
From the resulting series of fits we attempt to extract a sensible
interval for the fitting range of $Q^2$. First we determine the upper
cut-off in $Q^2$ on the $28^3$ lattice. The results are listed in
Tab.~\ref{tab:ga-sse-vart}.
\begin{table}[htb]
  \centering
  \begin{tabular}[c]{*{3}{c|}c}
    \hline\hline
    $Q^2$ max [GeV$^2$] & $\chi^2$/dof &
    $\tilde{B}_3$ & $\langle r_A^2\rangle(\mps=356{\rm MeV})$ [fm$^2$]
    \\ \hline
    1.5 & 4.82 & 0.725(23) & 0.1269(38) \\
    0.5 & 2.66 & 0.848(33) & 0.1485(46) \\
    0.4 & 2.67 & 0.901(39) & 0.1577(57) \\
    0.3 & 2.52 & 0.949(47) & 0.1660(71) \\
    0.2 & 1.50 & 1.094(70) & 0.1914(11) \\
    \hline\hline
  \end{tabular}
  \caption{Chiral fit to $G_A(Q^2,\mps=356{\rm MeV})$ with varying
    upper cut-off in $Q^2$.}
  \label{tab:ga-sse-vart}
\end{table}
Since the fit expression is only linear, the exclusion of points at
larger $Q^2$ does indeed improve the quality of the fit, although the
overall result is not as satisfactory as for the dipole fit. Both for
$g_A$ and for $\langle r_A^2\rangle$ we find that the chiral fit lies
below the results from the dipole fits. Furthermore, the fits in this
section possess larger uncertainties and the resulting $\chi^2$/dof is
worse.

Now, we apply the SSE fit simultaneously to the entire available data
set. From the previous section we have learned that $Q^2=[0,0.4]$
GeV$^2$ is a sensible fitting interval which we keep in the
following. When varying the upper cut-off in $\mps$, we obtain the fit
results in Tab.~\ref{tab:ga-chiral-varmpi}.
\begin{table}[htb]
  \centering
  \begin{tabular}[c]{*{3}{c|}c}
    \hline\hline
    $\mps$ max [MeV] & $\chi^2$/dof & $\tilde{B}_3$ & $\langle
    r_A^2\rangle(\mps=293{\rm MeV})$ [fm$^2$] \\ \hline
    300 & 0.71 & 0.993(130) & 0.1707(193) \\
    400 & 1.73 & 0.908(32)  & 0.1560(60)  \\
    500 & 1.54 & 0.891(20)  & 0.1531(49)  \\
    600 & 1.76 & 0.847(12) & 0.1457(42) \\
    \hline\hline
  \end{tabular}
  \caption{Chiral fits to $G_A(Q^2)$ at fixed interval $Q^2=[0,0.4]$
    GeV$^2$ with varying upper cut-off in $\mps$.}
  \label{tab:ga-chiral-varmpi}
\end{table}

\begin{figure}[htb]
  \centering
  \includegraphics[scale=0.3,clip=true]{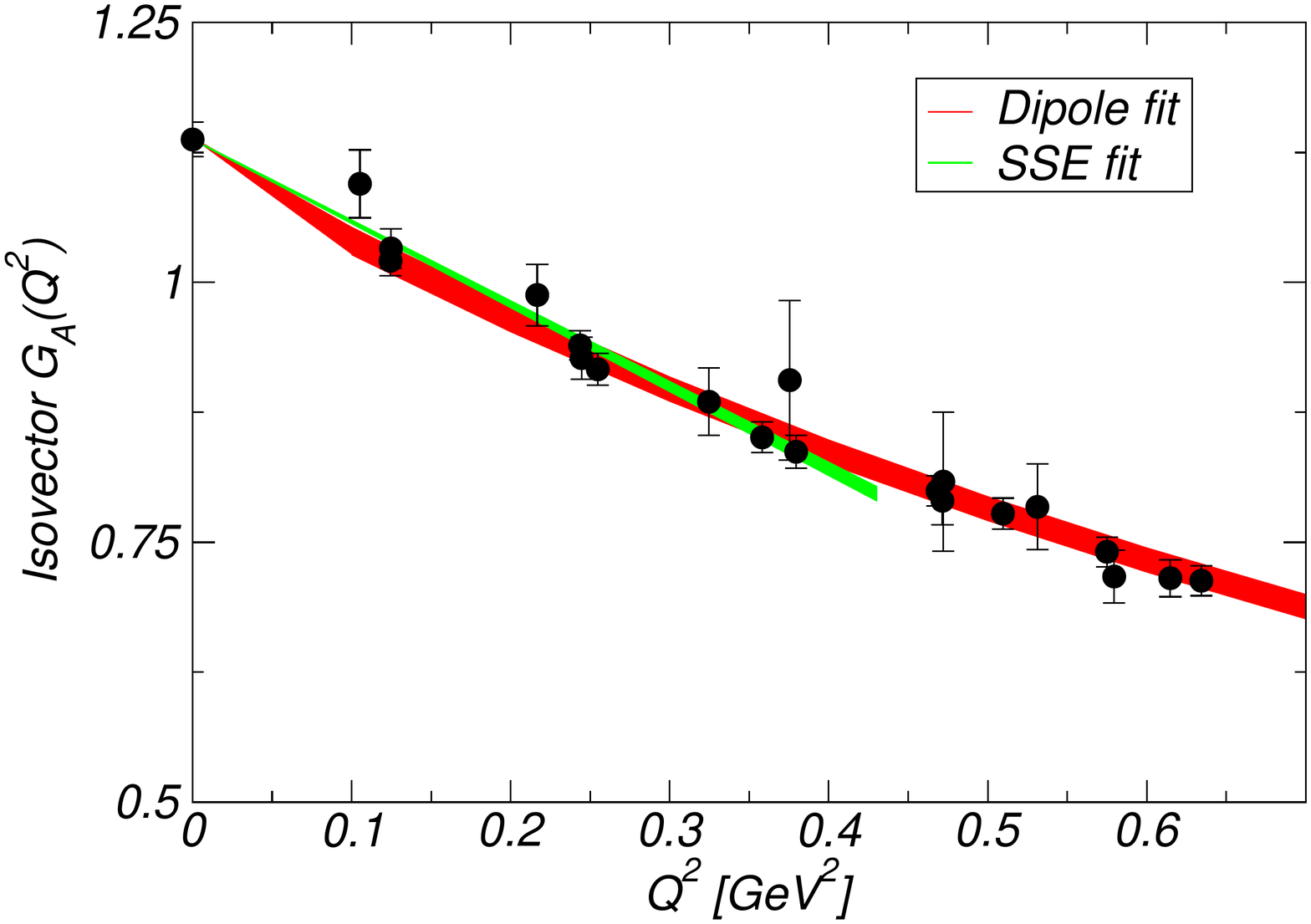}
  \caption{Comparison of dipole and chiral fit to $G_A(Q^2)$ for the
    $28^3$ lattice at $\mps=356$MeV.}
  \label{fig:ga-sse-dip-comp}
\end{figure}

Graphically, \fig\ref{fig:ga-sse-dip-comp} shows the results of the
chiral fit together with a dipole fit and the data set for the $28^3$
lattice at $\mps=356$MeV. The fitting range is $Q^2=[0,0.4]$ GeV$^2$
and $\mps<400$MeV for the SSE expression and all $Q^2$ values for the
dipole fit. The plot illustrates that the dipole fit works well and
that the SSE NLO result is linear in $Q^2$ and provides a good
description of lattice data only for relatively small values of $Q^2$.
There is a rather weak sensitivity to the pion mass, but we need to
point out that we do not at all reproduce the value
$\tilde{B}_3$=3.08(27) from Ref.~\cite{Bernard:1998gv} determined from
the experimental value of the axial radius $\langle
r_A^2\rangle=0.409(12)$fm$^2$ (from electroproduction) at the physical
pion mass.

To summarize, we find that $G_A(Q^2)$ is described excellently using a
dipole-type fit, but the axial radius is substantially smaller than
the experimental one. The SSE expansion simply gives a linear
dependence in $Q^2$ --- it can thus only be applied to small values of
$Q^2$, where the functional behavior is approximately linear. At fixed
$\mps$, the fit quality is acceptable. When performing a combined fit
in $Q^2$ and $\mps$, we still find that the axial radius is
substantially underestimated; we also detect a small residual pion
mass dependence that is not present in the SSE expression at the given
order since at this order one should take $g_A(\mps)=g_A=1.2$,
cf.~Tab.~\ref{tab:sse-pars}, in the denominator of
Eq.~\eqref{eq:ra-extract}.

\subsubsection{\label{sec:isov-pseud-form}Isovector pseudoscalar form
  factor}
This section discusses fit results to the induced pseudo-scalar form
factor, $G_P(Q^2,\mps)$. We will first focus on the isovector
case. Unlike the other form factors, the chiral expansion of
$G_P(Q^2,\mps)$ includes a pion-pole term~\cite{Bernard:1998gv,
  Gockeler:2007hj}:
\begin{equation}
  \label{eq:gp-pion-pole}
  G_P(Q^2,\mps) = \frac{4 m_N^2}{\mps^2+Q^2} \left( g_A - \frac{2\mps^2
      \tilde{B}_2}{(4\pi f_\pi)^2}\right) - \frac{2}{3} g_A m_N^2
  \langle r_A^2\rangle\,.
\end{equation}
This NLO expression has two fit parameters, $\tilde{B}_2$, and
$\langle r_A^2\rangle$. The former parameter, $\tilde{B}_2$, is a
correction to the residue of the pion pole. The parameter $\langle
r_A^2\rangle$ is the axial radius encountered above in the axial form
factor and induces an overall shift.

As discussed previously, we first perform a series of fits on the
$28^3$ lattice at fixed $\mps=356$MeV and vary the upper cut-off in
$Q^2$. For the input parameters to expression~\eqref{eq:gp-pion-pole}
see Tab.~\ref{tab:sse-pars}. This will give us an understanding of how
far the chiral expansion can be expected to hold, although we always
have to keep in mind that any agreement for $Q^2$ larger than
$0.5$GeV$^2$ should be considered merely
accidental. Table~\ref{tab:gp-sse-vart} summarizes the results.
\begin{table}[htb]
  \centering
  \begin{tabular}[c]{*{3}{c|}c}
    \hline\hline
    $Q^2$ max [GeV$^2$] & $\chi^2$/dof &
    $\tilde{B}_2$ & $\langle r_A^2\rangle$ [fm$^2$] \\
    \hline
    1.5 & 1.26 & -0.96(13) & 0.107(8)  \\
    0.5 & 1.08 & -1.01(17) & 0.108(14) \\
    0.4 & 1.22 & -0.84(22) & 0.088(21) \\
    0.3 & 0.84 & -0.55(31) & 0.051(35) \\
    \hline\hline
  \end{tabular}
  \caption{Pion pole fits to the isovector pseudoscalar form factor
    $G_P(Q^2,\mps=356{\rm MeV})$ on the $28^3$ lattice.}
  \label{tab:gp-sse-vart}
\end{table}

The overall quality of the fit is good and $\chi^2/$dof
acceptable. However, we notice that the axial radius, $\langle
r_A^2\rangle$, is smaller than the one obtained using the axial form
factor, $G_A(Q^2,\mps)$. It appears that the NLO SSE formula is not
able to connect our data to experiment for that observable.

With a fixed fit interval $Q^2=[0,0.5]$ GeV$^2$ we also performed a
combined fit in $Q^2$ and $\mps$ and varied the upper cut-off for the
latter. Note that in this case, the location of the pion pole varies
and is set to the appropriate value for each ensemble under
consideration. Table~\ref{tab:gp-chiral-varmpi} summarizes our
findings.
\begin{table}[htb]
  \centering
  \begin{tabular}[c]{*{3}{c|}c}
    \hline\hline
    $\mps$ max [MeV] & $\chi^2$/dof & $\tilde{B}_2$ & $\langle
    r_A^2\rangle$ [fm$^2$] \\
    \hline
    300 & 1.13 & -0.10(80)  & 0.058(46) \\
    400 & 1.72 & -0.94(15)  & 0.089(12) \\
    500 & 2.85 & -1.544(87) & 0.113(9)  \\
    600 & 3.14 & -1.875(56) & 0.139(7)  \\
    \hline\hline
  \end{tabular}
  \caption{Chiral fits to $G_P(Q^2)$ at fixed interval $Q^2=[0,0.5]$
    GeV$^2$ with varying upper cut-off in $\mps$.}
  \label{tab:gp-chiral-varmpi}
\end{table}
We find the fit formula applicable with the cuts $Q^2=[0,0.5]$ GeV$^2$
and $\mps<400$MeV.

\begin{figure}[htb]
  \centering
  \includegraphics[scale=0.3,clip=true]{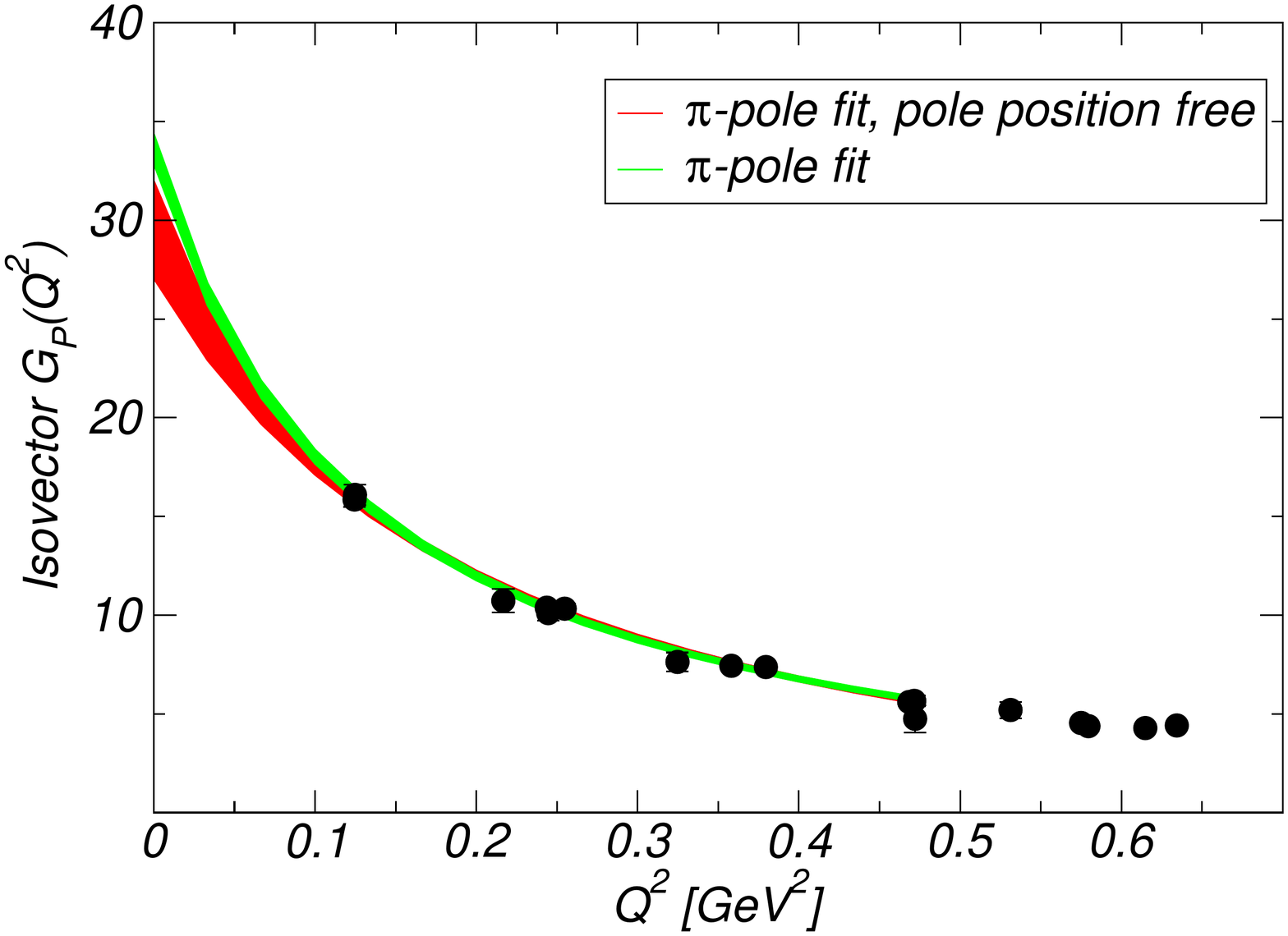}
  \caption{Comparison of pion-pole fits to isovector $G_P(Q^2)$ on the
    $28^3$ lattice with fixed pion pole and with the pion pole as a
    free parameter.}
  \label{fig:gp-pole-comp}
\end{figure}

Next, we study the location of the pole in
Eq.~\eqref{eq:gp-pion-pole}. Whereas in the previous fits the measured
pion mass was used to fix the dependence, we can also treat the pole
position as a free parameter. For this study, we choose the fit
interval $Q^2=[0,0.5]$ GeV$^2$ on the $28^3$ lattice and obtain the
results in Tab.~\ref{tab:gp-pole-fit}. It is evident that there is an
uncertainty of about $10\%$, but within this uncertainty, we find that
the location of the pion pole is reproduced.
\begin{table}[htb]
  \centering
  \begin{tabular}[c]{*{2}{c}}
    \hline\hline
    & Fit result \\ \hline
    $\chi^2$/dof & 0.94 \\
    $\tilde{B}_2$  & -1.98(61) \\
    $\langle r_A^2\rangle$ [fm$^2$] & 0.192(65) \\
    Fit result $\mps$ [MeV] & 417(43) \\
    Actual ensemble $\mps$ [MeV] & 356 \\
    \hline\hline
  \end{tabular}
  \caption{Result from fit of isovector $G_P(Q^2)$ to pion-pole form
    with variable pole mass parameter.}
  \label{tab:gp-pole-fit}
\end{table}
We compare the fits with varying pole position in
Tab.~\ref{tab:gp-pole-fit} and with fixed pole position in
Tab.~\ref{tab:gp-sse-vart} for the interval $Q^2=[0,0.5]$ GeV$^2$
graphically in \fig\ref{fig:gp-pole-comp}.

Our overall conclusion is that the isovector $G_P(Q^2)$ form factor
can be well described by the pion pole expression. We find that by
applying the cuts $Q^2=[0,0.5]$ GeV$^2$ and $\mps<400$MeV we obtain
acceptable results. Allowing the pion pole to vary freely we obtain a
result compatible with the ``true'' pion mass, albeit slightly higher.

\subsubsection{\label{sec:isoscalar-axial-form}Isoscalar axial form
  factors}
When considering the forward case of the isoscalar axial form factors,
we recover the connected part of the first moment of the
spin-dependent isoscalar parton distribution which we discuss
thoroughly in \sect\ref{sec:gener-form-fact}. Their $Q^2$-dependence
is not known experimentally, but they have been studied in the
framework of chiral perturbation theory. Reference~\cite{Ando:2006sk}
finds a counter-term with a linear $Q^2$-dependence, and
Ref.~\cite{Diehl:2006js} does not list any $Q^2$-dependence at the
order considered. In a previous paper~\cite{Diehl:2006ya}, however,
the same authors find a counter-term with linear $Q^2$-dependence, in
agreement with Ref.~\cite{Ando:2006sk}. In the following we adopt the
notation from Eq.~\eqref{eq:lattice-ax-gff-def} when referring to the
two form factors, i.e.~we denote them by $\tilde{A}_{10}(Q^2)$ and
$\tilde{B}_{10}(Q^2)$.

\begin{figure}[htb]
  \centering
  \includegraphics[scale=0.3, clip=true]{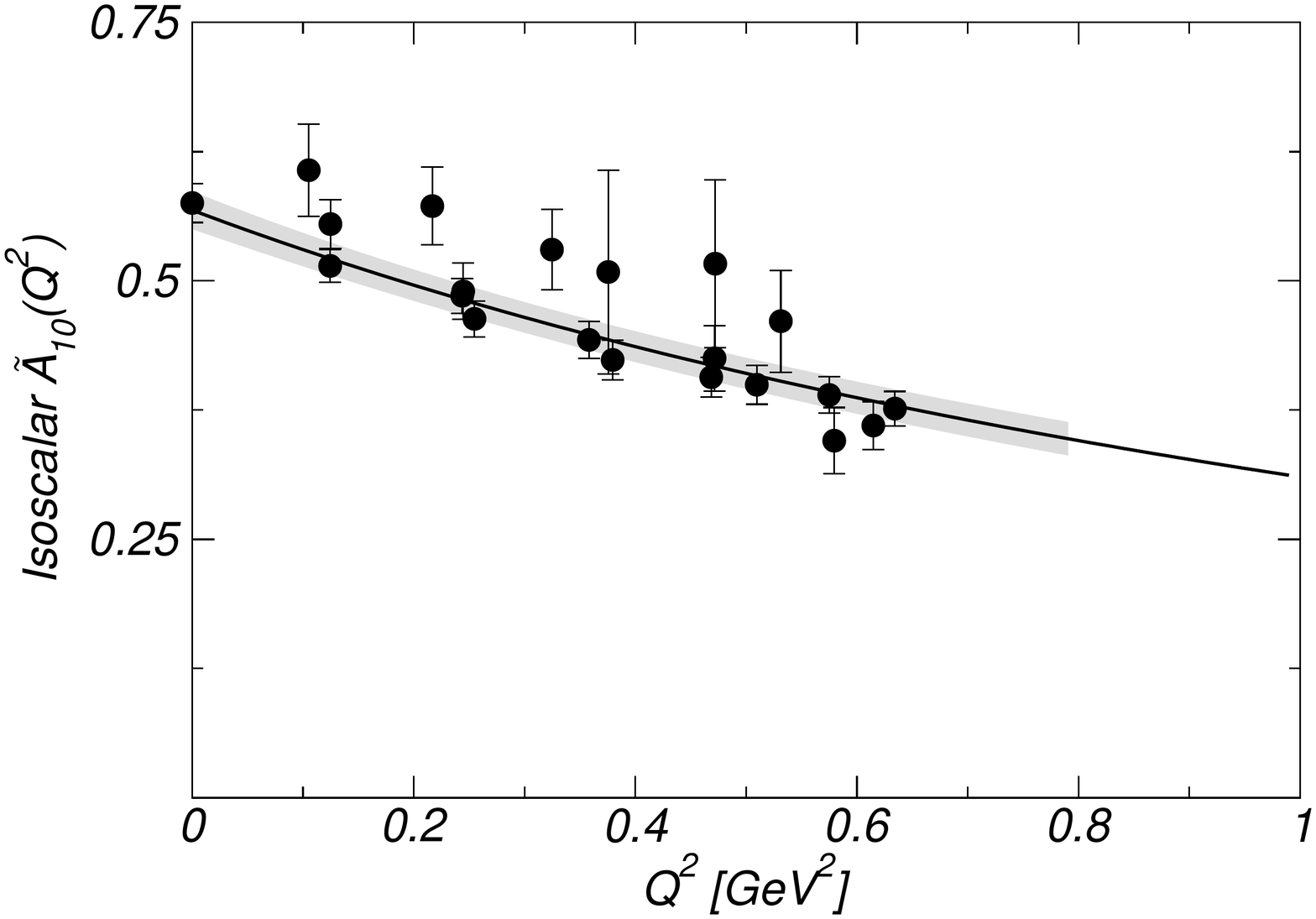}
  \caption{Dipole fit to the isoscalar axial form factor
    $\tilde{A}_{10}(Q^2)$ on the $28^3$ lattice at $\mps=356$MeV.}
  \label{fig:at10vsqQ}
\end{figure}

\paragraph{Isoscalar axial form factor $\tilde{A}_{10}$}
Since the $\mps$-dependence of the forward matrix element is already
covered in \sect\ref{sec:gener-form-fact}, we focus on the
$Q^2$-dependence at each ensemble. First, we study the case
$\mps=356$MeV on the $28^3$ lattice. As a generic fit formula we use
the dipole expression. Figure~\ref{fig:at10vsqQ} displays the result
of the fit. Notice that this figure contains an example for six data
points that appear superficially high as discussed previously in
Ref.~\cite{Bratt:2008uf}, cf.\ also
Sec.~\ref{sec:super-jackkn-analys}.
The dipole form actually provides a decent description of the data. We
find a dipole mass of $M_d=1683(89)$MeV with
$\chi^2/$dof=1.87. Unlike the isovector form factors of the vector
current, we cannot refer to previous phenomenological fits of
experimental data. Our results for the dipole fits to the different
ensembles are listed in Tab.~\ref{tab:at10-allfits}. The radii thus
obtained are denoted by $\langle \tilde{r}_1^2\rangle$.
\begin{table}[htb]
  \centering
  \begin{tabular}[c]{*{4}{c|}c}
    \hline\hline
    $\mps$ [MeV] & $\chi^2$/dof & $A_0$ & $M_d$ [GeV] & $\langle
    \tilde{r}_1^2\rangle$ [fm$^2$] \\ \hline
    293 & 1.45 & 0.560(31) & 1.626(143) & 0.176(31) \\
    356 on $28^3$ & 1.87 & 0.568(18) & 1.683(89) & 0.165(17) \\
    356 on $20^3$ & 1.58 & 0.542(17) & 1.903(97) & 0.129(13) \\
    495 & 1.10 & 0.586(9) & 1.711(40) & 0.1596(75) \\
    597 & 0.75 & 0.605(6) & 1.691(26) & 0.1635(51) \\
    \hline\hline
  \end{tabular}
  \caption{Dipole fits to isoscalar $\tilde{A}_{10}(Q^2)$ for all data
    sets.}
  \label{tab:at10-allfits}
\end{table}

\begin{figure}[htb]
  \centering
  \includegraphics[scale=0.3, clip=true]{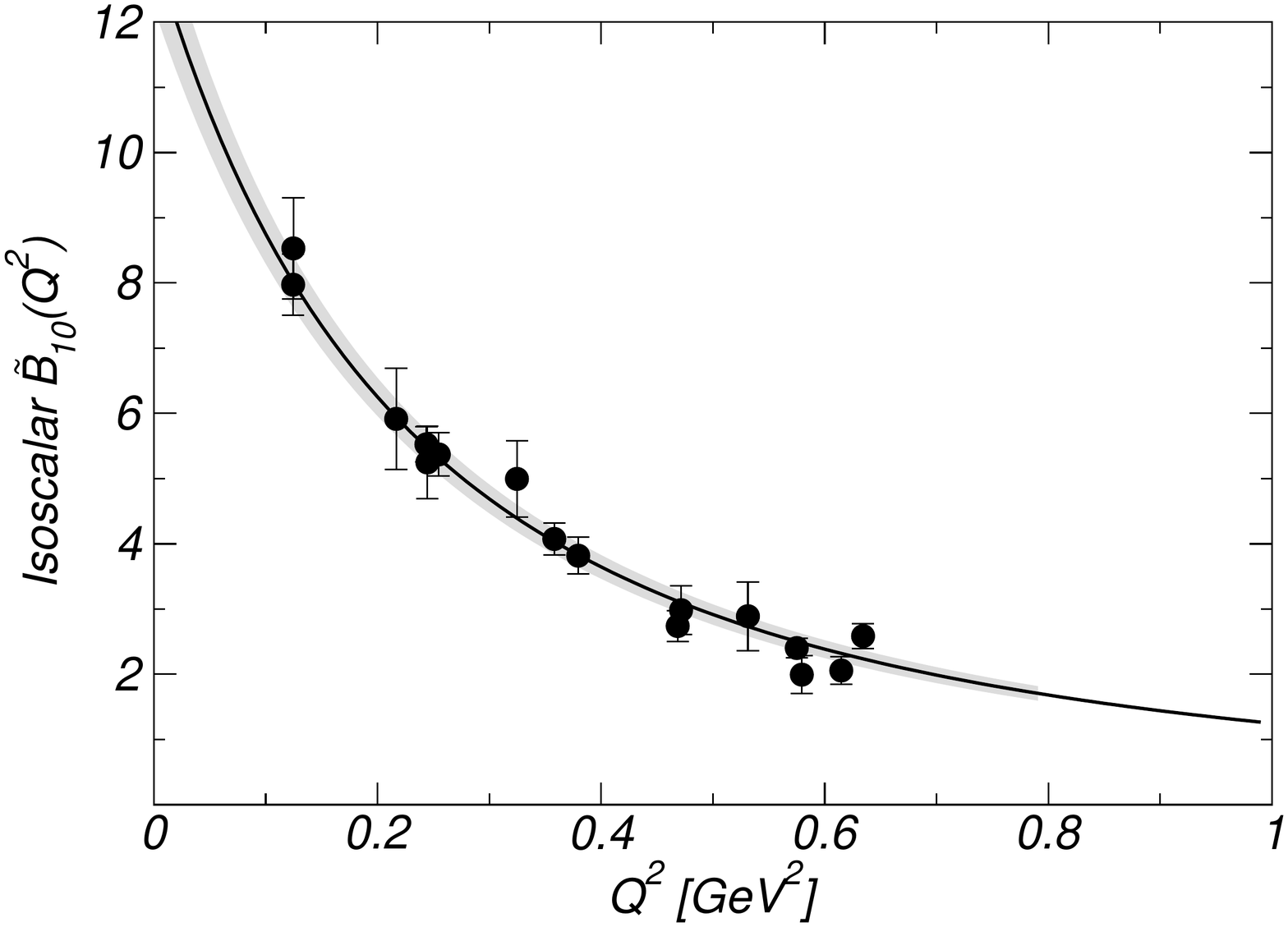}
  \caption{Dipole fit to the isoscalar axial form factor
    $\tilde{B}_{10}(Q^2)$ on the $28^3$ lattice at $\mps=356$MeV.}
  \label{fig:bt10vsQ}
\end{figure}

\paragraph{Isoscalar axial form factor $\tilde{B}_{10}$}
For the connected part of the isoscalar form factor
$\tilde{B}_{10}(Q^2)$ of the axial current we also find a non-trivial
$Q^2$ dependence. In this case, however, we observe that the form
factor has quite a large magnitude and a rather strong
fall-off. Again, we attempt a dipole-type fit and display the result
in \fig\ref{fig:bt10vsQ}. The dipole appears to fit the connected part
well with $\chi^2/$dof=1.88. However, since the functional form is
suggestive of the pion pole, cf.~Eq.~\eqref{eq:gp-pion-pole} in
\sect\ref{sec:isoscalar-axial-form}, we have also attempted a fit to
that form. If we keep the pion pole fixed to the actual pion mass of
the sample, $\mps=356$MeV, we obtain a $\chi^2/$dof$=1.86$, whereas
leaving it a free parameter gives $\chi^2/$dof$=1.97$ with a measured
position of $\mps=394(59)$MeV. This result is certainly in agreement
with the actual pion mass of the underlying ensemble. This indicates,
that the connected part of the isoscalar $\tilde{B}_{10}(Q^2)$ form
factor is indeed compatible with a pion pole form, although the data
does not favor it over a dipole form.

The resulting parameters from dipole fits to all data sets are
assembled in Tab.~\ref{tab:bt10-allfits}.
\begin{table}[htb]
  \centering
  \begin{tabular}[c]{*{4}{c|}c}
    \hline\hline
    $\mps$ [MeV] & $\chi^2$/dof & $A_0$ & $M_d$ [GeV] & $\langle
    \tilde{r}_2^2\rangle$ [fm$^2$] \\ \hline
    293 & 0.65 & 13.3(37) & 0.671(95) & 1.04(30) \\
    356 on $28^3$ & 1.88 & 13.14(85) & 0.668(22) & 1.05(7) \\
    356 on $20^3$ & 1.22 & 11.4(11) & 0.757(36) & 0.82(8) \\
    495 & 1.15 & 10.44(54) & 0.802(23) & 0.73(4) \\
    597 & 0.89 & 10.27(34) & 0.840(17) & 0.66(3) \\
    \hline\hline
  \end{tabular}
  \caption{Dipole fits to isoscalar $\tilde{B}_{10}(Q^2)$ for all data
    sets.}
  \label{tab:bt10-allfits}
\end{table}
It is evident that there is a notable pion mass dependence, with the
dipole mass being affected strongest. This is in line with the
functional form in Ref.~\cite{Ando:2006sk}, although the strength of
the effect is quite surprising. At $\mps=293$MeV the mean squared
radius, $\langle\tilde{r}_2^2\rangle$, is larger than $1$fm, making it
the largest radius of all observables. Further studies of the chiral
behavior for this observable and the computation of disconnected
diagrams are interesting directions for future work to shed further
light on these observations.

\subsubsection{\label{sec:summary-axial-form}Summary of axial form
  factors}
The isovector axial form factor $G_A(Q^2,\mps)$ at present lattice
pion masses systematically is much flatter as a function of $Q^2$ than
experiment and this result is not explained by one-loop chiral
perturbation theory. This is a qualitative and significant
quantitative mismatch between lattice and phenomenology that still
remains to be understood. The corresponding pseudoscalar form factor
$G_P(Q^2,\mps)$ is described very well by a pion-pole form and both
the functional form and the location of the pion pole are in excellent
agreement with theory. Still, the SSE can not explain the discrepancy
in the axial radii we extract from $G_A$ and $G_P$. The connected
parts of the isoscalar form factors of the axial current have a strong
$Q^2$ dependence and we have successfully used dipole forms in each
case.

\subsection{\label{sec:gener-form-fact}Generalized Form Factors}
In this section we present a survey of our results for the generalized
form factors, emphasizing their main qualitative features and
describing the progress made since our previous publication on the
subject~\cite{Hagler:2007xi}.

We start with the form factors of the twist-two quark bilinear
operator $\bar q D_{\{\mu}\gamma_{\nu\}} q$.  This operator is
particularly important since its forward matrix elements determine the
quark momentum fractions; in the isosinglet case, it is one of the
terms appearing in the energy-momentum tensor. Furthermore its forward
matrix elements contribute to the nucleon mass \cite{Ji:1994av} and
spin decompositions, see \sect\ref{sec:decomp-nucl-spin}.  The three
form factors $A_{20}$, $B_{20}$, $C_{20}$, which determine its matrix
elements between two arbitrary one-nucleon states via
Eq.~\eqref{eq:lattice-vec-gff}, are displayed in
\fig\ref{fig:ABCunpol_vs_t} both for the isovector and isosinglet
combinations. In the latter case, our computation lacks contributions
from the disconnected diagrams.

Similarly, the form factors $\tilde A_{20}$ and $\tilde B_{20}$
associated with the corresponding twist-two axial operator $\bar q
D_{\{\mu}\gamma_{\nu\}}\gamma_5 q$, see \eqref{eq:lattice-ax-gff}, are
displayed in \fig\ref{fig:ABCpol_vs_t}. The forward matrix elements of
this operator determine the first moments of the polarized structure
functions, in other words the polarized momentum fraction.  As is
clear from Eqs.~\eqref{eq:lattice-vec-gff}
and~\eqref{eq:lattice-ax-gff}, only the $A_{20}$ and $\tilde A_{20}$
form factors can be directly obtained at zero momentum transfer; the
others require an extrapolation, much as the Pauli form factor.
\begin{figure}[htb]
  \centerline{\includegraphics[width=14cm,angle=0]{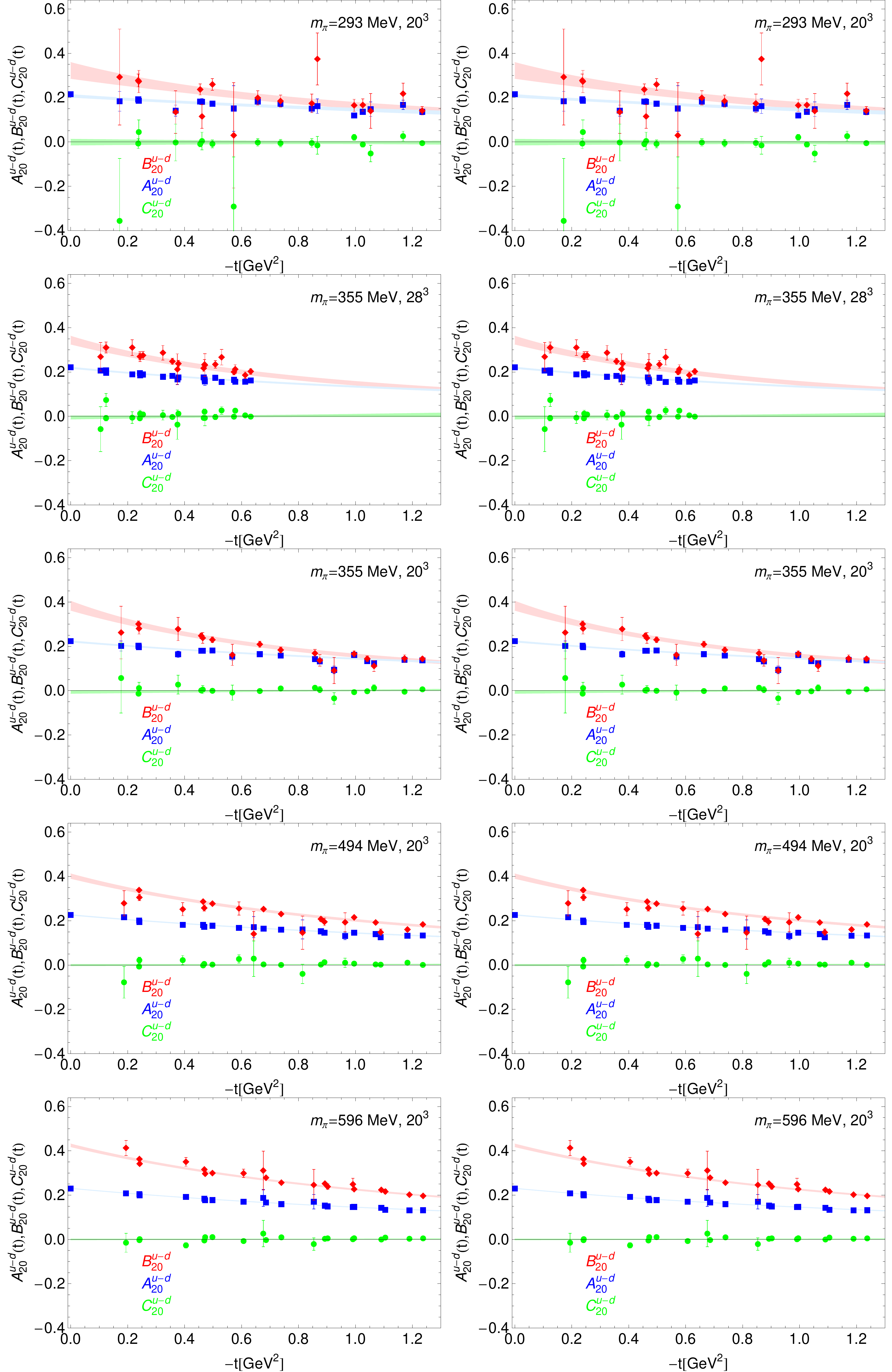}}
  \caption{The unpolarized isovector and isosinglet GFFs $A_{20}$,
    $B_{20}$, $C_{20}$.}
  \la{fig:ABCunpol_vs_t}
\end{figure}
\begin{figure}[htb]
  \centerline{\includegraphics[width=14cm,angle=0]{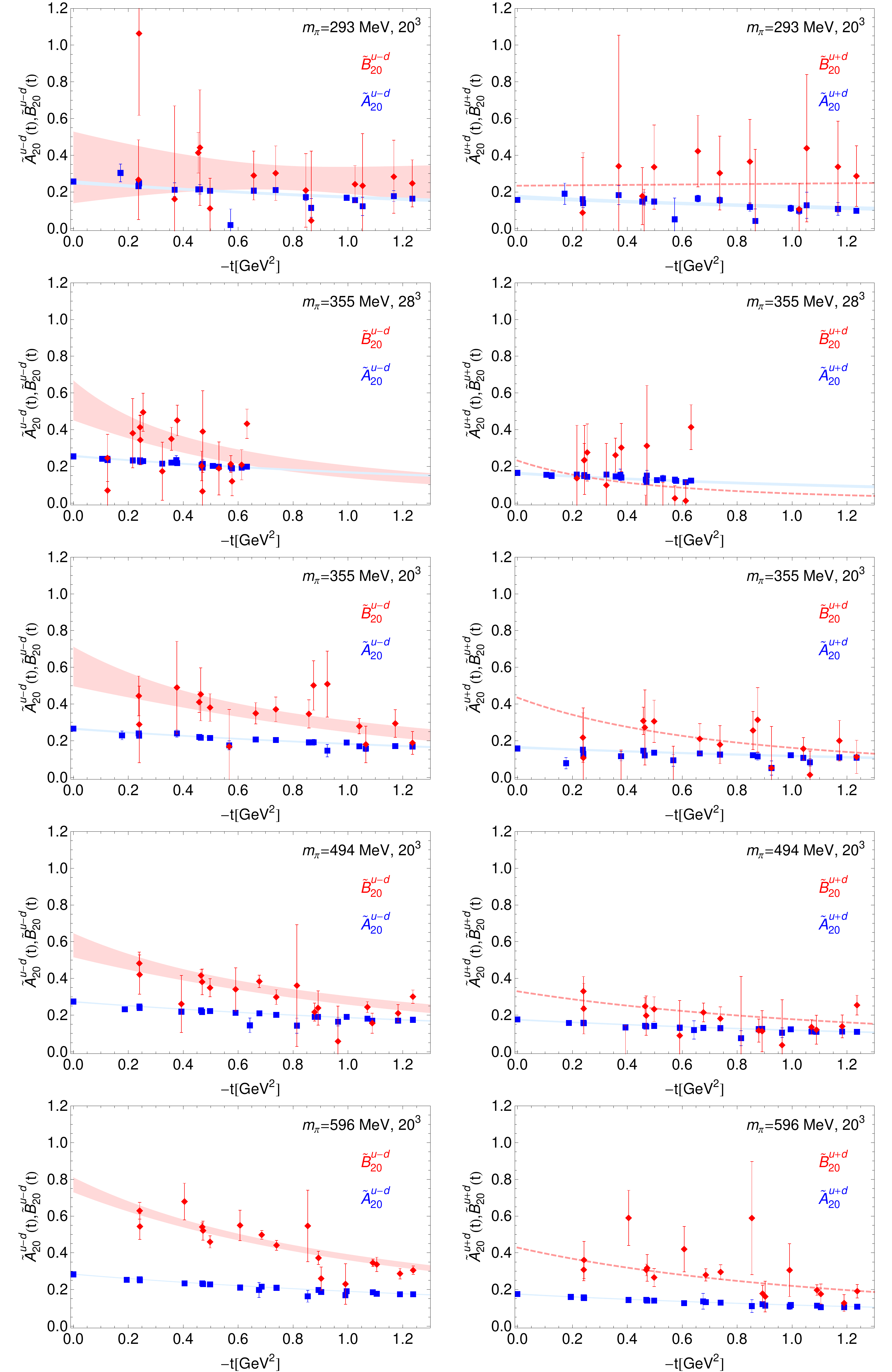}}
  \caption{The polarized isovector and isosinglet GFFs $\tilde A_{20}$
    and $\tilde B_{20}$.}
  \la{fig:ABCpol_vs_t}
\end{figure}

The statistical uncertainties on $A_{20}$, $B_{20}$, $C_{20}$ at
$m_\pi= 356$, 495 and 597MeV are reduced by about a factor three as
compared to our previous publication~\cite{Hagler:2007xi}.
Furthermore we have data at one lighter pion mass, 293MeV, with an
accuracy comparable to our previous data in the 600-700MeV range.
Given the expected rapid deterioration of the signal-to-noise ratio
when $m_\pi\to0$, this represents significant progress.

Similar remarks apply to the axial GFFs $\tilde A_{20}$ and $\tilde
B_{20}$.  The signal-to-noise ratio of the latter GFF is overall quite
poor and worsens visibly with the pion mass. While previously we were
essentially unable to obtain a signal below $m_\pi=500$MeV, we now
dispose of some information down to about $m_\pi=350$MeV.

As in our previous work, we observe a qualitative agreement of the
relative magnitudes of the GFFs with predictions from large $N_c$
counting rules, e.g.
\begin{equation}
  |A_{20}^{u+d}|\sim N_c^2 \gg |A_{20}^{u-d}|\sim N_c, 
  \qquad |B_{20}^{u-d}|\sim N_c^3 \gg |B_{20}^{u+d}|\sim N_c^2
  , \qquad |C_{20}^{u+d}|\sim N_c^2 \gg |C_{20}^{u-d}|\sim N_c\,.
  \label{eq:largeN-unpol}
\end{equation}
As a matter of fact, we remark that $C_{20}$ is consistent with zero
in the isovector channel, whereas it is clearly non-vanishing and
negative in the isosinglet channel. This is of interest, since
$C_{20}$ entirely determines the longitudinal momentum transfer $\xi$
dependence of the functions $H^{n=2}(\xi,t)$ and $E^{n=2}(\xi,t)$, see
Eq.~\eqref{eq:lattice-vec-gff-def}.  In the case of $B_{20}$, it is
the opposite: $B_{20}$ is very small in the isosinglet channel, but it
is positive and quite large (compared to $A_{20}$) in the isovector
channel. However, $B^{u-d}_{20}$ falls off faster than $A^{u-d}_{20}$,
so that the two form factors are practically equal by the time
$|t|\approx 1$GeV$^2$ is reached, for pion masses $\le 356$MeV. Their
$t$-dependence is thus qualitatively similar to the $Q^2$-dependence
of the electromagnetic form factors, for which chiral perturbation
theory predicts that $(r_1^v)^2\sim \log m_\pi$, and $(r_2^v)^2\sim
1/m_\pi$, and experimentally, it is indeed the case that $(r_1^v)^2 <
(r_2^v)^2$.

In the polarized case, the large-$N_c$ counting rules predict
\begin{equation}
  |\widetilde A_{20}^{u-d}|\sim N_c^2 \gg |\widetilde A_{20}^{u+d}|\sim N_c,
  \qquad |\widetilde B_{20}^{u-d}|\sim N_c^4 \gg |\widetilde
  B_{20}^{u+d}|\sim N_c^3\,.
  \label{eq:largeN-pol}
\end{equation}
Our lattice data shows that the isovector GFFs are only marginally
larger in magnitude than the corresponding isosinglet GFFs.  We
further observe in both the isovector and isosinglet channel that
$\tilde B_{20}(t)>\tilde A_{20}(t)$.  Noting that $\tilde
H^{n=2}=\tilde A_{20}$ and $\tilde E^{n=2}=\tilde B_{20}$, see
\eq\eqref{eq:lattice-ax-gff-def}, this is compatible with a dominance
of the GPD $\tilde E$ over $\tilde H$ at small $-t$ as indicated in a
recent HERMES study~\cite{HERMES:2009ua}.

It is also interesting to compare the size of corresponding polarized
and unpolarized form factors. The large-$N_c$ ordering is indeed
observed if one compares $A_{20}^{u-d}$ with $\tilde A_{20}^{u-d}$, or
$B_{20}^{u-d}$ with $\tilde B_{20}^{u-d}$, but the differences in
magnitude are not pronounced enough to speak of a hierarchy.
\begin{figure}[htb]
  \centerline{\includegraphics[width=14cm,angle=0]{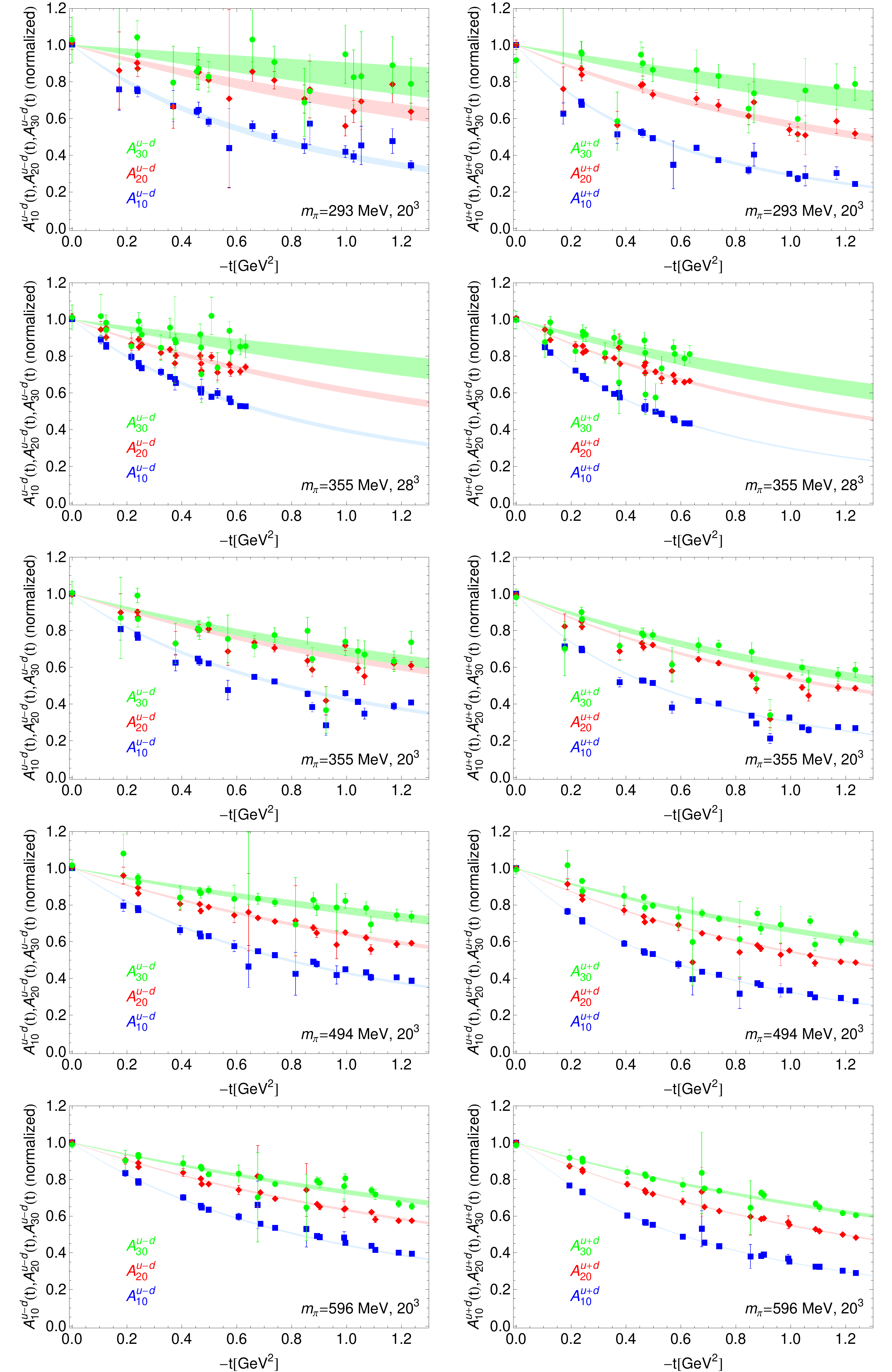}}
  \caption{Generalized form factors of twist-two operators of
    dimension 3, 4 and 5, in the isovector and isosinglet channels.
    Notice in particular the flattening of the slope of the $A_{n0}$
    GFF with increasing $n$.}
  \la{fig:A1230}
\end{figure}
\begin{figure}[htb]
  \centerline{\includegraphics[width=8.4cm,angle=0]{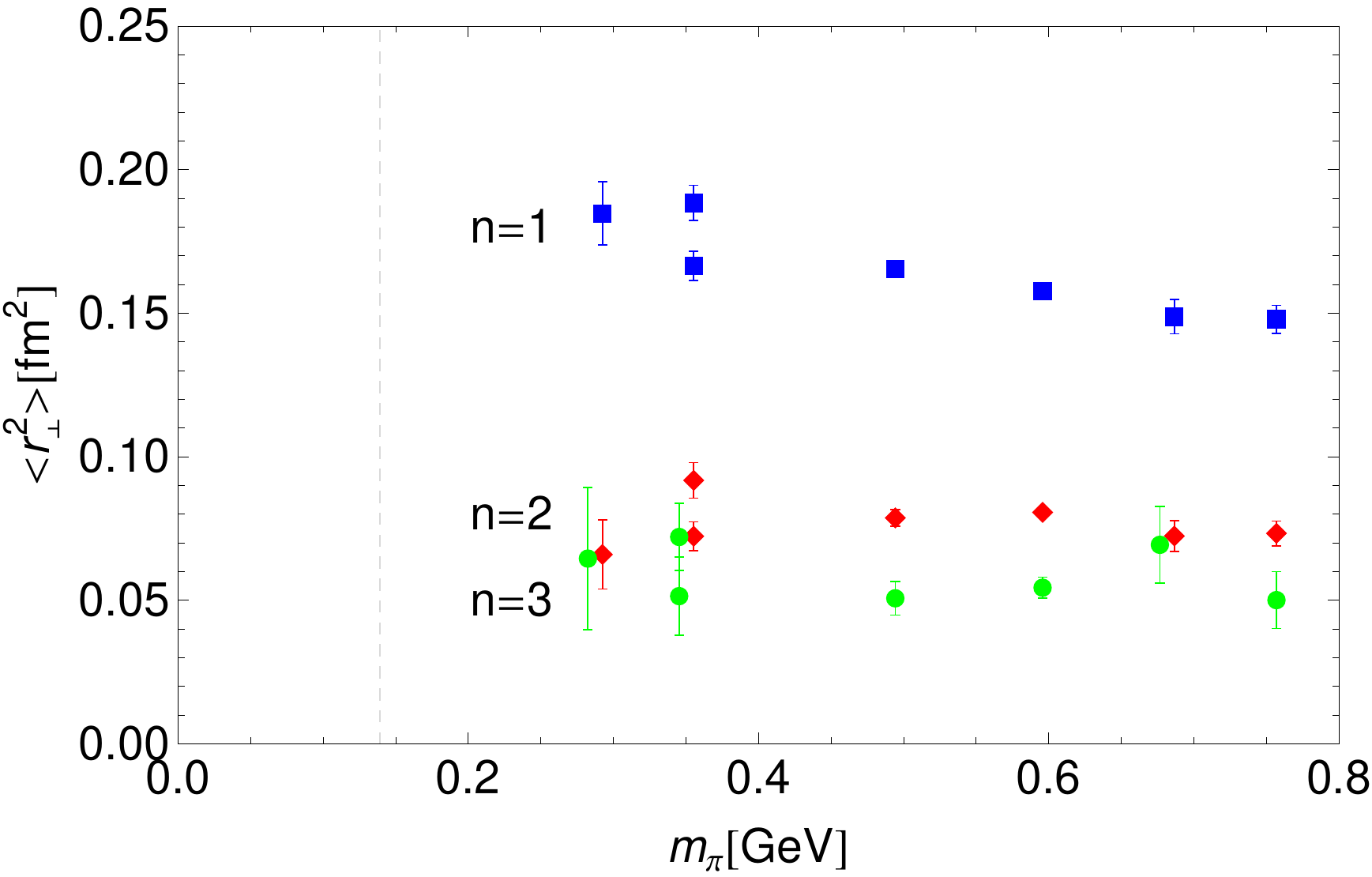}
    {\includegraphics[width=8.4cm,angle=0]{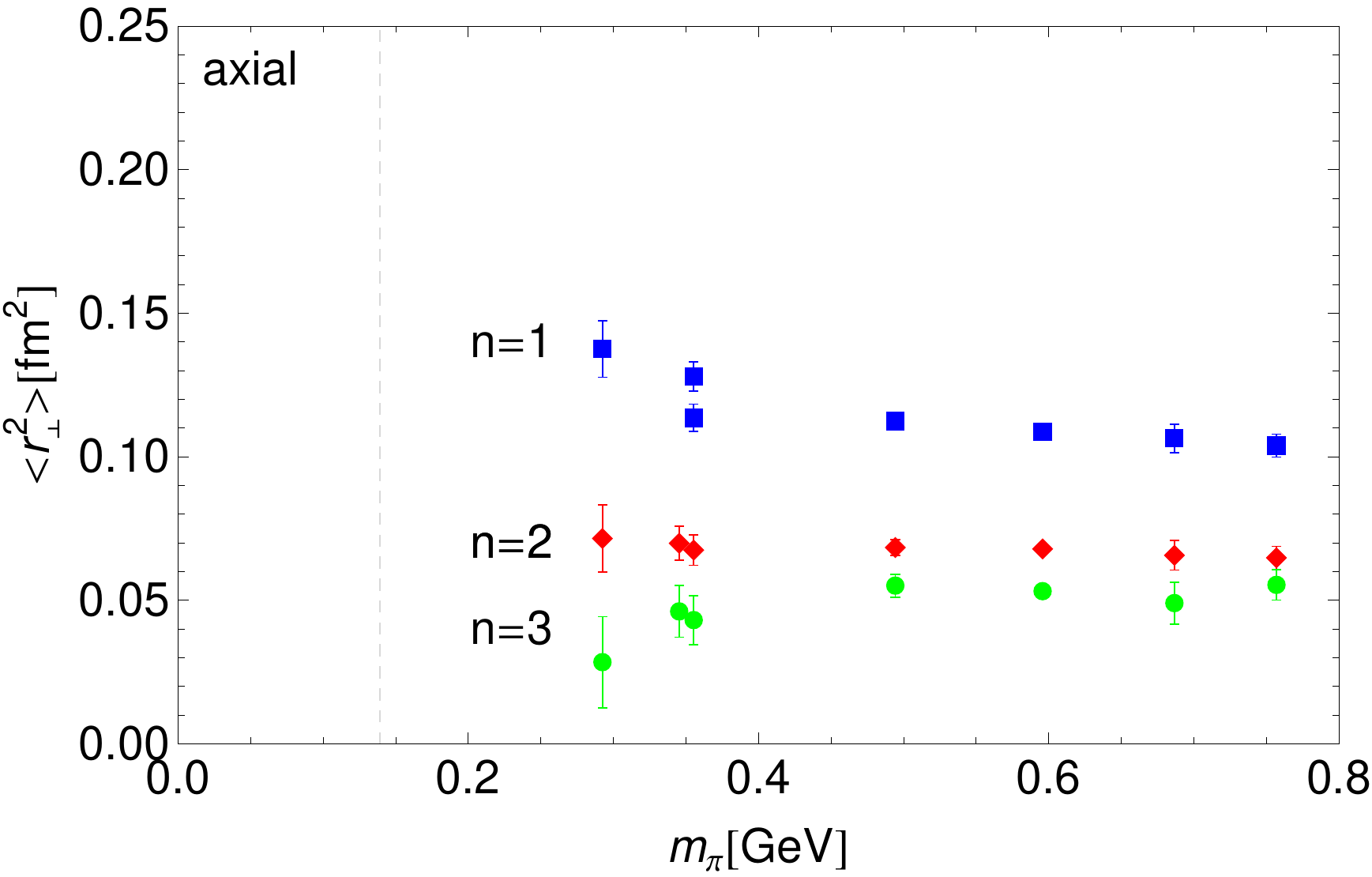}}}
  \caption{Transverse isovector radii as extracted from a dipole fit
    with momentum cut $|t|<0.5$GeV$^2$.  The left panel corresponds to
    the unpolarized case, and the right panel to the polarized
    case. At $m_\pi=356$MeV, the radius on the $28^3$ lattice is in
    all cases larger than on the $20^3$ lattice.}
  \la{fig:radii}
\end{figure}

\begin{figure}[htb]
  \centerline{\includegraphics[width=14cm,angle=0]{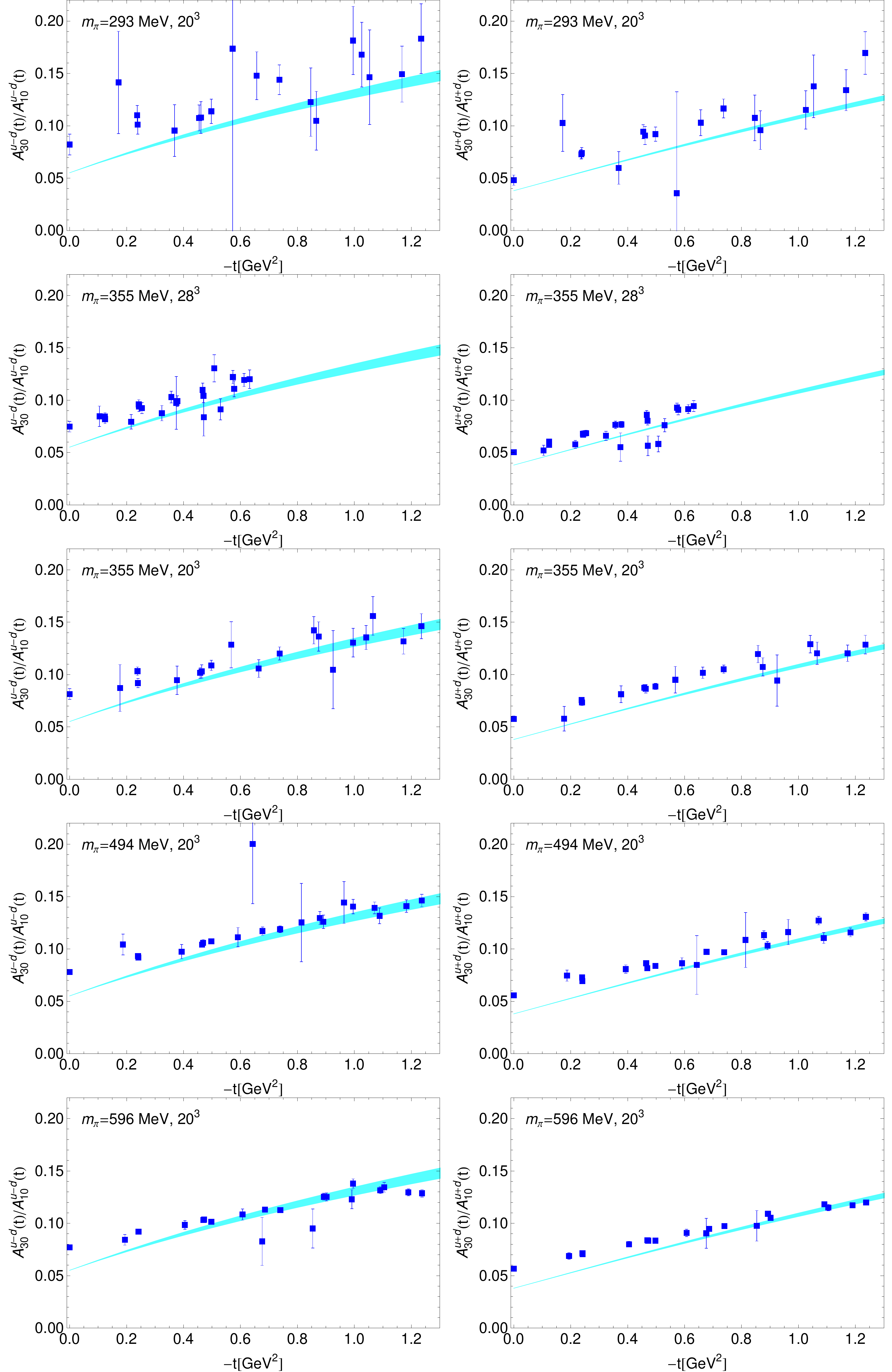}}
  \caption{Ratio of the $A_{30}$ to the $A_{10}$ unpolarized GFFs,
    isovector and isosinglet. Disconnected contributions have been
    omitted.}
  \la{fig:A30_ov_A10unpol}
\end{figure}
\begin{figure}[htb]
  \centerline{\includegraphics[width=14cm,angle=0]{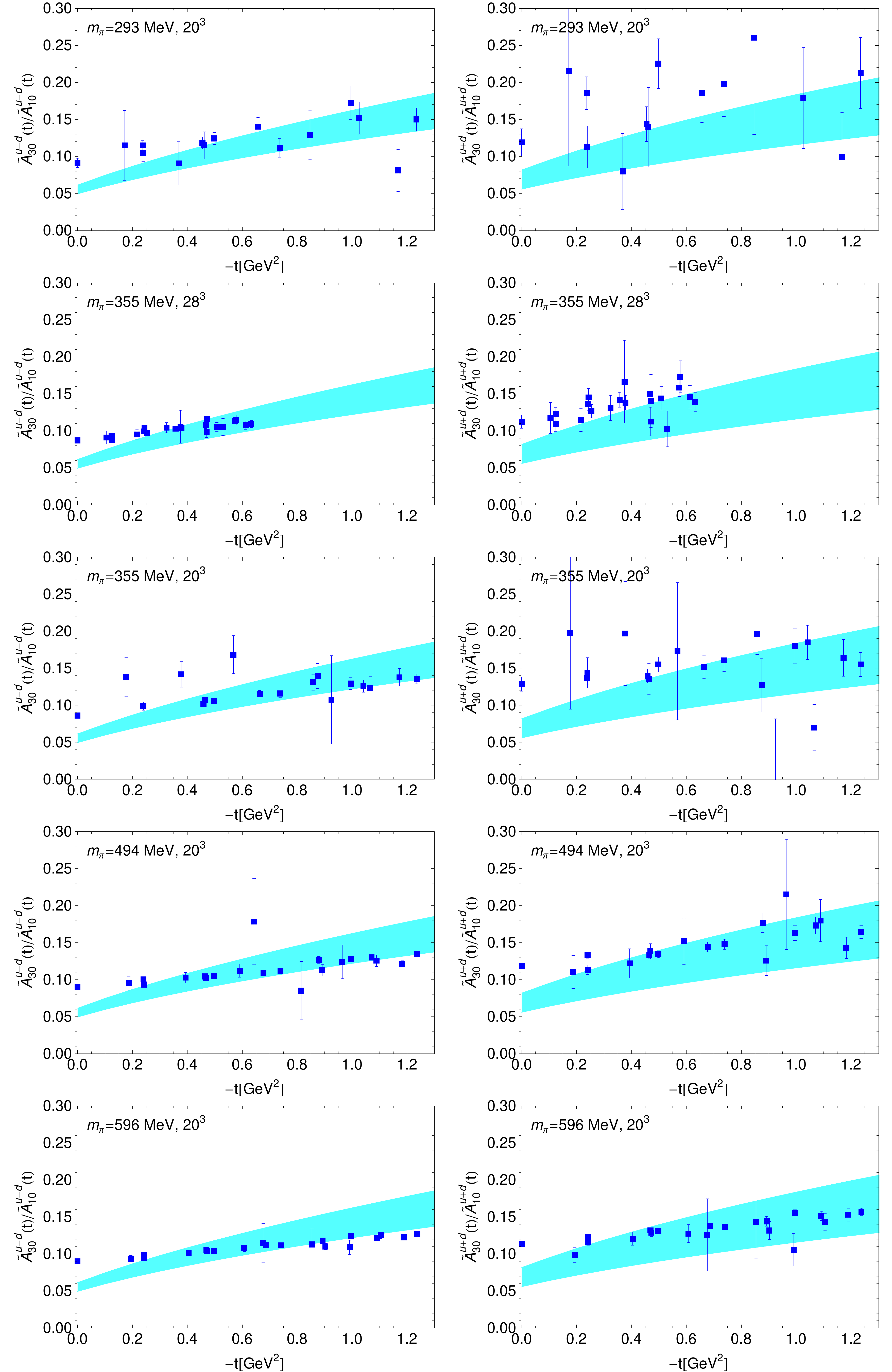}}
  \caption{Ratio of the $\tilde A_{30}$ to the $\tilde A_{10}$
    polarized GFFs, isovector and isosinglet. Disconnected
    contributions have been omitted.}
  \la{fig:A30_ov_A10pol}
\end{figure}

The momentum transfer dependence of the GFFs $A_{10}$, $A_{20}$,
$A_{30}$ is compared in \fig\ref{fig:A1230} for both the isovector and
isoscalar channels.  For this purpose they all have been rescaled to
be equal to unity at $t=0$. It is clear that in both channels, the
form factor flattens when the moment $n$ increases.  The effect is
observed at all pion masses, and it is larger when increasing $n$ from
1 to $2$ than when going from $n=2$ to $n=3$.

This observation can be made more quantitative by fitting the GFFs
with a dipole ansatz for $|t|$ ranging from zero to $0.5$GeV$^2$.  In
the isovector channel this leads, via the analogue of
Eq.~\eqref{eq:rms-radius-def}, to the squared radii displayed in
\fig\ref{fig:radii}.  The left panel corresponds to unpolarized GFFs,
the right panel to the polarized ones. In the infinite momentum frame,
the transverse radii correspond to the rms transverse distance (i.e.\
the impact parameter) of the active parton to the center of momentum
of the nucleon.  At each pion mass, the higher the moment of the
structure function, the smaller the radius associated with it. This
effect had been anticipated~\cite{Burkardt:2000za} and was seen by
direct lattice calculation for the first time in~\cite{Hagler:2003jd}.
Indeed most of the contribution to higher moments comes from the
large-$x$ region; a parton carrying by itself most of the nucleon's
momentum must be located near the center of mass in the transverse
plane.

On the $V=20^3$ lattices, corresponding to physical volumes of
$(2.5{\rm fm})^3$, it is only for the electromagnetic radius, $n=1$,
that we find a statistically significant increase when the pion mass
is reduced.  This corresponds to the idea that the growth of the
nucleon radius with decreasing quark mass is due to the low-$x$
partons, while the large-$x$ `valence' partons have a transverse
distribution that depends only weakly on the quark mass.  However,
these qualitative lessons are questionable due to the statistically
significant finite-size effect observed in several of these radii. The
larger lattice volume $V=28^3$, corresponding to $(3.5{\rm fm})^3$,
leads to larger radii. The largest finite-size effect is seen in the
vector and axial-vector form factors, see
\sect\ref{sec:finite-volume-effect}, but a similar effect is also seen
in the $A_{20}$ form factor.  It appears that the nucleon is somewhat
`squeezed' by the periodic box in which we study it, leading to an
underestimation of the radii.  This finite-size effect is not
statistically significant if one simply extracts an effective square
radius from the smallest available momentum transfer $q=2\pi/L$, as we
already pointed out in~\cite{Syritsyn:2009mx}.  The finite-size effect
affects the intermediate values of $-t$ most strongly, as discussed in
\sect\ref{sec:finite-volume-effect}.  We believe that this apparently
$t$-dependent finite-size effect on form factors is interesting and
deserves further investigation.

Regarding \fig\ref{fig:radii} we also note that the $n=2$ and 3
polarized radii at $m_\pi=$356MeV are statistically compatible with
the corresponding unpolarized ones.  This is in contrast with the
$n=1$ case, where the Dirac radius is significantly larger than the
axial vector radius.  In the chiral limit, the Dirac radius diverges
logarithmically, while the axial vector radius remains finite, see
Secs.~\ref{sec:form-factors}
and~\ref{sec:axial-form-factors}. Therefore, the former must become
larger than the latter below a certain pion mass.  What we see on
\fig\ref{fig:radii} is that this hierarchy survives up to very large
pion masses, far beyond the range of validity of the argument based on
chiral effective theory.
\begin{figure}[htb]
  \centerline{\includegraphics[width=11cm,angle=0]%
    {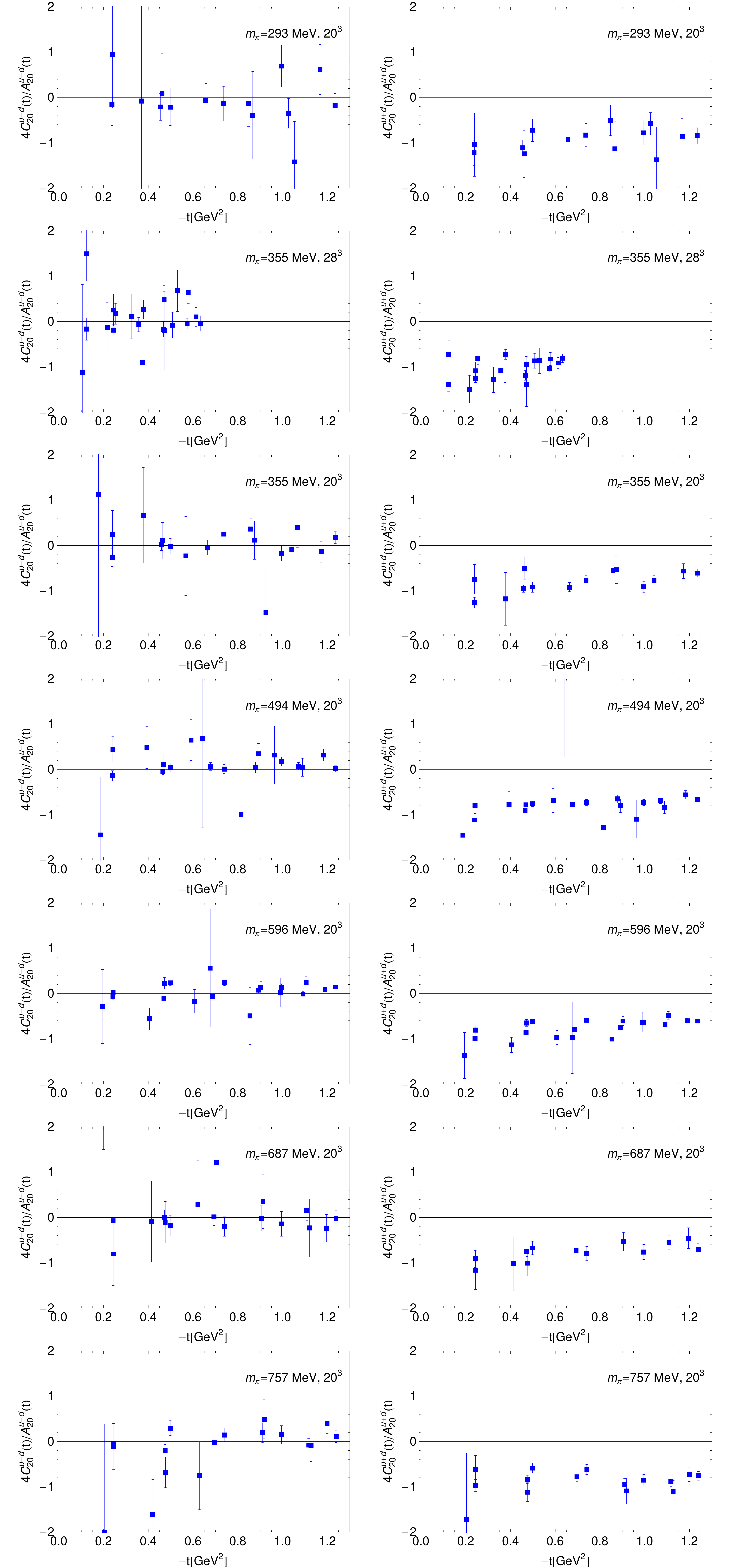}}
  \caption{Ratios of $n=2$ generalized form factors.}
  \la{fig:xi-dep2}
\end{figure}
\begin{figure}[htb]
  \centerline{\includegraphics[width=11cm,angle=0]%
    {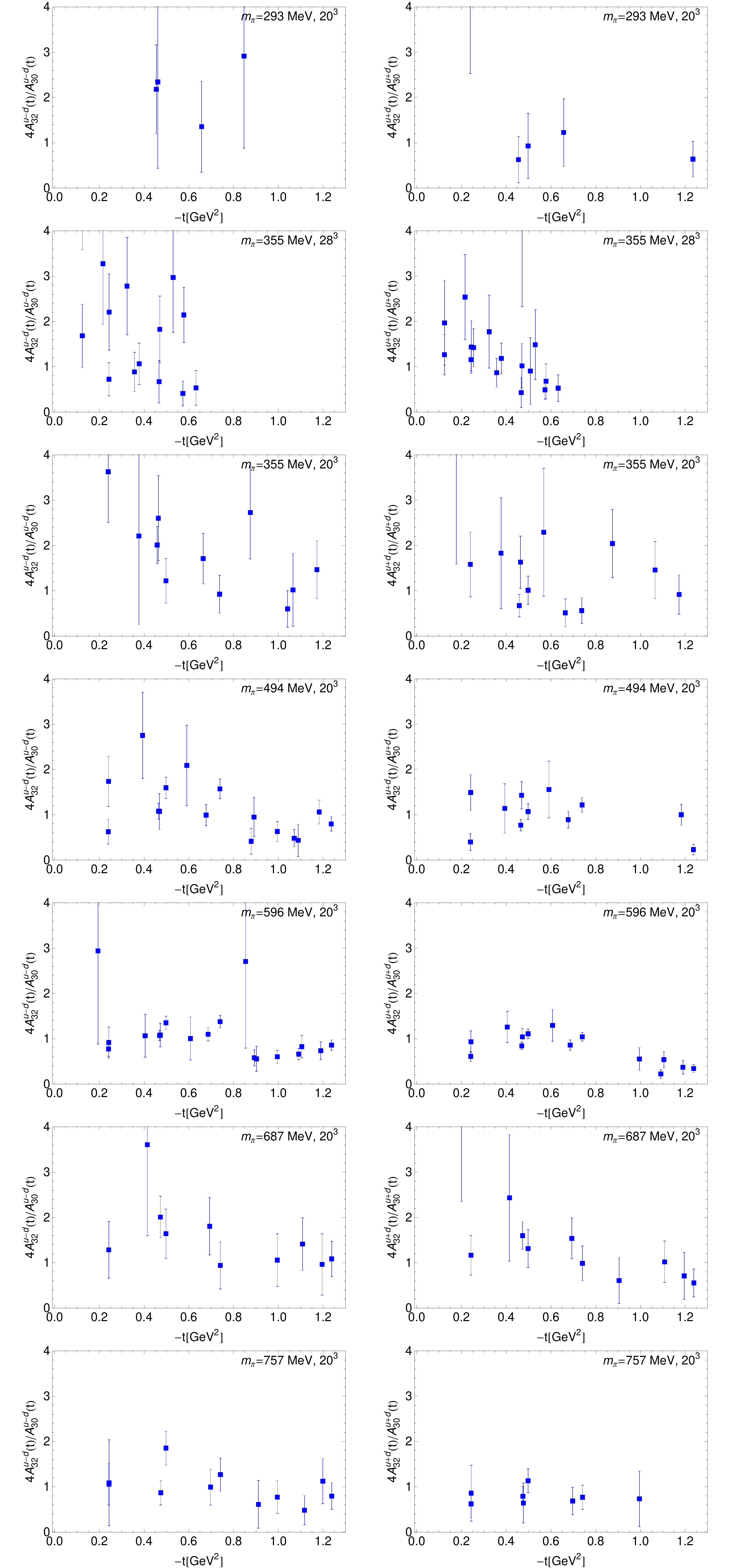}}
  \caption{Ratios of $n=3$ generalized form factors.}
  \la{fig:xi-dep3}
\end{figure}

As in \cite{Hagler:2007xi}, we have compiled data on the ratios
$A_{30}(t)/A_{10}(t)$ and $\tilde A_{30}(t)/\tilde A_{10}(t)$ in
\fig\ref{fig:A30_ov_A10unpol} and Fig.~\ref{fig:A30_ov_A10pol}
respectively, where we compare them to predictions based on
experimental data for the nucleon form factors and PDFs in combination
with a model-dependent ansatz for the combined $(x,t)$-dependence of
the GPDs, displayed by the error bands \cite{Diehl:2004cx}.  With
increased statistics, even at the lightest pion mass there is good
evidence that these ratios are generically not flat.  A flat ratio
would be a direct consequence of the factorization of the GPDs into an
$x$-dependent function and a $t$-dependent function.  In spite of a
semi-quantitative agreement, we see that the predictions of Diehl et
al.~\cite{Diehl:2004cx} are systematically steeper than the lattice
data, in particular at small $-t$. This is most probably related to
the fact the lattice data for the radii at the accessible pion masses
are systematically below the experimental values.  We further observe
that the ratio of polarized GFFs is flatter than the unpolarized one
(see for instance the $m_\pi\simeq 600$MeV graph, where the data is
rather accurate).

Figures~\ref{fig:xi-dep2} and~\ref{fig:xi-dep3} display respectively
the ratios of GFFs $C_{20}/A_{20}$ and $A_{32}/A_{30}$, both in the
isovector and isosinglet channels.  We have already remarked that
$C_{20}$ is consistent with zero in the isovector channel; however in
the isosinglet channel, the ratio $4C_{20}/A_{20}$ is of the order of
-1.  This makes the $\xi$ dependence of $H^{n=2}(\xi,t)$ and
$E^{n=2}(\xi,t)$ an order unity effect, see
Eq.~\eqref{eq:lattice-vec-gff-def}, which is of great interest due to
its direct relation to the frequently discussed $D$-term
\cite{Polyakov:1999gs}. For the next higher moment, $n=3$, the data is
rather noisy, but at $\mps=495$MeV and $\mps=597$MeV, it is
nevertheless clear that $4A_{32}/A_{30}$ is positive, and most likely
of order +1. Thus the $\xi$ dependence of $H^{n=3}(\xi,t)$ is also
substantial, and goes in the other direction compared to the $n=2$
sector.

\subsubsection{\label{sec:A20Isovector}BChPT extrapolation of
  $A_{20}^{u-d}, B_{20}^{u-d}, C_{20}^{u-d}$}
In this section we discuss the forward and small $-t$ behavior of the
generalized form factors $A_{20}^{u-d}, B_{20}^{u-d}$ and
$C_{20}^{u-d}$.  In particular $A_{20}^{u-d}(t=0)$ is the isovector
momentum fraction $\langle x \rangle_{u-d}$.  The main novelty as
compared to \cite{Hagler:2007xi} is that we now have sufficiently
accurate data below $m_\pi=500$MeV to test the applicability of
covariant baryon chiral perturbation theory (BChPT) in that regime
exclusively. In Ref.~\cite{Hagler:2007xi}, the fit range extended up
to $m_\pi=700$MeV.
\begin{figure}[htb]
  \centerline{\includegraphics[width=8.0cm,angle=0]%
    {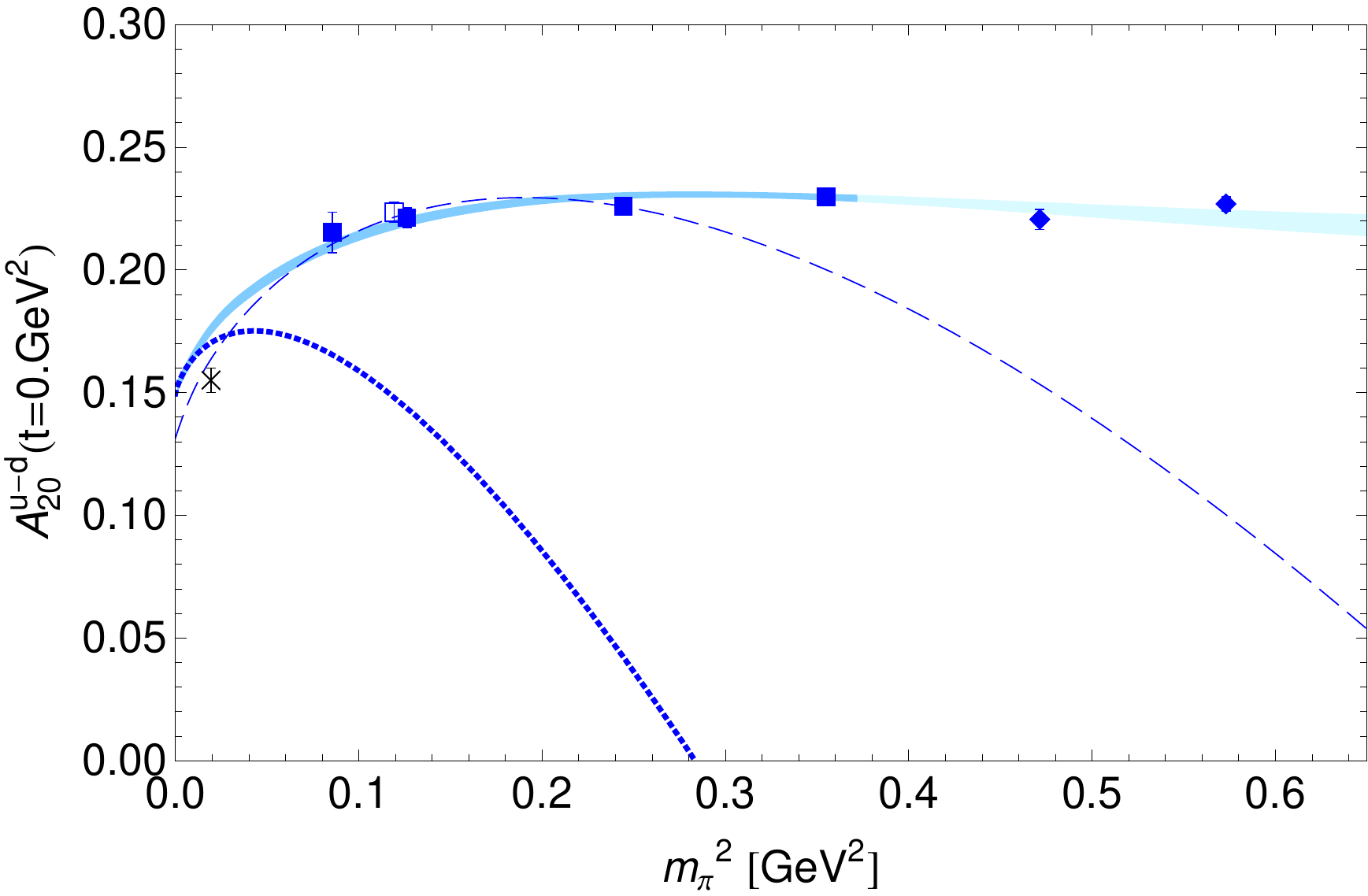}
    \includegraphics[width=8.0cm,angle=0]%
    {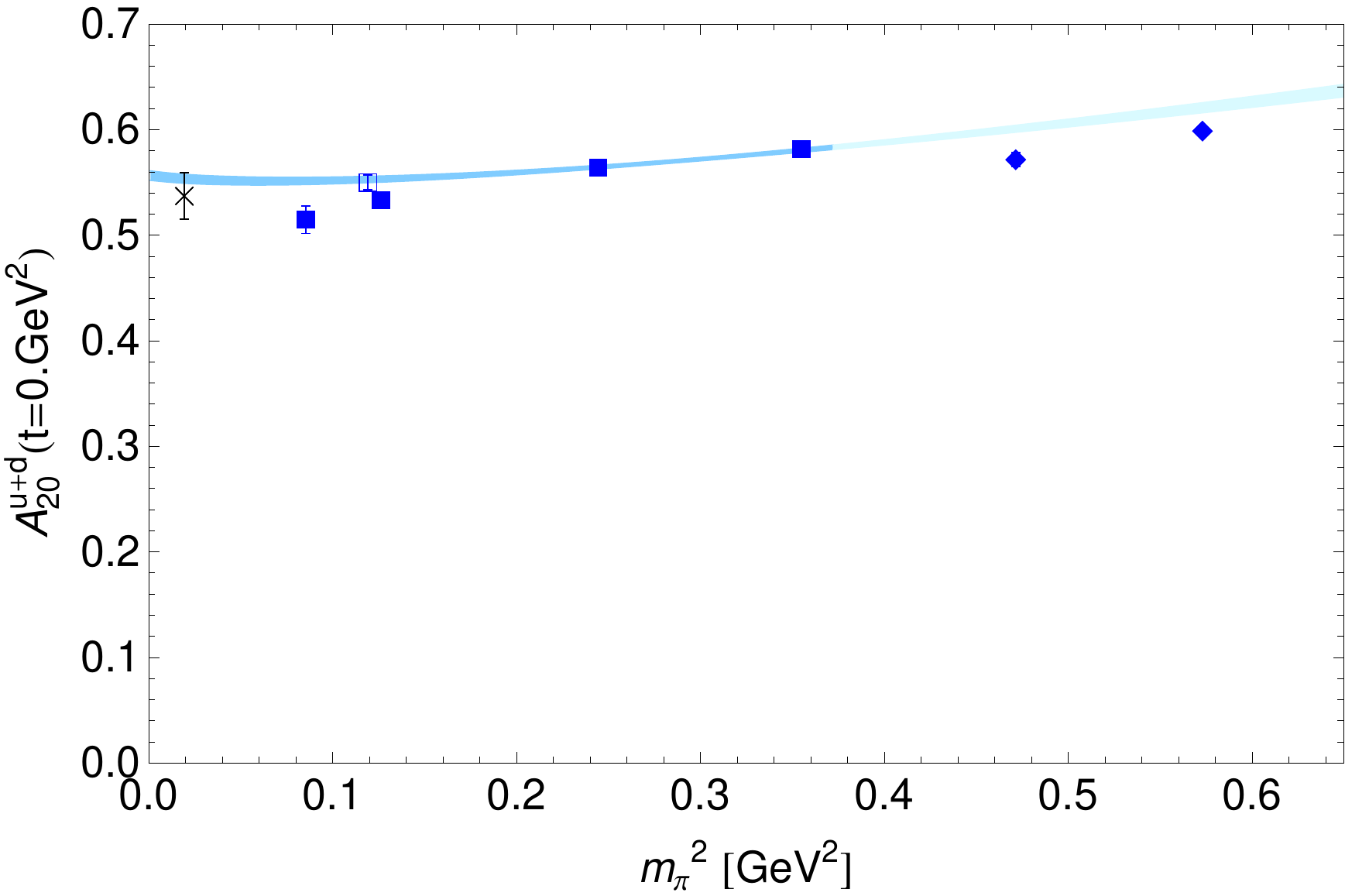}}
  \centerline{\includegraphics[width=8.0cm,angle=0]%
    {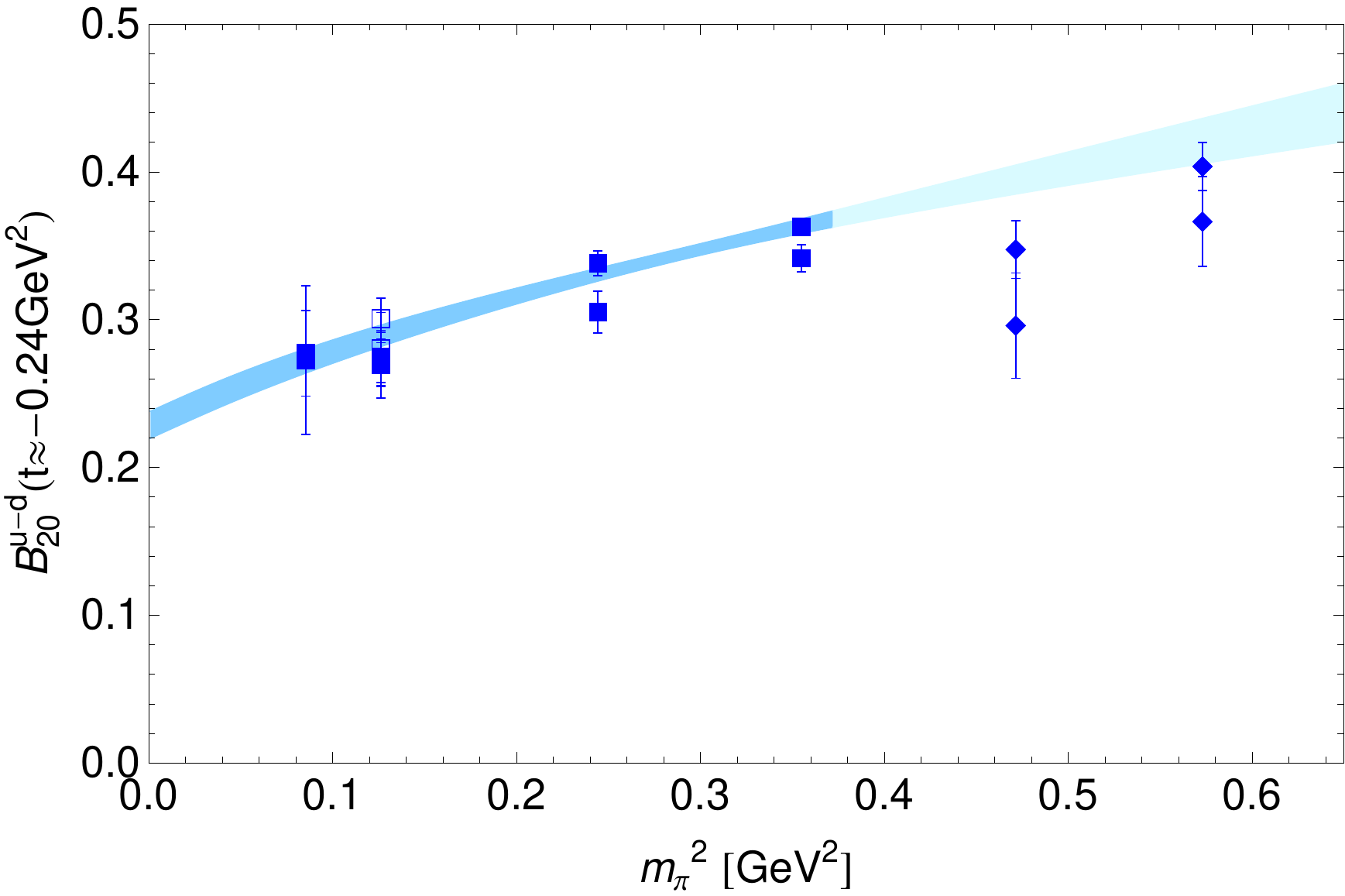}
    \includegraphics[width=8.0cm,angle=0]%
    {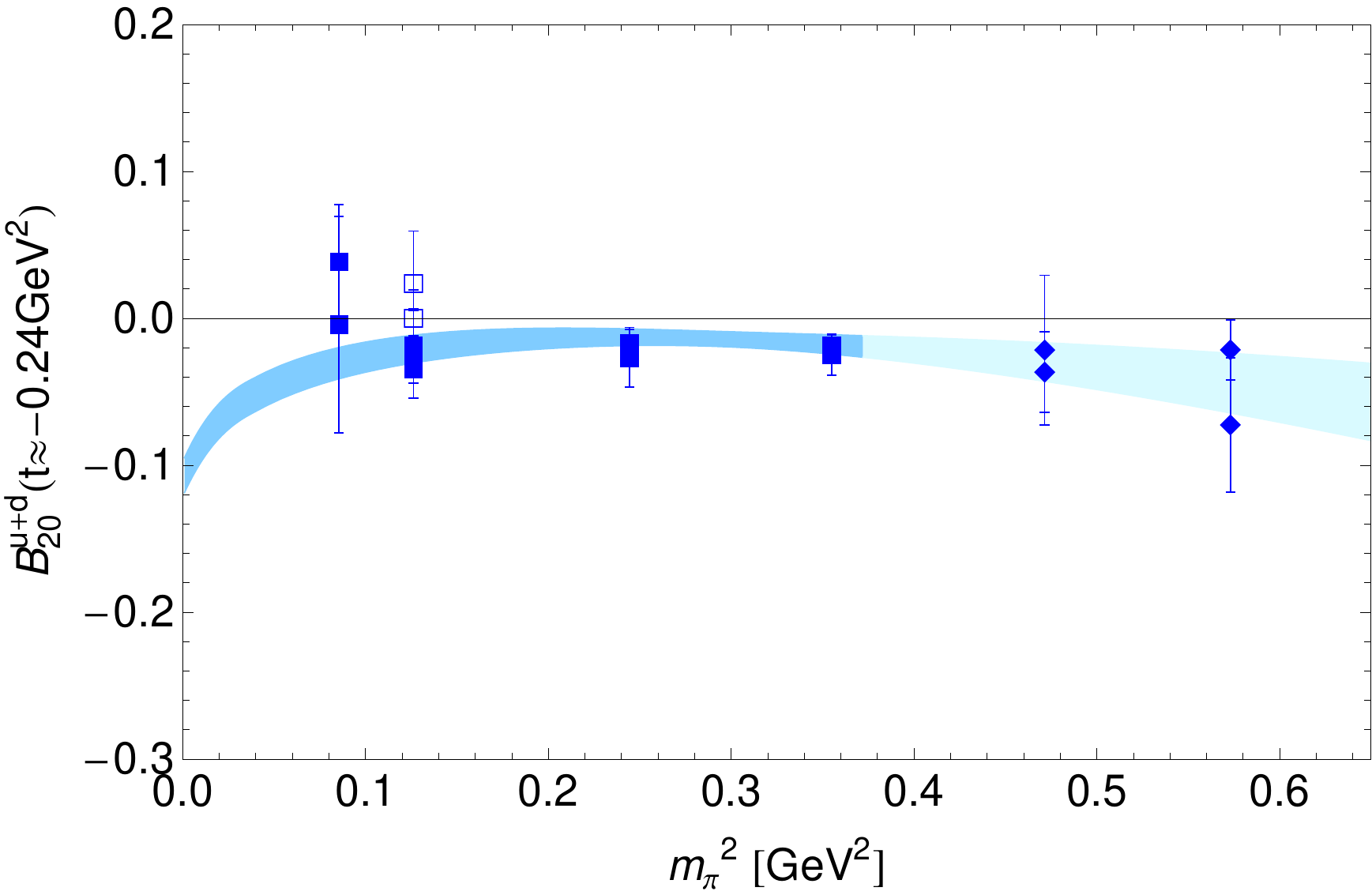}}
  \centerline{\includegraphics[width=8.0cm,angle=0]%
    {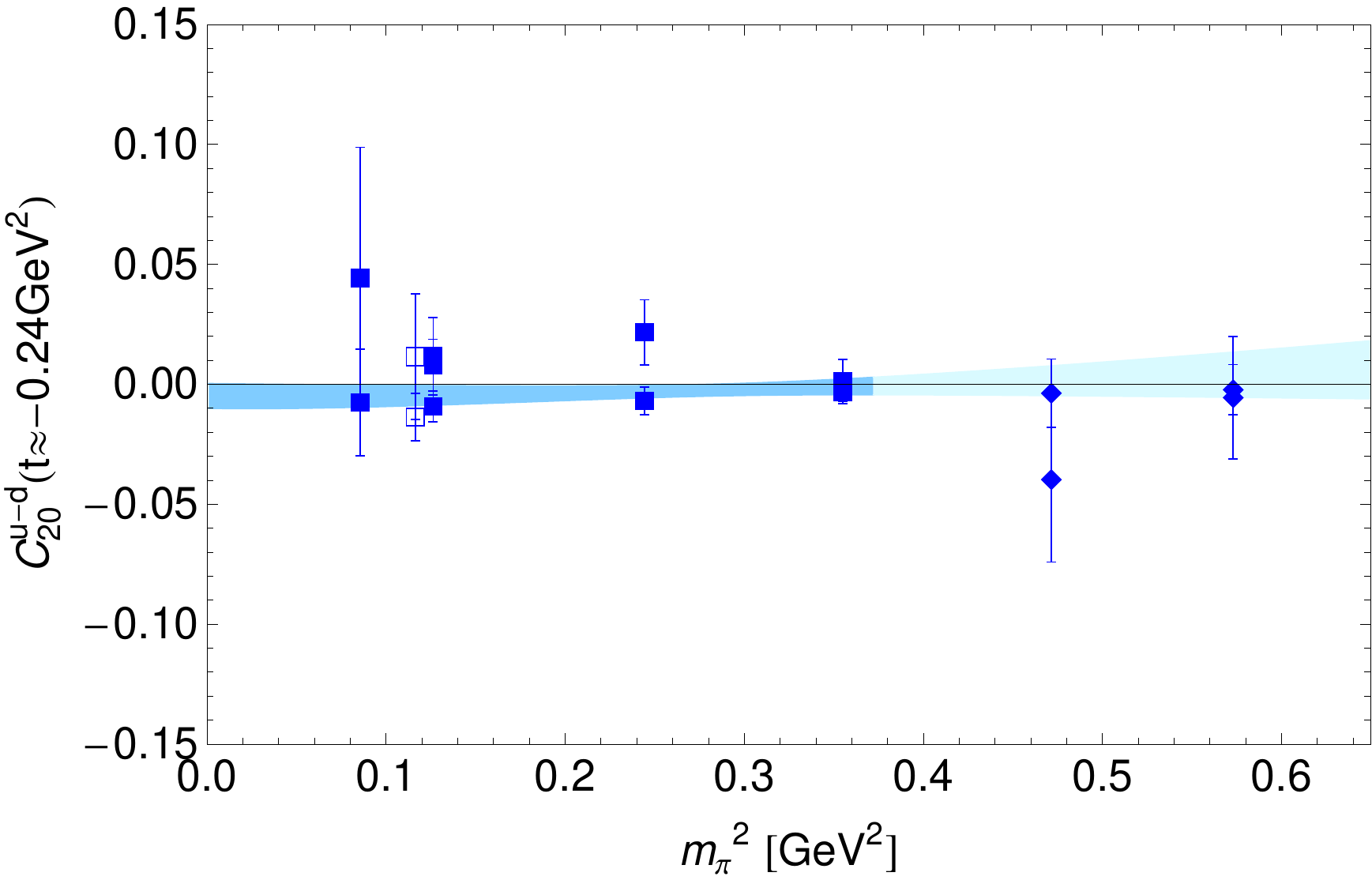}
    \includegraphics[width=8.0cm,angle=0]%
    {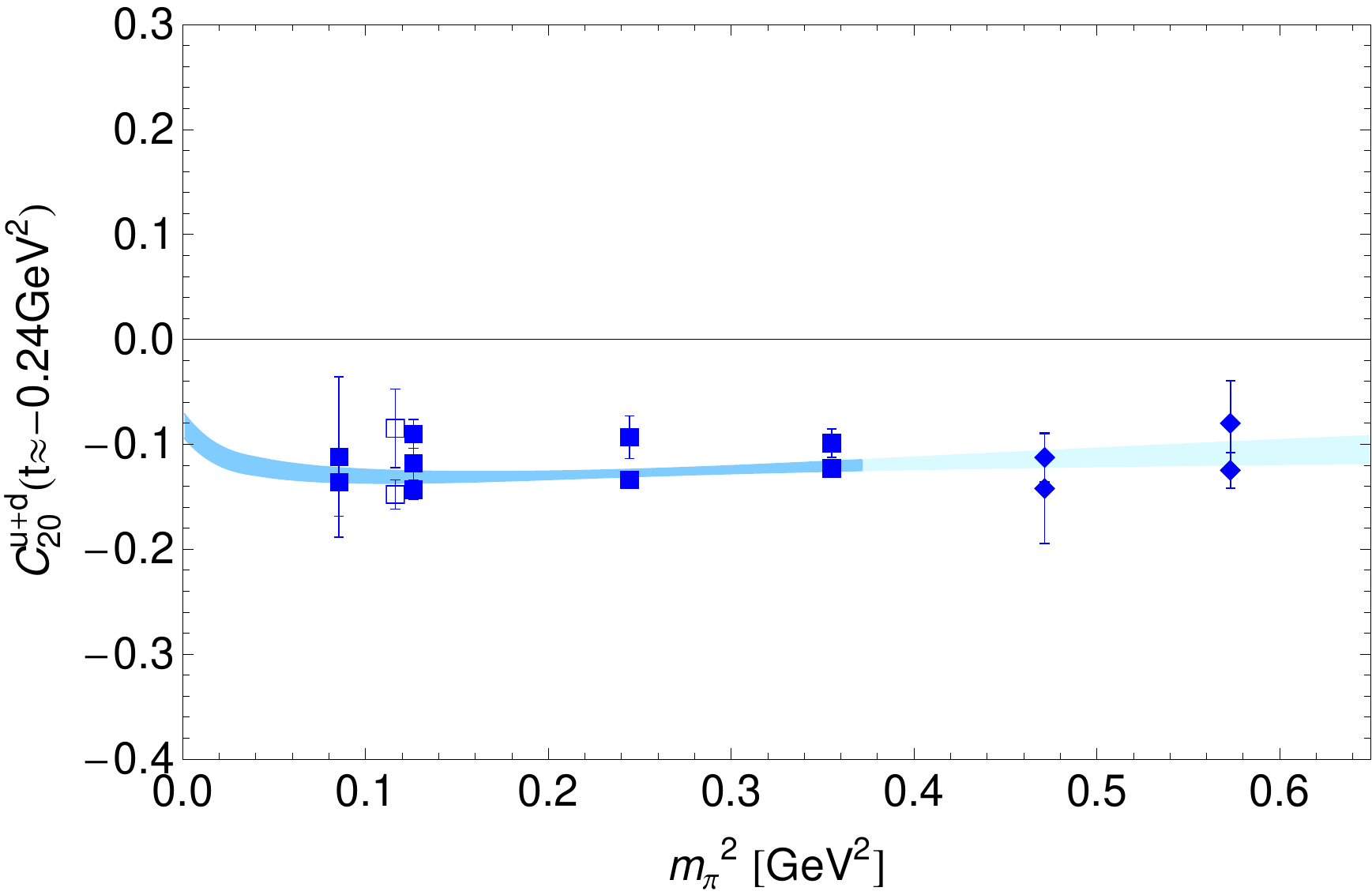}}
  \caption{Simultaneous BChPT fit to isovector (left) and isosinglet
    (right) lattice data for $m_\pi<580$MeV. The dotted line is the
    heavy-baryon limit of the BChPT fit. A HBChPT fit to the lattice
    data for $|t|<0.3$GeV$^2$ and $m_\pi<500$MeV is shown by the
    dashed line.  The phenomenological value of $\langle
    x\rangle_{u\pm d}$ (CTEQ6) is indicated by the cross.
    Disconnected contributions are omitted in the isosinglet case.
    Notice that $B_{20}$ and $C_{20}$ are displayed for
    $t\approx-0.24$GeV$^2$.  At $m_\pi^2\approx 0.12$GeV$^2$, the
    larger volume ($28^3$) is the filled symbol, the smaller volume
    ($20^3$) the open symbol; for all other pion masses, the volume is
    $20^3$.  The $\chi^2/{\rm d.o.f.}$ is 1.5 and 2.8 respectively in
    the isovector and isosinglet cases.}
  \la{fig:A20upmd}
\end{figure}
\begin{figure}[htb]
  \centerline{\includegraphics[width=11.5cm,angle=0]{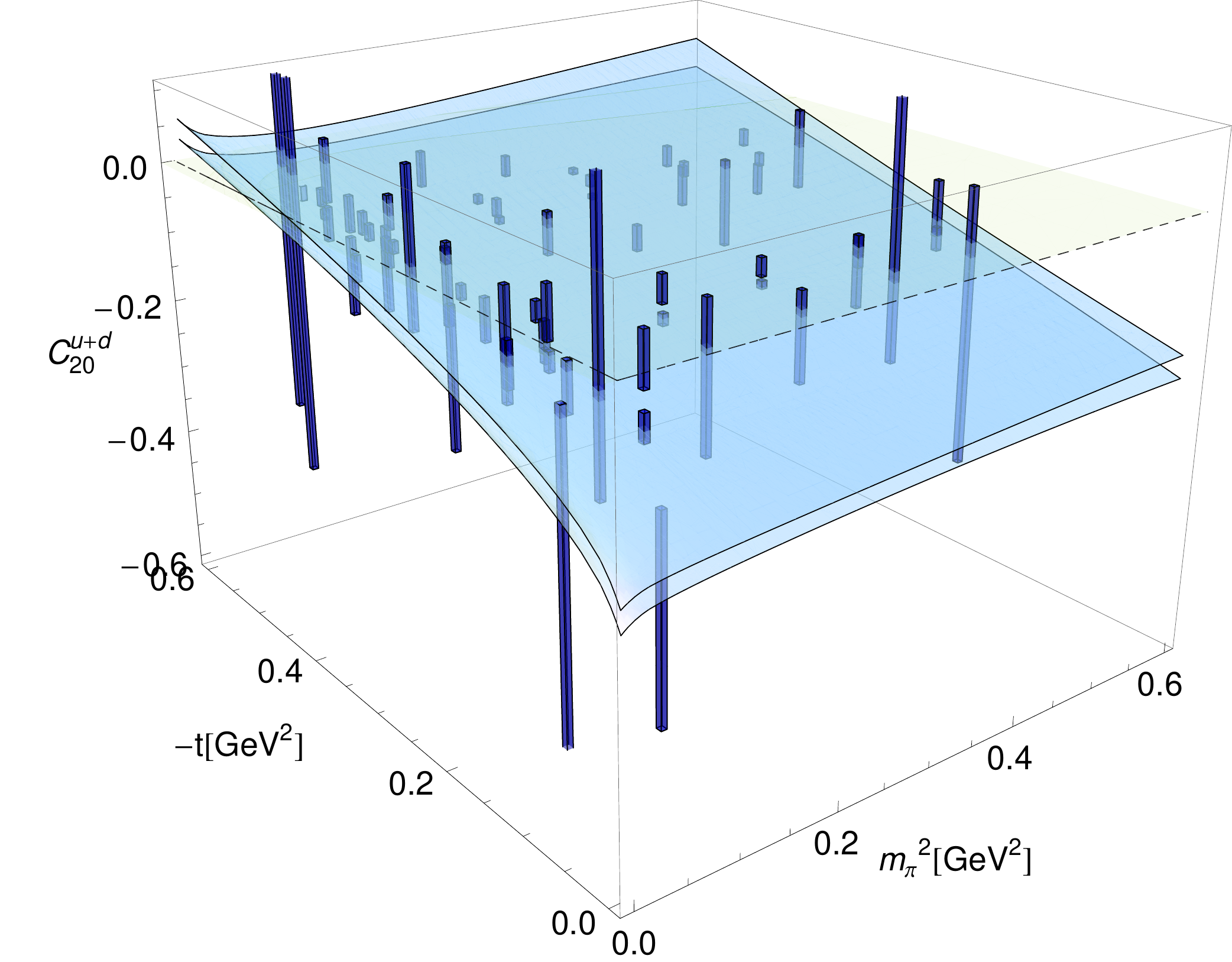}}
  \caption{The $t$ and $m_\pi^2$ dependence of the isosinglet
    generalized form factor $C_{20}$.  The two sheets indicate the
    statistical uncertainty of the BChPT fit.}
  \la{fig:C20-3D-BChPT}
\end{figure}

The $\mathcal{O}(p^2)$ BChPT result~\cite{Dorati:2007bk} for the
isovector GFFs $\{A,B,C\}_{20}^{u-d}(t)$ is
\begin{eqnarray}
  A_{20}^{u-d}(t,m_\pi) &=&  A_{20}^{0,u-d}\bigg(f_A^{u-d}(m_\pi) 
  + \frac{g_A^2}{192 \pi^2f_\pi^2}h_A(t,m_\pi)\bigg)
  \nn
  && \quad + \widetilde A_{20}^{0,u-d} j_A^{u-d}(m_\pi) + 
  A_{20}^{m_\pi,u-d} m_\pi^2 + A_{20}^{t,u-d} t\,,
  \label{ChPTA20umdp4} 
  \\
  B_{20}^{u-d}(t,m_\pi)&=&\frac{m_N(m_\pi)}{m_N} B_{20}^{0,u-d} + A_{20}^{0,u-d} h_B^{u-d}(t,m_\pi) 
  + \frac{m_N(m_\pi)}{m_N}\bigg\{\delta_{B}^{t,u-d}\,t + \delta_{B}^{m_\pi,u-d}\,m_\pi^2  \bigg\}\,,
  \label{ChPTB20umdp4}
  \\
  C_{20}^{u-d}(t,m_\pi)&=&\frac{m_N(m_\pi)}{m_N} C_{20}^{0,u-d} + A_{20}^{0,u-d} h_C^{u-d}(t,m_\pi) 
  + \frac{m_N(m_\pi)}{m_N}\bigg\{\delta_{C}^{t,u-d}\,t + \delta_{C}^{m_\pi,u-d}\,m_\pi^2 \bigg\}\,.
  \label{ChPTC20umdp4}
\end{eqnarray}
Here $f_A^{u-d}(m_\pi)$, $h_{A,B,C}(t,m_\pi)$ and $j_A^{u-d}(m_\pi)$
contain the non-analytic dependence on the pion mass and momentum
transfer squared (see for instance Eq. (28), (40) and (41)
of~\cite{Dorati:2007bk}), while $A_{20}^{0,u-d} \equiv
A_{20}^{u-d}(t=0,m_\pi=0)$, $A_{20}^{m_\pi,u-d}$ and $A_{20}^{t,u-d}$
are low-energy constants.  Similarly $B_{20}^{0,u-d} \equiv
B_{20}^{u-d}(t=0,m_\pi=0)$, $C_{20}^{0,u-d} \equiv
C_{20}^{u-d}(t=0,m_\pi=0)$, and we have included estimates of
$\mathcal{O}(p^3)$-corrections in form of $(\delta_{B}^{t,u-d}\,t)$,
$(\delta_{B}^{m_\pi,u-d}\,m_\pi^2)$, $(\delta_{C}^{t,u-d}\,t)$ and
$(\delta_{C}^{m_\pi,u-d}\,m_\pi^2)$. The associated low-energy
constants are treated as free parameters and may be obtained from a
fit to the lattice data.  Because of the small prefactor, the term
$\propto h_A(t,m_\pi)$ is of $\mathcal{O}(10^{-3})$ for
$m_\pi\le700$MeV, $|t|<1$GeV$^2$ and therefore numerically
negligible. Also, $h_{B,C}(t,m_\pi)$ are very weakly dependent on $t$,
so that the Ansatz for each form factor is practically linear in $t$.
The parameter $m_N$ in the denominator of Eqs.~\eqref{ChPTB20umdp4}
and~\eqref{ChPTC20umdp4} is the nucleon mass in the chiral limit and
was set to $890$MeV, as given in \tab\ref{tab:sse-pars}.

\begin{figure}[htb]
  \centerline{\includegraphics[width=8.5cm,angle=0]{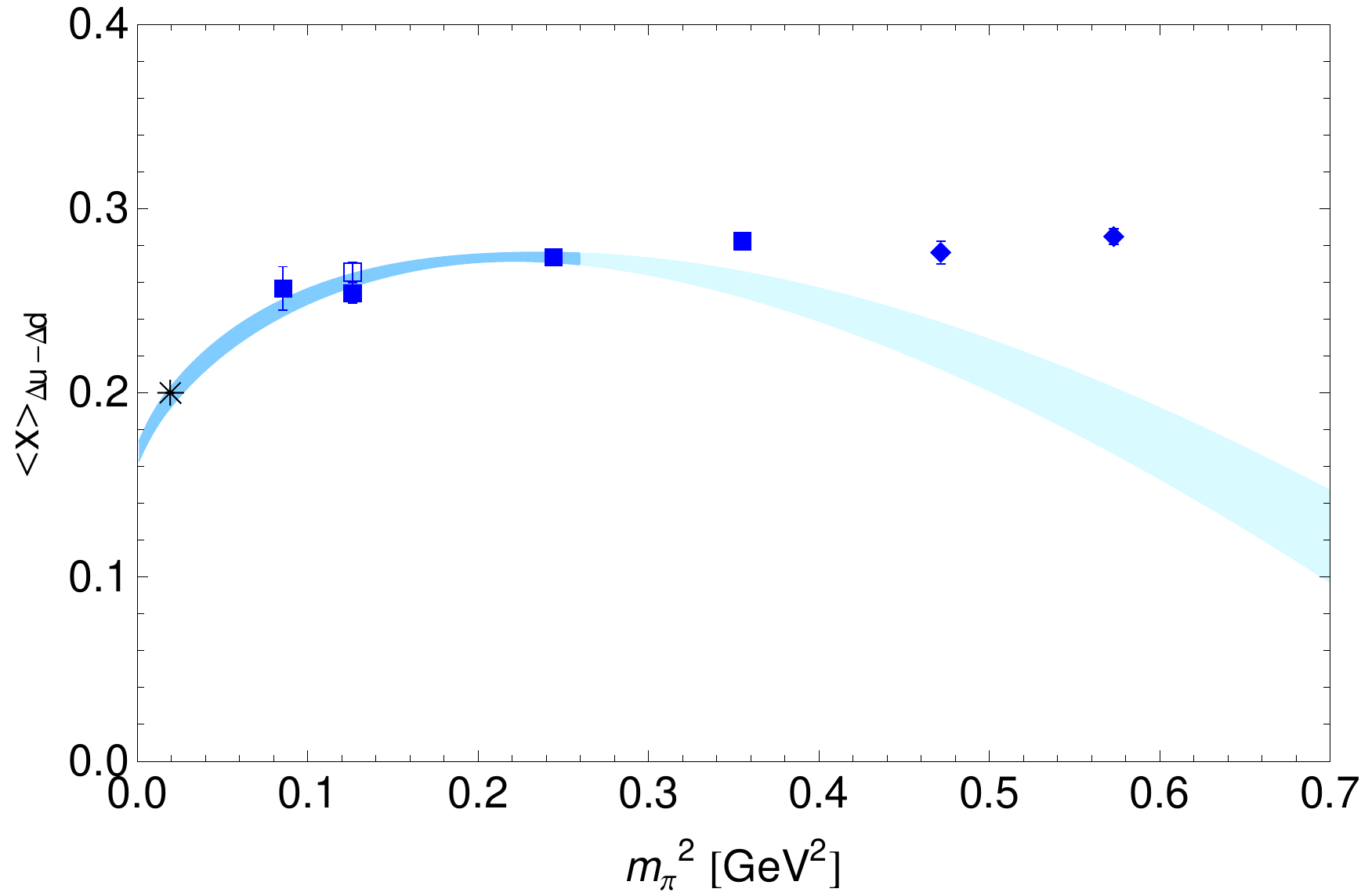}}
  \caption{The extrapolation of the polarized momentum fraction using
    HBChPT\@. The value obtained in the chiral limit is $0.165(8)$,
    the experimental value is from HERMES~\cite{Airapetian:2007aa}.}
  \label{fig:A20u-dfit}
\end{figure}

We use the value $\widetilde A_{20}^{0,u-d}=0.17$ obtained from a
heavy baryon chiral perturbation theory (HBChPT) fit to our
$m_\pi<500$MeV lattice results for $\widetilde
A_{20}^{u-d}(t=0)=\langle x\rangle_{\Delta u-\Delta d}$.  This value
is consistent with an earlier determination \cite{Edwards:2006qx},
which we used for a similar fit in \cite{Hagler:2007xi}.  The fit is
displayed in \fig\ref{fig:A20u-dfit}.

Since the low energy constant $A_{20}^{0,u-d}$ is a common parameter
in the BChPT-formulae for the GFFs $A_{20}^{u-d}$, $B_{20}^{u-d}$ and
$C_{20}^{u-d}$, we performed a simultaneous fit based on
Eqs.~\eqref{ChPTA20umdp4} to \eqref{ChPTC20umdp4} with a total of $9$
(1 common and 8 separate) fit parameters to over 83 lattice data
points. The fit is displayed in the left panel of
\fig\ref{fig:A20upmd} and the results in the chiral limit and at the
physical pion mass are listed below in \tab\ref{tab:BChPTtab}. The
${\chi^2}/{\rm dof}=1.5$ is good and gives us confidence that the fit
works well.  Among the most important results is
$\<x\>_{u-d}=0.1758(20)$, to be compared with $\langle
x\rangle_{u-d}^{\text{CTEQ6}}=0.155(5)$.  Our result is thus $10\%$ to
$15\%$ higher than the phenomenological value, a statistically
significant difference.  Compared to \cite{Hagler:2007xi}, it has
increased from 0.157(10) by a little less than two standard
deviations. The main reason is that $A_{20}^{u-d}(t=0)$ at our
lightest pion mass is bending down less than predicted by the chiral
expansion compared to the heavier pion data points.

We obtain $B_{20}^{u-d}(t=0,m_\pi^{\rm phys})=0.293(12)$, a value
compatible with our previous result 0.273(63) \cite{Hagler:2007xi},
but with a much reduced uncertainty.  The quantity
$C_{20}^{u-d}(t=0,m_\pi^{\rm phys})=-0.0157(86)$ still comes out much
smaller than $A_{20}$ and $B_{20}$, but there is now a hint that it is
in fact negative.

To study the difference between HBChPT and BChPT, we take the heavy
baryon limit of BChPT while keeping the same values of the fit
parameters, and obtained the dotted line in \fig\ref{fig:A20upmd}.
This curve only overlaps with the BChPT curve for $m_\pi<m_{\pi}^{\rm
  phys}$ and drops off sharply for $m_\pi>m_{\pi}^{\rm phys}$,
indicating the quantitative importance of the truncated terms when
using the coefficients from the BChPT fit.  The dashed curve in
\fig\ref{fig:A20upmd} shows the result of fitting our lattice data for
$|t|<0.3$GeV$^2$ and $m_\pi<500$MeV directly with the HBChPT
expression, and indicates that the latter describes the behavior of
our lattice data over a significantly smaller range of pion masses
than the BChPT expression.

\begin{figure}[thbp]
  \centerline{\includegraphics[width=8.5cm,clip=true,angle=0]{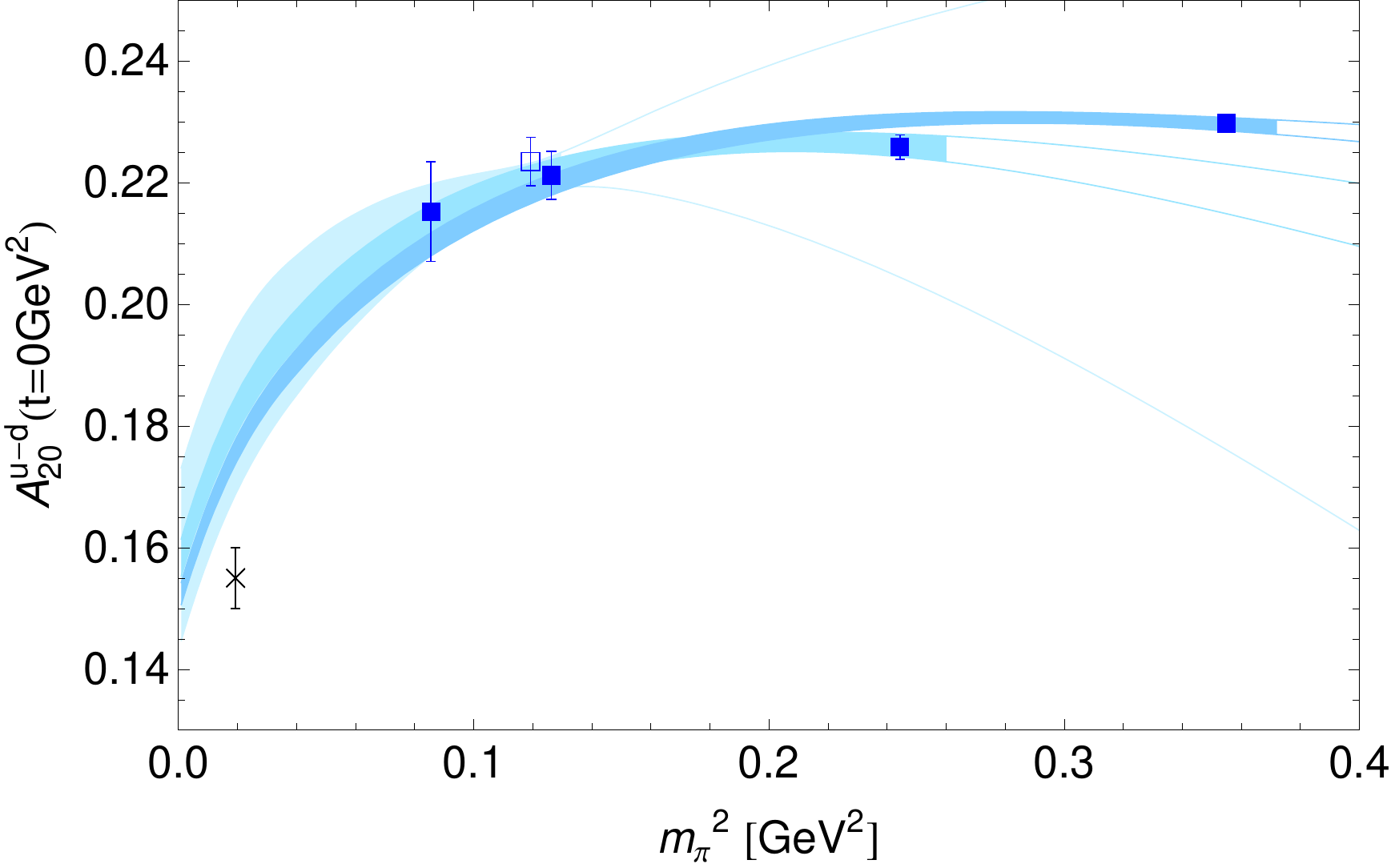}}
  \caption{A study of the stability of the BChPT extrapolation of
    $A_{20}^{u-d}$ at $t=0$GeV$^2$ based on Eqs.~\eqref{ChPTA20umdp4},
    \eqref{ChPTB20umdp4}, and \eqref{ChPTC20umdp4}.  The three
    different error bands represent chiral fits to lattice results for
    pion masses in the regions $m_\pi<400$, $500$ and $600$MeV,
    respectively.}
\label{BChPTstudy}
\end{figure}

As already mentioned above, one of the main achievements of this work
is the substantial reduction of the statistical uncertainties of the
data points at $m_\pi=356$, $495$ and $597$MeV compared to our
previous work~\cite{Hagler:2007xi}, and the inclusion of an additional
ensemble at $m_\pi=293$MeV. This allows us in particular to lower the
cut in $m_\pi$ in the chiral fits from $700$MeV to $600$MeV.  However,
it is also important to note that a full BChPT analysis of $A_{20},
B_{20}$ and $C_{20}$ consistently including all terms of
$\mathcal{O}(p^3)$ is still not available. In view of this, it is
interesting to perform a first check of the stability and potential
uncertainty of the chiral extrapolations by repeating the fit for
different maximal values of the included pion masses.
Figure~\ref{BChPTstudy} shows a comparison of the BChPT extrapolations
of $A_{20}^{u-d}$ from fits to the lattice data in the regions
$m_\pi<400$, $500$ and $600$MeV. Most importantly, we find that the
error bands from all three extrapolations are consistent and do
overlap in the region below $m_\pi^2\sim 0.25$MeV$^2$ down to the
chiral limit. Apparently, the bending towards the physical point is
not overly sensitive to the large pion mass region, where
$\mathcal{O}(p^3)$ corrections would have the strongest impact.  A
quantitative estimate of the uncertainties of the chiral extrapolation
must therefore be based on an improved ChPT analysis (e.g.~including
higher order effects) and higher precision lattice data at even lower
pion masses, which is beyond the scope of the present work.

\subsubsection{\label{sec:A20Isosinglet}BChPT extrapolation of
  $A_{20}^{u+d}, B_{20}^{u+d}, C_{20}^{u+d}$}
In this section we treat the GFFs in the isosinglet channel.  The
quantity $A_{20}^{u+d}(t=0)=\langle x\rangle_{u+d}$ is not only an
important hadron structure observable on its own but in addition plays
an important role for the computation of the total angular momentum
contribution of quarks to the nucleon spin,
$J^{u+d}=1/2(A_{20}^{u+d}(0)+B_{20}^{u+d}(0))$, a discussion that we
postpone to \sect\ref{sec:decomp-nucl-spin}. The combined
$(t,m_\pi)$-dependence in BChPT is given by \cite{Dorati:2007bk}:
\begin{eqnarray}
  A_{20}^{u+d}(t,m_\pi)&=&
  A_{20}^{0,u+d}\bigg(f^{u+d}_A(m_\pi) - \frac{g_A^2}{64
    \pi^2f_\pi^2}h_A(t,m_\pi)\bigg)
  \nonumber \\ && \quad + 
  A^{m_\pi,u+d}_{20} m_\pi^2 + A^{t,u+d}_{20} t + \Delta
  A_{20}^{u+d}(t,m_\pi) + \mathcal{O}(p^3)\,, 
  \label{ChPTA20umdp4new} \\
  B_{20}^{u+d}(t,m_\pi)&=&
  \frac{m_N(m_\pi)}{m_N} B_{20}^{0,u+d} + A_{20}^{0,u+d}
  h_B^{u+d}(t,m_\pi)
  + \Delta B_{20}^{u+d}(t,m_\pi)
  \nonumber \\ && \quad +
  \frac{m_N(m_\pi)}{m_N}\bigg\{\delta_{B}^{t,u+d}\,t 
  + \delta_{B}^{m_\pi,u+d}\,m_\pi^2\bigg\}  + \mathcal{O}(p^3) \,,
  \label{ChPTB20updp4} \\
  C_{20}^{u+d}(t,m_\pi)&=&\frac{m_N(m_\pi)}{m_N} C_{20}^{0,u+d} +
  A_{20}^{0,u+d} h_C^{u+d}(t,m_\pi)  + 
  \Delta C_{20}^{u+d}(t,m_\pi) + \mathcal{O}(p^3)\,,
  \label{ChPTC20updp4}
\end{eqnarray}
where $A_{20}^{0,u+d} \equiv A_{20}^{u+d}(t=0,m_\pi=0)$, and
$f^{u+d}_A(m_\pi)$ and $h_A(t,m_\pi)$ contain the non-analytic
dependence on the pion mass and momentum transfer squared.  Also,
$B_{20}^{0,u+d} \equiv B_{20}^{u+d}(t=0,m_\pi=0)$, and the terms
$\Delta B_{20}$, $\delta_{B}^{t,u+d}\,t$ and
$\delta_{B}^{m_\pi,u+d}\,m_\pi^2$ are of $\mathcal{O}(p^3)$ and
represent only a part of the full $\mathcal{O}(p^3)$ contribution.
Similarly, $C_{20}^{0,u+d} \equiv C_{20}^{u+d}(t=0,m_\pi=0)$, and the
term $\Delta C_{20}^{u+d}$, proportional to the pion momentum fraction
carried by quarks in the chiral limit $\langle
x\rangle^{\pi,0}_{u+d}$, is a part of the full
$\mathcal{O}(p^3)$-corrections \cite{Dorati:2007bk}.  As in the
isovector case, the parameter $m_N$ in the denominator of
Eqs.~\eqref{ChPTB20updp4} and~\eqref{ChPTC20updp4} was set to
$890$MeV.

The constants $A^{m_\pi,u+d}_{20}$, $A_{20}^{t,u+d}$,
$B_{20}^{0,u+d}$, $\delta_{B}^{t,u+d}$ and $\delta_{B}^{m_\pi,u+d}$
may be obtained from a fit to the lattice data.  In this counting
scheme, contributions from operator insertions in the pion line,
proportional to $\langle x\rangle^{\pi,0}_{u+d}$, are of order
$\mathcal{O}(p^3)$.  Counter terms of the form $\delta_{C}^{t,u+d}\,t$
and $\delta_{C}^{m_\pi,u+d}\,m_\pi^2$ first appear at
$\mathcal{O}(p^4)$.  Since in \cite{Hagler:2007xi} the term $\Delta
B_{20}^{u+d}(t,m_\pi)$ was found to lead to unstable fits, we perform
a fit dropping this contribution but keeping the counter terms
$\propto t$ and $\propto m^2_\pi$.  
We have also included the formally
higher order counter terms $\delta_{C}^{t,u+d}\,t$ and
$\delta_{C}^{m_\pi,u+d}\,m_\pi$ in the fit to our lattice data.
Furthermore, in order to see if such contributions could be relevant
for the pion masses and values of the momentum transfer squared
accessible in our calculation, we also include the estimate of the
$\mathcal{O}(p^3)$-contribution $\Delta A_{20}^{u+d}$ provided
in~\cite{Dorati:2007bk} in the fit to the lattice data points.

Similar to the isovector case discussed in the previous sections, the
low energy constant $A_{20}^{0,u+d}$ is a common parameter in the
chiral extrapolation formulae for the isosinglet GFFs $A_{20}^{u+d}$,
$B_{20}^{u+d}$ and $C_{20}^{u+d}$.  Using $\langle
x\rangle^{\pi,0}_{u+d}=0.5$ as an input
parameter~\cite{Sutton:1991ay,Gluck:1999xe,Meyer:2007tm,Guagnelli:2004ga},
we performed a simultaneous fit to 83 lattice data points for these
three GFFs, based on Eqs.~\eqref{ChPTA20umdp4new},
\eqref{ChPTB20updp4} and~\eqref{ChPTC20updp4}, with 1 common and 8
separate low energy constants as fit parameters. The results are also
summarized in Tab.~\ref{tab:BChPTtab}.
\begin{table}[htb]
  \centerline{\begin{tabular}{c@{~~~}c@{~~~}c}
      \hline\hline
      &   $m_\pi=0$     &  $m_\pi=m_\pi^{\rm phys}$  \\
      \hline
      $A^{u-d}_{20}$   &   0.1496(18)  &0.1758(20) \\
      $B^{u-d}_{20}$   &   0.282(12)   & 0.293(12) \\
      $C^{u-d}_{20}$   &   -0.0150(84) &-0.0157(86) \\
      \hline
      $A^{u+d}_{20}$   &  0.5567(45)  &  0.5534(43) \\
      $B^{u+d}_{20}$   & -0.126(16)  &  -0.077(16)  \\
      $C^{u+d}_{20}$   &  -0.277(15) &  -0.255(13) \\
      \hline\hline
    \end{tabular}}
  \caption{Chirally extrapolated GFFs $A_{20}, B_{20}, C_{20}$ using 
    BChPT.}
  \la{tab:BChPTtab}
\end{table}

The result $\chi^2/{\rm d.o.f.}$ for this fit is 2.8. The result in
the chiral limit is $A_{20}^{0,u+d}=0.5567(45)$, and $\langle
x\rangle_{u+d}= A_{20}^{u+d}(t=0,m_{\pi}^{\rm phys})=0.5534(43)$ at
the physical point.  Incidentally, this value is in very good
agreement with phenomenological results from CTEQ and MRST
\cite{DurhamDatabase} parameterizations, $\langle
x\rangle_{u+d}^{\text{MRST2001}}=0.538(22)$ and $\langle
x\rangle_{u+d}^{\text{CTEQ6}}=0.537(22)$.  A variation of the input
parameter $\langle x\rangle^{\pi,0}_{u+d}$ by $\pm10\%$ only leads to
a small change in $A_{20}^{0,u+d}(t=0)$ of a few percent.  The results
of the fit are shown in \fig\ref{fig:A20upmd}.  As already noted
in~\cite{Hagler:2007xi}, the slight upward bending in
\fig\ref{fig:A20upmd} at low $m_\pi$, is due to the
$\mathcal{O}(p^3)$-contribution $\Delta A_{20}^{u+d}$.  The fact that
the lightest pion mass data point does not exhibit this upward bending
is partly responsible for the large value of the $\chi^2$.  Further
calculations are required to see whether, for example, this point is
pushed downward by finite-size effects.  The inclusion of
contributions from disconnected diagrams could also lead to a
different $m_\pi$ dependence.

In \fig\ref{fig:C20-3D-BChPT}, the simultaneous dependence of
$C_{20}^{u+d}$ on $t$ and $m_\pi^2$ is displayed.  The error bars of
the lattice data points are illustrated by the stretched cuboids.  The
two surfaces represent the BChPT discussed in this section with its
uncertainty. The figure illustrates that $C^{u+d}_{20}$ becomes
relatively large near the origin, which implies that the $\xi$
dependence of $H^{n=2}(\xi,t)$ and $E^{n=2}(\xi,t)$ is far from being
negligible.
\begin{figure}[t]
  \centerline{\includegraphics[width=8.4cm,angle=0]{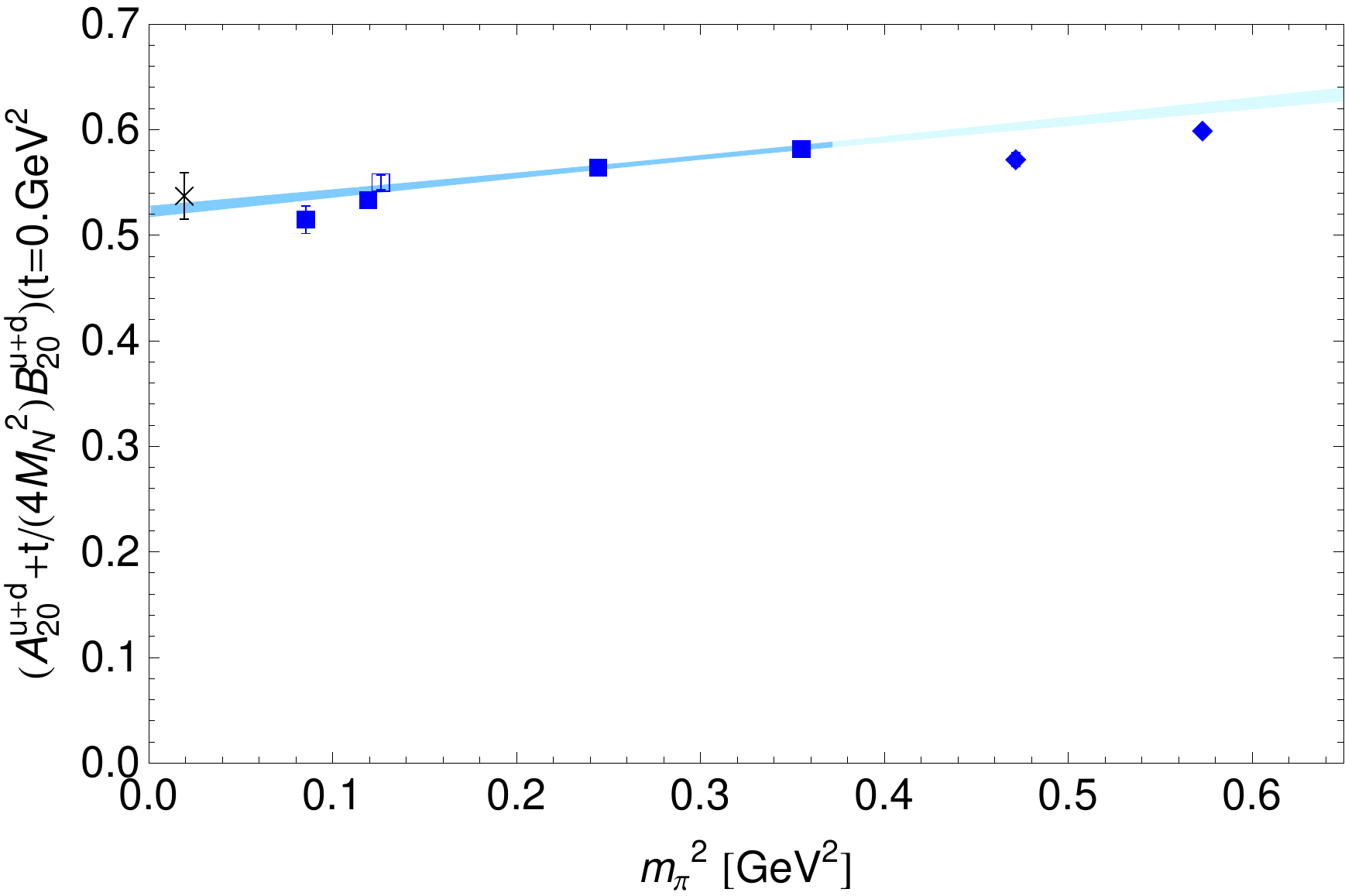}
    \includegraphics[width=8.4cm,angle=0]{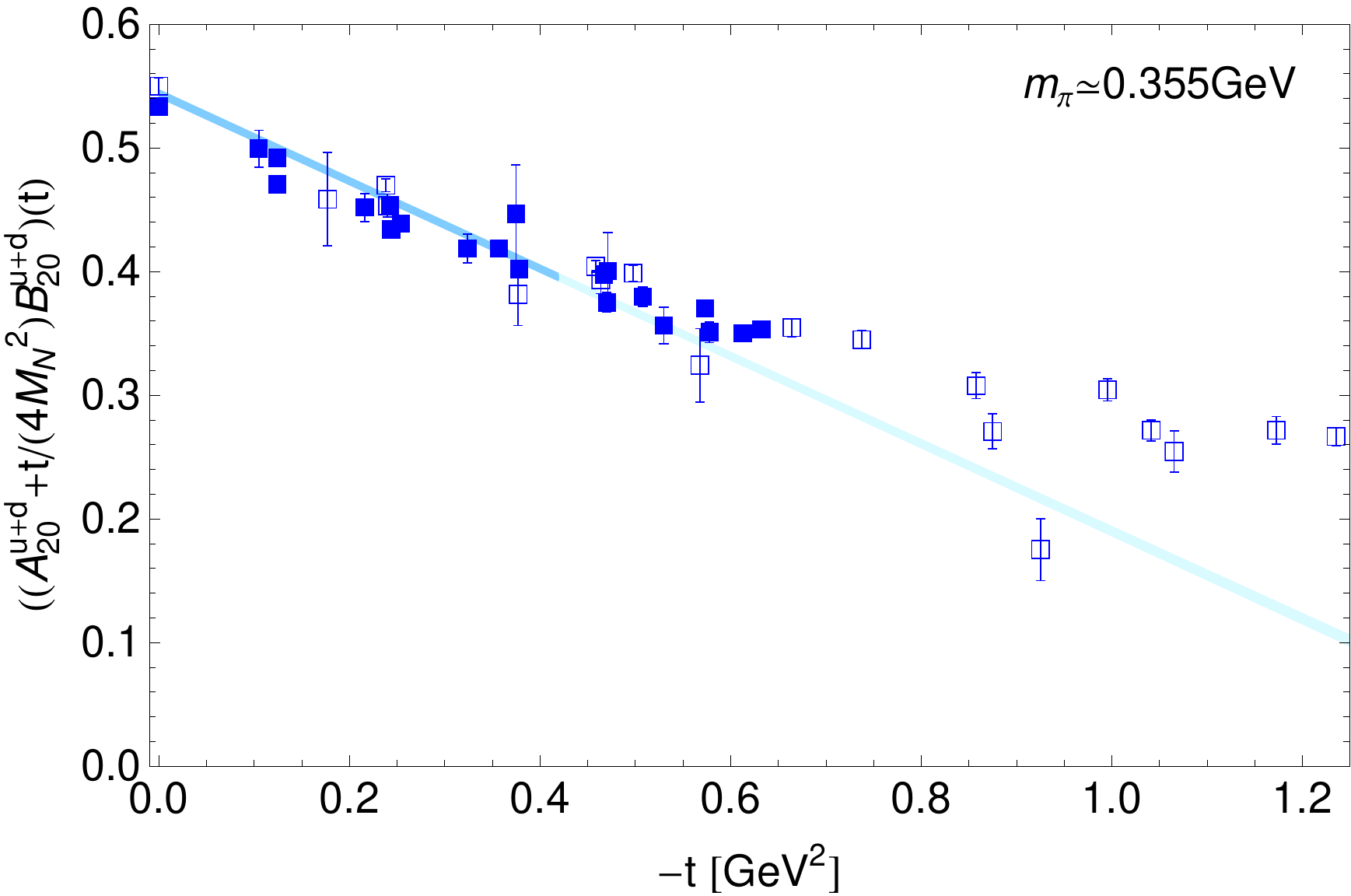}}
  \caption{Pion mass and momentum transfer dependence of $E_{20}$.  At
    $m_\pi=356$MeV, the open symbol corresponds to $20^3$, the filled
    symbol to $28^3$.  At all other pion masses, the volume is $20^3$.
    The extrapolation in $m_\pi$ at $t=0$ can be compared to the
    extrapolation of $A_{20}^{u+d}$ in \fig\ref{fig:A20upmd}.}
  \la{fig:E20-upd}
\end{figure}
\begin{figure}[htb]
  \centerline{\includegraphics[width=11.5cm,angle=0]{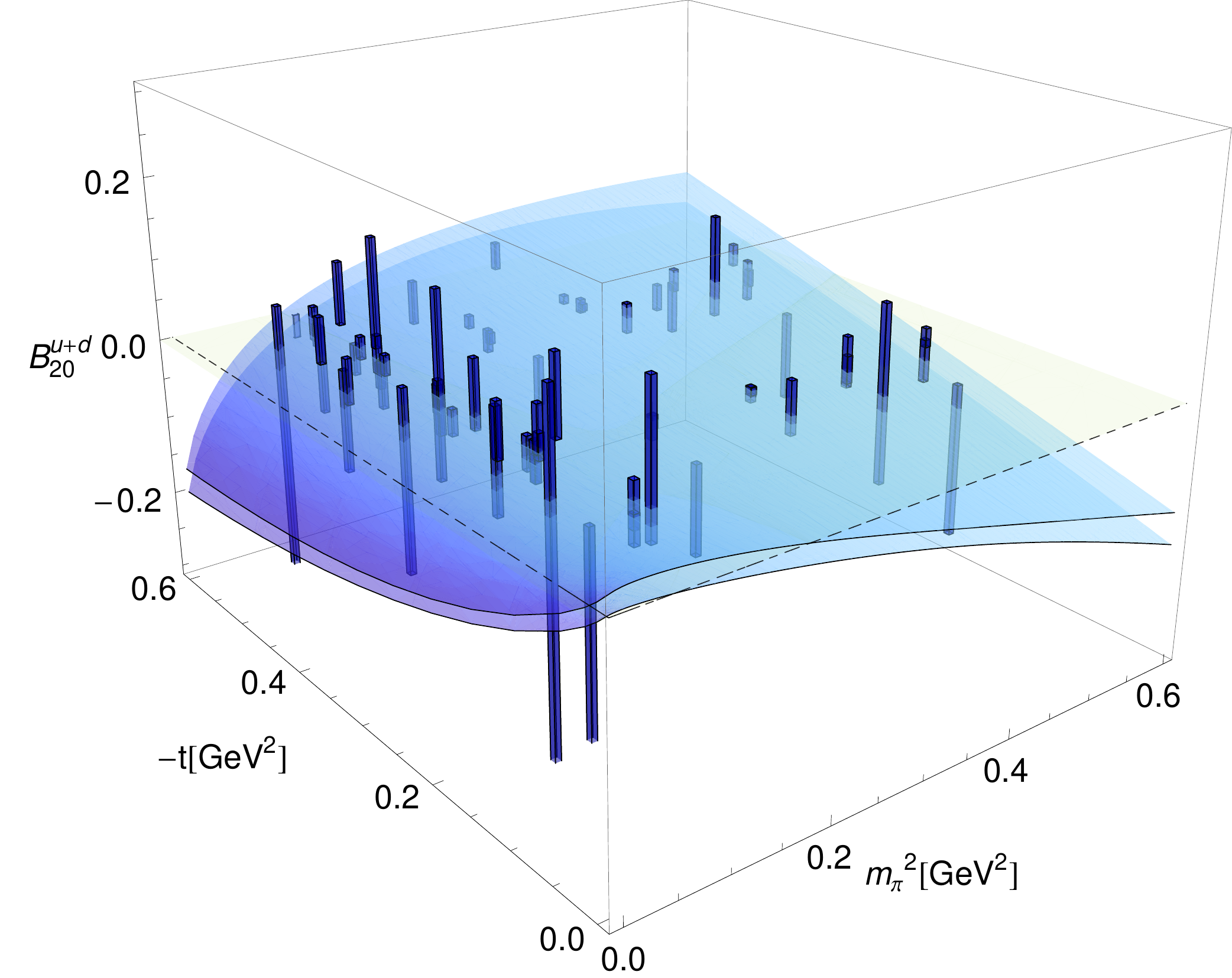}}
  \centerline{\includegraphics[width=11.5cm,angle=0]{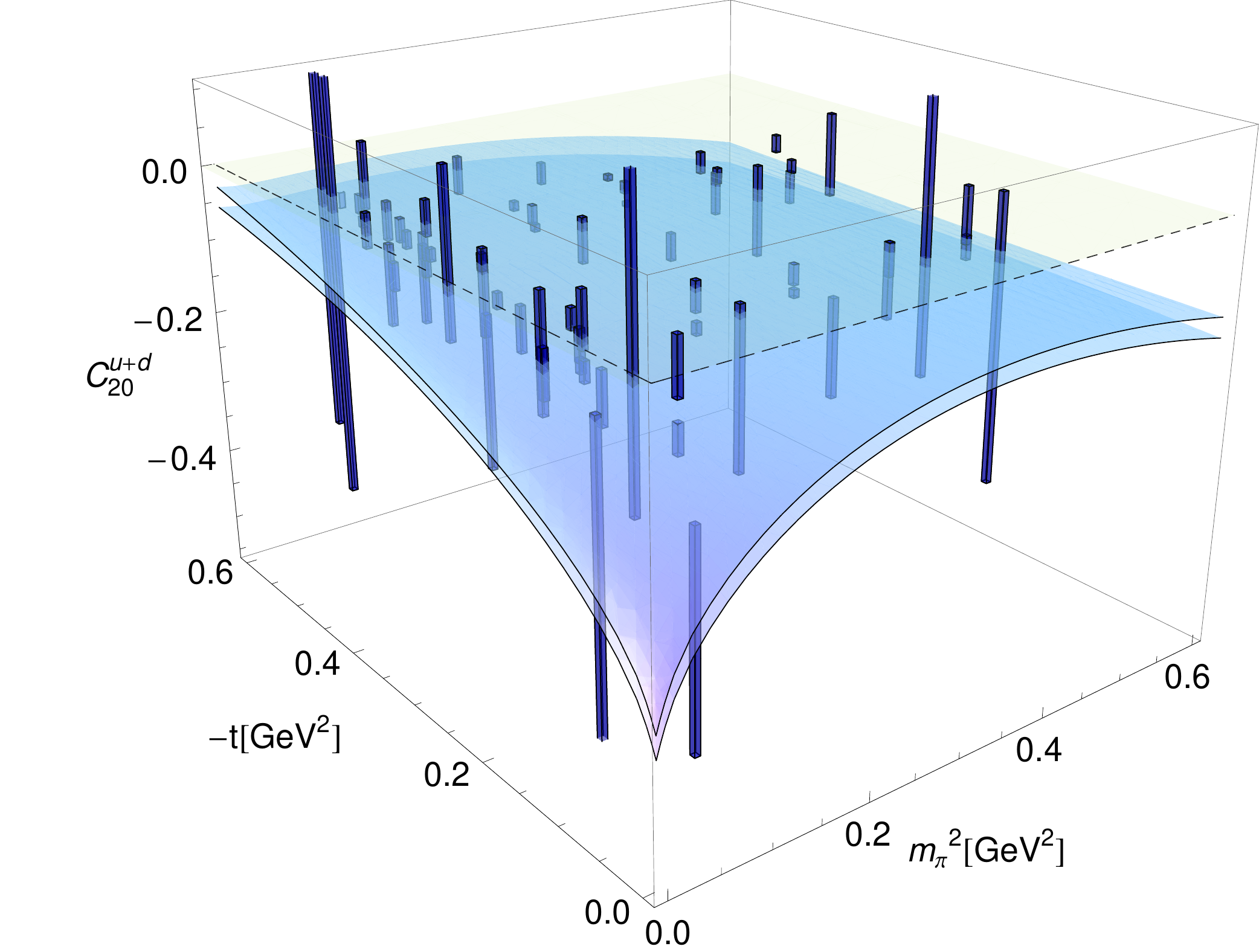}}
  \caption{The $t$ and $m_\pi^2$ dependence of the isosinglet
    generalized form factors $\frac{m_N^{\rm phys}}{m_N(m_\pi)}B_{20}$
    and $\frac{m_N^{\rm phys}}{m_N(m_\pi)}C_{20}$, and a HBChPT fit.
    Its uncertainty is indicated by the two surfaces.}
  \la{fig:BC20-3D}
\end{figure}
\begin{figure}[t]
  \centerline{\includegraphics[width=8.4cm,angle=0]{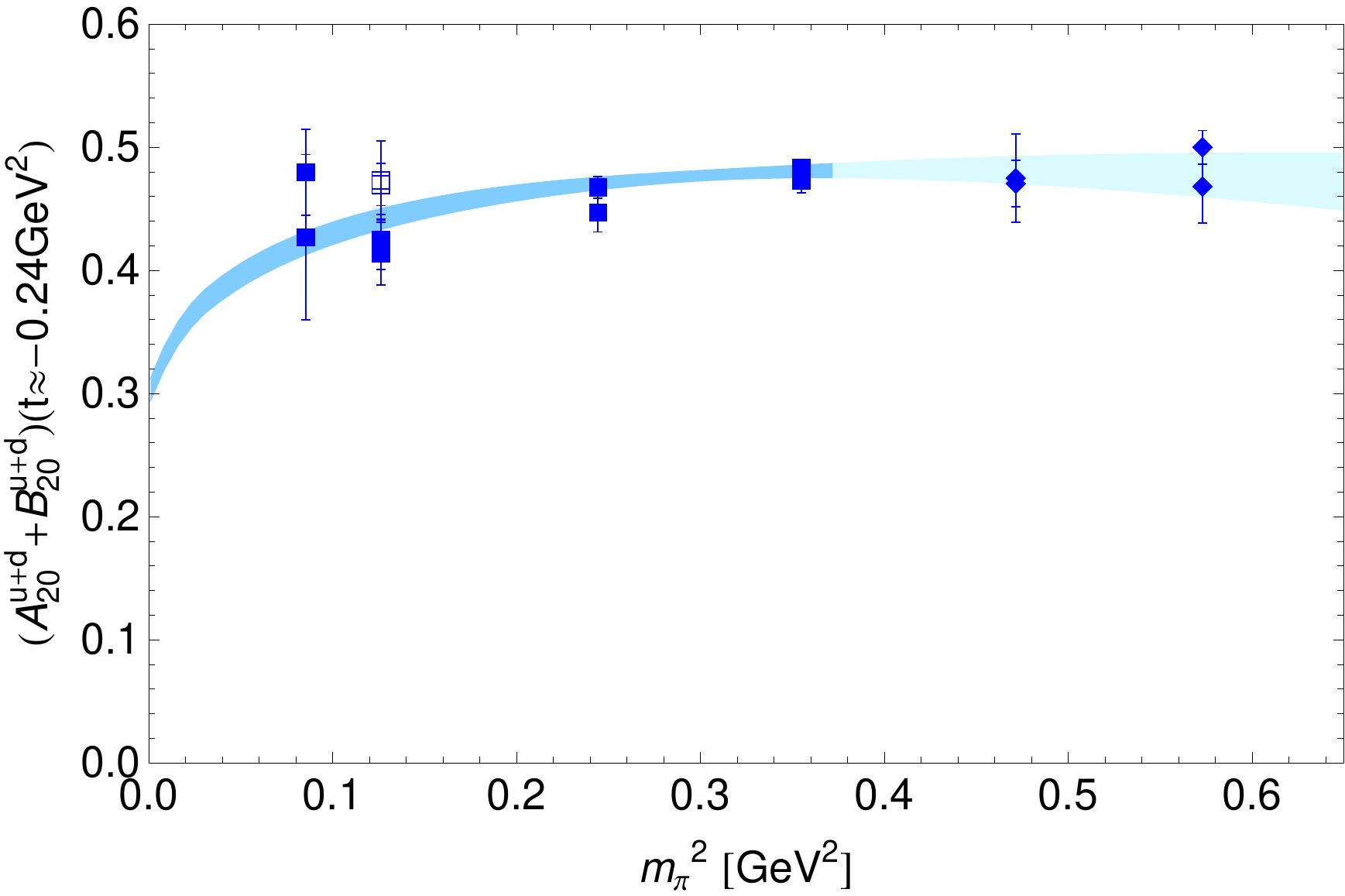}
    \includegraphics[width=8.4cm,angle=0]{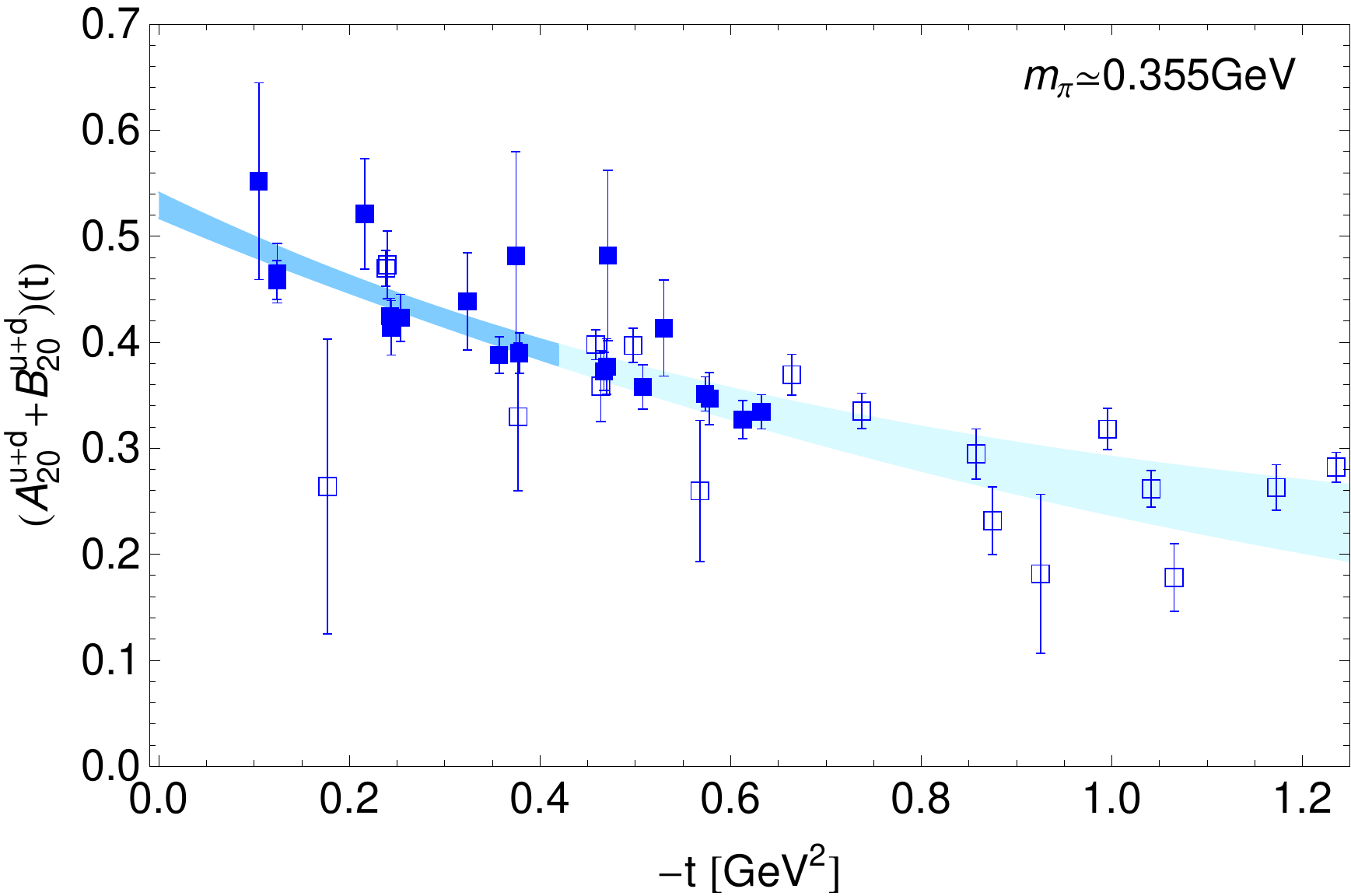}}
  \caption{Pion and momentum transfer dependence of $M_{20}$.  At
    $m_\pi=356$MeV, the open symbol corresponds to $20^3$, the filled
    symbol to $28^3$.  At all other pion masses, the volume is $20^3$.
  }
  \la{fig:M20-upd}
\end{figure}

\subsubsection{\label{EMIsoSingletextr}HBChPT extrapolation of
  $E_{20}^{u+d}$ and $M_{20}^{u+d}$}
In heavy baryon chiral perturbation theory
(HBChPT)~\cite{Diehl:2006ya, Diehl:2006js} to $\mathcal{O}(p^2)$, the
combined ($t,m_\pi$)-dependence of the GFF-combination
$E_{20}^{u+d}(t)= A_{20}^{u+d}(t)+t/(4m_N^2)B_{20}^{u+d}(t)$ is quite
different from that of
$M_{20}^{u+d}(t)=A_{20}^{u+d}(t)+B_{20}^{u+d}(t)$, which in the
forward limit is equal to two times the total quark contribution to
the nucleon spin, $2J_q=M_{20}^{u+d}(t=0)$.  The notation is chosen by
analogy with the Sachs form factors in the $n=1$ sector, $E$ being the
analog of $G_E$ and $M$ of $G_M$.  Our results for $B_{20}^{u+d}$ are
in units of the quark-mass dependent nucleon mass; to be consistent
with the conventions of the chiral expansion Eq.~\eqref{M20chPT1}
below, we multiply the $B_{20}^{u+d}$ data by $m_N^{\rm
  phys}/m_N(m_\pi)$ before performing the fit.  In
\cite{Hagler:2007xi}, this conversion had been omitted.

While at this order $M_{20}^{u+d}$ shows a non-analytic dependence on
$t$ and $m_\pi$ as discussed below, $E_{20}^{u+d}$ is constant up to
analytic tree-level contributions,
\begin{equation}
  E_{20}^{u+d}(t,m_\pi)=E_{20}^{0,u+d} + E_{20}^{m_\pi,u+d} m_\pi^2 +
  E_{20}^{t,u+d} t\,.
  \label{ChPTE20}
\end{equation}
A fit to our lattice results based on Eq.~\eqref{ChPTE20} is shown in
Fig.~\ref{fig:E20-upd}. The linear dependence of $E_{20}$ on $t$ and
$m_\pi^2$ works well within the fit range, while the higher mass
points do not quite lie on a common smooth curve.  In contrast to the
covariant approach, the functional form of $E_{20}(t=0)=A_{20}(t=0)$
does not exhibit a term that could lead to an upward bending as seen
on the right panel of \fig\ref{fig:A20upmd}.

The pion mass dependence of $M_{20}^{u+d}(t)$ for non-zero $t$ is
given by \cite{Diehl:2006ya,Diehl:2006js}
\begin{equation}
  M_{20}^{u+d}(t,m_\pi)=M_{20}^{0,u+d}
  \bigg\{ 1-\frac{3g_A^2 m_\pi^2}
  {(4 \pi f_\pi)^2}\ln\left(\frac{m_\pi^2}{\Lambda_\chi^2}\right)\bigg\} 
  + M_2^{(2,\pi)}(t,m_\pi) + M_{20}^{m_\pi,u+d} m_\pi^2 +
  M_{20}^{t,u+d} t\,,
  \label{M20chPT1}
\end{equation}
with new counter terms $ M_{20}^{m_\pi,u+d}$ and $M_{20}^{t,u+d}$.
The non-analytic dependence on $t$ and $m_\pi$ in
$M_2^{(2,\pi)}(t,m_\pi)$ results from pion-operator insertions and is
directly proportional to the (isosinglet) momentum fraction of quarks
in the pion in the chiral limit, $\langle x\rangle^{\pi,0}_{u+d}$.  We
use $\langle x\rangle^{\pi,0}_{u+d}=0.5$ for the fit.  The results of
chiral fits based on Eq.~\eqref{M20chPT1} are presented in
\fig\ref{fig:M20-upd}. We note that the data for $M_{20}(t=-0.24{\rm
  GeV}^2,m_\pi)$ has a tendency of bending downward as a function of
$m_\pi$, and the fit is able to describe this behavior. We find
$M_{20}^{u+d}(t=0,m_\pi^{\rm phys}) = 0.528(11)$.  We will use this
result in \sect\ref{sec:decomp-nucl-spin}.

\subsubsection{HBChPT extrapolation of $B_{20}^{u+d}$}
Since the total anomalous gravitomagnetic moment of quarks and gluons
in the nucleon has to vanish, $\sum_{q,g} B_{20}(t=0)=0$, an
interesting question is whether the individual quark and gluon
contributions to $B_{20}$ are separately zero or very small.  The GFF
$B_{20}^{u+d}$ can be written as a linear combination of
Eqs.~\eqref{ChPTE20} and~\eqref{M20chPT1}.  A separate fit to the data
with fixed $E_{20}^{0,u+d}=0.522 $ 
gives $B_{20}^{u+d}(t=0,m_{\pi}^{\rm phys})= 0.015(11)$ 
which is compatible with the fits to $M_{20}^{u+d}$ and $E_{20}^{u+d}$
above that in combination give $(M-E)_{20}^{u+d}(t=0,m_{\pi}^{\rm
  phys})= 0.003(12)$. Although the absolute value of
$B_{20}^{u+d}(t=0)$ is again rather small, we note that the sign is
different from that found in \sect\ref{sec:A20Isosinglet} based on the
BChPT fit, where it was negative. A more accurate calculation and the
full ${\cal O}(p^3)$ BChPT expression will help resolve the sign of
$B_{20}^{u+d}(t=0)$. The simultaneous dependence of $B_{20}^{u+d}$ on
$m_\pi$ and $t$ is shown in \fig\ref{fig:BC20-3D}.  We recall that the
sum of $B_{20}^{u+d}$ and the corresponding quantity for gluons
vanishes in the forward direction.  The very non-trivial interplay of
the $t$ and $m_\pi$ dependence results in a small value of
$B_{20}^{u+d}$ near the origin of the $(t,m_\pi)$ plane, suggesting
also a small value of $B_{20}^{g}(t=0)$ in the real world. However,
near the chiral limit the form factor $B_{20}^{u+d}$ becomes sizable,
about -0.1 at $-t\approx 0.24$GeV$^2$.

\subsubsection{\label{subsecHBC20}HBChPT extrapolation of
  $C_{20}^{u+d}$ }
At order $\mathcal{O}(p^2)$, the pion mass dependence of the GFF
$C^{u+d}_{20}(t)$ is given by
\cite{Ando:2006sk,Diehl:2006ya,Diehl:2006js}
\begin{equation}
  C^{u+d}_{20}(t,m_\pi) = \frac{1}{1-t/(4m_N^2)}\bigg\{ C^{0,u+d}_{20}
  + E_2^{(1,\pi)}(t,m_\pi) + E_2^{(2,\pi)}(t,m_\pi) 
  + C^{m_\pi,u+d} m_\pi^2 + C^{t,u+d} \,t \bigg\}\,,
  \label{C20chPT1}
\end{equation}
where $C^{0,u+d}_{20} \equiv C^{u+d}_{20}(t=0,m_\pi=0)$.  The terms
$E_2^{(1,\pi)}(t,m_\pi)$ and $E_2^{(2,\pi)}(t,m_\pi)$ contain
non-analytic terms in $t$ and $m_\pi$ that come from insertions of
pion operators proportional to $\langle x\rangle^{\pi,0}_{u+d}$.
Additionally, $E_2^{(2,\pi)}(t,m_\pi)$ depends on the low energy
constants $c_1$, $c_2$ and $c_3$, which are set to the same values as
in \cite{Hagler:2007xi}. To be consistent with the chiral expansion
Eq.~\eqref{C20chPT1}, we rescale the $C_{20}^{u+d}$ lattice data by
$m_N^{\text{phys}}/m_N(m_\pi)$ before performing the fit. We find a
value of $C_{20}^{0,u+d}=-0.407(14)$ in the chiral limit, and
$C_{20}^{u+d}(t=0,m_\pi^{\text{phys}})=-0.325(14)$ at the physical pion
mass. The simultaneous dependence of $C_{20}^{u+d}$ on $m_\pi$ and $t$
is shown in \fig\ref{fig:BC20-3D}, along with the HBChPT fit.
Comparing the figure with \fig\ref{fig:C20-3D-BChPT}, one sees that
the fits differ significantly at larger values of $m_\pi$ and
$-t$. The result is also shown as a function of the pion mass squared
for fixed $t\approx -0.24$GeV$^2$ in \fig\ref{fig:C20_upd_HB} (left
panel), and the $t$-dependence at a pion mass of $356$MeV is presented
in the right panel.
\begin{figure}[t]
  \centerline{\includegraphics[width=8.5cm,angle=0]%
    {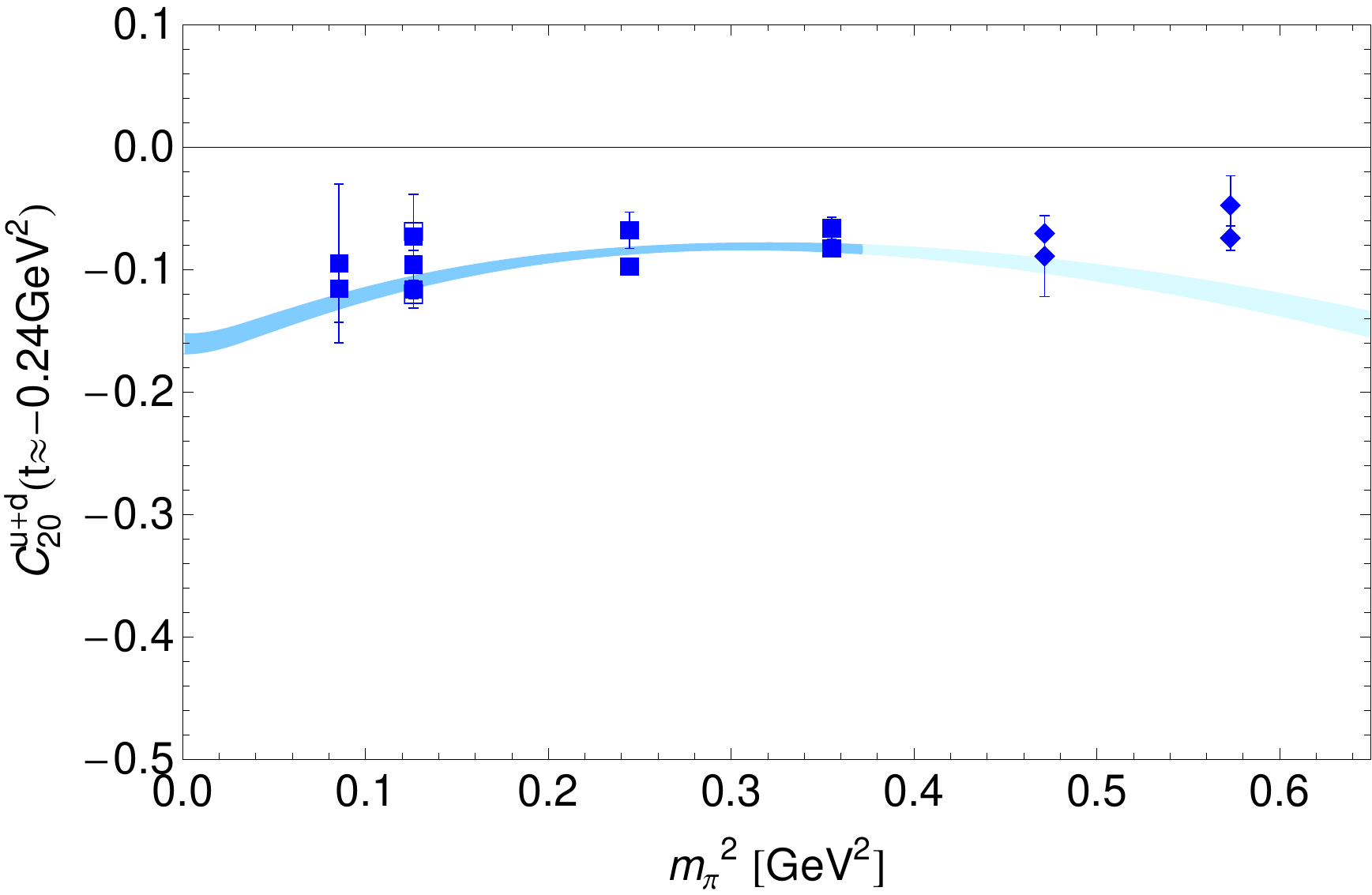}
    \includegraphics[width=8.5cm,angle=0]%
    {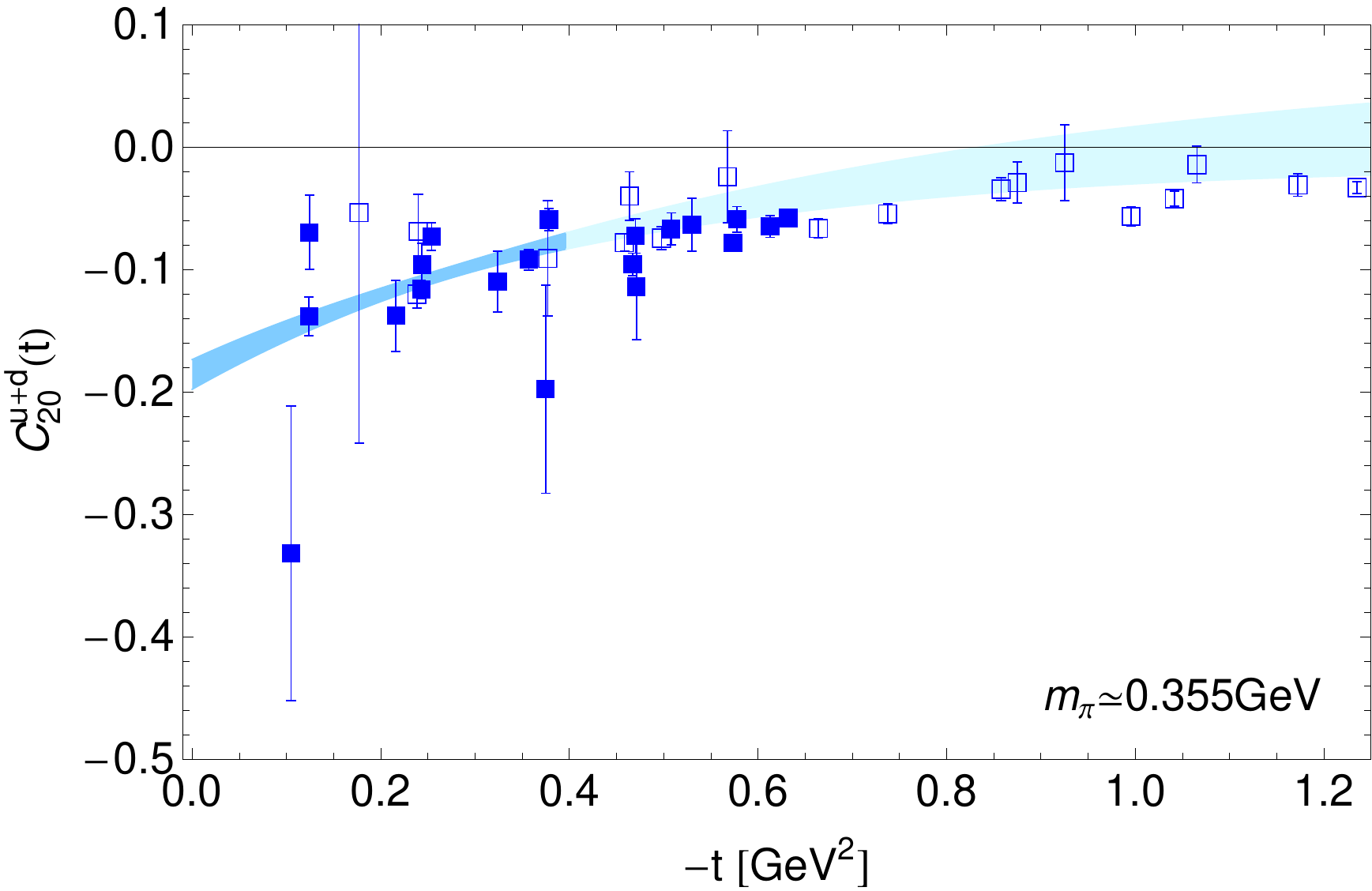}}
  \caption{Projections of $C_{20}^{u+d}$ --- as depicted in
    \fig\ref{fig:BC20-3D} --- on a constant $-t$ plane and on a
    constant $m_\pi^2$ plane. The HBChPT fit appears as a shaded
    band.}
  \la{fig:C20_upd_HB}
\end{figure}

\subsection{\label{sec:decomp-nucl-spin}Quark contributions to the proton spin}
In the following, we will present and discuss our results for the
quark spin and orbital angular momentum contributions (OAM) to the
proton spin. We first remind the reader that the form factors of the
energy momentum tensor, which in this section we denote by
$A^{q,g}(t)\equiv A^{q,g}_{20}(t)$ and $B^{q,g}(t)\equiv
B^{q,g}_{20}(t)$, in the forward limit $t=0$ give direct access to the
quark and gluon angular momenta in Ji's nucleon spin sum rule,
\begin{equation}
  \frac{1}{2}=\sum_{q=u,d,\ldots}J^q + J^g
  =\frac{1}{2}\Bigg\{\sum_{q=u,d,\ldots}\bigg(A^{q}(0) 
  + B^{q}(0)\bigg) + \bigg(A^g(0) + B^{g}(0)\bigg)\Bigg\} \,.
  \label{Ji}
\end{equation}
Furthermore, the quark angular momentum may be decomposed in terms of
the quark spin, $\Delta\Sigma^q$, and orbital angular momentum, $L^q$,
such that
\begin{equation}
  \frac{1}{2}=\sum_{q=u,d,\ldots}\bigg(\frac{1}{2}\Delta\Sigma^q +
  L^q\bigg) + J^g\,.
  \label{Ji2}
\end{equation}
It is important to note that the decompositions in Eqs.~\eqref{Ji}
and~\eqref{Ji2} are fully gauge-invariant, and that the individual
terms will in general be renormalization scale and scheme dependent.
Since the momentum fractions carried by quarks and gluons have to add
up to one, i.e.~the total nucleon momentum,
\begin{equation}
  1 =\sum_{q=u,d,\ldots}A^{q}(0) + A^g(0) = \sum_{q=u,d,\ldots}\langle
  x\rangle^q + \langle x\rangle^g\,,
  \label{momsumrule}
\end{equation}
one also finds that the anomalous gravitomagnetic moments,
$B^{q,g}(0)$, have to cancel exactly in sum,
\begin{equation}
  0 =\sum_{q=u,d,\ldots}B^{q}(0) + B^g(0)\,.
  \label{anomgravmom}
\end{equation}
It is a prominent goal of future lattice hadron structure calculations
to study the above sum rules in great detail.  Since gluonic
observables suffer in general from very low signal-to-noise ratios and
have so far not been studied on the lattice with sufficient precision,
for the moment we will have to concentrate on the (connected)
contributions from up- and down quarks.

In \sect\ref{sec:Jq} below, we will begin with a discussion of $J^q$
based on our results for the GFFs $A^{u,d}_{20}(t)$ and
$B^{u,d}_{20}(t)$.  There, the results of the covariant and heavy
baryon chiral extrapolations of Secs.~\ref{sec:A20Isovector},
\ref{sec:A20Isosinglet}, and~\ref{EMIsoSingletextr} will be
supplemented and compared with an extrapolation of $J^{u+d}$ using a
ChPT-formalism that includes explicitly the $\Delta$-resonance as an
additional degree of freedom. A decomposition of $J^q$ in quark spin
and OAM contributions, together with corresponding chiral
extrapolations, will be presented in \sect\ref{sec:SqLq}.

\subsubsection{\label{sec:Jq}Quark angular momentum $J$}
From the covariant baryon chiral perturbation theory (BChPT)
extrapolation in \sect\ref{sec:A20Isosinglet} of the isosinglet GFFs
$A_{20}(t)$ and $B_{20}(t)$, we find a value of
$J_{BChPT}^{u+d}=0.238(8)$ for the total quark angular momentum
contribution at the physical pion mass.  This corresponds to
$\simeq48\%$ of the total nucleon spin $S=1/2$, which is somewhat
larger than our result in \cite{Hagler:2007xi}, although the
difference is clearly not significant within statistical errors.  We
note again that these values have to be considered with some caution,
since contributions from disconnected diagrams have not been included.

\begin{figure}[htb]
  \centerline{\includegraphics[width=8.5cm,clip=true,angle=0]{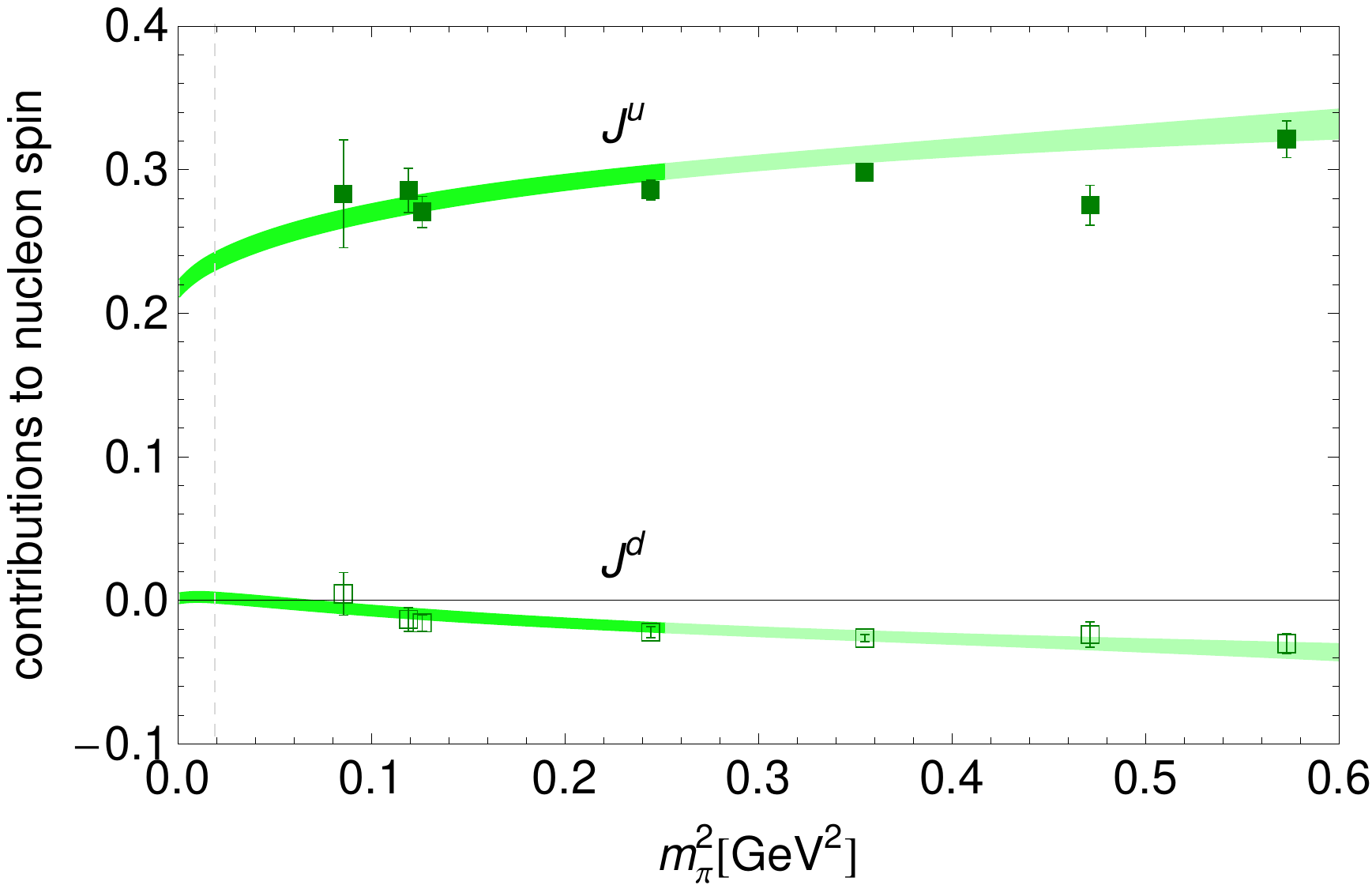}}
  \caption{Chiral extrapolations of $J^{u,d}$ using BChPT\@. Note that
    the displayed lattice data points were not directly employed in
    the chiral fits. Details are given in the text.}
  \label{figJq1}
\end{figure}

A result that is more accurate regarding the systematics can be given
for the isovector, $u-d$, channel where disconnected diagrams cancel
out exactly. From the chiral extrapolations in
\sect\ref{sec:A20Isovector}, we obtain $J_{BChPT}^{u-d}=0.234(6)$ at
$m_{\pi}^{\rm phys}$.  That this value is so close to the $u+d$-quark
angular momentum already points to a small contribution from down
quarks.  Indeed, combining the isovector and isosinglet results, we
find that the up quarks carry a substantial amount of angular
momentum, $J_{BChPT}^{u}=0.236(6)$, while the contribution from down
quarks is very small and even negligible within the small statistical
errors, $J_{BChPT}^{d}=0.0018(37)$.  We will see in
\sect\ref{sec:SqLq} below that the smallness of $J_{}^{d}$ can be
traced back to a remarkably precise cancellation between spin and
orbital angular momentum of quarks.  The results we have just
discussed are illustrated in \fig\ref{figJq1}, showing $J_{}^{u,d}$ as
a function of $m_\pi^2$, together with the corresponding BChPT
extrapolations indicated by the error bands.  We note that the lattice
data points were obtained from separate dipole extrapolations of the
GFFs $B_{20}^{u,d}(t)$ to $t=0$.
\begin{figure}[htb]
  \centerline{\includegraphics[scale=0.48,clip=true,angle=0]{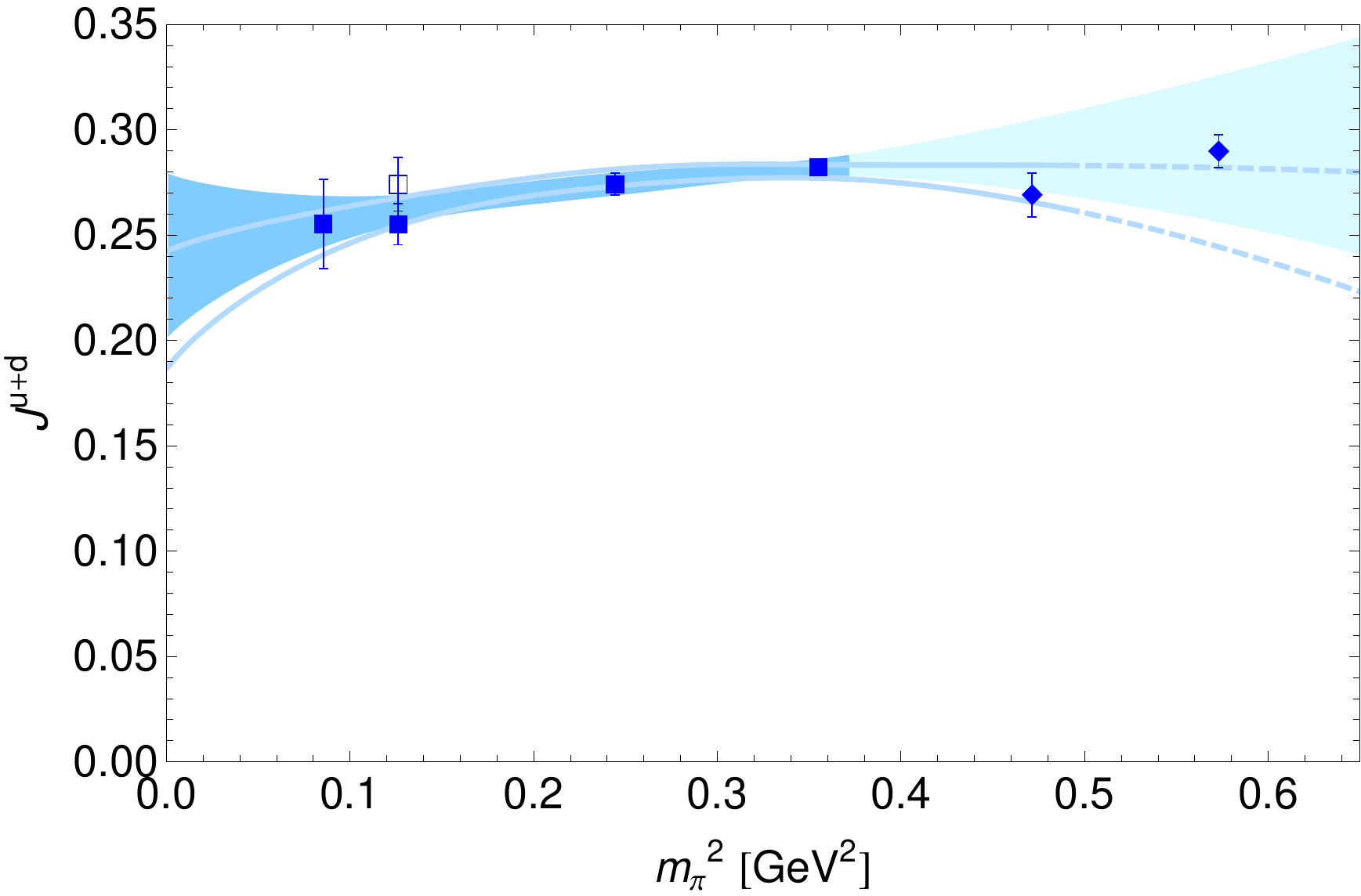}}
  \caption{Chiral extrapolation of $J^{u+d}$ using HBChPT including
    the $\Delta$ resonance, Eq.~\eqref{ChPTJqDelta}.  The fit and
    error bands are explained in the text.}
  \label{fitJq2}
\end{figure}

In order to study possible systematic uncertainties in the chiral
extrapolation of quark orbital angular momentum just described, we now
briefly discuss a completely different approach to the chiral
extrapolation of $J^{u+d}$.  This will be based on HBChPT including
the $\Delta$ resonance as an explicit degree of freedom, predicting at
leading-one-loop order a pion mass dependence of $J^{u+d}$ of the form
\cite{Chen:2001pv}
\begin{eqnarray}
  J^{u+d}_{\text{HBChPT}+\Delta}(m_\pi)&=&
  J_{\text{HBChPT}}^{u+d}(m_\pi) - \frac{1}{2}\left(\frac{9}{2}
    (A+B)^{0,u+d}_{20}+ 3 \langle x\rangle^{\pi,0}_{u+d} -
    \frac{15}{2}b_{q\Delta}\right)
  \nonumber \\ && \quad \times
  \frac{8 g_{\pi N\Delta}^2}{9(4 \pi
    f_\pi)^2}\left\{(m_\pi^2-2\Delta^2)
    \ln\left(\frac{m_\pi^2}{\Lambda_\chi^2}\right)
    +2\Delta \sqrt{\Delta^2 - m_\pi^2} \ln\left(\frac{\Delta -
        \sqrt{\Delta^2 - m_\pi^2}}{\Delta + \sqrt{\Delta^2 -
          m_\pi^2}}\right)\right\}\label{ChPTJqDelta} \,,
\end{eqnarray}
where $\Delta=m_\Delta-m_N$ denotes the $\Delta$-nucleon mass
difference, and $g_{\pi N\Delta}$ is the pion-nucleon-$\Delta$
coupling.  The $m_\pi$-dependent $J^{u+d}_{\text{HBChPT}}$ in
Eq.~\eqref{ChPTJqDelta} corresponds to the HBChPT result without
explicit $\Delta$ intermediate states as obtained from
Eq.~\eqref{ChPTC20updp4} for $t=0$ and is given by~\footnote{In the
  notation of \cite{Chen:2001pv},
  $b_{qN}=(A+B)^{0,u+d}_{20}=M^{0,u+d}$.}
\begin{equation}
  J^{u+d}_{\text{HBChPT}}(m_\pi)=\frac{1}{2}\left\{(A+B)^{0,u+d}_{20}
    + 3\bigg(\langle
    x\rangle^{\pi,0}_{u+d}-(A+B)^{0,u+d}_{20}\bigg)\frac{g_A^2
      m_\pi^2}{(4 \pi
      f_\pi)^2}\ln\left(\frac{m_\pi^2}{\Lambda_\chi^2}\right)\right\}
  + J^{m_\pi,u+d} m_\pi^2 \,.
  \label{JqchPT1}
\end{equation}
Since the GFF $B_{20}(t)$ cannot be extracted directly at $t=0$, we
have first performed separate dipole extrapolations of $B^u_{20}(t)$
and $B^d_{20}(t)$ to $t=0$, and combined this with our values for
$\langle x\rangle^{u+d}=A^{u+d}_{20}(0)$ to obtain
$J^{u+d}=(A^{u+d}_{20}(0)+B^{u+d}_{20}(0))/2$.  The resulting lattice
data points, including the full jackknife errors from the
extrapolations of the $B_{20}(t)$ to the forward limit, are displayed
in \fig\ref{fitJq2}.  Chiral fits based on Eq.~\eqref{ChPTJqDelta},
with the three free parameters $b_{qN}\equiv (A+B)^{0,u+d}_{20}$,
$b_{q\Delta}$ and $J^{m_\pi,u+d}$, to the data with $m_\pi\le 600\MeV$
and $m_\pi\le 700\MeV$ are represented by the shaded error band and
the curves (representing the upper and lower bounds of an error band)
respectively. In both cases, we have fixed $\Delta=0.3$GeV and used
the large-$N_c$ relation $g_{\pi N\Delta} = 3/(2^{3/2}) g_A$ as given
in \tab\ref{tab:sse-pars}.

The fit to our lattice results with $m_\pi\le 600$ MeV gives
$(A+B)^{0,u+d}_{20}=b_{qN}=0.514(41)$, $b_{q\Delta}=0.486(55)$ and
$J^{u+d}_{\text{HBChPT}+\Delta}(m_{\pi}^{\rm phys})=0.245(30)$ at the
physical pion mass.  Including the data point at $m_\pi=687\MeV$ in
the fit, we find consistent values with somewhat smaller errors,
$(A+B)^{0,u+d}_{20}=b_{qN}=0.546(24)$ and $b_{q\Delta}=0.449(39)$ and
$J^{u+d}_{\text{HBChPT}+\Delta}(m_{\pi}^{\rm phys})=0.226(22)$.  It is
encouraging to see that these values fully agree within statistical
errors with the results from the global simultaneous BChPT
extrapolations of the GFFs $A_{20}(t)$, $B_{20}(t)$ and $C_{20}(t)$
discussed above.

\subsubsection{\label{sec:SqLq}Quark spin and orbital angular momentum
  contributions}
For a consistent decomposition of the quark angular momentum, $J^q$,
into quark spin, $\Delta\Sigma^q$, and orbital angular momentum,
$L^q$, contributions, we need in addition lattice results for
$\widetilde A^{u+d}_{10}(t=0)=\Delta\Sigma^{u+d}$ and $\widetilde
A^{u-d}_{10}(t=0)=\Delta\Sigma^{u-d}$.

\begin{figure}[htb]
  \begin{minipage}{0.48\textwidth}
    \centering
    \includegraphics[scale=0.48,clip=true,angle=0]{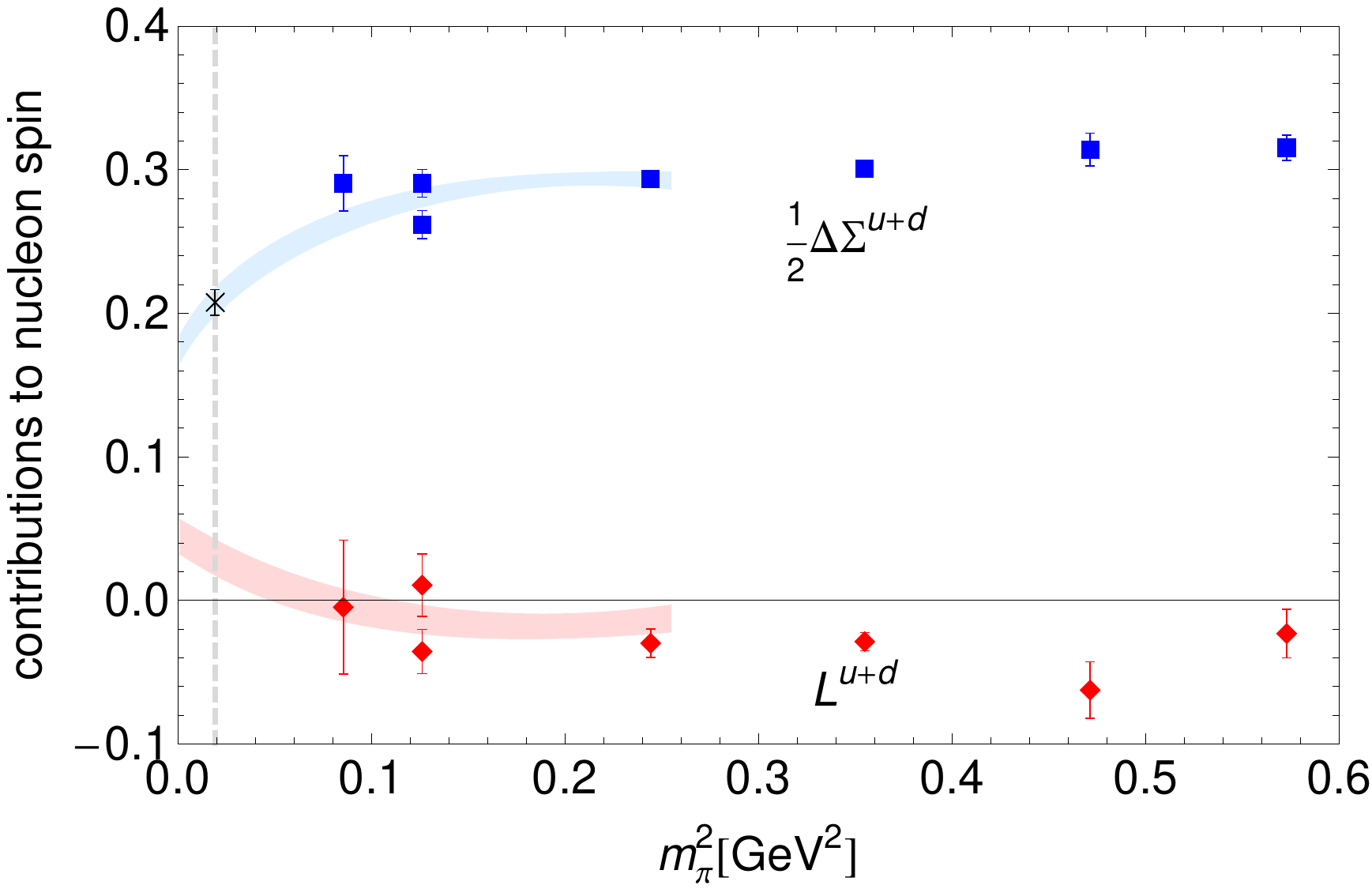}
    \caption{Total quark spin and orbital angular momentum
      contributions to the spin of the proton. The cross represents
      the value from the HERMES 2007 measurement
      \cite{Airapetian:2007aa}. The error bands are explained in the
      text. Disconnected contributions are not
      included.\label{Lq1}}
  \end{minipage}
  \hspace{0.3cm}
  \begin{minipage}{0.48\textwidth}
    \centering
    \includegraphics[scale=0.48,clip=true,angle=0]{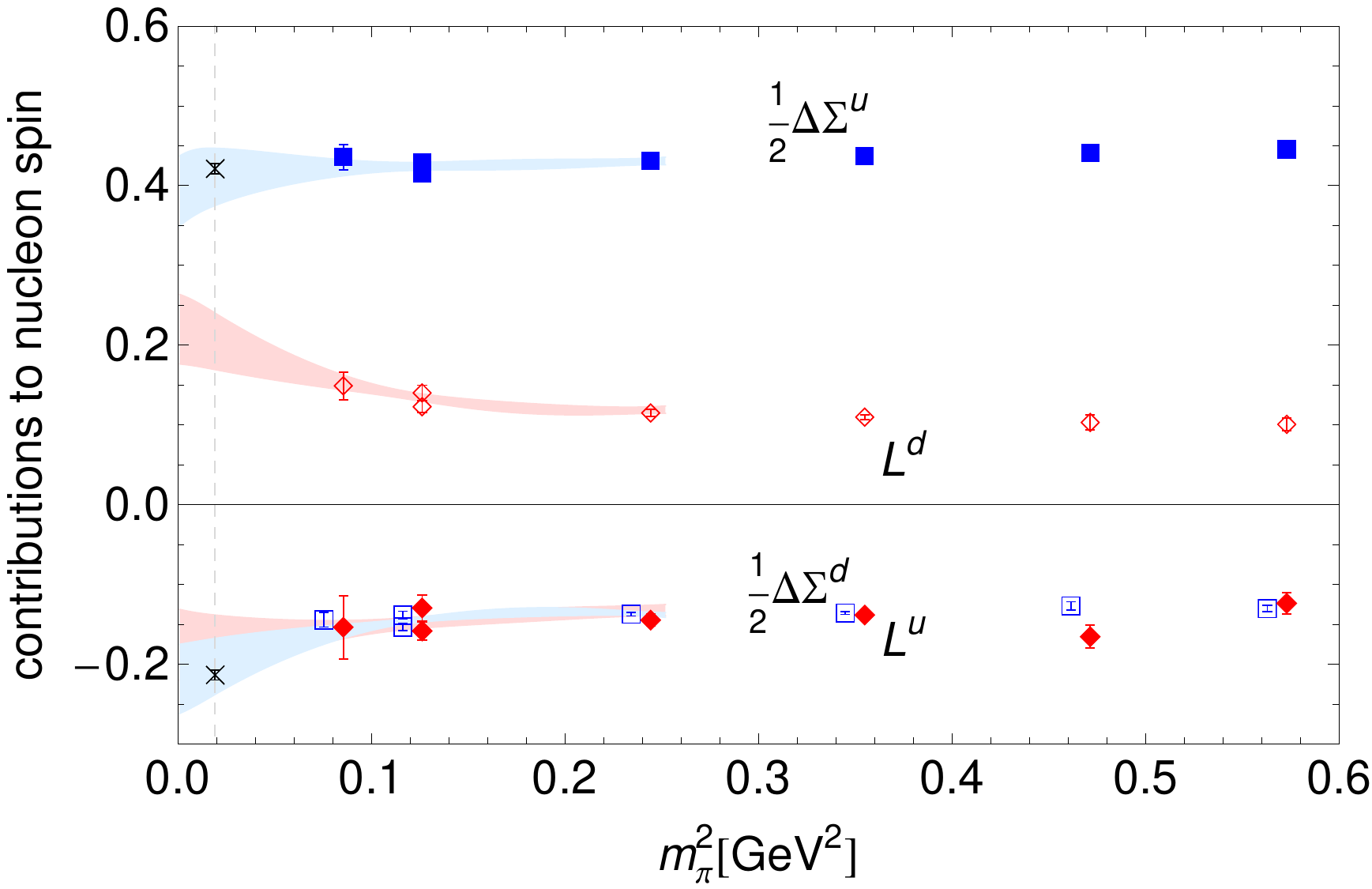}
    \caption{Quark spin and orbital angular momentum contributions to
      the spin of the proton for up and down quarks.  Filled and open
      squares denote $\Delta \Sigma^u/2$ and $\Delta \Sigma^d/2$, and
      filled and open diamonds denote $L^u$ and $L^d$, respectively.
      The crosses represent the values from the HERMES 2007
      measurement \cite{Airapetian:2007aa}. The error bands are
      explained in the text. Disconnected contributions are not
      included.\label{Lq2}}
  \end{minipage}
\end{figure}

Our lattice data for $\Delta\Sigma^{u+d}/2$ is displayed in
\fig\ref{Lq1}, together with a 2-parameter HBChPT-fit represented by
the upper shaded error band.  The chiral extrapolation leads to a
value of $\widetilde A^{u+d}_{10}/2(t=0)=\Delta\Sigma^{u+d}
/2=0.208(10)$ at the physical pion mass, perfectly matching the recent
results from HERMES~\cite{Airapetian:2007aa} indicated by the cross.
However, since this is a leading 1-loop HBChPT fit at comparatively
large pion masses, the agreement with the experimental value should be
considered with great caution and seen as indicative.  Combining this
with the results from the previous section for $J^{u+d}$ and the
corresponding BChPT extrapolation, we find a remarkably small quark
orbital angular momentum $L^{u+d}=J^{u+d}-\Delta\Sigma^{u+d}/2$
contribution to the nucleon spin for a wide range of pion masses, as
indicated by the filled diamonds and the lower error band in
\fig\ref{Lq1}.  From the combined covariant and heavy baryon chiral
extrapolations, we obtain a value of $L^{u+d}=0.030(12)$ at the
physical pion mass.

Superficially seen, such a small OAM contribution from $u+d$ quarks of
only $\approx6\%$ to the nucleon spin is in clear conflict with
general expectations from relativistic quark models, which suggest
that $L^{u+d}=30-40\%$ of $1/2$.  Moreover, from quite general
arguments, e.g.\ based on light cone wave function representations of
hadrons, substantial quark orbital motion is essential for the Pauli
form factor $F_2$ to be non-vanishing in general, and also for the
formation of azimuthal single spin asymmetries in semi-inclusive deep
inelastic scattering related to, e.g., the Sivers effect
\cite{Brodsky:2000ii,Brodsky:2002cx,Burkardt:2003je}.  As we will see
in the following, these apparent inconsistencies may be explained by
studying on the one hand the renormalization scale dependence of quark
OAM, and on the other the contributions from individual quark flavors.

We begin with the latter by noting that a study of the separate up-
and down-quark OAM contributions requires in addition knowledge of the
spin and angular momentum in the isovector, $u-d$, channel.  To this
end, we identify the $u-d$ quark spin contribution with the axial
vector coupling constant, $g_A=\widetilde
A^{u-d}_{10}(t=0)=\Delta\Sigma^{u-d}$, which we have discussed above
in \sect\ref{sec:gA}, cf.~\fig\ref{fig:gA-SSE-600}.  Using our lattice
data for $g_A$, $\Delta\Sigma^{u+d}$, and $(A+B)^{u\pm d}_{20}$, we
have computed the individual spin and OAM contributions from up- and
down quarks to the nucleon spin, which are displayed in \fig\ref{Lq2}
as functions of $m_\pi^2$.  We stress that the chiral extrapolations
represented by the shaded error bands in \fig\ref{Lq2} were not
obtained from direct fits to the shown lattice data points, but are
the result of combining the super-jackknife error bands of the heavy
baryon, SSE, and covariant baryon chiral extrapolations in
Figs.~\ref{Lq1}, \ref{fig:gA-SSE-600}, and~\ref{fig:A20upmd}. The
very good overlap of the bands with the data points may be seen as a
first consistency-check of our approach, in particular with respect to
the different types of extrapolations in the squared momentum transfer
and the pion mass that we have employed throughout this work.

Most remarkably, \fig\ref{Lq2} shows that the individual up- and
down-quark OAM contributions are sizable and of similar magnitude over
a wide range of pion masses, but opposite in sign, and therefore only
cancel to a large extent in the sum, $L^{u+d}\approx0$, as already
observed in \fig\ref{Lq1}.  At the physical pion mass, we find from
the chiral extrapolations that $|L^u|\approx |L^d|\approx 33\%$ of
$1/2$.  A more accurate result can be given for the isovector channel
where disconnected contributions cancel out, $L^{u-d}=-0.379(71)$.

These observations may be seen in analogy to a corresponding analysis
of the nucleon anomalous magnetic moment, $\kappa=F_2(0)=B_{10}(0)$:
Although the proton and neutron anomalous magnetic moments are sizable
(and, as noted in \cite{Brodsky:1980zm, Brodsky:2000ii}, related to
non-zero quark orbital motion), $\kappa^p=1.79$ and $\kappa^n=-1.91$,
they largely cancel in the sum, $\kappa^{u+d}=3\kappa^{p+n}=-0.36$,
see also Sec.~\ref{sec:isosc-form-fact}. However, as for $L^{u+d}$,
this doesn't imply that the orbital motion of individual quarks in the
nucleon is necessarily small.

Furthermore, from a direct comparison of Figs.~\ref{figJq1}
and~\ref{Lq2}, we find that the smallness of the angular momentum of
down-quarks can be seen as the result of another remarkable
cancellation, in this case between spin and OAM\@.  While both types
of contributions for the down-quarks are similarly large in magnitude,
$|\Delta\Sigma^d|/2\approx |L^d|\approx20-30\%$ of $1/2$, they are
again of opposite sign and hence cancel out in the sum,
$J^d=\Delta\Sigma^d/2+L^d\approx0$.

\begin{figure}[htbp]
  \centerline{\includegraphics[width=8.5cm,clip=true,angle=0]{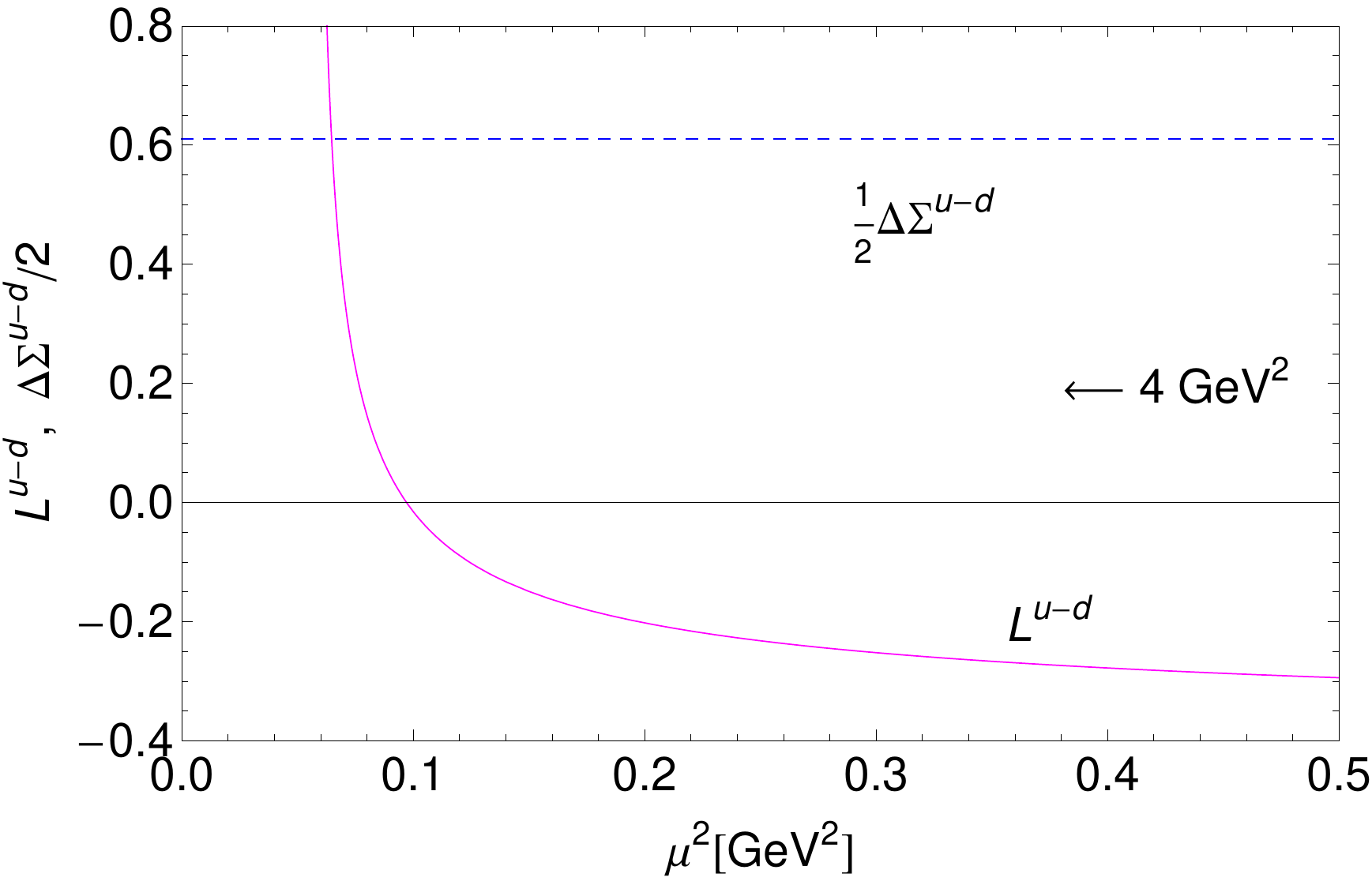}}
  \caption{Leading order evolution of lattice results for $L^{u-d}$
    (solid line) and $\Delta\Sigma^{u-d}$ (dashed line) from
    $\mu_0^2=4$GeV$^2$ down to low hadronic scales.}
  \label{fig:LqEvo}
\end{figure}

Even though the small value for $L^{u+d}$ in our lattice calculation
can be understood to arise from a cancellation between the different
quark flavors, it would still be important to understand the striking
discrepancy with results from relativistic quark model calculations,
$L^{u+d}\approx30-40\%$, as noted above.  In principle, this could be
related to a severe deficiency of the model, or unexpectedly large,
systematic uncertainties in the lattice calculation.  A more likely
solution has been proposed in \cite{Wakamatsu:2007ar, Thomas:2008ga},
just by noting that quark model calculations generically correspond to
a low, hadronic scale $\mu\ll1$GeV, while the resolution scale chosen
here is $\mu_{\MSbar}^2=4$GeV$^2$. Since quark OAM is not conserved
under (leading and higher order) evolution \cite{Ji:1995cu}, it is
therefore in general pointless to directly compare the lattice results
with the model expectations at the different scales.  For a more
sensible comparison, one might instead try to evolve the model results
up to the lattice scale --- or, alternatively, transform the lattice
values down to the hadronic scale.  In order to show that the
different scales might be in principle responsible for the observed
discrepancy, we follow Refs.~\cite{Thomas:2008ga,Bratt:2008uf} and
consider the LO scale dependence of $L^{u-d}$, given by
\cite{Ji:1995cu}
\begin{equation}
  L^{u-d}(t)=\bigg(\frac{t}{t_0}\bigg)^{\frac{2}{\beta_0}
    \big(\frac{-16}{9}\big)}\bigg\{L^{u-d}(t_0)
  + \frac{1}{2}\Delta\Sigma^{u-d} \bigg\}-
  \frac{1}{2}\Delta\Sigma^{u-d}\,,
  \label{eq:LqEvo}
\end{equation}
where $t_{(0)}=\ln(\mu_{(0)}^2/\Lambda_{QCD}^2)$ with an initial scale
denoted by $\mu_{0}$, and where the isovector spin contribution,
$\Delta\Sigma^{u-d}/2=g_A/2$, is exactly conserved under QCD
evolution.  Working in the isovector channel allows us to avoid issues
related to disconnected diagrams (that were not included in our
lattice calculation), and in particular the mixing with gluon
operators under evolution. Figure~\ref{fig:LqEvo} shows the result of
the LO evolution according to Eq.~\eqref{eq:LqEvo} from an initial
scale of $\mu_0=2$GeV, with lattice starting values $L^{u-d}=-0.38$
and $\Delta\Sigma^{u-d}/2=0.61$, down to very low (model) scales, as
indicated by the left pointing arrow.

Although the application of LO evolution equations at such low scales
cannot be quantitatively trusted, we note at least that qualitatively
a strong scale dependence sets in as $\mu\rightarrow \Lambda_{QCD}$,
eventually leading to a change of sign of $L^{u-d}$.  We note that
similar observations have been made in the isosinglet channel
\cite{Thomas:2008ga}, even allowing for an approximate quantitative
agreement of the lattice value for $L^{u+d}$ with the expectations
from relativistic quark models.  One should keep in mind, however,
that the renormalization scheme (e.g. $\MSbar$) of a model is in
general indeterminate, which is a fundamental limitation in the
comparison with results from the full theory.  To sum up, the above
exercise shows very clearly that a one-to-one confrontation of hadron
structure observables from the lattice and in models must be
considered with great caution.

For convenience, we present in \tab\ref{TabSpinDecomp} an overview of
our results for the (connected) quark contributions to the proton
spin, in the $\MSbar$ scheme at $\mu^2=4$GeV$^2$, at the physical pion
mass, as obtained from the different chiral extrapolations discussed
in this {section} and the previous sections.  In the table, we also
include estimates of systematic uncertainties due to the
renormalization of the 1-derivative operators, as discussed in
Sec.~\ref{sec:renorm-latt-oper}. At this point we note again that no
phenomenological values for $\Delta\Sigma=\langle 1\rangle_{\Delta
  q}$, $\langle x\rangle_q$ and $\langle x\rangle_{\Delta q}$ have
been used in the chiral fits, and that we have so far only included
contributions from connected diagrams in the lattice calculations.
\begin{sidewaystable}[htbp]
  \begin{tabular}{*{12}{c|}c}
    \hline\hline
    &  $J^{u-d}$ & $J^{u+d}$ & $J^{u}$ & $J^{d}$ & $g_A=\Delta\Sigma^{u-d}$ & $\Delta\Sigma^{u+d}/2$ & $\Delta\Sigma^{u}/2$ & $\Delta\Sigma^{d}/2$ &
    $L^{u-d}$ & $L^{u+d}$ & $L^{u}$ & $L^{d}$\\ \hline
    BChPT & $0.234(6)(16)$ & $0.238(8)(17)$ & $0.236(6)(17)$ & $0.0018(37)(1)$ &  & & &  & & & &\\
    HBChPT &  & $0.264(6)(18)$ &  &  & & $0.208(10)$ & & & & $0.056(11)(18)$ & & \\
    HBChPT + $\Delta$ & &  $0.226(22)(16)$ & &  &  $1.21(17)$ &   &   & & & & &  \\
    mixed ChPT&  &  &  &  &   &  &  $0.411(36)$ & $-0.203(35)$ &  $-0.379(71)(16)$ & $0.030(12)(17)$ & $-0.175(36)(17)$ & $0.205(35)(0)$ \\ \hline
    experiment &  &  &  &  & $1.2670(35)$  & $0.208(9)$ &  $0.421(6)$ & $-0.214(6)$ &   &  &  & \\  
    \hline\hline
  \end{tabular}
  \caption{Overview of proton spin observables at $m_{\pi}^{\rm phys}$
    from chiral extrapolations of the lattice results, in the
    $\overline{\text{MS}}$ scheme at a scale of $\mu^2=4$GeV$^2$.
    Statistical and estimated systematic uncertainties due to the
    renormalization of the 1-derivative operators are given in
    brackets in the form
    $(\ldots)_{\text{stat}}(\ldots)_{\text{ren}}$, cf.\
    Sec.~\ref{sec:renorm-latt-oper}. For the other observables we
    quote only the statistical uncertainty. We refer to
    Fig.~\ref{BChPTstudy} and the adjacent discussion for a study of
    potential uncertainties due to the chiral extrapolations.  The
    experimental numbers for $\Delta\Sigma$ are from
    \cite{Airapetian:2007aa} for a scale of $\mu^2=5$GeV$^2$. Details
    are given in the text.}
  \label{TabSpinDecomp}
\end{sidewaystable}

As a final note, we look forward to different angles of attack from
the experimental and phenomenological sides becoming available. These
will supplement the currently available DVCS measurements at JLab and
HERMES and help to narrow down the statistical and systematic
uncertainties~\cite{JLab:2007vj,HERMES:2008jga}. Stronger constraints
might come from experimental data on, e.g., the transverse target-spin
asymmetry $A_{UT}$ for the electroproduction of vector mesons, which
is in particular sensitive to the GPD $E$.  In this respect, it is
interesting to note that the (model dependent) phenomenological
analysis in~\cite{Kroll08}, which is compatible with recent data from
HERMES~\cite{HERMES:2009uta} and preliminary results from
COMPASS~\cite{Schill:2008iy} on $A_{UT}$, has found a similar pattern
of valence quark (orbital) angular momentum contributions as compared
to our result in \tab\ref{TabSpinDecomp}.

%
%

\section{\label{sec:summary-conclusions}Summary and conclusions}
We have performed a comprehensive lattice QCD study of the observables
characterizing the structure of the nucleon. The moments of forward
parton distributions, the electromagnetic and axial form factors, and
a set of generalized form factors, both for the isovector and the
isoscalar flavor combinations were calculated. We have applied various
kinds of chiral expansion schemes to the data and compared our results
to experimental values. Determining a wide set of nucleon properties
simultaneously allows us to assess the range of applicability of these
schemes, and allows us to gauge how sensitive the nucleon properties
are to changes in the up and down quark masses.

The picture that emerges is that a nucleon composed of quarks
corresponding to pion masses in the range $290{\rm MeV}<m_\pi<760{\rm
  MeV}$ is significantly more compact than the nucleon in nature, when
probed by a local vector current, as realized experimentally through
photon-nucleon interaction. Both the Dirac and Pauli radii are less
than a factor 2/3 of their physical values,
cf.~Figs.~\ref{fig:r1v-varmpi} and~\ref{fig:r2vkv-final}. A similar
observation holds for the axial radius, see Tab.~\ref{tab:ga-dip-vart}
--- consequently, a nucleon composed of heavier up and down quarks
would also appear much smaller than the physical nucleon if probed by
a $W$ boson. When performing fits to our data using the small scale
expansion (SSE) and covariant baryon chiral perturbation theory
(BChPT) schemes we find that the pion mass dependence of our data is
weaker than what we expect from those chiral expansions. Still, our
findings are qualitatively compatible with the picture of a nucleon
growing substantially in size when approaching the chiral limit. We
also find that the Dirac form factor, at $\mps=356$MeV and for
$0.3<Q^2/{\rm GeV}^2<0.7$, exhibits a statistically significant
finite-size effect in the femto-universe in which we study the
nucleon, see Fig.~\ref{fig:f12v-finvol}. The latter appears to be
`squeezed' when observed on the hypertorus of size (2.5fm)$^3$. For
this reason the chiral schemes might well be applicable to the pion
mass regime which we probe if we had quantitative control of finite
size effects. Hence, an important task for future lattice calculations
will be to determine at what level such finite-size effects are
influencing the extraction of the nucleon's infinite-volume
properties. For the axial charge, we find the finite-size dependence
to be at most $(4-5)\%$ at $\mps=356$MeV,
cf.~Fig.~\ref{fig:ga-plateau}, but the effect is expected to grow in
the chiral regime.

We find the axial charge of the nucleon to be $(8-10)\%$ lower than in
the physical world, and its pion mass dependence is very weak in the
explored range, see Fig.~\ref{fig:gA-SSE-600}.  The latter observation
is in qualitative agreement with the prediction of chiral effective
theory, and continues up to pion masses far outside its range of
validity.  It is an important challenge to show that $g_A$ really
rises by $(8-10)\%$ as $m_\pi$ is decreased from 300MeV down to its
physical value.

We have also presented extensive data on the generalized form factors
(GFFs), which are related to the $x$-moments of the generalized parton
distributions (GPDs).  The higher the spin of the twist-two operator,
the smaller the radii of the associated GFFs\@. This observation holds
throughout the explored pion mass range, see Fig.~\ref{fig:A1230}, and
is in accord with the expectation of a decreasing transverse size of
the nucleon, as measured by the impact parameter of an active quark
relative to the nucleon center of momentum, with increasing
longitudinal momentum fraction $x$.  Since only the vector and
axial-vector radii (i.e.~for the lowest moment $n=1$) exhibit a
statistically significant growth when the pion mass is reduced,
cf.~Fig.~\ref{fig:radii}, the hierarchy between these radii and those
associated with higher spin operators increases towards the physical
point.  We have also found that the longitudinal momentum transfer
$\xi$ dependence of the GPDs is significant, as the ratios
$C^{u+d}_{20}/A^{u+d}_{20}$ and $A^{u+d}_{32}/A^{u+d}_{30}$ show quite
clearly, cf.~Eq.~\eqref{eq:lattice-vec-gff-def} and
Figs.~\ref{fig:xi-dep2} and~\ref{fig:xi-dep3}. This is in fact
consistent with large-$N_c$ counting rules, which predict in
particular that $C^{u-d}_{20}\sim N_c\sim A^{u-d}_{20}$.

The GFFs in the forward limit, namely the moments of parton
distributions, exhibit a rather mild pion mass dependence in the
accessible range of pion masses.  Specifically, both the unpolarized
and polarized isovector momentum fractions, shown in
Figs.~\ref{fig:A20upmd} and~\ref{fig:A20u-dfit} respectively, will
have to bend down as a function of $m_\pi$ beyond our range, if the
lattice data is to make contact with experiment.

Finally, we applied our results for the moments of the GPDs to the
decomposition of the proton spin --- the final results are listed in
Tab.~\ref{TabSpinDecomp}. Thus, we extend our previous calculation in
Ref.~\cite{Hagler:2007xi} down to $m_\pi=290$MeV, and increase its
accuracy. The decomposition of quark angular momentum in terms of
quark spin and quark orbital angular momentum is displayed in
Fig.~\ref{Lq2}.  The spin contributions are in agreement with the
HERMES 2007 data~\cite{Airapetian:2007aa}, while orbital contributions
are our predictions.  We remind the reader, however, that our
calculation did not include the disconnected graphs which contribute
to all but isovector quantities, and that the mixing of the isosinglet
quark operators with the gluonic operators was not taken into account.
These systematic uncertainties should be kept in mind in the
interpretation of the non-isovector results.  The isovector quantities
however, such as $\frac{1}{2}\Delta\Sigma^{u-d}$ and $L^{u-d}$, do not
suffer from these systematic uncertainties.  One of the surprises,
from the relativistic quark-model point of view, is the negative sign
of $L^{u-d}$.  Various interpretations of this result have been
proposed~\cite{Thomas:2008ga, Burkardt:2008ua}.  As pointed out in
Ref.~\cite{Thomas:2008ga}, $L^{u-d}$ evolves rapidly at low
renormalization scales, because
$J^{u-d}=L^{u-d}+\frac{1}{2}\Delta\Sigma^{u-d}$ renormalizes
multiplicatively and $\frac{1}{2}\Delta\Sigma^{u-d}$ is
scale-invariant.  In any case, lattice studies of the Ji sum rule,
Eq.~\eqref{Ji}, have led to a renewed interest in the problem of
`decomposing the total angular momentum of an interacting
multiconstituent system into contributions from various constituents'
(cited from \cite{Burkardt:2008ua}; see also \cite{Wang:2009dq}).
 
Given that lattice QCD provides a systematically improvable way of
solving QCD, there should be eventual agreement between nucleon
properties calculated in this framework and those measured by
experiment. Thus, an extension of the present calculation down to
$m_\pi=200$MeV is almost certain to uncover the dramatic effects
predicted by chiral effective theory near the chiral limit.  To be
specific, the isovector Dirac radius $\langle r_1^2\rangle$ and the
Pauli radius $\langle r_2^2\rangle$ can provide important benchmarks
of our understanding of nucleon structure --- they both need to grow
particularly fast as $m_\pi$ decreases below $250$MeV, if our lattice
data are to make contact with the phenomenological values, see
Figs.~\ref{fig:r1v-comb-final} and~\ref{fig:F2v-final}. The
calculation of these and similar observables would thus test our
understanding of both chiral effective field theory and of lattice
calculations as applied to nucleon structure.

To achieve this goal we must be sure that statistical uncertainties,
cut-off effects, and finite-size effects are well understood and under
control. We believe that our current work provides an important step
in this direction, as it demonstrates the improvement we have achieved
in reducing statistical uncertainties and the application of chiral
perturbation theory. In view of the success of our methods and
techniques and given the recent increase in the computing power
available to lattice QCD practitioners, we are optimistic that this
program can be carried out successfully in the course of the next few
years.

%
%

\begin{acknowledgments}
  The authors wish to thank George T.~Fleming, Dru B.~Renner, and
  Andre P.~Walker-Loud for their contributions to this project and to
  the LHPC Collaboration and for valuable discussions of the physics
  and presentation of this work. This work was supported in part by
  U.S.~DOE Contract No.~DE-AC05-06OR23177 and DE-FG03-97ER4014, by the
  DOE Office of Nuclear Physics under grants DE-FG02-94ER40818,
  DE-FG02-04ER41302, DE-FG02-96ER40965, DE-FG02-05ER25681 and
  DE-AC02-06CH11357 and the EU (I3HP) under contract
  No.~RII3-CT-2004-506078. This article is coauthored by Jefferson
  Science Associates, LLC under U.S.~DOE Contract
  No.~DE-AC05-06OR23177, and by The Southeastern Universities Research
  Association, Inc.~under U.S.~DOE Contract No.~DE-AC05-84ER40150; the
  U.S.~Government retains a non-exclusive, paid-up, irrevocable,
  world-wide license to publish or reproduce this manuscript for
  U.S.~Government purposes. Ph.H.~and B.M.~acknowledge support by the
  Emmy-Noether program and the cluster of excellence ``Origin and
  Structure of the Universe'' of the DFG\@. W.S.~acknowledges support
  by the National Science Council of Taiwan under grants
  NSC96-2112-M002-020-MY3 and NSC96-2811-M002-026, by NuAS in Germany,
  and wishes to thank the Institute of Physics at Academia Sinica for
  their kind hospitality and support. W.S.~particularly thanks
  Jiunn-Wei Chen at National Taiwan University and Hai-Yang Cheng and
  Hsiang-nan Li at Academia Sinica for their hospitality and for
  valuable physics discussions and suggestions. K.O.~acknowledges
  support from the Jeffress Memorial Trust grant J-813 and Ph.H.,
  M.P.~and W.S.~acknowledge support by the Alexander von
  Humboldt-foundation through the Feodor-Lynen program.  It is a
  pleasure to acknowledge the use of resources provided by the New
  Mexico Computing Applications Center (NMCAC) on Encanto, and of
  computer resources provided by the DOE through the USQCD project at
  Jefferson Lab and through its support of the MIT Blue Gene/L. These
  calculations were performed using the Chroma software
  suite~\cite{Edwards:2004sx}. We are indebted to members of the MILC
  Collaboration for providing the dynamical quark configurations that
  made our full QCD calculations possible.
\end{acknowledgments}

%
%

\bibliography{General,GPD_Refs,FF_Refs,LHPC_mixed_2009}

%
%

\end{document}